\documentclass[fleqn,usenatbib]{mnras}

\usepackage{newtxtext,newtxmath}

\usepackage[T1]{fontenc}

\DeclareRobustCommand{\VAN}[3]{#2}
\let\VANthebibliography\thebibliography
\def\thebibliography{\DeclareRobustCommand{\VAN}[3]{##3}\VANthebibliography}

\usepackage{newtxtext,newtxmath}

\usepackage[T1]{fontenc}
\usepackage{ae,aecompl}
\usepackage{graphicx}	
\usepackage{amsmath}	
\usepackage{mathtools}
\usepackage{cases}
\usepackage{cuted}
\usepackage{bigstrut}
\usepackage{subfig}
\usepackage{paralist}
\usepackage{scrextend}
\usepackage{lscape}
\usepackage{pdflscape}
\usepackage{wrapfig}
\usepackage{adjustbox} 
\usepackage{underscore}
\usepackage{float}
\usepackage{amsfonts}
\usepackage{multicol} 
\usepackage{booktabs}
\usepackage{longtable}
\usepackage[maxfloats=256]{morefloats}
\usepackage[flushleft]{threeparttable}
\usepackage[english]{babel}
\usepackage{stmaryrd}
\usepackage{subfig}
\usepackage{geometry}
\usepackage{makecell}
\usepackage{rotating}
\newcommand{\NI}{\mbox{N\,{\sc i}}}

\newcommand{\ovi}{\mbox{O\,{\sc vi}}}
\newcommand{\ovii}{\mbox{O\,{\sc vii}}}
\newcommand{\oviii}{\mbox{O\,{\sc viii}}}

\newcommand{\SI}{\mbox{S\,{\sc i}}}
\newcommand{\SII}{\mbox{S\,{\sc ii}}}

\newcommand{\SiIV}{\mbox{Si\,{\sc iv}}}

\newcommand{\mnii}{\mbox{Mn\,{\sc ii}}}

\newcommand{\di}{\mbox{D\,{\sc i}}}
\newcommand{\neviii}{\mbox{Ne\,{\sc viii}}}

\newcommand{\civ}{\mbox{C\,{\sc iv}}}
\newcommand{\ciii}{\mbox{C\,{\sc iii}}}
\newcommand{\cii}{\mbox{C\,{\sc ii}}}
\newcommand{\ci}{\mbox{C\,{\sc i}}}

\newcommand{\nitri}{\mbox{N\,{\sc i}}}
\newcommand{\nii}{\mbox{N\,{\sc ii}}}
\newcommand{\niii}{\mbox{N\,{\sc iii}}}
\newcommand{\niv}{\mbox{N\,{\sc iv}}}
\newcommand{\feii}{\mbox{Fe\,{\sc ii}}}

\newcommand{\caii}{\mbox{Ca\,{\sc ii}}}
\newcommand{\siiv}{\mbox{Si\,{\sc iv}}}
\newcommand{\siiii}{\mbox{Si\,{\sc iii}}}
\newcommand{\siii}{\mbox{Si\,{\sc ii}}}

\newcommand{\oi}{\mbox{O\,{\sc i}}}
\newcommand{\oii}{\mbox{O\,{\sc ii}}}
\newcommand{\oiii}{\mbox{O\,{\sc iii}}}
\newcommand{\oiv}{\mbox{O\,{\sc iv}}}
\newcommand{\oxyv}{\mbox{O\,{\sc v}}}
\newcommand{\nv}{\mbox{N\,{\sc v}}}

\newcommand{\alii}{\mbox{Al\,{\sc ii}}}

\newcommand{\CLOUDY}{\textsc{cloudy}}
\newcommand{\cai}{\mbox{Ca\,{\sc i}}}
\newcommand{\nai}{\mbox{Na\,{\sc i}}}
\newcommand{\fei}{\mbox{Fe\,{\sc i}}}
\newcommand{\mgii}{\ifmmode {\rm Mg}{\textsc{ii}} \else Mg\,{\sc ii}\fi}
\newcommand{\heii}{\ifmmode {\rm He}{\textsc{ii}} \else He\,{\sc ii}\fi}

\newcommand{\msun}{M$_{\odot}$} 
\newcommand{\hi}{\mbox{H\,{\sc i}}}
\newcommand{\mgi}{\ifmmode {\rm Mg}{\textsc{i}} \else Mg\,{\sc i}\fi}

\newcommand{\sn}{S/N}

\newcommand{\hst}{\it HST}
\newcommand{\lya}{Ly$\alpha$}
\newcommand{\lyb}{Ly$\beta$}
\newcommand{\lyg}{Ly$\gamma$}
\newcommand{\lyd}{Ly$\delta$}
\newcommand{\rvir}{$R_{\rm vir}$}

\newcommand{\xh}{\ensuremath{\rm [X/H]}}

\newcommand{\degree}{\ensuremath{^\circ}}
\def\kms{\hbox{km~s$^{-1}$}}
\def\cmsq{\hbox{cm$^{-2}$}}
\def\cc{\hbox{cm$^{-3}$}}
\newcommand{\mstar}{$\log M_{\star}/M_{\odot}$}
\newcommand{\metallicity}{$\log Z/Z_{\sun}$}
\newcommand{\hdenu}{$\log n_{\H}/\cc$}
\newcommand{\coldenu}{$\log N(\hi)/\cmsq$}
\newcommand{\totalcoldenu}{$\log N(\rm H)/\cmsq$}
\newcommand{\thicknessu}{$\log L/kpc$}
\newcommand{\tempu}{$\log T/K$}
\newcommand{\hden}{$\log n_{\rm H}$}
\newcommand{\colden}{$\log N(\hi)$}
\newcommand{\totalcolden}{$\log N(\rm H)$}
\newcommand{\thickness}{$\log L$}
\newcommand{\temp}{$\log T$}
\newcommand{\lsfr}{$\log {\rm sSFR}$}
\newcommand{\nodata}{$\cdots$}
\newcommand{\mhalo}{$\log M_{\rm halo}/M_{\odot}$}

\newcommand{\vabsscaled}{|$V_{\rm abs}$/$V_{\rm c}$(D)|}
 
\newcommand{\btherm}{b$_{\rm {T}} (\rm {\hi})$}
\newcommand{\bturb}{b$_{n\rm {T}}$}

\newcommand{\bnet}{b$(\rm {\hi})$}

\def\H{\hbox{{\rm H}}}
\def\HI{\hbox{{\rm H~}\kern 0.1em{\sc i}}}

\defcitealias{Pointon2019}{P19}
\defcitealias{Gnat2017}{G17}
\defcitealias{Hafen2024}{Hafen, Sameer et al., 2024}

\title[Multiphase CGM of Low-$z$ Galaxies]{Cloud-by-cloud Multiphase Investigation of the Circumgalactic Medium of Low-redshift Galaxies}

\author[Sameer et al.]
{Sameer,$^{1,2}$\thanks{E-mail: sameer@nd.edu} 
Jane C. Charlton,$^{1}$
Bart P. Wakker,$^{3}$
Glenn G. Kacprzak,$^{4,5}$
Nikole M. Nielsen,$^{4,5}$
\newauthor
Christopher W. Churchill,$^{6}$
Philipp Richter,$^{7}$
Sowgat Muzahid,$^{8}$
Stephanie H. Ho,$^{6}$
\newauthor
Hasti Nateghi,$^{4,5}$
Benjamin Rosenwasser,$^{9}$
Anand Narayanan,$^{10}$
Rajib Ganguly$^{11}$
\\
$^{1}$Department of Astronomy \& Astrophysics, 525 Davey Lab,
The Pennsylvania State University, University Park, PA 16802, USA\\
$^{2}$Department of Physics \& Astronomy, Nieuwland Science Hall, The University of Notre Dame, Notre Dame, IN 46556, USA\\
$^{3}$Department of Astronomy, University of Wisconsin-Madison, 475 N. Charter Street, Madison, WI 53706, USA\\
$^{4}$Centre for Astrophysics and Supercomputing, Swinburne University of Technology, Hawthorn, Victoria 3122, Australia\\
$^{5}$ ARC Centre of Excellence for All Sky Astrophysics in 3 Dimensions (ASTRO 3D)\\
$^{6}$Department of Astronomy, New Mexico State University, Las Cruces, NM 88003, USA\\
$^{7}$ Institut für Physik und Astronomie, Universität Potsdam, Haus 28, Karl-Liebknecht-Str. 24/25, D-14476, Potsdam, Germany\\
$^{8}$ IUCAA, Post Bag-04, Ganeshkhind, Pune, India - 411007\\
$^{9}$Department of Physics \& Astronomy, The University of Toledo, Toledo, OH 43606, USA\\
$^{10}$Department of Earth and Space Sciences, Indian Institute of Space Science \& Technology, Thiruvananthapuram 695547, Kerala, INDIA\\
$^{11}$College of Innovation \&\ Technology, University of Michigan-Flint, 196 Murchie Science Building, 303 Kearsley Street, Flint MI 48502, USA\\
}

\date{Accepted 2024 April 3. Received 2024 March 31; in original form 2024 February 12}

\pubyear{2024}

\begin{document}
\maketitle

\begin{abstract}

The pervasive presence of warm gas in galaxy halos suggests that the circumgalactic medium (CGM) is multiphase in its ionization structure and complex in its kinematics. Some recent state-of-the-art cosmological galaxy simulations predict an azimuthal dependence of CGM metallicities. We investigate the presence of such a trend by analyzing the distribution of gas properties in the CGM around 47 $z <$ 0.7 galaxies from the Multiphase Galaxy Halos Survey determined using a cloud-by-cloud, multiphase, ionization modelling approach. We identify three distinct populations of absorbers: cool clouds ($T \sim$ 10$^{4.1}$ K) in photoionization equilibrium, warm-hot
collisionally ionized
clouds ($T \sim$ 10$^{4.5-5}$ K) affected by time-dependent photoionization, and hotter clouds ($T \sim$ 10$^{5.4-6}$ K) with broad {\ovi} and {\lya} absorption consistent with collisional ionization. We find that fragmentation can play a role in the origin of cool clouds, that warm-hot clouds are out of equilibrium due to rapid cooling, and that hotter clouds are representative of virialized halo gas in all but the lowest mass galaxies. The metallicities of clouds do not depend on the azimuthal angle or other galaxy properties for any of these populations. At face value, this disagrees with the simplistic model of the CGM with bipolar outflows and cold-mode planar accretion. However, the number of clouds per sightline is significantly larger close to the minor and major axes. This implies that the processes of outflows and accretion are contributing to these CGM cloud populations, and our sightlines are probing gas of mixed origins at all azimuthal angles in these low redshift galaxies.


\end{abstract}

\begin{keywords}
galaxies: evolution — galaxies: haloes — galaxies: individual … — quasars: absorption lines — methods: statistical
\end{keywords}



\section{Introduction}
\label{sec:introduction}

The circumgalactic medium (CGM) of a galaxy is the tenuous gaseous medium permeating the space beyond the interstellar medium (ISM) and extending out to its virial radius~\citep{kacprzak2008halo,kacprzak2010,Chen2010,steidel2010structure,Tumlinson2011,tumlinson2017circumgalactic,Rudie2012,Burchett2013,Nielsen20131,Nielsen20132,werk2013cos,johnson2015possible,lehner2020}. The CGM plays a vital role in galaxy evolution by moderating gas flows between the ISM and the intergalactic medium (IGM). A diverse array of processes such as pristine inflows from the intergalactic medium, galactic-scale outflows, recycled accretion via the galactic fountain mechanism, tidal interactions via satellite galaxy mergers, and effects of cosmic rays and magnetic fields, influence the properties of the CGM. Understanding how these processes manifest in the observed properties of the CGM is a crucial step in understanding galaxy formation and evolution. 

\smallskip

The chief ingredients in galaxy formation and evolution models include: 1) accretion of metal-poor inflowing gas from the IGM through filaments onto galaxies (e.g.,~\citealt{Oort1970,Wakker1997,Wakker1999Natur,Fumagalli2011,Stewart2011,Stewart2013,Stewart2017,Danovich2012,Danovich2015,Oppenheimer2012,Voort2012,shen2013circumgalactic,Kacprzak2016,Borthakur2022}).  2) recycled accretion from previously ejected metal-rich large-scale galactic outflows~\citep{Springel2003,Keres2005,Dekel2009,Oppenheimer2010,Dave2011a,Dave2011b,Dave2012,Stewart2011,Rubin2012,Ford2014,alcazar2017,Zheng2017,Hafen2019,Lochhaas2020,Pandya2020,Pandya2021,Fielding2022}. Accretion, in these forms, is necessary to sustain the formation of stars over billions of years~(e.g., \citealt{vandenBergh1962,Pagel1975,Pagel1989,Chiappini1997,Maller2004,Dekel2006}). 3) Outflows from galaxies are needed to regulate the rate of star formation within galaxies by removing gas from the ISM (e.g., \citealt{Shapiro1976,Bregman1980,kacprzak2008halo,Oppenheimer2010,Faucher2011,Oppenheimer2012}). Outflows carry enriched material out of the galaxy and into the CGM, and are characterized by relatively high metallicities compared with material reaccreting onto the galaxy. On the other hand, pristine accretion is expected to have even lower metallicities.

\smallskip

Recent cosmological magnetohydrodynamical simulations of galaxy formation (e.g., Illustris TNG50, \citealt{Nelson2019,Pillepich2019,peroux2020}) find that gas with the highest metallicities preferentially aligns with the minor axis, and the highest CGM gas metallicities are associated with high-velocity outflowing gas from the concerted feedback action of supernovae as well as from AGN. The lower metallicities, on the other hand, are associated with accreting/inflowing gas along the major axis. Thus CGM properties, and particularly the distribution of metal-rich gas in the CGM, are anisotropic according to these high-resolution simulations, with metallicity differences of $\sim$ 0.5 dex.

\smallskip

Contrary to these theoretical expectations, previous studies, e.g., \citet{Pointon2019} (hereafter \citetalias{Pointon2019}), have not found a dependence of metallicity on galaxy azimuthal angle. This was unexpected given that a bimodality is found in distribution of the azimuthal angle of {\mgii} absorbers~\citep{rongmon2011,rongmon2014,bouche2012,kacprzak2012,kacprzak2015azimuthal,rubin2014,lan2018,martin2019,Dutta2021,Lundgren2021}. However, consistent with the results on outflow and accretion in hydro-dynamical simulations, \citet{wendt2021} identify metal enrichment of $\lesssim$ 1 dex for sightlines along the minor axis compared with sightlines along the major axis, using dust depletion as a proxy for metallicity. They argue that metals deplete onto dust grains preferentially along the minor axis, thus needing to account for the dust when estimating metallicities. Their results are based on only 13 galaxy-{\mgii} absorber pairs (9$-$81 kpc distance) from the MusE GAs FLOw and Wind (MEGAFLOW) survey at 0.4 $< z <$ 1.4~\citep{wendt2021}. However, this work is only applicable to damped {\lya} absorbers (DLAs) limiting the usefulness of this approach to other lower column density {\hi}-absorbers.

\smallskip

The findings of an association between galaxy azimuthal angle and metal absorption properties are based primarily on observations using {\mgii} absorption. However, CGM gas is observed to be multiphase and must be investigated accounting for the contribution of all the gas phases traced by low, intermediate, and high ionization gas, to the {\hi} Lyman series lines in the metallicity determination. Information is lost by assuming an integrated single-valued metallicity per sightline. For simulated absorption profiles, \citet{Churchill2015} and \citet{peeples2019figuring} found that gas arises in a wide range of structures and physical locations. In addition, there might be little physical overlap between multi-phase gas structures that contribute to the same absorption components~\citep{marra2024}. 

\smallskip

Even a single line of sight through the CGM of a galaxy would be expected to have complex multiphase structure with a range of properties. The latest efforts to model observed systems find just that, with metallicities often varying by one or two orders of magnitude along the line of sight through a single absorption system~\citep{Lehner2019,Wotta2019,zahedy2019,sameer21,narayanan2021,haislmaier2021, lehner2022}. Observational investigations of individual absorbers have identified a variety of metallicities within a single absorption system, indicating that CGM characteristics vary along each sight-line \citep[e.g.][]{Churchill2012,Crighton2015,muzahid2015extreme,rosenwasser2018understanding,zahedy2019,Nielsen2022}. Therefore it is important to move beyond observational modelling efforts that typically average together properties (e.g., \citealt{Pointon2019}) and conduct a large statistical study to determine if the metallicities of any of the multiple structures in the CGM depend on orientation. This will allow us to disentangle the complex nature of the CGM and possibly identify a hidden metallicity/azimuthal bimodality.

\smallskip

Ionization modelling in observational work is often limited to the characterization of the low ionization absorption, which contributes the most to the absorption seen in {\hi}. The properties of the cooler gas phases which typically dominate the {\hi} optical depth are captured by single-phase ionization models~\citep{Marra2021}, while the warm-hot phases are often left unaccounted for. Observationally, constraining the properties of high ionization gas traced by {\ovi} has been challenging because the {\ovi} exhibits a significantly broader velocity profile compared with the nearby {\hi} absorbers, and often with misaligned {\ovi} kinematics~(e.g., \citealt{fox2013,savage2014properties,Werk2016}). Thus it is often excluded from single-phase ionization modelling. The multiphase nature of the CGM can only be fully understood by analyzing, in combination, the absorption line diagnostics from low, intermediate, and high ions.

\smallskip

In this study, we reanalyze the \citetalias{Pointon2019} sample using a cloud-by-cloud, multiphase, Bayesian ionization modelling method. This approach will allow us to delineate the phase structure along the line of sight and ascertain the contribution of multiphase gas to the {\hi} absorption. The \citetalias{Pointon2019} sample is unique because all 47 galaxies at redshift $z < 0.7$: 1) have galaxy imaging with {\hst}, which is essential for the determination of azimuthal angles; 2) have UV {\hst}/COS spectra in the MAST archive mostly from PIDs 13398 (P.I. Christopher Churchill) and 11598 (P.I. Jason Tumlinson), with 36/47 absorption systems supplemented with high-resolution optical spectra, covering a range of ionization states including the {\hi} series, {\ovi}, {\civ}, and {\mgii} using optical spectra. This paper is organized as follows: in Section~\ref{sec:observations} we describe the spectral observations that are analysed in this work; in Section~\ref{sec:methodology} we describe the methodology used to determine the physical conditions of an absorber; in Section~\ref{sec:ionization modelling} we present the ionization modelling of all the absorbers in our sample; in Section~\ref{sec:results} we present the results from our ionization modelling; in Section~\ref{sec:discussion} we discuss the implications based on the inferred absorption properties seen along these sightlines. We conclude in Section~\ref{sec:summary}. Throughout this work, we assume a $\Lambda$CDM cosmology with $H_{0}\,=\,70$~km~s$^{-1}$~Mpc$^{-1}$, $\Omega_{\rm M}\,=\,0.3$, and $\Omega_{\Lambda}\,=\,0.7.$  Metallicities are given in the notation {\metallicity} with solar relative abundances taken from \citet{grevesse2011chemical}. All the distances given are in physical units. All the logarithmic values are presented in base-10.

\section{OBSERVATIONS AND DATA}
\label{sec:observations}

 \onecolumn
\begin{landscape}
\begin{table*}
\begin{threeparttable}
  \setlength{\tabcolsep}{0.025in}
  \def\colhead#1{\multicolumn{1}{c}{#1}}
  \caption{Quasar Observations \label{tab:obsqso}}
  \begin{tabular}{llcllcccccccc}
    \hline\hline
    \colhead{(1)}           &
    \colhead{(2)}     &
    \colhead{(3)}        &
    \colhead{(4)}       &
    \colhead{(5)} &
    \colhead{(6)} &
    \colhead{(7)} &
    \colhead{(8)}&
    \colhead{(9)}&
    \colhead{(10)}&
    \colhead{(11)}&
    \colhead{(12)}&
    \colhead{(13)}\\
    \colhead{J-Name}           &
    \colhead{Proper Name}           &
    \colhead{$z_{\rm qso}$}     &
    \colhead{RA}        &
    \colhead{DEC}       &
    \colhead{COS} &
    \colhead{COS} &
    \colhead{STIS} &
    \colhead{STIS} &
    \colhead{$FUSE$} &
    \colhead{$FUSE$} &
    \colhead{Optical} &
    \colhead{Optical} \\
    &
    &
    &
    \colhead{(J2000)}        &
    \colhead{(J2000)}       &
    \colhead{Gratings} &
    \colhead{PID(s)} &
    \colhead{Gratings} &
    \colhead{PID(s)} &
    \colhead{Gratings} &
    \colhead{PID(s)} &
    \colhead{Spectrograph} &
    \colhead{PID(s)} \\

    \hline
    J0125		& PKS0122$-$00	& 	1.074		&	$		01$:$25$:$28.84		$	&	$		-00$:$05$:$55.93		$	&		G160M, G185M, G225M	        	&		13398		&		{\nodata}		&		{\nodata}		&		{\nodata}		&		{\nodata}		&		UVES		&		075.A-0841(A)	\\
J0351		&3C95	&	0.616		&	$		03$:$51$:$28.54		$	&	$		-14$:$29$:$08.71 		$	&		G130M, G160M, G185M		&		13398		&		{\nodata}		&		{\nodata}		&		{\nodata}		&		{\nodata}		&		UVES		&		076.A-0860(A)	\\
J0407		&PKS0405$-$12	&	0.572		&	$		04$:$07$:$48.43		$	&	$		-12$:$11$:$36.66		$	&	G130M, G160M 	&	11508, 11541 	&	E140M	&	7576	&	SiC2A, LiF1, LiF2	&	B087, D103	&		HIRES		&		G01H, U68H	\\
J0456		&PKS0454$-$22&	0.533		&	$		04$:$56$:$08.92		$	&	$		-21$:$59$:$09.40		$	&	\makecell{G160M, G185M, \\G225M, G200M\tnote{a}} 	&	\makecell{12466, 12252, \\ 12536, 13398} 	&	E230M	&	8672	&		{\nodata}		&		{\nodata}		&		UVES 		&		076.A-0463(A)	\\
J0853		&US1867	&	0.514		&	$		08$:$53$:$34.25		$	&	$		+43$:$49$:$02.33		$	&	\makecell{G130M, G160M, \\ G185M, G225M}		&		13398		&		{\nodata}		&		{\nodata}		&		{\nodata}		&		{\nodata}		&		{\nodata}		&		{\nodata}	\\
J0914		& SDSSJ091440.40$+$282330.0	&	0.735		&	$		09$:$14$:$40.38		$	&	$		+28$:$23$:$30.62		$	&		G130M, G160M		&		11598		&		{\nodata}		&		{\nodata}		&		{\nodata}		&		{\nodata}		&		HIRES		&		U059Hb	\\
J0943		&	SDSSJ094331.60$+$053131.0	& 0.564		&	$		09$:$43$:$31.61		$	&	$		+05$:$31$:$31.49		$	&		G130M, G160M		&		11598		&		{\nodata}		&		{\nodata}		&		{\nodata}		&		{\nodata}		&		HIRES		&		U066Hb	\\
J0950		&	HS0946$+$4845 &	0.589		&	$		09$:$50$:$00.73		$	&	$		+48$:$31$:$29.38		$	&		G130M, G160M		&	 11598, 13033	&		{\nodata}		&		{\nodata}		&		{\nodata}		&		{\nodata}		&		HIRES		&		U059Hb	\\
J1004		&	PG1001$+$291 &	0.327		&	$		10$:$04$:$02.61		$	&	$		+28$:$55$:$35.39		$	&	 G130M, G160M 	&		12038		&	E140M	&	9184	&	SiC2A, LiF1, LiF2	&	P207	&		{\nodata}		&		{\nodata}	\\
J1009		&	SDSSJ100902.10$+$071344.0 &	0.456		&	$		10$:$09$:$02.06		$	&	$		+07$:$13$:$43.87		$	&		G130M, G160M		&	 11598, 14708	&		{\nodata}		&		{\nodata}		&		{\nodata}		&		{\nodata}		&		HIRES		&		U066Hb	\\
J1041		&	4C06.41 &	1.270		&	$		10$:$41$:$17.16		$	&	$		+06$:$10$:$16.92		$	&		G160M	        	&		12252		&		{\nodata}		&		{\nodata}		&		{\nodata}		&		{\nodata}		&		HIRES		&		C17H	\\
J1119		&	PG1116$+$215	& 0.176		&	$		11$:$19$:$08.67		$	&	$		+21$:$19$:$18.01		$	&	 G130M, G160M 	&		12038		&	E140M, E230M	&	8097, 8165	&	SiC2A, LiF1, LiF2	&	P101	&		HIRES		&		U152Hb	\\
J1133		&	SDSSJ113327.78$+$032719.1 &	0.524		&	$		11$:$33$:$27.78		$	&	$		+03$:$27$:$19.17		$	&		G130M, G160M		&		11598		&		{\nodata}		&		{\nodata}		&		{\nodata}		&		{\nodata}		&		HIRES		&		U059Hb	\\
J1139		& HE1136$-$1334	&	0.556		&	$		11$:$39$:$10.70		$	&	$		-13$:$50$:$43.63		$	&		G130M	        	&		12275		&		{\nodata}		&		{\nodata}		&		{\nodata}		&		{\nodata}		&		{\nodata}		&		{\nodata}	\\
J1219 	& PG1216$+$069	&	0.331		&	$		12$:$19$:$20.93		$	&	$		+06$:$38$:$38.52		$	&	 G130M, G160M 	&		12025		&	E140M	&	9184	&	SiC2A, LiF1, LiF2	&	P107	&		{\nodata}		&		{\nodata}	\\
J1233		& LBQS1230$-$0015	&	0.470		&	$		12$:$33$:$04.05		$	&	$		-00$:$31$:$34.20		$	&		G130M, G160M		&	 11598, 12486 	&		{\nodata}		&		{\nodata}		&		{\nodata}		&		{\nodata}		&		HIRES		&		U059Hb	\\
J1241		&	SDSSJ124154.00$+$572107.0 &	0.583		&	$		12$:$41$:$54.02		$	&	$		+57$:$21$:$07.38		$	&		G130M, G160M		&	 11598, 13033	&		{\nodata}		&		{\nodata}		&		{\nodata}		&		{\nodata}		&		HIRES		&		U059Hb	\\
J1244		&PG1241$+$176	&	1.273		&	$		12$:$44$:$10.82		$	&	$		+17$:$21$:$04.52		$	&	 G160M, G185M       	&		12466		&	E230M	&	8672	&		{\nodata}		&		{\nodata}		&		HIRES		&	 \tnote{b}	\\
J1301		&PG1259$+$593	&	0.477		&	$		13$:$01$:$12.93		$	&	$		+59$:$02$:$06.75		$	&	 G130M, G160M 	&		11541		&	E140M	&	8695	&	SiC2A, LiF1, LiF2	&	P108	&		{\nodata}		&		{\nodata}	\\
J1319		&TON153	&	1.014		&	$		13$:$19$:$56.23		$	&	$		+27$:$28$:$08.22		$	&	 G160M, G185M       	&		11667		&	E230M	&	8672	&		{\nodata}		&		{\nodata}		&		HIRES		&		U074	\\
J1322		&SDSSJ132222.70$+$464535.0&	0.374		&	$		13$:$22$:$22.68		$	&	$		+46$:$45$:$35.22		$	&		G130M, G160M		&	 11598, 13033 	&		{\nodata}		&		{\nodata}		&		{\nodata}		&		{\nodata}		&		HIRES		&		U066Hb	\\
J1342		&	HE1340$-$0038	& 0.326		&	$		13$:$42$:$51.60		$	&	$		-00$:$53$:$45.31		$	&		G130M, G160M		&	 11598, 13033 	&		{\nodata}		&		{\nodata}		&		{\nodata}		&		{\nodata}		&		HIRES		&		U059Hb	\\
J1357		&	PKS1354$+$19	& 0.720		&	$		13$:$57$:$04.43		$	&	$		+19$:$19$:$07.37		$	&	 G160M, G185M, G225M	&		13398		&		{\nodata}		&		{\nodata}		&		{\nodata}		&		{\nodata}		&		UVES		&		076.A-0860(A)	\\
J1547		&3C323.1	&	0.264		&	$		15$:$47$:$43.53		$	&	$		+20$:$52$:$16.61		$	&	 G130M, G160M, G185M 	&		13398		&		{\nodata}		&		{\nodata}		&		{\nodata}		&		{\nodata}		&		{\nodata}		&		{\nodata}	\\
J1555		&SDSSJ155504.40$+$362847.0	& 0.714		&	$		15$:$55$:$04.40		$	&	$		+36$:$28$:$48.04		$	&		G130M, G160M		&		11598		&		{\nodata}		&		{\nodata}		&		{\nodata}		&		{\nodata}		&		HIRES		&		U059Hb	\\
J1704		&	3C351.0 &	0.371		&	$		17$:$04$:$41.37		$	&	$		+60$:$44$:$30.50		$	&		{\nodata}		&		{\nodata}		&		E140M	    	&		8015		&	SiC2A, LiF1, LiF2	&	Q106	&		HIRES		&		G400H, U019Hb	\\
J2131 	&	PKS2128$-$12 &	0.501		&	$		21$:$31$:$35.26		$	&	$		-12$:$07$:$04.79 		$	&		\makecell{G130M, G160M,\\ G185M, G225M}		&	 13398, 12536	&		{\nodata}		&		{\nodata}		&		{\nodata}		&		{\nodata}		&		HIRES		&		C54H, U51H, C99H	\\
J2137		&	PKS2135$-$14 &	0.200		&	$		21$:$37$:$45.17		$	&	$		-14$:$32$:$55.81		$	&	 G130M, G160M, G185M 	&		13398		&		{\nodata}		&		{\nodata}		&	SiC2A, LiF1, LiF2	&	C030	&		{\nodata}		&		{\nodata}	\\
J2253		&	3C454.3	& 0.859		&	$		22$:$53$:$57.74		$	&	$		+16$:$08$:$53.56		$	&	 G130M, G160M, G185M 	&		13398		&		{\nodata}		&		{\nodata}		&		{\nodata}		&		{\nodata}		&		UVES		&		075.A-0841(A)	\\[-5pt]
\\ 
    \hline
  \end{tabular}
  \begin{tablenotes}
  \item [a] For the J0456 sightline, we also use the GHRS G200M spectrum from the PID = 5961.
    \item[b] Spectra from \citet{churchill01}.
  \end{tablenotes}
\end{threeparttable}
\end{table*}
 \end{landscape}
 \twocolumn

We study the distribution of CGM properties using the ``Multiphase Galaxy Halos'' Survey, which consists of the {\hst} program with PID 13398~\citep{kacprzak2015,Kacprzak2019ApJ,muzahid2015extreme,Muzahid2016,nielsen2017highly,ng2019,Nateghi2021} and data from the literature~\citep{Yuan2002,Danforth2010,Meiring2011,Churchill2012,Shull2012,Tilton2012,werk2013cos,fox2013,Tilton2013,mathes2014}. {\hst} imaging and UV spectra are available for all 29 quasar fields.

\smallskip

This sample consists of 47 galaxies with spectroscopic redshifts between 0.07 $< z <$ 0.66, which have an impact parameter range of 21 kpc $< D <$ 276 kpc from a background quasar. Two absorption systems in \citetalias{Pointon2019}, the $z =$ 0.2198 absorber towards the quasar J1139 and the $z =$ 0.2013 absorber towards the quasar J1342, are not considered in our analysis, as for these systems the strongest observed potential {\hi} line ({\lyb} and {\lya, respectively) cannot be disentangled from a complex set of lines from higher-redshift systems, and there are no associated metal lines. Two new systems not presented in \citetalias{Pointon2019} are presented in this work. These are the $z =$ 0.1538 and $z =$ 0.3900 absorber towards the quasar J2253. The properties of the galaxies associated with these two systems were determined by and presented in~\citet{GF1}. The properties of the remaining galaxies are adopted from \citetalias{Pointon2019}. For the convenience of the reader, we reproduce Table 2 from \citetalias{Pointon2019} and present it in Table~\ref{tab:obsgal}. Additionally, we have added the columns of virial radii of the galaxies estimated using the halo masses, impact parameter scaled by the virial radius, absorber velocity scaled by the circular velocity of the galaxy (see \S~\ref{sec:absorber_velocities}), and information on specific star formation. The propagation of asymmetric uncertainties is carried out using the \textsc{asymmetric\_uncertainty}~\citep{Gobat2022} Python package. The galaxies in this sample are isolated such that they have no neighbors within 100~kpc and have a line-of-sight velocity separation $>500$~{\kms} from the nearest galaxy~\citetalias{Pointon2019}. This selection minimizes the effect of mergers on the CGM. The halo mass range of the galaxies in the sample is $10.8 < \log M_{h}/M_{\odot} < 12.5$, which is typical for $L_{\star}$ galaxies. Galaxy morphologies and other properties for each of the galaxy-absorber pairs in the ``Multiphase Galaxy Halos'' survey are from~\citealt{kacprzak2015, Kacprzak2019ApJ,muzahid2015extreme,Muzahid2016,nielsen2017highly,Pointon2017,Pointon2019,ng2019}, and from the literature~(\citealt{chen2001origin,chen2009,Prochaska2011,werk2012,werk2013cos,Johnson2013}). We define $\Phi=$ 0{\degree} and $\Phi=$90{\degree} as the projected major and minor axes. The survey reaches an average limiting magnitude of $\approx$ 26 for a {\sn} of $\approx$ 10, in an average exposure time of $\approx$ 800 s. 

\smallskip

\subsection{UV Quasar Spectra}

The UV spectra in this study were acquired with the COS and STIS instruments onboard the {\hst} telescope, and the LiF and SiC2 channels onboard the $FUSE$ telescope. While \citetalias{Pointon2019} have used only the COS G130M and/or G160M when available, and the STIS/E140M for the sightline towards J1704, we have complemented {\hst}/COS observations with $FUSE$ and {\hst}/STIS observations whenever available. The data from these additional gratings often provide better constraints on the inferred absorber properties. The UV \textit{$HST$}/COS spectra in this study are downloaded from the Barbara A. Mikulski Archive for Space Telescopes (MAST). We use COS data when both STIS and COS are available, owing to the significantly higher {\sn} of COS data. We use STIS data when a COS observation does not cover a transition of interest. In Table~\ref{tab:obsqso}, we list the background quasars and various instruments/gratings they have been observed with.

\smallskip

The COS G130M/G160M spectra have an average resolving power of $R \approx 20 ,000$ and cover a range of ions including the {\hi} Lyman series, {\cii}, {\ciii}, {\civ}, {\siii}, {\siiii}, {\siiv}, {\nii}, {\niii}, {\nv}, and {\ovi}. The {\hst}/COS spectra were reduced following the procedures in \citet{wakker2015}. In essence, spectra are processed with the \textsc{calcos} v3.2.1 pipeline and are aligned using a cross-correlation, and then shifted to ensure that (1) the velocities of the interstellar lines match the 21cm {\hi} profile, and (2) the velocities of the lines in a single absorption system are aligned properly. The exposures are then collectively combined by summing total counts per pixel prior to converting to flux. The $FUSE$ observations were processed with the \textsc{calfuse} pipeline (ver 3.2.3). A zero point offset correction was applied using the methods in \citet{wakker2006fuse}. $FUSE$ spectra have an average resolving power of $R \approx 15,000$. In some cases, where the absorption of interest is affected by the {\oi}~$\lambda$1302 geocoronal emission, we use the \textsc{timefilter} module from the \textsc{costools} package to filter unwanted data (\hbox{SUN\_ALT $>$ 0}) and rerun the \textsc{calcos} pipeline on the filtered data to generate the airglow emission corrected spectrum for the affected portion of the spectrum. The STIS archival observations are available for some of the sightlines with the E230M and E140M echelle gratings. STIS data were processed using the \textsc{calstis} ver.~3.4.2 pipeline to reduce the raw data into 1-D spectrum. E230M and E140M spectra have an average resolving power of $R \approx$ 30,000 and 45,000, respectively. 

\smallskip

Continuum normalization was done by fitting a cubic spline to the spectrum. The statistical uncertainties in the continuum fits were determined by using ``flux randomization'' Monte Carlo simulations (e.g., \citealt{peterson1998optical}), varying the flux in each pixel of the spectrum by a random Gaussian deviate based on the spectral uncertainty. The pixel-error weighted average and standard deviation of 1000 iterations were adopted as the flux and uncertainty, respectively. 

\subsection{Optical Quasar Spectra}

The UV spectra are complemented with the optical spectra, which cover the low ionization transitions of {\mgi}, {\cai}, {\nai}, {\fei}, {\mgii}, {\caii}, {\feii}, {\mnii}, and {\alii}. Some of these transitions are especially useful in resolving the component structure. Optical spectra from Keck/HIRES or VLT/UVES are available for 34 absorption systems with an average resolving power of $R \simeq 40,000$. The observation IDs and instruments are listed in Table~\ref{tab:obsqso}. The HIRES spectra were reduced using the Mauna Kea Echelle Extraction (\textsc{makee}) package~\footnote{\url{https://sites.astro.caltech.edu/~tb/makee/index.html}}. The UVES spectra were reduced using the European Southern Observatory (ESO) pipeline \citep{dekker2000optical} and the UVES Post-Pipeline Echelle Reduction (UVES POPLER) code~\citep{murphy2016,murphy2019}.

\section{Cloud-by-cloud multiphase Bayesian ionization modelling}
\label{sec:CMBM}

To extract the properties of CGM absorbers, we employ a novel forward-modelling Bayesian inference suite, Cloud-by-cloud, Multiphase, Bayesian, ionization Modelling method \citep[CMBM;][]{sameer21,Sameer2022}, which couples ionization modelling from {\CLOUDY} with nested sampling Bayesian inference techniques. This modelling approach has been tested using both observations~\citep{sameer21,Sameer2022} and large cosmological boxes to cosmological zoom-in simulations of the CGM~\citepalias{Hafen2024}, and found to obtain observational estimates accurate to within 0.1 dex of the true source values. Identifying and characterizing truly individual, physically distinct clouds may prove challenging, possibly due to the gas being too diffuse and/or because the sightline intersects shorter pathlength resulting in the absence of detected absorption in the observed profiles, as highlighted in modeling of simulations. For example, \citet{Stephanie2021} found that in low redshift, star-forming galaxies, up to half of the observed {\ovi} absorption can originate outside the galaxy virial radius. For low ionization clouds with small cloud sizes, \citetalias{Hafen2024} did find that more localized regions, with derived sizes consistent with the simulations, are responsible for the absorption. This method improves the efficiency of component by component modelling that has been successful in recovering the physical conditions for various individual absorbers~\citep[e.g.][]{churchill1999multiple,charlton2000anticipating,ding2003quadruple,charlton2003high,ding2003multiphase,zonak2004absorption,ding2005absorption,masiero2005models,lynch2007physical,misawa2008supersolar,lacki2010z,jones2010bare,muzahid2015extreme,Crighton2015,richter2018,rosenwasser2018understanding,zahedy2019characterizing,Norris2021,Qu2023}. CMBM allows for self-consistently modelling the multicomponent, multiphase absorption profiles seen in multiple ionization states. It uses the observed profile shapes in all the available line transitions, allowing us to better capture the multiphase nature of the absorption systems. This procedure eliminates the need for assuming ionization corrections in the estimation of the metallicities of absorbers as it marginalizes over the unknown amounts of ionization corrections to be applied during parameter inference.

\smallskip

Low-density diffuse gas is seldom found in galaxy halos where there are no substantial stellar or extragalactic sources of ionizing radiation. This photoionizing radiation modulates the ionization states, and thus affects the absorption line signatures from CGM gas~\citep[e.g.,][]{Mallik2023}. A variety of extragalactic photoionization background radiation (EBR) models exist in the literature. The KS19~\citep{KS19} EBR model uses the latest QSO emissivities, galaxy star formation rates, estimates of {\hi} distribution in the IGM, dust attenuation in galaxies, and the escape fraction of ionizing photons from galaxies. An investigation of the systematic uncertainties in the inferred parameters due to the choice of the radiation field models is presented in \citet{acharya21} and \citet{Gibson2022}. \citet{acharya21} find a variation of 4--6.3 and 1.6--3.2 times in the inferred density and metallicity, respectively, depending on whether the gas is high or low density. For $z<1$, \citet{Gibson2022} find that the absorber metallicities increase on average by $\sim 0.3$~dex as the EUV slope is increased from  $\alpha_{\rm EUVB} = -2.0$ to $-1.4$. We adopt the KS19 default EUV slope of $\alpha_{\rm EUVB} = -1.8$.

\subsection{Methodology}
\label{sec:methodology}

 We first generate a grid of converged photoionization thermal equilibrium (PIE) {\CLOUDY} models on a 3--D grid, adopting the KS19 EBR model. The gas conditions of CGM absorbers, which are subject to photoionization by the EBR, are functions of the metallicity, density, and temperature. The axes of the PIE grid are {\metallicity}, {\hden}, and {\colden}. The {\colden} functions as a stopping criterion for a given {\CLOUDY} model. For a PIE model, {\CLOUDY} determines an equilibrium temperature, $T$, such that the absorber is in thermal equilibrium with the background radiation field. The ranges for these parameters are $\log Z/Z_{\sun}$ $\in$ [$-3.0,1.5$], $\log n_{\H}$ $\in$ [$-6.0, 2.0$], and $\log N(\hi)$ $\in$ [11.0, 21.0], and obtained with a 0.1 dex step-size. 

\smallskip

The galaxies in our sample have spectroscopic redshifts between 0.07 < $z$ < 0.66. Therefore, we construct the PIE grids at each $z \in$ [0.05, 0.65], with a 0.05 step size. To create large grids of {\CLOUDY} calculations, we use the grid command in {\CLOUDY} and generate grids on a distributed system--Stampede~\footnote{\url{https://www.tacc.utexas.edu/systems/stampede2}} cluster.  We adopt the {\CLOUDY} default solar abundance pattern in model generation. We do not account for depletion due to dust or deviations from solar abundance pattern in our model generation. We report any deviations from the solar abundance pattern.

\smallskip

We start by inspecting the spectrum for low ionization transitions \textit{viz.}\ {\mgii}, {\siii} or {\cii} that can serve as constraining ions to determine the component structure. We obtain a Voigt Profile (VP) fit to the constraining ion to obtain the redshifts, z$_{i}$, and total Doppler broadening, $b_{i}$, parameters for each, $i$th, component of an absorption system. The VPs are modelled using an analytic approximation~\citep{garcia2006voigt} and implemented using the VoigtFit package~\citep{krogager2018voigtfit}. With this assumed component structure, we optimise over five parameters in each component, the five parameters being {\metallicity}, {\hden}, {\colden}, b$_{nt}$, and $z$. Here b$_{nt}$ is the broadening due to non-thermal processes, and $z$ is the redshift of the absorption component. We use uniform priors for all five parameters, in the range mentioned above for {\metallicity}, {\hden}, and {\colden}. For b$_{nt}$ we adopt the range $\in$ [0, $b+3\sigma_{b}$], where $b$ is the value determined from the VP fit, and $\sigma_{b}$ is the uncertainty on $b$ value, and for $z$ in the range $\in$ [$z-3\sigma_{z}$, $z+3\sigma_{z}$], where z is the absorption centroid for the component.

\smallskip

Typically, an absorption component in an absorption system is composed of one or more ``clouds''. Each cloud represents a parcel of gas in a specific spatial location with its own unique properties. These different clouds could potentially contribute to the absorption at a given velocity, albeit potentially being spatially at disparate locations~(e.g., \citealt{marra2024}). In some cases, asymmetry in the component profile can reveal the presence of two clouds. While in other cases, the need for different ionization parameters to explain the range of transitions observed at a given velocity can reveal the need for two clouds to explain the component. It is generally the case that a model that explains the absorption in low-ionization transitions cannot fully account for the absorption seen in intermediate/high ionization states (e.g., {\siiii}, {\ciii}, {\siiv}, {\civ}, {\nv}, {\ovi}, {\neviii}). When it is found that there is absorption unaccounted for by the low-ionization gas phases, we include higher ionization transitions as additional constraining ions. This unaccounted absorption may arise in environments where particle-particle interactions dominate over energy exchange with the radiation field. Earlier works (e.g., \citealt{Tripp2001,Tripp2011, Narayanan2011,Narayanan2012}) which have undertaken the ionization modelling of high-ionization absorbers have shown that this gas phase is dominated by collisional ionization, primarily based upon the large Doppler broadening parameters of the absorption profiles. 

\smallskip

Furthermore, in some instances collisional ionization equilibrium (CIE) cannot sufficiently explain the ratios of observed column densities and/or profile shapes~(e.g., \citealt{savage2014properties}). For highly ionized gas at temperatures $\lesssim 5 \times 10^{6}$~K, and in the presence of external photoionizing radiation, recombination can slow down in comparison to cooling causing the plasma to remain over-ionized compared with what would be predicted from CIE. Departures from CIE become more prominent for high-metallicity gas. \citet{Tchernyshyov2022} using the CGM$^{2}$ survey~\citep{wilde2021} found some evidence for departures from a hydrostatic equilibrium temperature to explain the observed {\ovi} amount, galaxy mass dependence, and extent. \citet{Gnat2017} (hereafter \citetalias{Gnat2017}) examined the non-equilibrium evolution of photoionized cooling gas and computed the fractional abundances of different ions as a function of temperature, metallicity, and hydrogen number density. We thus model higher ionization gas, by assuming cooling in the presence of external photoionizing radiation, a more realistic scenario than CIE, using the time-dependent photoionized (TDP) ion fractions from Table 11 of~\citetalias{Gnat2017}. In TDP cooling gas, the abundances of the major coolants may be considered to be either in PIE or affected by time-dependent collisional (TDC) processes, depending on the density. However, the abundance of some specific species may differ by a few factors from the TDC/PIE values, and thus we use the TDP ion fractions. We optimize over five parameters in each cloud modelled as a TDP cloud, the five parameters being {\metallicity}, {\hden}, {\temp}, b$_{nt}$, and $z$. We use uniform priors on all parameters in the range - $\log Z/Z_{\sun}$ $\in$ [$-3.0,0.3$], {\hdenu}$\in$ [$-6.0, 2.0$], and {\tempu} $\in$ [4.0, 7.0]. The bounds on these priors are set by the model limits obtained by~\citetalias{Gnat2017}. The priors for b$_{nt}$ and $z$ are determined in a way similar to discussed earlier. We do not adopt TDP models to model low ionization phases (e.g., {\mgii}, {\siii}, {\cii}) because the~\citetalias{Gnat2017} models have a lower bound on temperature of 10$^{4}$ K, and, a posteriori in a number of cases we observe narrow {\mgii} absorption suggestive of even lower temperatures. 

\smallskip

\begin{figure*}
  \centering
  \subfloat{\includegraphics[width=0.55\linewidth]{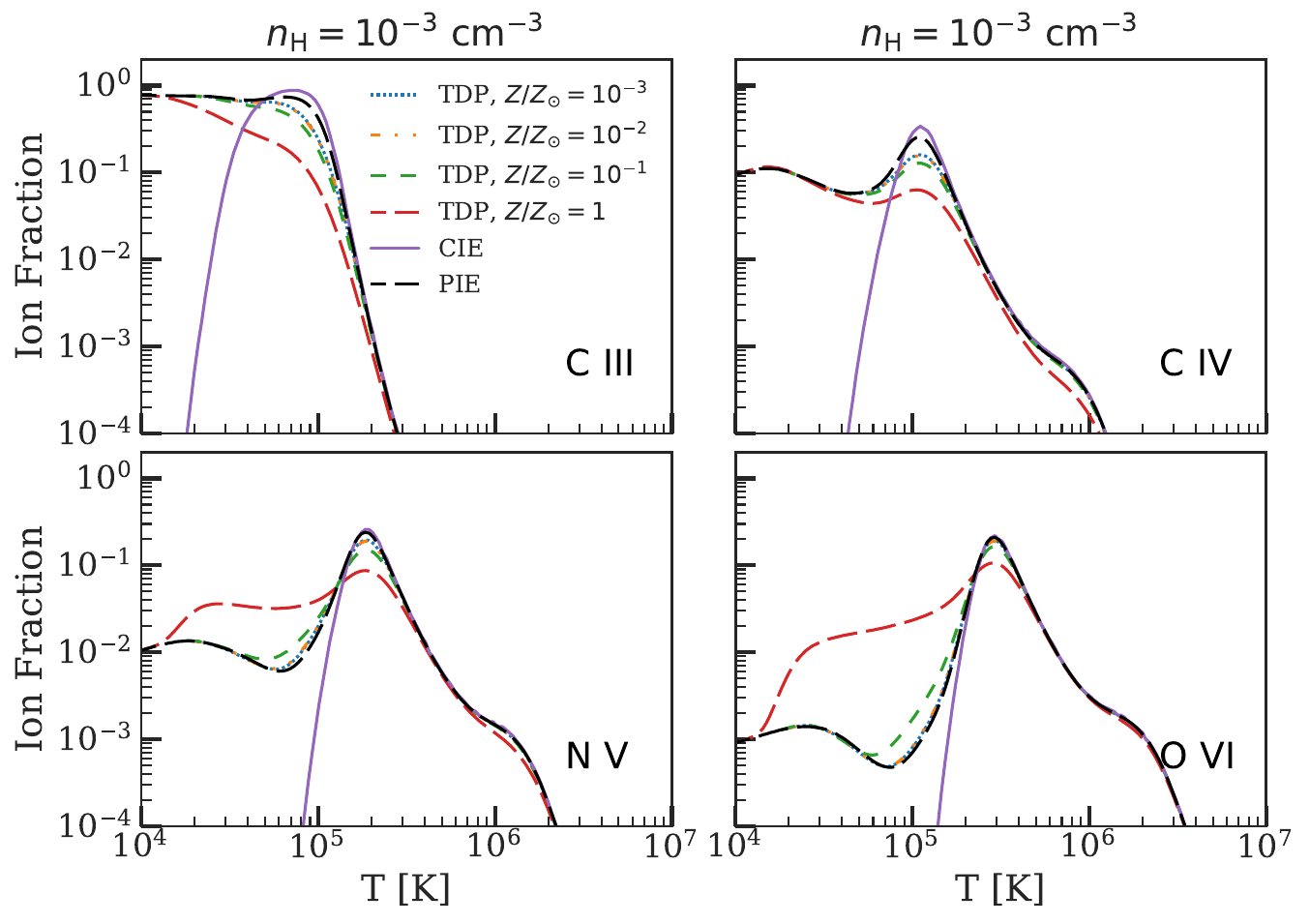}\label{fig:subfig1}}
  \hfill
  \subfloat{\includegraphics[width=0.55\linewidth]{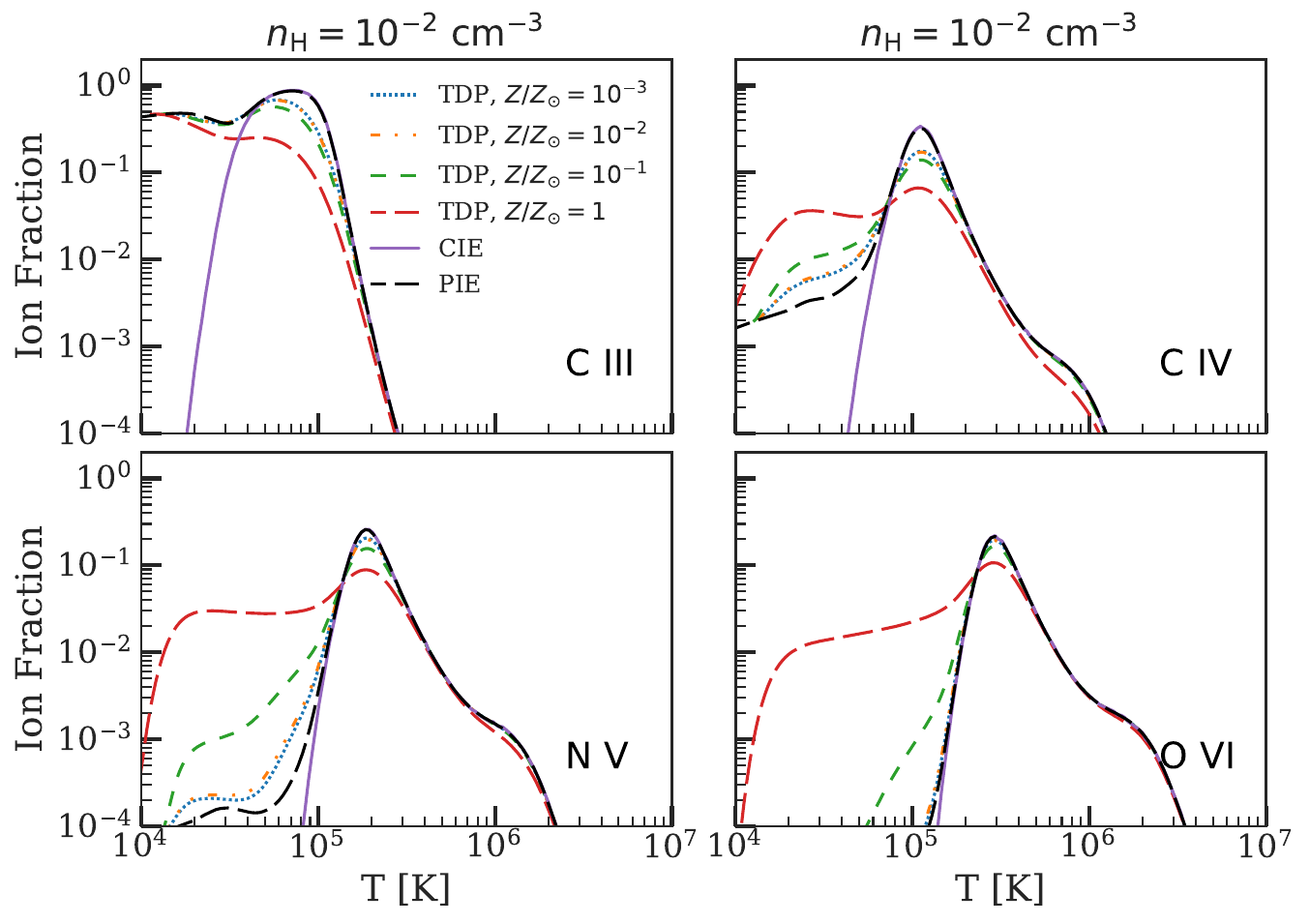}\label{fig:subfig2}}
  \hfill
  \subfloat{\includegraphics[width=0.55\linewidth]{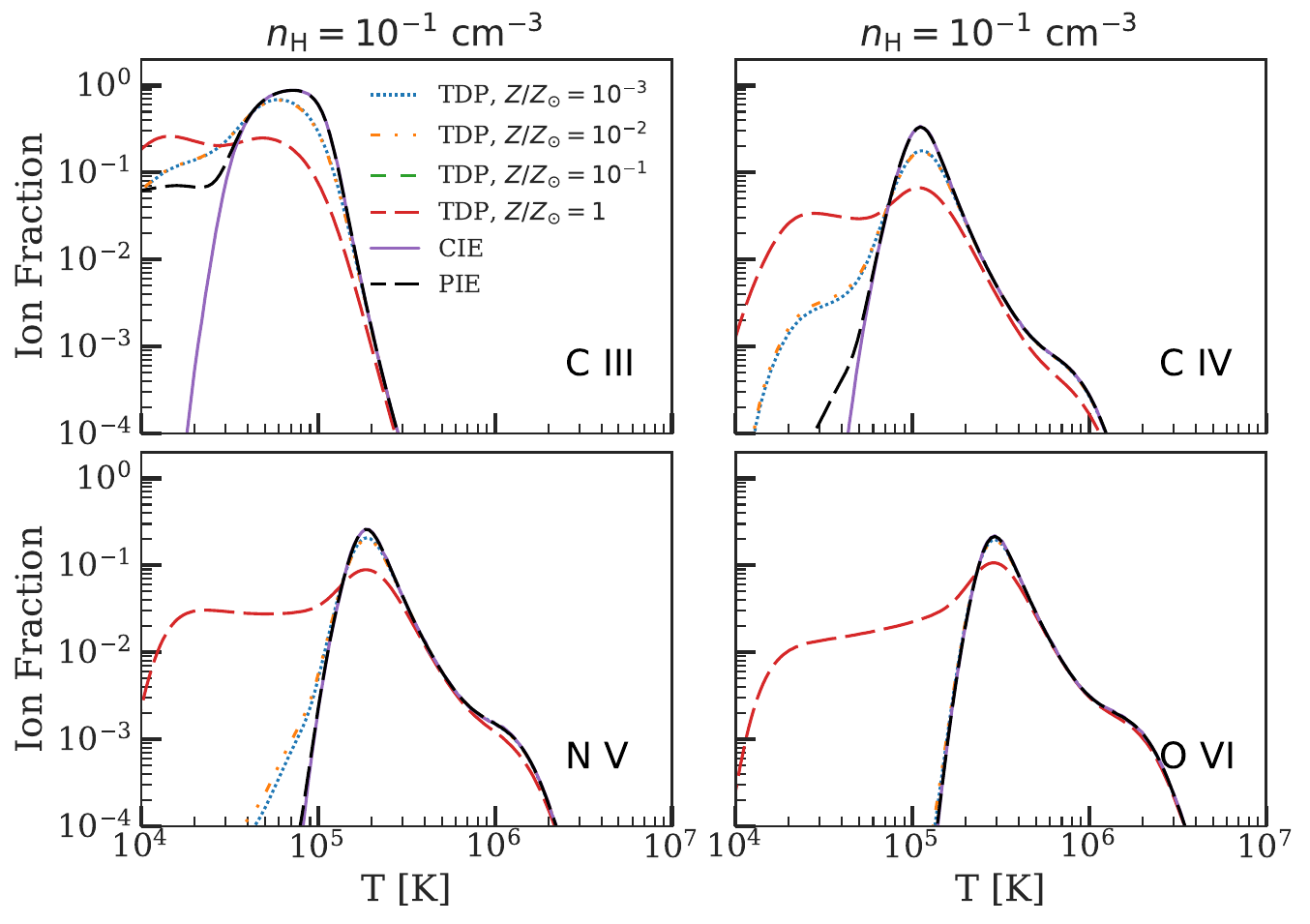}\label{fig:subfig3}}
  \caption{Ionization fractions of chosen low, intermediate, and higher ionization ions from the models in \citetalias{Gnat2017} for the densities of {\hdenu} = $-3$, $-2$, and $-1$, from top to down. CIE indicates collisional ionization equilibrium. PIE indicates photoionization equilibrium, and TDP indicates time-dependent photoionization. For CIE gas, cooling times are much longer than the ionization and recombination timescales. On the other hand, for gas out of ionization equilibrium, the cooling time is shorter than the ionization and recombination timescales. Cooling is more efficient for higher metallicity gas, where the departures from CIE are noticeably marked. For {\ovi}, at temperatures greater than  $T \simeq$ 2.5 $\times$ 10$^{5}$ K cooling times are longer than the ionization timescales and independent of the gas metallicity such that the CIE and non-equilibrium ion fractions are identical.}
  \label{fig:ionfracs}
\end{figure*}

In Fig.~\ref{fig:ionfracs}, we compare the temperature dependence of ionization fractions for gas traced by intermediate and high ionization ions under the various conditions of photoionization equilibrium, collisional ionization equilibrium, and time-dependent photoionization for metallicities $Z/Z_{\odot}$ = 10$^{-3}$, 10$^{-2}$, 10$^{-1}$, and 1, at the densities of 10$^{-3}$ \cmsq, 10$^{-2}$ \cmsq, 10$^{-1}$ \cmsq. For the intermediate and higher ionization ions of {\ciii}, {\civ}, {\nv}, and {\ovi}, we observe that the departure between equilibrium models (PIE or CIE) and time-dependent photoionization models is smallest for lower metallicity gas, where the cooling times are longest. The PIE and TDP fractions diverge for metal-enriched gas at temperature, $T \gtrsim$ 10$^{4}$ K. The ionizaton fractions in Fig.~\ref{fig:ionfracs} show that non-equilibrium ionization can produce significant intermediate and higher ions in a temperature regime often associated with photoionization. While it is possible for the intermediate and higher ionization phases to arise at lower temperatures, by using profile shapes in multiple ionization states in conjunction with the column densities we are able to break the ionization fraction degeneracy. For {\ovi}, at $T \gtrsim$ 4$\times$10$^{5}$ K, the different ionization models are consistent with one another and independent of the density and metallicity, suggesting that CIE is a good approximation for the ionization of {\ovi} gas.

\smallskip

For parameter estimation and model comparison, we use the nested sampling algorithm, PyMultiNest~\citep{buchner2014}. PyMultiNest is well-suited for tackling multi-dimensional and multimodal posteriors. During the log-likelihood evaluation, the column densities and temperatures from the {\CLOUDY} grids at the sampled parameter values are calculated on the fly by interpolating over the preconstructed {\CLOUDY} grid. The interpolant is constructed by triangulating the grid data using the Quick hull algorithm~\citep{barber1996quickhull}, and on each triangle linear barycentric interpolation is carried out. We create a ``mask'' over the regions contaminated by interlopers and over noisy continuum-level regions of the spectrum. The masked regions do not influence the parameter estimation. We perform deblending in cases where there is substantial contribution from {\siii}~$\lambda$989 line to the absorption in {\niii}~$\lambda$989 line. For {\hi}-only clouds with no metals detected we adopt 3$\sigma$ upper limits on the metallicity measurement. To confirm the detection (or lack thereof) of the absorption of interest at the location of {\hi} absorption, we calculate the limiting equivalent width and column density using the pyND\footnote{https://github.com/jchowk/pyND} software.

\smallskip

We compute the Bayes factor to compare models of increasing complexity (more clouds), adopting the model with the highest log-evidence and the least model complexity. In models of increasing complexity, the gain in likelihood does not compensate for the increase in parameters of the model and is found to yield clouds with unconstrained posteriors. We synthesize the expected absorption profiles and compare them to the observed profiles in order to infer the column densities, Doppler parameters, and physical conditions of the absorbing gas. By using the shapes of the absorption profiles and centering of individual components of constraining ions, the CMBM method allows for more robust constraints on inferred model parameters than methods that average together components and perform {\CLOUDY} modelling of total column densities derived from the data.

\section{Absorption profiles and physical conditions}
\label{sec:ionization modelling}

Here, we present the CMBM analysis of the ``Multiphase Galaxy Halos'' survey. For each absorption system, we present a system plot (e.g., Fig.~\ref{fig:J0125_0.3985}) showing the absorption in various radiative transitions and the Maximum Likelihood Estimate (MLE) ionization models are overlaid on the observed absorption profiles. The observed profiles are presented as a function of velocity relative to the zero-point defined by the redshift of the associated galaxy. The detailed notes on the ionization modelling and system plots for all the systems are presented in appendices~\ref{appendix:notes_ionizationmodelling} and~\ref{appendix:systemplots}, respectively. A summary of the cloud-by-cloud properties of all the individual absorption systems determined by using our CMBM analysis is presented in Table~\ref{tab:cloudproperties}. 

\smallskip

Based on the models that explain the observed absorption in various phases, we classify clouds into PIE clouds and TDP clouds. Furthermore, for the TDP clouds, we subclassify them into two types--those that show absorption in the high ionization ions of at least the {\ovi}, and possibly {\nv} and/or {\civ}, but not in the low or intermediate ions, which we refer to as the TDP--High clouds, and those that show absorption in intermediate and high ionization states, and possibly also in low ionization states, which we refer to as the TDP--Low clouds.

\smallskip

In the rightmost panel of every system plot, the inferred marginalised posterior distributions representing the probability distributions of the parameters of interest are presented. The parameters are {\metallicity}, {\hden}, {\colden}, {\temp}, {\thickness}, and b$_{nt}$. The distributions are presented in the form of violin plots to visualise any skewness and multi-modality of the distributions. The region enclosing 99.7\% of the distribution is shown as a dotted boundary, and the region enclosing the $16-84$ percentiles is shaded using a colour corresponding to the modelled cloud. The MLE values are indicated as a $\star$ on top of the distributions. In many cases, the MLE provides a good point estimate of the parameters of the distribution. However, it is possible for the MLE to be outside the peak of the distribution.  The MLE can be affected by noise in the data which can pull the estimate away from the center of the distribution towards the tail.  

\smallskip

For each absorption system, we also discuss the various blends that affect the absorption of interest, and these blended regions are excluded from the modelling. Pixels that are contaminated by absorption from other intervening systems are blanked out and shaded in grey. However, we do not discuss contaminating blends that might appear to be present in a weaker transition but absent in the stronger one, where the line strength is proportional to the product of the oscillator strength and the rest wavelength of the line. We exclude {\caii} from the modeling. \citet{Robertson1988}, in the first census of extragalactic {\caii} absorption, found that the {\caii} is severely underabundant compared to both {\feii} and {\mgii}, typically smaller by a factor of 100. They argued that the underabundance suggests that {\caii} condenses more readily onto dust grains than {\mgii} and {\feii}. \citet{Richter2011} observed a column-density-dependent dust depletion of calcium, finding enhanced dust depletion of {\caii} in high {\hi} column density systems.

\onecolumn
\begin{landscape}
\begin{table*}

\begin{threeparttable}
  \setlength{\tabcolsep}{0.06in}
 \def\colhead#1{\multicolumn{1}{c}{#1}}
 \caption{Cloud-by-cloud properties of the absorption systems 
 \label{tab:cloudproperties}}

\begin{tabular}{llllllllllllc}
    \hline\hline
      \colhead{(1)}     &
          \colhead{(2)}     &
          \colhead{(3)}     &
          \colhead{(4)}     &
          \colhead{(5)}     &
          \colhead{(6)}	    &
          \colhead{(7)}	    &
          \colhead{(8)}     &
           \colhead{(9)}     &
            \colhead{(10)}    &
            \colhead{(11)}     &
             \colhead{(12)}     &
            \colhead{(13)}     \\
          \colhead{J-Name}                  &
          \colhead{$z_{\rm gal}$}           &
          \colhead{$V$}           &
          \colhead{{\metallicity}}           &
          \colhead{{\hdenu}}           &
        \colhead{{\totalcoldenu}}           &
          \colhead{{\coldenu}}           &
          \colhead{{\tempu}}           &
          \colhead{{\thicknessu}} &
          \colhead{{\bturb}} &
          \colhead{{\btherm}} &
          \colhead{{\bnet}} &
          \colhead{{Ionization Model}} \\
          \colhead{}                &
          \colhead{}                &
        \colhead{{\kms}}                &
          \colhead{}                &
          \colhead{}                &
          \colhead{}                &
          \colhead{}                &
          \colhead{}                &
          \colhead{} & 
          \colhead{{\kms}} & 
          \colhead{{\kms}} & 
          \colhead{{\kms}} &
          \colhead{} \\
         
    \hline 
          J0125 & $0.3985$ & $-114.8_{-8.7}^{+8.3}$ & $<0.03$ & $>-2.85$ & $14.79_{-0.47}^{+0.83}$ & $14.00_{-0.12}^{+0.11}$ & $4.02_{-0.06}^{+0.10}$ & $<-1.93$ & $37.89_{-5.52}^{+4.40}$ & $13.14_{-0.86}^{+1.62}$ & $40.28_{-5.03}^{+4.26}$ & PIE\\[2pt]
 &  & $-64.2_{-3.4}^{+3.4}$ & $<0.30$ & $>-4.00$ & $19.83_{-0.49}^{+0.34}$ & $13.40_{-0.30}^{+0.18}$ & $5.84_{-0.13}^{+0.13}$ & $<2.34$ & $51.43_{-7.09}^{+6.86}$ & $106.97_{-14.86}^{+16.62}$ & $119.08_{-12.91}^{+14.66}$ & TDP--High\\[2pt]
& & $-63.3_{-2.4}^{+2.3}$ & $-0.42_{-0.19}^{+0.19}$ & $-2.75_{-0.21}^{+0.22}$ & $17.77_{-0.20}^{+0.23}$ & $15.16_{-0.19}^{+0.21}$ & $4.32_{-0.10}^{+0.12}$ & $-1.0_{-0.4}^{+0.4}$ & $8.6_{-2.7}^{+3.0}$ & $18.5_{-2.0}^{+2.6}$ & $20.7_{-2.1}^{+2.6}$ & PIE \\[2pt]
& & $-9.6_{-0.3}^{+0.2}$ & $-1.36_{-0.11}^{+0.11}$ & $-2.36_{-0.05}^{+0.06}$ & $19.40_{-0.13}^{+0.13}$ & $17.19_{-0.10}^{+0.11}$ & $4.26_{-0.02}^{+0.02}$ & $0.3_{-0.2}^{+0.2}$ & $8.8_{-0.3}^{+0.4}$ & $17.3_{-0.3}^{+0.3}$ & $19.4_{-0.3}^{+0.3}$ & PIE \\[2pt]
 &  & $5.4_{-2.3}^{+2.0}$ & $<0.30$ & $>-4.00$ & $18.30_{-0.24}^{+0.51}$ & $12.38_{-0.18}^{+0.30}$ & $5.55_{-0.12}^{+0.13}$ & $<0.81$ & $12.60_{-5.13}^{+5.59}$ & $76.72_{-10.07}^{+12.64}$ & $77.91_{-9.49}^{+12.44}$ & TDP--High\\[2pt]
& & $86.7_{-0.9}^{+0.8}$ & $-2.06_{-0.46}^{+0.48}$ & $-3.60_{-0.09}^{+0.09}$ & $19.39_{-0.06}^{+0.06}$ & $14.18_{-0.12}^{+0.12}$ & $5.16_{-0.03}^{+0.03}$ & $1.5_{-0.1}^{+0.1}$ & $26.4_{-1.4}^{+1.4}$ & $48.8_{-1.8}^{+2.0}$ & $55.6_{-1.6}^{+1.7}$ & TDP--Low \\[2pt]
& & $88.1_{-1.0}^{+0.9}$ & $0.02_{-0.17}^{+0.16}$ & $-2.60_{-0.13}^{+0.15}$ & $17.24_{-0.18}^{+0.19}$ & $14.94_{-0.09}^{+0.09}$ & $4.15_{-0.07}^{+0.08}$ & $-1.7_{-0.3}^{+0.3}$ & $10.0_{-2.7}^{+2.6}$ & $15.2_{-1.3}^{+1.4}$ & $18.4_{-1.1}^{+1.2}$ & PIE \\[2pt]
& & $184.2_{-0.2}^{+0.2}$ & $-0.03_{-0.02}^{+0.02}$ & $-0.66_{-0.04}^{+0.03}$ & $17.78_{-0.02}^{+0.03}$ & $17.37_{-0.03}^{+0.02}$ & $3.65_{-0.01}^{+0.02}$ & $-3.1_{-0.0}^{+0.1}$ & $4.4_{-0.2}^{+0.2}$ & $8.6_{-0.1}^{+0.2}$ & $9.7_{-0.1}^{+0.1}$ & PIE \\[2pt]
 &  & $197.8_{-1.2}^{+1.1}$ & $<0.30$ & $>-4.00$ & $19.15_{-0.39}^{+0.53}$ & $13.07_{-0.22}^{+0.31}$ & $5.63_{-0.12}^{+0.13}$ & $<1.66$ & $38.63_{-2.75}^{+2.87}$ & $83.43_{-10.70}^{+13.33}$ & $92.09_{-9.18}^{+11.55}$ & TDP--High\\[2pt]
& & $198.4_{-0.3}^{+0.2}$ & $0.16_{-0.04}^{+0.03}$ & $-2.25_{-0.06}^{+0.06}$ & $18.28_{-0.05}^{+0.05}$ & $16.41_{-0.04}^{+0.04}$ & $3.97_{-0.02}^{+0.02}$ & $-1.0_{-0.1}^{+0.1}$ & $5.4_{-0.3}^{+0.4}$ & $12.4_{-0.2}^{+0.2}$ & $13.5_{-0.2}^{+0.2}$ & PIE \\[2pt]
& & $216.3_{-1.2}^{+1.1}$ & $-0.42_{-0.16}^{+0.13}$ & $-3.04_{-0.05}^{+0.07}$ & $19.11_{-0.13}^{+0.14}$ & $16.04_{-0.10}^{+0.13}$ & $4.45_{-0.02}^{+0.02}$ & $0.7_{-0.1}^{+0.1}$ & $5.7_{-1.8}^{+1.9}$ & $21.7_{-0.5}^{+0.6}$ & $22.5_{-0.4}^{+0.4}$ & PIE \\[2pt]
& & $307.0_{-1.0}^{+1.1}$ & $-0.15_{-0.23}^{+0.23}$ & $-4.62_{-0.06}^{+0.06}$ & $18.50_{-0.04}^{+0.04}$ & $14.49_{-0.03}^{+0.03}$ & $4.56_{-0.03}^{+0.03}$ & $1.6_{-0.1}^{+0.1}$ & $26.0_{-1.4}^{+1.4}$ & $24.5_{-0.7}^{+0.9}$ & $35.7_{-1.0}^{+1.1}$ & TDP--Low \\[2pt]
\hline
J0351 & $0.2617$ & $-91.7_{-10.8}^{+6.5}$ & $<0.24$ & $>-3.29$ & $14.48_{-0.41}^{+1.69}$ & $13.91_{-0.09}^{+0.06}$ & $4.01_{-0.06}^{+0.20}$ & $<0.28$ & $56.17_{-7.46}^{+6.81}$ & $13.02_{-0.81}^{+3.42}$ & $58.51_{-7.23}^{+6.65}$ & PIE\\[2pt]
& & $-45.8_{-5.0}^{+3.5}$ & $-0.21_{-1.23}^{+0.53}$ & $-2.74_{-0.26}^{+0.51}$ & $16.26_{-0.63}^{+0.53}$ & $14.00_{-0.08}^{+0.10}$ & $4.19_{-0.22}^{+0.23}$ & $-2.5_{-1.1}^{+0.8}$ & $31.6_{-5.5}^{+5.2}$ & $15.9_{-3.6}^{+4.9}$ & $35.8_{-4.0}^{+5.0}$ & PIE \\[2pt]
\hline
J0351 & $0.3570$ & $-149.3_{-1.1}^{+1.0}$ & $-0.40_{-0.16}^{+0.15}$ & $-2.58_{-0.10}^{+0.12}$ & $17.33_{-0.15}^{+0.15}$ & $15.03_{-0.02}^{+0.02}$ & $4.24_{-0.06}^{+0.06}$ & $-1.6_{-0.3}^{+0.3}$ & $20.4_{-1.4}^{+1.5}$ & $17.0_{-1.2}^{+1.1}$ & $26.6_{-0.9}^{+0.8}$ & PIE \\[2pt]
& & $-74.6_{-0.2}^{+0.2}$ & $0.07_{-0.04}^{+0.04}$ & $-2.24_{-0.08}^{+0.07}$ & $17.10_{-0.10}^{+0.10}$ & $15.31_{-0.04}^{+0.04}$ & $4.00_{-0.03}^{+0.03}$ & $-2.1_{-0.2}^{+0.2}$ & $1.3_{-0.6}^{+0.7}$ & $12.9_{-0.4}^{+0.5}$ & $13.0_{-0.4}^{+0.5}$ & PIE \\[2pt]
& & $-48.3_{-1.7}^{+1.8}$ & $0.21_{-0.04}^{+0.04}$ & $-2.67_{-0.34}^{+0.32}$ & $19.29_{-0.02}^{+0.01}$ & $14.18_{-0.03}^{+0.03}$ & $5.13_{-0.01}^{+0.01}$ & $0.5_{-0.3}^{+0.3}$ & $56.3_{-2.8}^{+2.7}$ & $47.1_{-0.7}^{+0.6}$ & $73.4_{-2.2}^{+2.1}$ & TDP--Low \\[2pt]
& & $-26.4_{-0.9}^{+0.9}$ & $-0.73_{-0.05}^{+0.05}$ & $-2.76_{-0.03}^{+0.03}$ & $18.68_{-0.05}^{+0.05}$ & $16.06_{-0.03}^{+0.02}$ & $4.36_{-0.02}^{+0.02}$ & $-0.1_{-0.1}^{+0.1}$ & $40.8_{-1.2}^{+1.1}$ & $19.4_{-0.4}^{+0.4}$ & $45.1_{-1.1}^{+1.0}$ & PIE \\[2pt]
& & $-13.2_{-0.5}^{+0.5}$ & $-0.63_{-0.22}^{+0.19}$ & $-0.77_{-0.41}^{+0.38}$ & $16.74_{-0.21}^{+0.31}$ & $16.17_{-0.02}^{+0.03}$ & $3.96_{-0.10}^{+0.06}$ & $-4.0_{-0.6}^{+0.7}$ & $8.8_{-1.6}^{+1.6}$ & $12.2_{-1.3}^{+0.9}$ & $15.0_{-0.9}^{+1.0}$ & PIE \\[2pt]
 &  & $-7.4_{-2.7}^{+2.4}$ & $<0.30$ & $>-4.00$ & $19.33_{-0.37}^{+0.30}$ & $13.08_{-0.23}^{+0.19}$ & $5.73_{-0.09}^{+0.08}$ & $<1.84$ & $13.03_{-6.20}^{+7.27}$ & $93.88_{-8.95}^{+9.15}$ & $95.03_{-8.78}^{+9.26}$ & TDP--High\\[2pt]
& & $39.6_{-0.1}^{+0.1}$ & $0.07_{-0.02}^{+0.02}$ & $-2.89_{-0.02}^{+0.02}$ & $19.21_{-0.02}^{+0.02}$ & $16.56_{-0.01}^{+0.01}$ & $4.24_{-0.00}^{+0.00}$ & $0.6_{-0.0}^{+0.0}$ & $2.9_{-0.1}^{+0.1}$ & $16.9_{-0.1}^{+0.1}$ & $17.2_{-0.1}^{+0.1}$ & PIE \\[2pt]
\hline
J0407 & $0.1534$ & $-536.2_{-1.2}^{+1.1}$ & $<0.63$ & $>-2.97$ & $13.63_{-0.50}^{+0.79}$ & $12.85_{-0.07}^{+0.06}$ & $4.02_{-0.06}^{+0.10}$ & $<-3.38$ & $11.48_{-3.13}^{+2.82}$ & $13.09_{-0.90}^{+1.55}$ & $17.64_{-1.78}^{+1.78}$ & PIE\\[2pt]
 &  & $-503.1_{-4.4}^{+4.5}$ & $<0.60$ & $>-3.12$ & $13.75_{-0.66}^{+0.89}$ & $12.79_{-0.08}^{+0.07}$ & $4.04_{-0.08}^{+0.13}$ & $<-2.99$ & $27.56_{-4.60}^{+4.06}$ & $13.44_{-1.17}^{+2.23}$ & $30.96_{-4.05}^{+3.62}$ & PIE\\[2pt]
 &  & $-314.9_{-0.2}^{+0.2}$ & $<-0.35$ & $>-2.43$ & $14.05_{-0.34}^{+0.58}$ & $13.39_{-0.00}^{+0.01}$ & $4.02_{-0.05}^{+0.08}$ & $<-3.85$ & $17.02_{-1.16}^{+0.86}$ & $13.12_{-0.74}^{+1.23}$ & $21.56_{-0.45}^{+0.45}$ & PIE\\[2pt]
 &  & $-175.7_{-0.9}^{+1.0}$ & $<0.17$ & $>-2.07$ & $12.89_{-0.20}^{+0.41}$ & $12.51_{-0.03}^{+0.03}$ & $3.97_{-0.05}^{+0.05}$ & $<-5.65$ & $4.14_{-2.29}^{+2.27}$ & $12.37_{-0.63}^{+0.69}$ & $13.10_{-0.76}^{+1.05}$ & PIE\\[2pt]
 &  & $-95.9_{-0.3}^{+0.3}$ & $<-0.14$ & $>-2.62$ & $14.47_{-0.42}^{+0.73}$ & $13.76_{-0.00}^{+0.00}$ & $4.02_{-0.06}^{+0.11}$ & $<-3.03$ & $43.91_{-0.67}^{+0.68}$ & $13.15_{-0.85}^{+1.72}$ & $45.95_{-0.51}^{+0.55}$ & PIE\\ 
          \hline

\end{tabular}
   
 Properties of the different gas phases that contribute to the absorption towards different sightlines. Notes: (1) J-Name of the quasar; (2) Redshift of the galaxy; (3) Velocity of the cloud in the rest frame of the galaxy; (4) cloud metallicity; (5) cloud total hydrogen number density; (6) cloud total hydrogen column density; (7) cloud hydrogen column density; (8) cloud temperature; (9) cloud line of sight thickness; (10) cloud non-thermal Doppler broadening parameter (11) cloud thermal Doppler broadening parameter measured for {\hi}; (12) cloud total Doppler broadening parameter measured for {\hi}; (13) type of ionization model for the cloud; PIE - Photoionized Equilibrium, TDP - Time Dependent Photoionized (see Section~\ref{sec:methodology}). The marginalised posterior values of model parameters are given as the median along with the upper and lower bounds corresponding to the 16--84 percentiles. The median and the 1-sigma credible interval are derived from fractional weighted quantiles. The weighting is done by sample probability following \citet{Buchner2016}. The lower and upper limits correspond to $\mu_{1/2} - 2\sigma$ and $\mu_{1/2} + 2\sigma$, respectively. The rest of the table is presented in appendix~\ref{appendix:cloudbycloud}.
 
\end{threeparttable}
 \end{table*}
\end{landscape}
\twocolumn

\section{Results}
\label{sec:results}

We present the metallicity analysis of the ``Multiphase Galaxy Halos'' Survey using CMBM analysis of the 47 absorption systems associated with 47 galaxies with well characterized properties. We resolved the absorption in these systems into 240 clouds. These absorbers were found to be explained by one of three kinds of models PIE (157 absorbers in number), TDP--Low (35 absorbers in number), or TDP--High (48 absorbers in number). We plot the various properties to visualize the range and trends, summarize them using the interquartile range (IQR) to capture the dispersion, and the median and the standard deviation of the median from bootstrapped estimates to capture the central tendency of the data. The IQR is calculated as the difference between the 75th and 25th percentiles of the data. Table~\ref{tab:summarystat} summarizes the summary statistics of the properties of different cloud types.

\smallskip

In addition to the distribution of metallicities, we also consider the relationship between metallicity and other intrinsic absorber properties including {\hi} column density, total hydrogen column density, temperature, total hydrogen number density, line of sight thickness, and Doppler parameter. We then compare cloud metallicities to galaxy properties including the impact parameter scaled by the galaxy virial-radius, galaxy azimuthal angle, galaxy colour, halo mass, and redshift. We then examine the connection between the azimuthal angle and the CGM metallicities.  

\subsection{Metallicity and its relationship with absorber intrinsic properties}
\label{sec:metallicity_intrinsicproperties}

\begin{table}
\begin{center}
\caption{\bf Summary statistics of the properties of different cloud types}
\label{tab:summarystat}

\begin{tabular}{ |c|c|c|c| } \hline

 Property & Cloud type & Median $\pm$ 1$\sigma$ & IQR\\\hline
 {\metallicity} & PIE & $-0.5 \pm 0.1$ & [$-1.1, 0.0$]\\
  & PIE--SLFSs  & $-0.3 \pm 0.2$ & [$-1.0, 0.1$]\\
  & PIE--pLLSs  & $-0.4 \pm 0.1$ & [$-0.9, -0.1$]\\
  & PIE--LLSs  & $-1.1 \pm 0.2$ & [$-1.8, -0.6$]\\
  & TDP--Low  & $-0.6 \pm 0.2$ & [$-1.2, 0.0$]\\
  & TDP--High  & {\nodata} & {\nodata}\\\hline
 {\coldenu} & PIE & $15.6 \pm 0.1$ & [$14.0, 16.6$]\\
  & TDP--Low  & $14.4 \pm 0.2$ & [$13.8, 15.5$]\\
  & TDP--High  & $12.9 \pm 0.1$ & [$12.5, 13.3$]\\\hline
 {\totalcoldenu} & PIE & $17.6 \pm 0.1$ & [$16.4, 18.5$]\\
  & TDP--Low  & $18.5 \pm 0.1$ & [$18.2, 18.8$]\\
  & TDP--High  & $19.2 \pm 0.1$ & [$18.6, 19.8$]\\\hline  
  {\tempu} & PIE & $4.11 \pm 0.03$ & [$4.0, 4.3$]\\
  & TDP--Low  & $4.62 \pm 0.06$ & [$4.4, 5.0$]\\
  & TDP--High  & $5.71 \pm 0.04$ & [$5.6, 5.9$]\\\hline  
  {\btherm} [{\kms}]  & PIE & $14.7 \pm 0.4$ & [$13.0, 18.2$]\\
  & TDP--Low  & $26.3 \pm 2.1$ & [$21.2, 42.5$]\\
  & TDP--High  & $89.8 \pm 5.6$ & [$76.5, 119.7$]\\\hline  
    {\bturb} [{\kms}]  & PIE & $11.4 \pm 1.5$ & [$5.7, 23.8$]\\
  & TDP--Low  & $25.8 \pm 2.3$ & [$16.0, 31.7$]\\
  & TDP--High  & $24.8 \pm 3.7$ & [$13.1, 45.4$]\\\hline
    {\bnet} [{\kms}]  & PIE & $21.3 \pm 1.2$ & [$15.5, 33.5$]\\
  & TDP--Low  & $38.0 \pm 2.1$ & [$33.9, 48.7$]\\
  & TDP--High  & $95.0 \pm 5.7$ & [$80.7, 123.9$]\\\hline
      {\hdenu}   & PIE & $-2.5 \pm 0.1$ & [$-2.8, -1.6$]\\
  & TDP--Low  & $-3.5 \pm 0.4$ & [$-4.4, -2.2$]\\
  & TDP--High  & {\nodata} & {\nodata}\\\hline
       {\thicknessu}   & PIE & $-1.4 \pm 0.2$ & [$-2.9, -0.4$]\\
  & TDP--Low  & $0.7 \pm 0.4$ & [$-1.2, 1.5$]\\
  & TDP--High  & {\nodata} & {\nodata}\\\hline
\end{tabular} \\
\end{center}
\end{table}

\subsubsection{Metallicity distributions of clouds, {\metallicity}}
\label{sec:metallicity_distribution_pie_tdp}

We present the histograms of the metallicities of different cloud types from our sample in Fig.~\ref{fig:metallicity_pie_tdp}. The summary statistics of metallicity distributions are summarized in Table~\ref{tab:summarystat}. The histogram for PIE clouds is shown in panel (a). The median of the metallicity distribution of PIE clouds is estimated to be {\metallicity} = $-0.47 \pm 0.10$.\footnote{We used the Kaplan-Meier estimator, implemented using the \textit{survfit} function in R language to determine the median, accounting for the upper limits on some of the metallicity measurements. To account for random variations in the observations, we obtained 1000 bootstrapped estimates of the median to ascertain the uncertainty on the median.} The metallicity distribution for PIE clouds is skewed left with a low fraction of very low metallicity clouds. The hatched distribution is the PDF of the upper limits and corresponds to the metallicity limits of {\hi}-only clouds. These clouds have just {\hi} detected with no associated metal absorption, and we adopt 2$\sigma$ upper limits for their metallicities. The upper limits on PIE clouds are modelled as uniform distributions because of the inferred broad posterior distributions for these clouds. 
 The range of measured metallicities for PIE clouds is in close agreement with the result of \citetalias{Pointon2019}. However, they obtain a median metallicity of [Si/H] $\approx -1.3$. \citetalias{Pointon2019} infer a metallicity value using only the low and intermediate ionization phases, by measuring total column densities and performing ionization corrections with the help of {\CLOUDY}. With the CMBM approach, we are able to measure how much of the {\hi} is associated with the various phases contributing to the absorption, and obtain separate metallicities for various clouds and phases. This difference in modelling, and potentially to some extent the choice of EBR, could explain the disagreement between \citetalias{Pointon2019} and our result. \citetalias{Pointon2019} adopted HM12~\citep{haardt2012radiative}, while we have used KS19 to model the PIE clouds.

\begin{figure}
\centerline{
\includegraphics[scale=0.55]{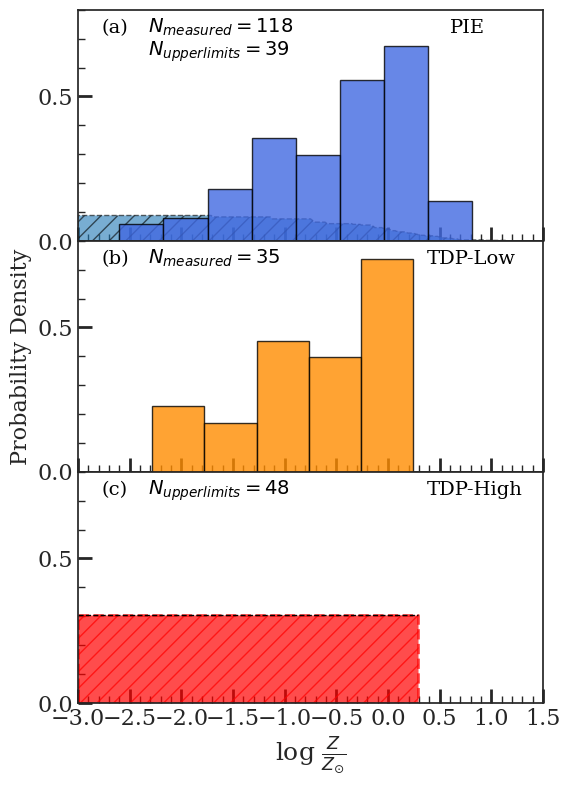}
}
\caption{The histograms of metallicity for various cloud types obtained from the CMBM analysis. The cloud types are (a) Photoionized Equilibrium (PIE), (b) Time-Dependent Photoionized - High (TDP--High), and (c) TDP--Low clouds. The upper limits (2$\sigma$) in PIE clouds are shown as lightly shaded hatched region. The metallicity distribution of TDP--High clouds only comprises upper limits, and is shown as a uniform distribution. The number of clouds with measurements and upper limits are also noted in each panel. The optimum bin width for the displayed histograms is determined using Knuth's rule~\citep{Knuth2006}.}
\label{fig:metallicity_pie_tdp}%
\end{figure}

\smallskip

The metallicity distribution of TDP--Low clouds is shown in panel (b) of Fig.~\ref{fig:metallicity_pie_tdp}, and falls in the range [$-1.1$, $0.0$] with a median metallicity of {\metallicity} = $-0.56 \pm 0.21$. For the TDP--High clouds we display a uniform distribution for their metallicities spanning between [$-3$, 0.3] due to the independency on metallicity for these clouds as indicated by their platykurtic posterior distributions.  

\smallskip

We compare the metallicity distributions in the context of previous studies that investigated the properties of cool CGM gas. \citet{prochaska2017cos} modelled the cool CGM gas of absorbers associated with $z \approx$ 0.2 $L^{\star}$ galaxies and found a median metallicity of [Si/H] = $-0.51$, with a range of $-1.9$ $\lesssim$ [Si/H] $\lesssim$ 1.0, adopting the HM12 ionizing background. \citet{zahedy2019} found a median metallicity of [M/H] = $-0.7 \pm 0.2$, with a 1$\sigma$ confidence interval of [M/H] = ($-1.6$, $-0.1$) for absorbers associated with the CGM of luminous red galaxies (LRGs) at redshifts, $0.21 \lesssim z \lesssim 0.55$. The median metallicity of PIE clouds is consistent with the inferred metallicities from these previous works.

\subsubsection{Neutral and total hydrogen column density distributions of clouds, {\colden} and {\totalcolden}}
\label{sec:metallicity_NHI}

\begin{figure*}
    \centering
    \subfloat{{\includegraphics[scale=0.465]{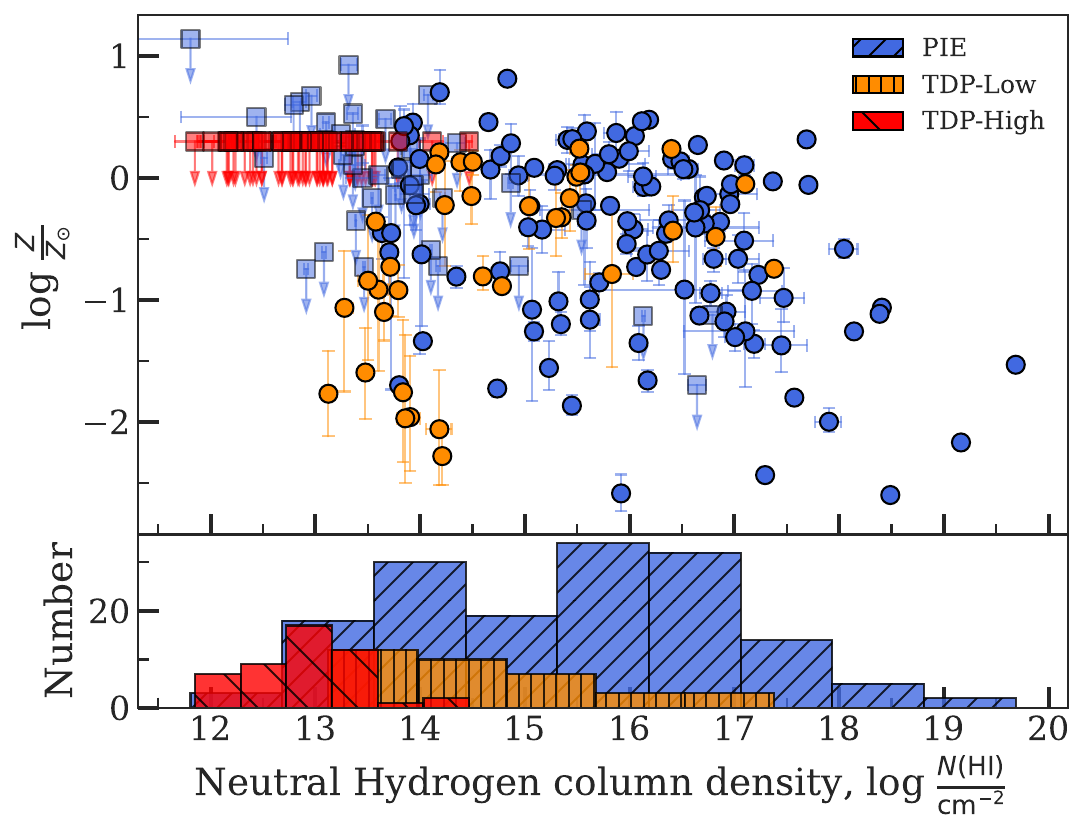} }}%
    \qquad
    \subfloat{{\includegraphics[scale=0.465]{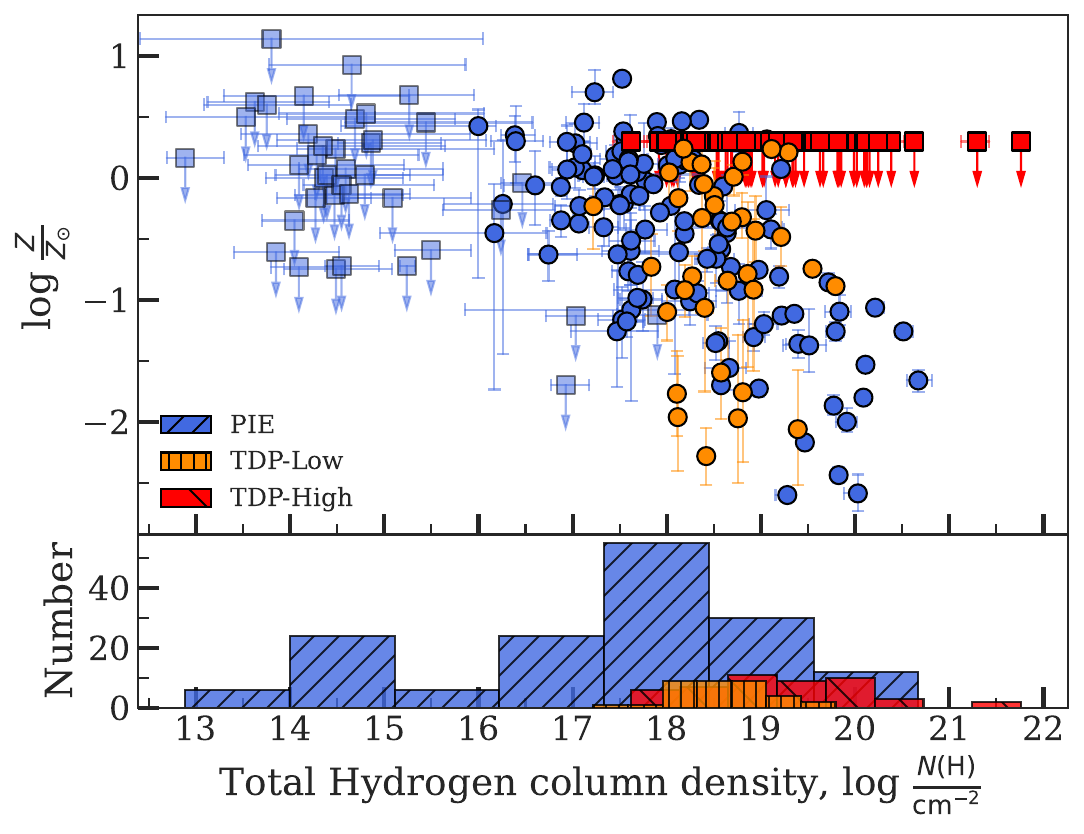} }}%
 
    \caption{(a): The metallicity as a function of neutral hydrogen column density for the three cloud types. The metallicity measurements are shown as filled circles, while the upper limits on metallicity are shown as filled squares with downward pointing arrows. The histogram shows the distribution of neutral hydrogen column densities for different cloud types.  The metallicity of PIE clouds shows an anti-correlation with the neutral hydrogen column density. (b): Same as on the left but for the total hydrogen column density. The metallicity of PIE clouds also shows an anti-correlation with the total hydrogen column density. Error bars show the 1$\sigma$ uncertainty.}%
    \label{fig:metallicity_NHI_D}%
\end{figure*}

The scatter plots in Figure~\ref{fig:metallicity_NHI_D}a show the inferred metallicities plotted against the inferred neutral hydrogen column densities for PIE, TDP--Low, and TDP--High clouds. The {\hi} column density of our sample spans several orders of magnitude:
$11.8 \lesssim$ {\colden} $\lesssim 19.7$. The distributions of neutral hydrogen column densities are presented in the bottom panel. The summary statistics of {\hi} column density distributions for the three cloud types are summarized in Table~\ref{tab:summarystat}.

\smallskip

The metallicity of PIE clouds appears to indicate an anti-correlation with the {\hi} column density. We test for the null hypothesis of no correlation between the {\hi} column density and metallicity by computing the Akritas–Theil–Sen (ATS) Kendall correlation coefficient~\citep{Akritas95} and the associated null-hypothesis probability\footnote{This test accounts for the upper limits present on some of the metallicity measurements.}. We obtain a $p$-value of $\approx$ 0.18, suggesting that the null hypothesis cannot be rejected. However, if we just consider those sightlines which are impacting within the virial radius of the galaxy, we find evidence for significant anticorrelation between metallicity and neutral hydrogen column density ($p$-value = 0.03). \citetalias{Pointon2019} using this same sample of absorbers identified a significant anti-correlation between metallicity and {\hi} column density.  However, they found that the correlation vanishes when the HM05~\citep{Haardt2001} EBR is used in ionization modelling instead of HM12~\citep{haardt2012radiative}. \citet{zahedy2019} investigated the [M/H] -- N(\rm {\hi}) relationship for luminous red galaxies and found no correlation when the  
HM05 background was used in ionization modelling. \citet{prochaska2017cos} also found significant evidence for an
anti-correlation between metallicity and {\hi} column density in their analysis of 32 absorption systems from the COS-Halos survey comprising of $L^{\star}$ galaxies using the HM12 EBR. An anticorrelation is also seen for a sample of weak-{\mgii} analogs at low-$z$~\citep{muzahid2018cos}; they adopted the KS15 EBR~\citep{Khaire2015}. The existence of an anticorrelation could be attributed to similarity in the EBR spectrum of KS19 and HM12, both of which have harder spectrum compared with HM05.

\smallskip

 \citet{Lehner2019} characterized the metallicities of cool, photoionized gas in the {\hi}-selected Cosmic Origins Spectrograph (COS) circumgalactic medium compendium (CCC) survey~\citep{lehner2018cos,Wotta2019,Lehner2019}, consisting of {\hi} absorbers with column densities ranging between 15 $\leq$ {\colden} $\leq$ 19 at $z \lesssim$ 1. Within our PIE sample, we separately tabulate the summary statistics for the metallicities of SLFSs (15 $<$ {\colden} $<$ 16.2), pLLSs (16.2 $\leq$ {\colden} $<$ 17.2), and LLSs (17.2 $\leq$ {\colden} $\leq$ 19) in Table~\ref{tab:summarystat}, as these groupings of {\hi} absorbers overlap with the {\hi}-selected absorbers from the CCC survey. We find that the observed range of our metallicity values are consistent with the range of values estimated in the CCC survey. However, our estimated mean values for SLFSs and pLLSs are higher compared to the corresponding values for the CCC survey. The CCC survey obtains mean values of {\metallicity} = $-1.2$ and $-1.3$ for SLFSs and pLLSs, respectively. The metallicities of our LLSs are lower compared with the LLSs in CCC; they obtain a value of {\metallicity} of $-0.6$. We attribute these differences to the sample selection, our sample is galaxy-selected while the CCC survey is {\hi}-selected. Systematic differences can also stem from the choice of the radiation field. The CCC survey adopted the HM05 radiation field. 

\smallskip

The observed relationship between {\metallicity} and neutral hydrogen column density can be directly compared with simulations. \citet{Hafen2017} find that the metallicity distribution of {\hi} absorbers with {\colden} $\leq$ 15, the {\lya} forest, extends to below {\metallicity} = $-2$, with many of these absorbers likely tracing IGM gas yet to be enriched by galaxies. We find that a significant percentage ($57 \pm 6$\%)
of our absorbers with {\colden} $\leq$ 15 are found beyond the virial radius, with no associated metals detected, consistent with results of \citet{Hafen2017}. They also find that simulated absorbers with 16.2 $\leq$ {\colden} $\leq$ 19 have a metallicity within 1 dex of the mean metallicity of {\metallicity} = $-0.9$, also consistent with our estimates for this sample.

\smallskip
  
We also explore the relationship between the metallicity and {\hi} column density of the TDP--Low clouds and find a significant positive correlation ($p$-value = 0.007). While we have modelled these clouds assuming time-dependent photoionization, many of these absorbers are found to be overlapping in metallicity with the PIE clouds. We do not interpret this positive correlation as it is quite possible that some of these clouds could be in PIE, depending on their density, complicating the interpretation of the correlation. We do not investigate the trend between the metallicity and neutral hydrogen column density of TDP--High clouds, as their metallicities are unconstrained.

\smallskip

In Fig.~\ref{fig:metallicity_NHI_D}b, the metallicity is plotted as a function of the total hydrogen column density for different absorber types. The total hydrogen column density of our sample spans several orders of magnitude: $12.9 \lesssim$ {\totalcolden} $\lesssim 22.0$. The distribution of total hydrogen column densities for PIE clouds seems to show a bimodality, with a smaller peak at {\totalcolden} $\approx$ 14.6, and a more prominent peak at $\approx$ 18.0. However, Hartigan's Dip test suggests that the null hypothesis of unimodality cannot be rejected ($p-$value = 0.09). We determine that the {\totalcolden} of {\hi}-only PIE clouds ranges from 12.9 to 17.9, with an IQR of [14.2, 15.0].

\smallskip

The {\totalcolden} distribution of TDP--Low clouds is quite flat and spans a narrow range of values. The predicted range and median for {\totalcolden} of TDP--High clouds are consistent with the properties of low redshift, $z < 0.5$, warm {\ovi} absorbers (5.4 $\lesssim$ {\temp} $\lesssim$ 6.2) from \citet{savage2014properties} who obtained a range of 18.4 $<$ {\totalcolden} $<$ 20.4, and median of {\totalcolden} = 19.35.

\subsubsection{Temperature, {\temp}}
\label{sec:metallicity_temp}

\begin{figure*}
    \centering
    \subfloat{{\includegraphics[scale=0.46]{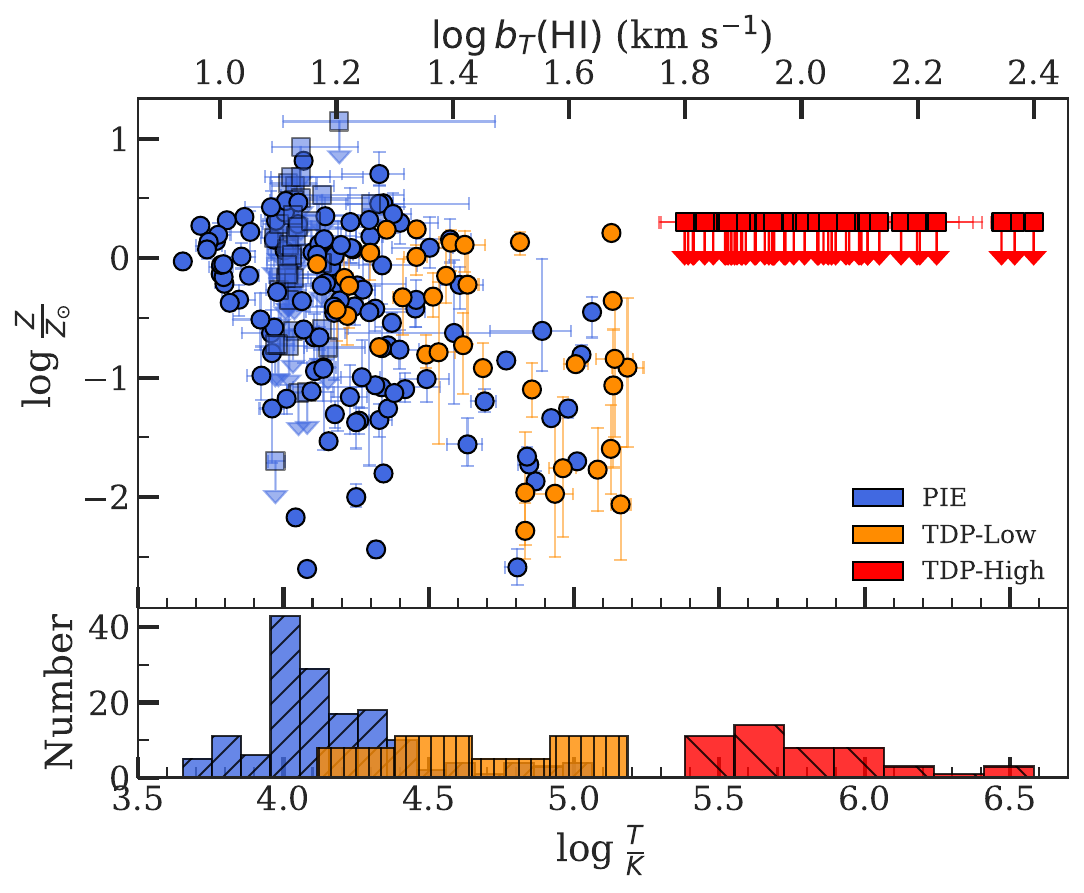} }}%
    \qquad
    \subfloat{{\includegraphics[scale=0.46]{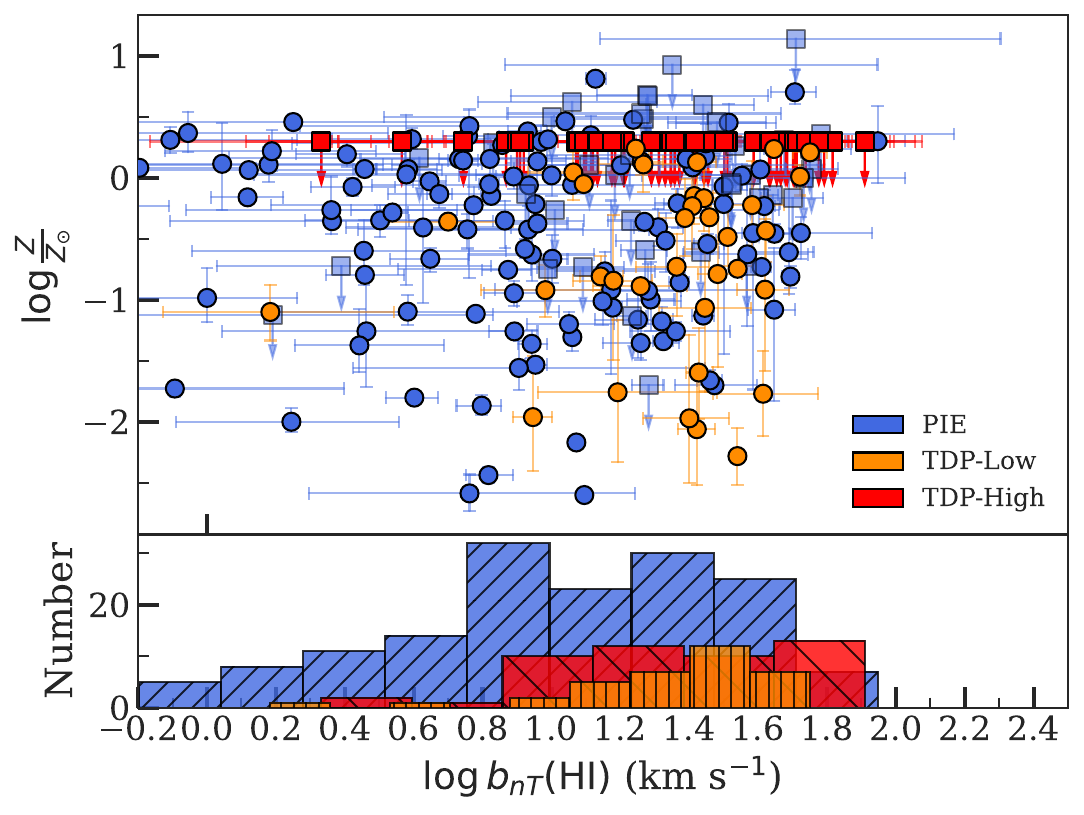} }}%
   \caption{(a): The metallicity as a function of temperature for the three cloud types. The metallicity measurements are shown as filled circles, while the upper limits on metallicity are shown as filled squares with downward pointing arrows. The histogram shows the distribution of cloud temperatures, which is reminiscent of the multiphase gas observed in high-resolution cosmological simulations. Clouds in photoionization equilibrium show lower temperatures compared to the clouds that are affected by time-dependent photoionization. The top-axis shows the thermal Doppler broadening parameter in log values. (b): The metallicity as a function of non-thermal Doppler broadening parameter for the three cloud types. The histogram shows the distribution of the non-thermal Doppler broadening parameter for these cloud types in log values. Error bars reflect the 1$\sigma$ uncertainty.}%
    \label{fig:metallicity_temp}%
\end{figure*}

In Figure~\ref{fig:metallicity_temp}a, we show metallicity as a function of temperature, {\temp}, plotted on the bottom x-axis. The summary statistics of the temperature distribution for the different cloud types is given in Table~\ref{tab:summarystat}. We find that the temperature ranges for these three cloud types agree with the multiphase CGM manifested in hydrodynamic simulations~\citep{Hummels2019,Nelson2020}, which exhibit a mixture of cool, warm-hot, and hot phases. The temperature distribution of PIE clouds appears to be right-skewed with a majority of the absorbers concentrated around the median of $T \sim$ 10$^{4}$ K, and a tail extending to $T \sim$ 10$^{5}$ K. The temperature distribution of TDP--Low and TDP--High clouds appears to be flat between $T \sim $10$^{4.2-5.2}$ K) and ($T \sim $10$^{5.4-6.0}$ K), respectively. The temperatures of some of the TDP--High clouds are consistent with Coronal Broad Lyman-$\alpha$ (CBLA) absorbers--the collisionally ionized, million-degree gas hypothesized to be pervading the extended, hot gaseous halos of low-redshift galaxies~(e.g., \citealt{Narayanan2010,Savage2010,Savage2011a,Savage2011b,richter2020}). 

\smallskip

Theoretical models of \citet{Faerman2023} infer the presence of a cold component at $\sim$\,10$^{4}$ K in photo-ionization and thermal equilibrium. The cold phase with a small volume filling fraction, and non-thermal pressure fraction higher than the hot $\sim$\,10$^{5.5}$ K phase, was found to predict roughly similar column densities of the ions tracing the cold phase. However, their model lacks a TDP-Low phase at $\sim$ 10$^{5}$ K, which we see in many systems. \citet{ADutta2023} using log-normal distributions to quantify the volume fraction of different CGM phases, and a more generalized prescription of \citet{Faerman2017,Faerman2020} models, find a good match with multi-wavelength observations, inferring the presence of gas akin to the cool, warm-hot, and hot phases seen in Figure~\ref{fig:metallicity_temp}a.

\smallskip

We perform an ATS test between the {\metallicity} and {\temp} measurements of the PIE clouds and TDP--Low clouds. For PIE clouds, we obtain a $p$-value of 0.01, indicating evidence for anti-correlation between these parameters. We also see evidence for anti-correlation between the metallicity and temperature of the TDP--Low clouds ($p$-value = 0.005). Higher metallicity allows the gas to cool efficiently, leading to a lower temperature--causing the anticorrelation. 

\subsubsection{Doppler parameters}
\label{sec:metallicity_Doppler}

The Doppler parameter, $b$, quantifies the line broadening mechanism, which is influenced by thermal and/or turbulent processes. The thermal motions of the particles result in Doppler broadening of the observed profiles. The thermal Doppler broadening and temperature are related by $b_{\rm T} = \sqrt{2kT/m}$ {\kms}. In Fig.~\ref{fig:metallicity_temp}a, we show the metallicity as a function of the thermal Doppler broadening parameter, {\btherm}, plotted on the top-axis in log values.

\begin{figure*}
    \centering
    \subfloat{{\includegraphics[scale=0.46]{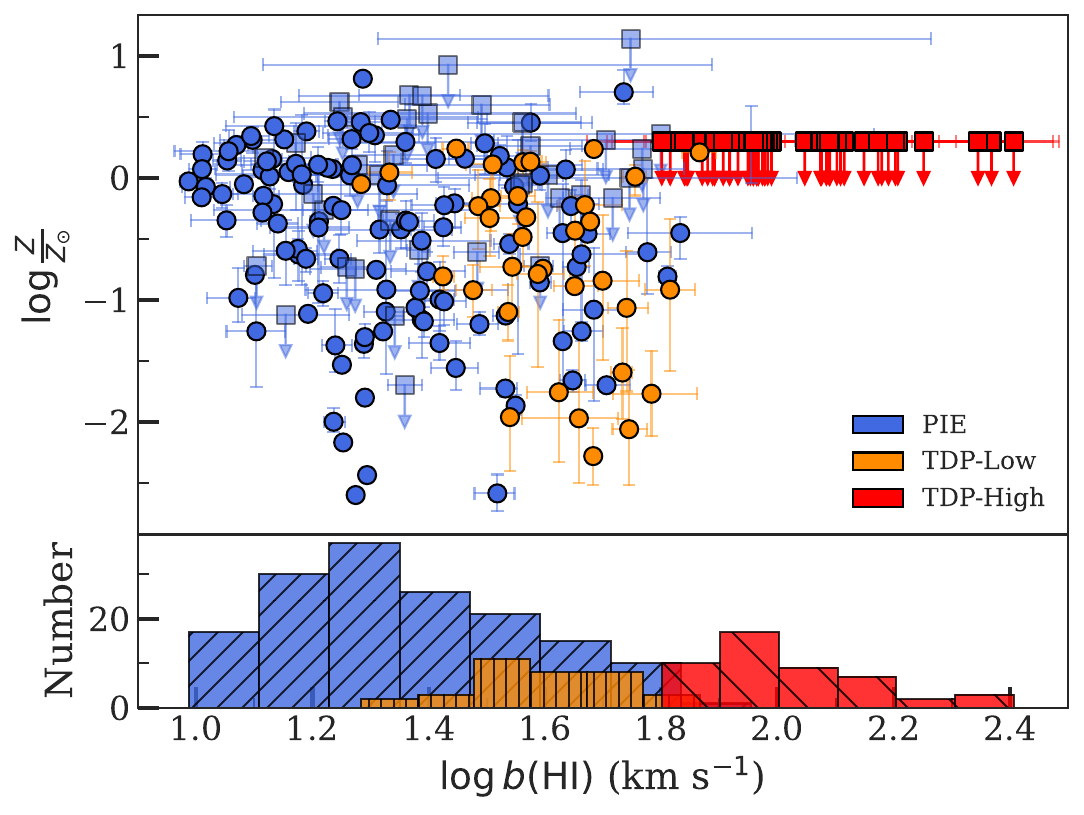} }}%
    \qquad
    \subfloat{{\includegraphics[scale=0.46]{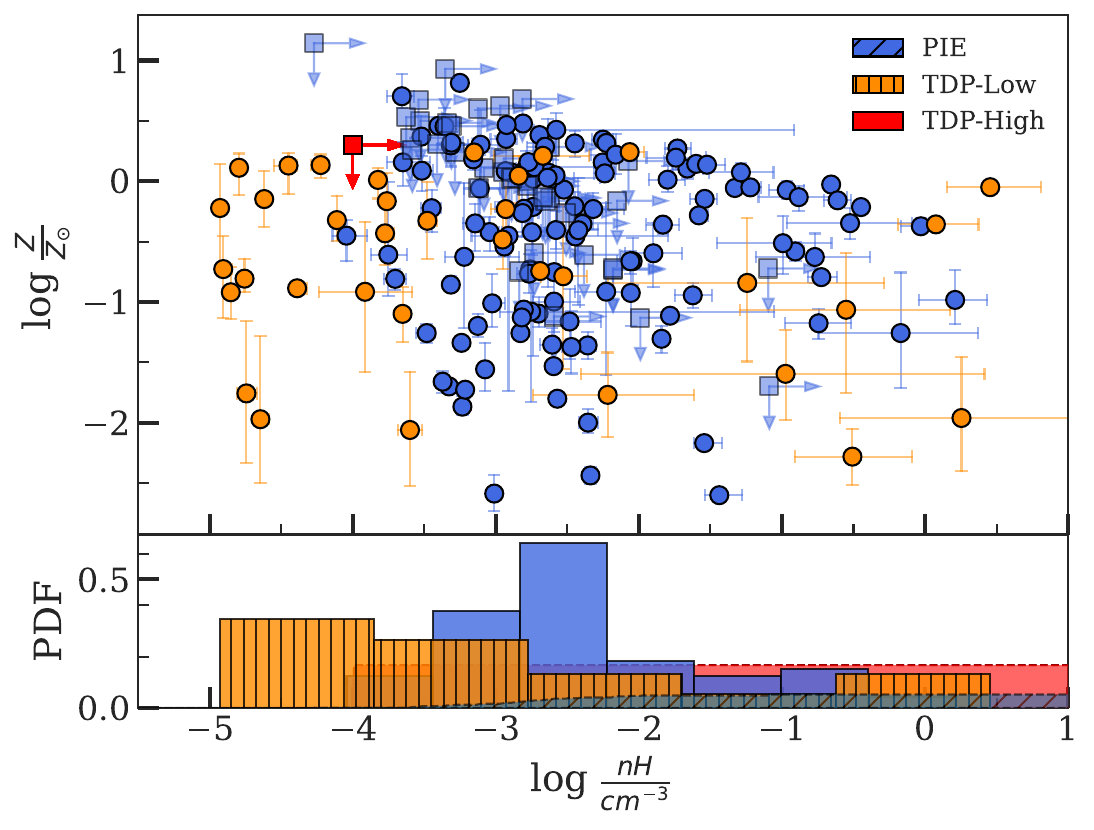} }}%
 \caption{(a): The metallicity as a function of total Doppler broadening parameter for the three cloud types. The metallicity measurements are shown as filled circles, while the upper limits on metallicity are shown as filled squares with downward pointing arrows. The histogram shows the distribution of the total Doppler broadening parameter for these cloud types in log values. (b): The metallicity as a function of total hydrogen number density for the three cloud types. The histogram shows the distribution of total hydrogen number densities. The lower limits in hydrogen number density for PIE clouds are shown as right-slanted hatching. The hydrogen density for TDP--High clouds is unconstrained and shown as a uniform distribution in red. Error bars reflect the 1$\sigma$ uncertainty.}%
    \label{fig:metallicity_doppler}%
\end{figure*}

\smallskip

 In Fig.~\ref{fig:metallicity_temp}b, we show the metallicity as a function of the non-thermal Doppler broadening parameter, {\bturb}, plotted in log values. The non-thermal Doppler broadening parameter in absorbing structures is potentially mediated by post-shock turbulence from outflows and/or accretion (e.g., \citealt{Cen2006,Draganova2015}). Magnetohydrodynamic turbulence produced by collisionless shocks could result in a significant non-thermal contribution to the observed line profiles, as observed in TNG100 simulations~\citep{Nelson2018}. The non-thermal Doppler broadening distribution of PIE clouds is relatively flat with a lack of pronounced mode, however, there are some clouds with smaller values, suggesting that some of these clouds are predominantly thermally broadened. The non-thermal Doppler broadening distribution of TDP clouds is also flat, but the TDP clouds always show some amount of contribution from non-thermal broadening to their line profiles.

\smallskip

In Figure~\ref{fig:metallicity_doppler}a, we show the metallicity as a function of the total Doppler broadening parameter, {\bnet}, and the {\bnet} distribution for different cloud types. The distribution of {\bnet} for PIE clouds appears broad and ranges between 10--90 {\kms} with a median of 21.3 $\pm$ 1.2. The COS CGM Compendium (CCC) survey~\citep{lehner2018cos} comprising of {\hi}-selected absorbers found a median Doppler broadening parameter of {\bnet} $\approx$ 28 {\kms} for gas consistent with being primarily photoionized. The distribution of {\bnet} for TDP--Low clouds overlaps with that of the right tail of PIE clouds. \citet{richter2020} who systematically studied the extended, hot gaseous halos of low-redshift with CBLAs, inferred Doppler parameters ranging from 70 -- 200 {\kms}. The total Doppler broadening parameter, {\bnet}, for TDP--High clouds in our sample is consistent with the range determined for CBLAs.

\subsubsection{Total hydrogen density, {\hden}}
\label{sec:metallicity_hden}

In our model clouds the total hydrogen density, {\hden}, allows the ionization parameter to be well-defined for the adopted EBR spectrum at the redshift of the absorber.  
In Fig.~\ref{fig:metallicity_doppler}b, we show the metallicity as a function of total hydrogen number density and the histogram of {\hden}. The summary statistics of the hydrogen density distribution for the different cloud types is given in Table~\ref{tab:summarystat}. The hydrogen density distribution of PIE clouds is right-skewed, with a pronounced peak at the median value of {\hden} $\approx -2.5$. The TDP--Low clouds show a range of hydrogen densities, but there is a higher frequency of these clouds with low densities. For the TDP--High clouds, the posterior distribution of {\hden} of the modelled clouds is unconstrained. For gas at {\temp} $\gtrsim$ 5.4, the ionization of the gas is most likely dominated by collisional ionization since heating by photoionizing radiation is unlikely to produce such high temperatures. For such gas, the radiative cooling efficiencies are functions of the temperature only and independent of density, thus resulting in large model uncertainties, provided the density is high enough ($\gtrsim$ 10$^{-4}$ \cc) for particle interactions to become more important than photoionization~\citep{Gnat2017}. Thus, for the TDP--High clouds we adopt a lower limit on their {\hden} and their {\hden} distribution is presented as a uniform distribution in [$-4.0, 2.0$].

\subsubsection{Line of sight thickness, {\thickness}}
\label{sec:metallicity_thickness}

\begin{figure}
\centerline{
\includegraphics[scale=0.5]{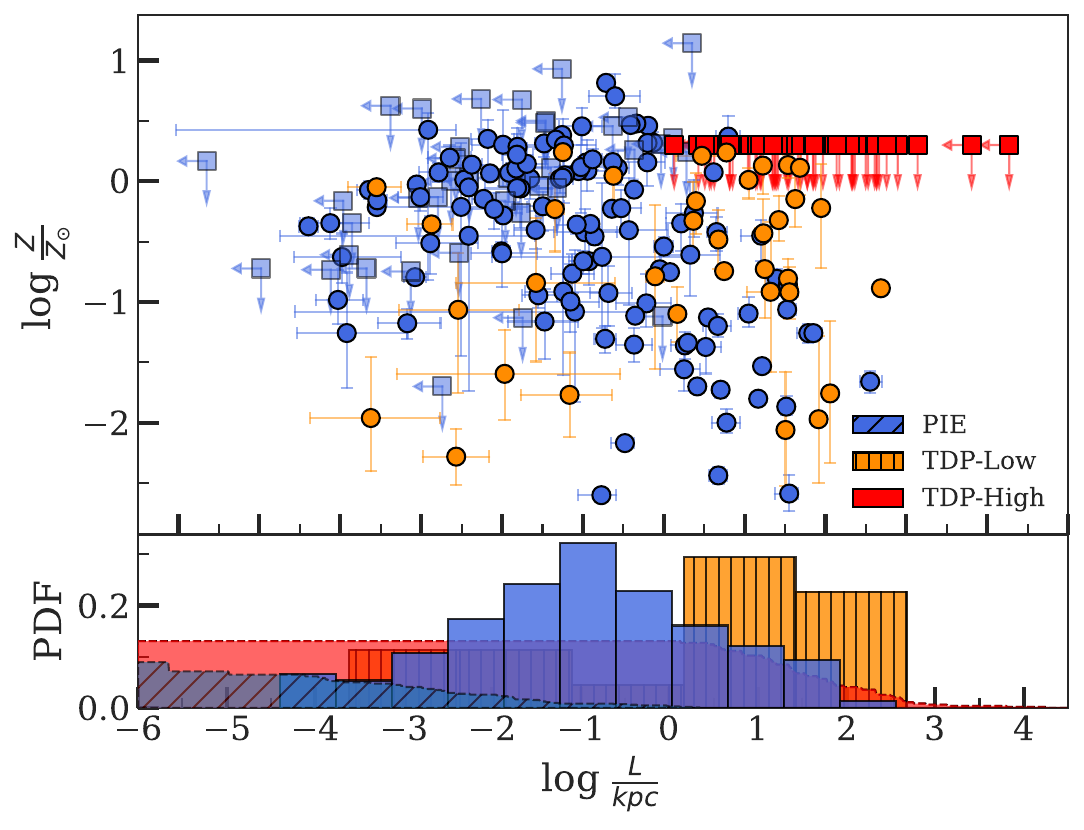}
}
\caption{The metallicity as a function of the inferred line of sight thickness. The metallicity measurements are shown as filled circles, while the upper limits on metallicity are shown as filled squares with downward pointing arrows. The histogram shows the distribution of the line of sight thickness. Error bars reflect the 1$\sigma$ uncertainty.}
\label{fig:metallicity_size}%
\end{figure}

 Our models assume that the cloud geometry is plane-parallel, and the ionization structure of the model cloud is determined using the boundary condition set by the neutral hydrogen column density. Using the predicted total hydrogen column density corresponding to the boundary condition, the thickness can be determined using the relation $L = \frac{N_{\rm H}}{n_{\rm H}}$, where N$_{\rm H}$ and n$_{\rm H}$ are the total hydrogen column density and total hydrogen number density, respectively. The summary statistics of the size distribution for the different cloud types is given in Table~\ref{tab:summarystat}.
In Fig.~\ref{fig:metallicity_size}, we show the metallicity as a function of cloud size for the three cloud types. The {\thickness} distribution of PIE clouds appears to be platykurtic, with a median value of {\thicknessu} = $-1.4 \pm 0.2$. The TDP--Low clouds show a range of sizes, but there is a higher frequency of these clouds with large sizes. For the TDP--High clouds, because of the dependence of {\thickness} on {\hden}, we obtain upper limits on the inferred {\thickness} values. We determine upper limits based on the lower limit of {\hden} $\geq -4$ and the model predicted {\totalcolden}.

\subsection{Metallicity and its relationship with galaxy properties}
\label{sec:metallicity_galaxyproperties}

We next examine the trends of metallicity with a set of galaxy properties. 

\subsubsection{M$_{\rm halo}$}
\label{sec:metallicity_mhalo}
\begin{figure*}
    \centering
    \subfloat{{\includegraphics[scale=0.46]{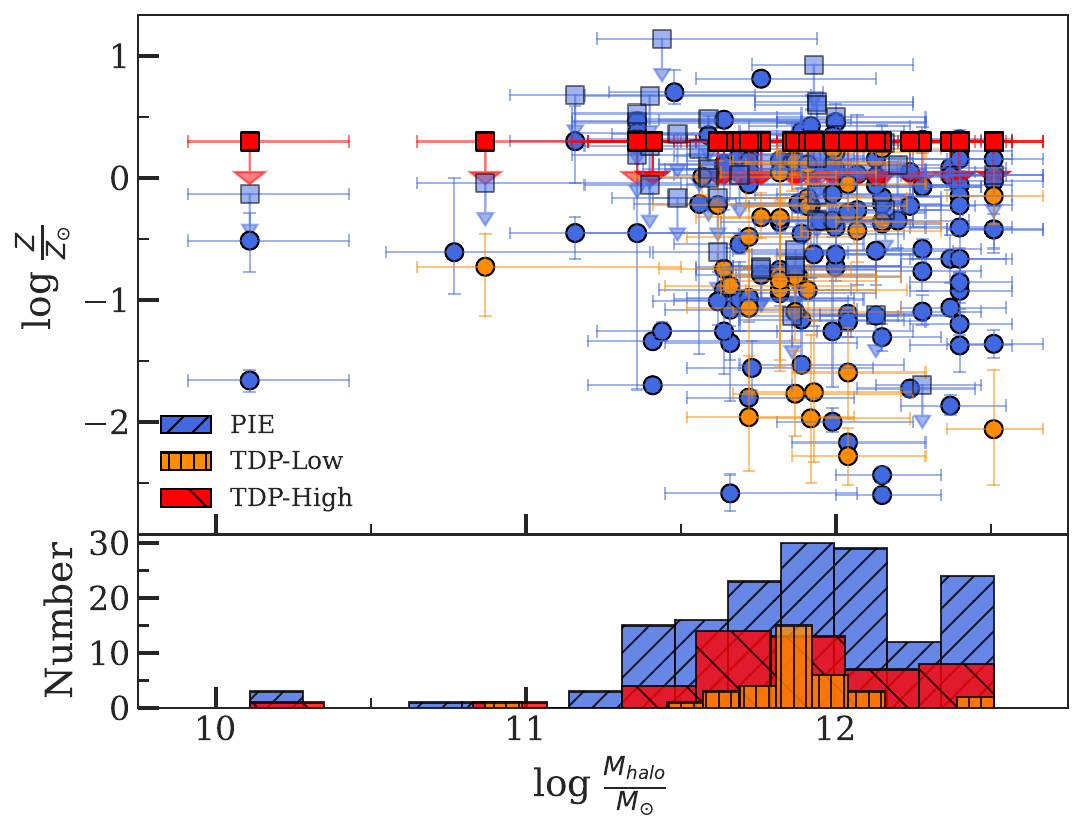} }}%
    \qquad
    \subfloat{{\includegraphics[scale=0.46]{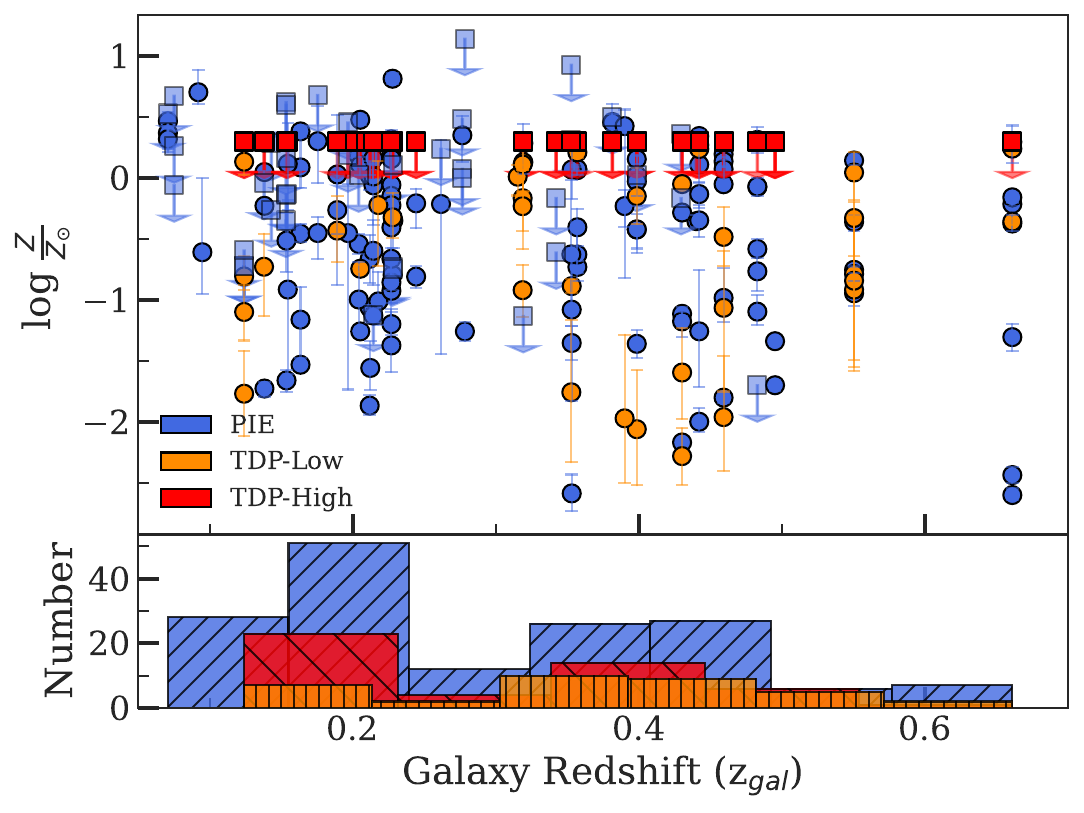} }}%
 \caption{(a): The metallicity as a function of galaxy halo mass for the three cloud types. The metallicity measurements are shown as filled circles, while the upper limits on metallicity are shown as filled squares with downward pointing arrows. The histogram shows the distribution of galaxy halo masses. (b): The metallicity as a function of galaxy redshift. The histogram shows the distribution of galaxy redshifts. No trends are evident for metallicity as a function of halo mass or galaxy redshift for any of the cloud types. Error bars reflect the 1$\sigma$ uncertainty.}%
    \label{fig:metallicity_galaxymass}%
\end{figure*}

 In Figure~\ref{fig:metallicity_galaxymass}a, we plot the metallicity, {\metallicity}, as a function of the galaxy halo mass, M$_{\rm halo}$. All the types of clouds are seen for all halo masses. We examine the relationship between metallicity and halo mass for any underlying trends. We do not find evidence to reject the null hypothesis of no correlation between metallicity and halo mass for PIE, TDP--Low, and TDP--High clouds ($p$-values $\approx$ 0.75, 0.92, and 0.82, respectively). We also do not find evidence for significant differences in the halo mass distributions of the three groupings.

 \smallskip

 \subsubsection{Galaxy Redshift}
\label{sec:metallicity_zgal}

Galaxies and quasars contribute most of the photoionizing radiation at all redshifts and their contribution is known to evolve with redshift~(e.g., \citealt{Gibson2022}). The galaxies in our sample span the redshift range of 0.07 $< z < $ 0.66. We investigate any trends due to the evolution of EBR on how various absorber types would be manifest in our sample. In Figure~\ref{fig:metallicity_galaxymass}b, we plot the metallicity as a function of galaxy redshift, z$_{\rm gal}$. We do not see an indication of trend between the metallicity and galaxy redshift for PIE and TDP--Low clouds. These cloud types are found at all redshifts. We do not expect and see the effect of EBR on the properties of TDP--High clouds as these are dominated by collisional ionization.


 \subsubsection{Impact Parameter}
\label{sec:metallicity_impact}

\begin{figure*}
    \centering
    \subfloat{{\includegraphics[scale=0.46]{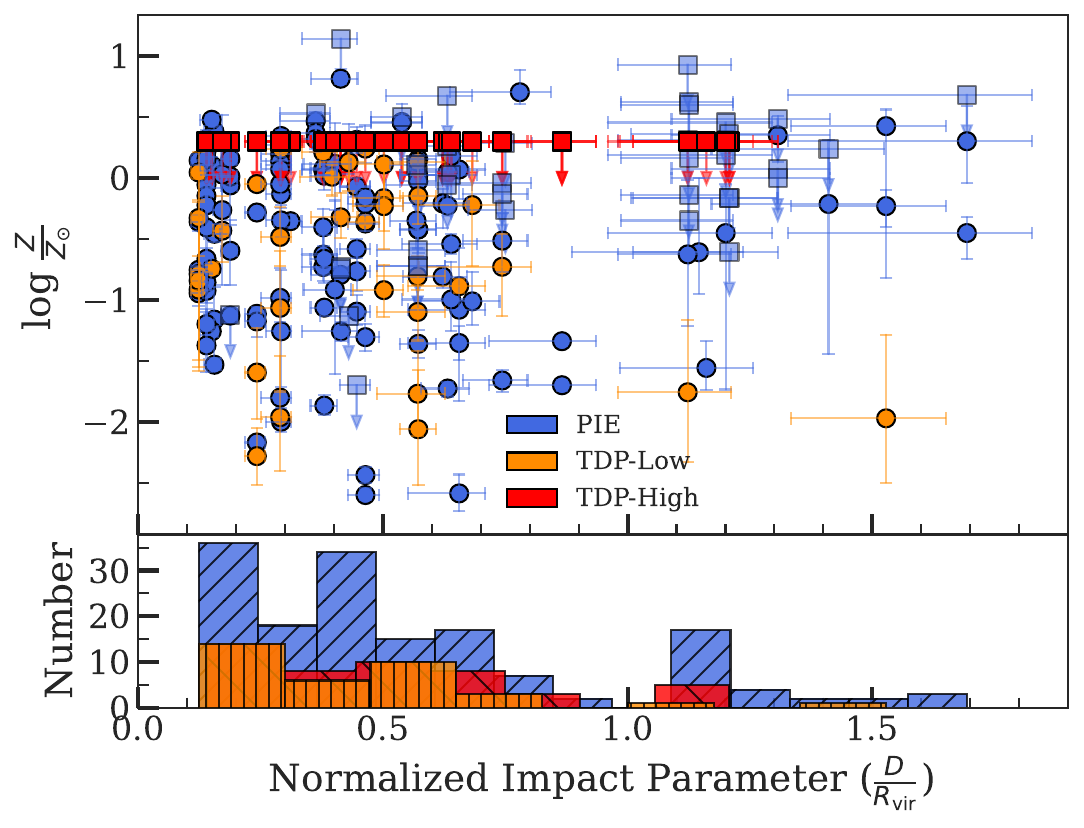} }}%
    \qquad
    \subfloat{{\includegraphics[scale=0.46]{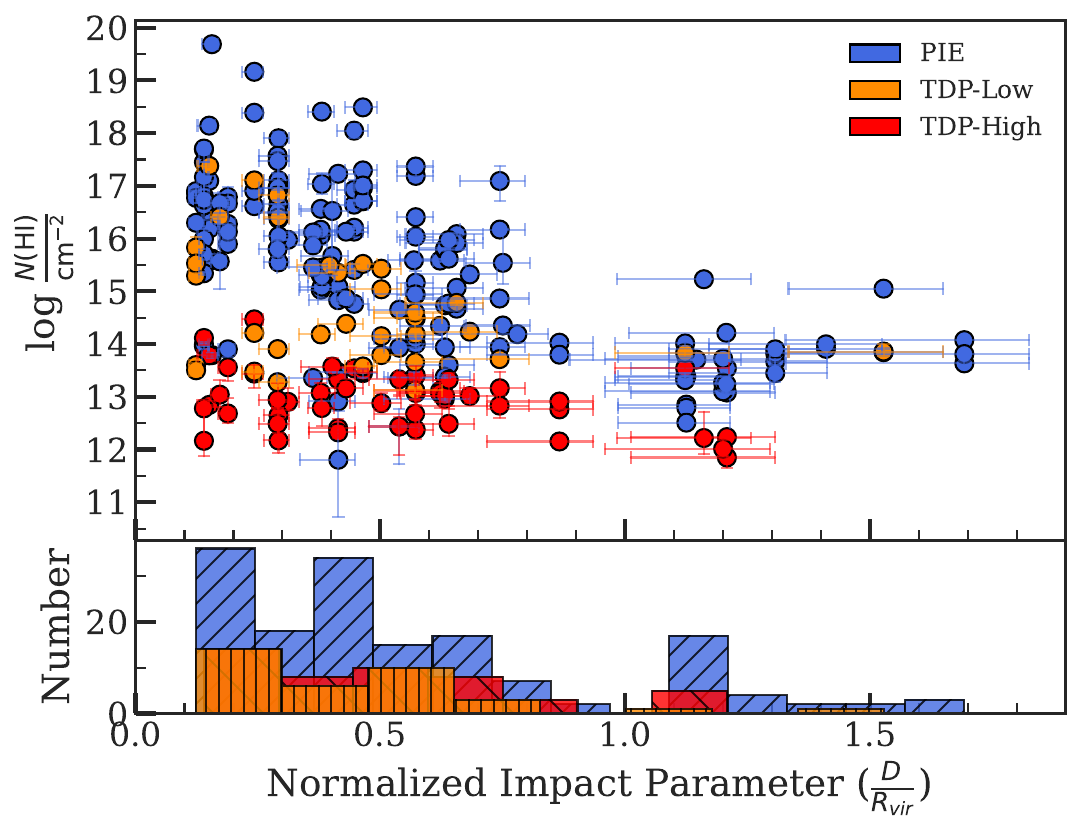} }}%
 \caption{(a): The metallicity as a function of the impact parameter scaled by the virial radius of galaxy for the three cloud types. The metallicity measurements are shown as filled circles, while the upper limits on metallicity are shown as filled squares with downward pointing arrows. The histogram shows the distribution of normalized impact parameters. (b): The neutral hydrogen column density as a function of normalized impact parameter. There is significant evidence for an anticorrelation between these parameters for PIE and TDP--Low clouds, but not for the TDP--High clouds. The histogram shows the distribution of impact parameters. Error bars reflect the 1$\sigma$ uncertainty.}%
    \label{fig:metallicity_impact}%
\end{figure*}

In Figure~\ref{fig:metallicity_impact}a, we plot the metallicity, {\metallicity}, as a function of the normalized impact parameter,  D/{\rvir}. The PIE clouds and TDP--clouds are found up to 1.7 {\rvir}. We determine that within {\rvir}, $84 \pm 3$\%  of the clouds show detected metals at the 2$\sigma$ level, but beyond {\rvir} only $36_{-8}^{+9}$\% of the clouds show detected metals at the 2$\sigma$ level. This is indicative of the fact that most of the metals are retained within the {\rvir} of galaxies. We investigate the existence of any trends between these two parameters and find that none of the groupings show evidence for a correlation. 

\smallskip

In Figure~\ref{fig:metallicity_impact}b, we investigate the relationship between neutral hydrogen column density and normalized impact parameter. Previous studies (e.g., \citealt{lanzetta1995,tripp1998,chen2001origin,wakker2009,rao2011,werk2014,Borthakur2015,Curran2016,prochaska2017cos,Kulkarni2022}) found that the {\hi} column density declines with increasing impact parameter for CGM absorbers. We also observe a similar trend, consistent with these previous works. The dispersion in neutral hydrogen column densities for PIE and TDP--Low clouds decreases with the normalized impact parameter, while the TDP--High clouds show a similar level of dispersion. We test for correlation for each of the three groupings using the ATS test. For the PIE clouds, mainly tracing the cool CGM gas, we see significant anti-correlation ($p$-value $\approx$ 10$^{-5}$) between {\colden} and D/{\rvir}. TDP--Low clouds also show significant anti-correlation ($p$-value $\approx$ 0.03), while the TDP--High clouds do not show evidence for anti-correlation ($p$-value $\approx$ 0.73).

\begin{figure}
\centerline{
\includegraphics[scale=0.7]{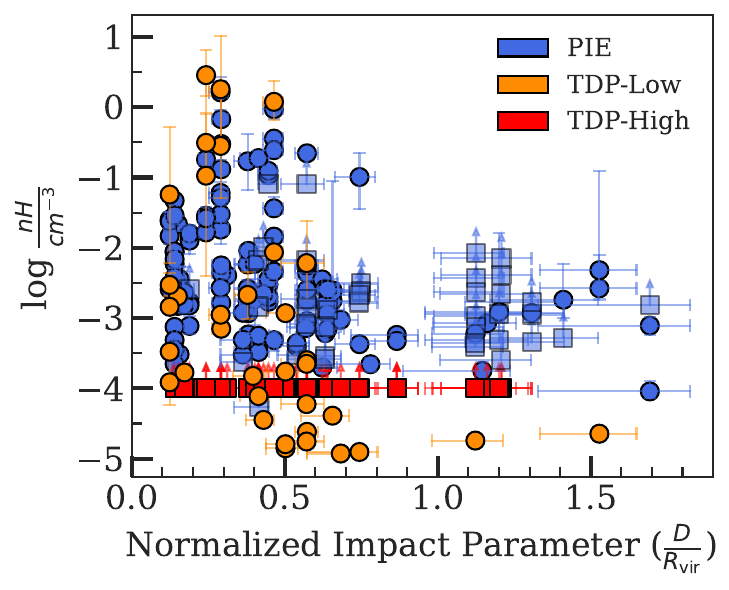}
}
\caption{The hydrogen number density, {\hden}, as a function of normalized impact parameter for the three cloud types. The {\hden} measurements are shown as filled circles, while the lower limits on density are shown as filled squares with upward pointing arrows.}
\label{fig:density_impact_par}%
\end{figure}

\smallskip

\citet{werk2014} found an indication of declining hydrogen density as a function of impact parameter scaled to the galaxy virial radius for cool, photoionized CGM gas using the COS-Halos~\citep{Tumlinson2011} gas column density measurements. In Fig.~\ref{fig:density_impact_par}, we plot the hydrogen number density as a function of the normalized impact parameter. For PIE clouds, we observe that within {\rvir}, a large range of hydrogen densities are seen spanning between {\hdenu} = $-3.7$ to 0.2. However, for impact parameters $>$ {\rvir} we find a narrower range of densities spanning between {\hdenu} = $-4.0$ to $-2.3$. \citet{Qu2023} also observe a similar trend with cool gas density showing a decline from the inner CGM to the outskirts in a sample of 19 unique galaxies and galaxy groups at redshifts $z = 0.89-1.21$ in six QSO fields from the Cosmic Ultraviolet Baryon Survey (CUBS).

\smallskip

We perform an ATS test to investigate the null hypothesis of no correlation between the hydrogen density and normalized impact parameter. We find significant evidence of a negative correlation ($p$-value = 0.01) between these parameters for PIE clouds, excluding the lower limits on hydrogen density. We also find significant evidence for a negative correlation between these parameters for TDP--Low clouds ($p$-value = 0.003).

\subsubsection{Absorber velocities}
\label{sec:absorber_velocities}
Cosmological hydrodynamical simulations (e.g., TNG50; \citealt{peroux2020}) predict that the highest metallicities in the CGM are associated with high-velocity outflowing gas expelled from the ISM. On the other hand, gas with lower metallicities is 
associated with lower velocity accreting or inflowing processes. Simulations (e.g., \citealt{Ford2014}) also predict that cool photoionized gas traces recycled accretion, warm-hot gas traces both recycled accretion and ancient outflows depending on impact parameter, while hotter gas probes ancient outflows. 

\begin{figure*}
    \centering
    \subfloat{{\includegraphics[scale=0.46]{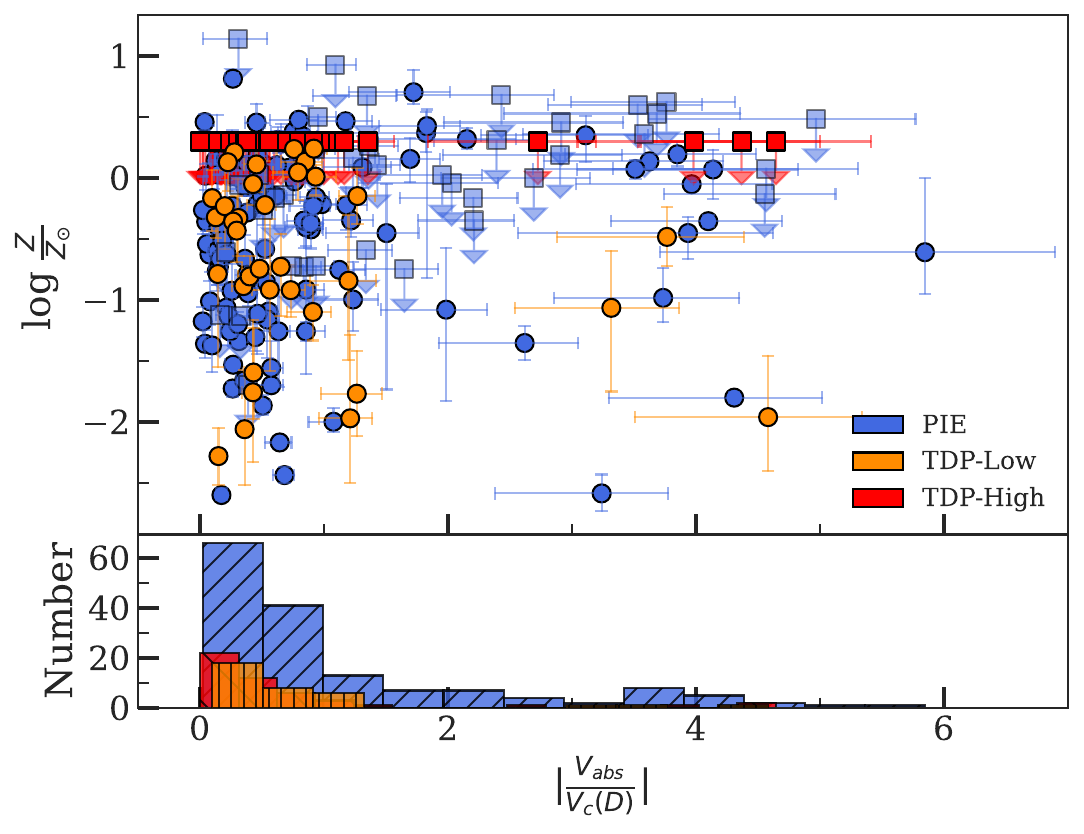}}}%
    \qquad
    \subfloat{{\includegraphics[scale=0.46]{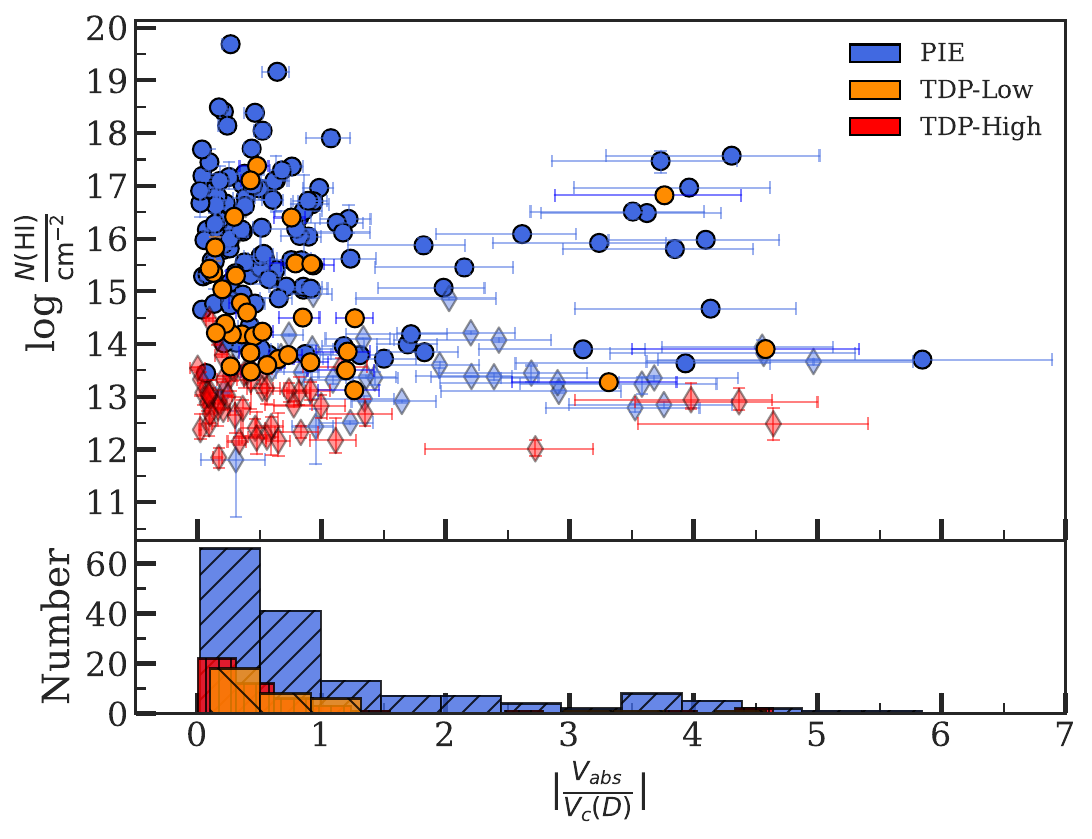} }}%
\caption{(a): The metallicity as a function of absorber velocities scaled by the circular velocity of galaxy for the three cloud types. The metallicity measurements are shown as filled circles, while the upper limits on metallicity are shown as filled squares with downward pointing arrows. (b): The neutral hydrogen column density as a function of absorber velocities scaled by the circular velocity of galaxy for the three cloud type. The histograms show the distribution of absorber velocities scaled by the circular velocity of galaxy. Error bars reflect the 1$\sigma$ uncertainty.}%
    \label{fig:NHI_absvel}%
\end{figure*}

In Figure~\ref{fig:NHI_absvel}a, we plot the metallicity, {\metallicity}, as a function of absorber velocity scaled by the galaxy circular velocity measured at the impact parameter, {\vabsscaled} for different cloud types. Galaxy circular velocities are calculated at the impact parameter of absorption using equation (5) of \citet{Navarro1996} and equation (B2) of \citet{Churchill2013}. The velocity zero point is the host galaxy systemic velocity. Accretion is expected to roughly match the rotation velocity of the host galaxy at a maximum, whereas outflows can sometimes exceed the escape velocity of galaxies, especially for lower mass galaxies. But also more massive galaxies can host absorption with higher velocities. Our sample of galaxies spans a range of halo masses; thus, we scale the absorber velocities shifted with respect to the galaxy by the circular velocity at the measured impact parameter of the host galaxy, V$_{c}$(D) to account for the galaxy halo mass~\citep[see e.g.,][]{ng2019}. We tabulate the V$_{c}$(D) values in Table~\ref{tab:obsgal}.

\smallskip

We observe that clouds with high {\vabsscaled} show both low and high metallicities; similarly clouds with low {\vabsscaled} also exhibit a range of metallicities.  We determine that the metal detection fraction for clouds with absolute absorber velocities $\leq$ V$_{c}$(D) is $85_{-4}^{+3}$\%, while for clouds with absorber velocities $>$ V$_{c}$(D) it is only $54 \pm 7$\%. This observation likely indicates that material beyond V$_{c}$(D) is not closely associated with the host galaxy and is likely tracing IGM material. We further investigate if there is an underlying trend by plotting the {\colden} as a function of {\vabsscaled} in Figure~\ref{fig:NHI_absvel}b, we see that the data points with upper limits on metallicity are the ones that have low {\colden} values, $\lesssim 14.5$, indicated by lightly shaded diamond markers. At these low {\colden} values the expected metal columns are quite low, which are difficult to measure at the resolution of COS and {\sn} of the observed data. These high-velocity clouds can be reasonably attributed to the fact that the material likely at the periphery of galaxies or in the IGM is unlikely to exhibit a direct connection with the kinematics of the galaxy. 

\subsubsection{Azimuthal Angle}
\label{sec:metallicity_azimuthal}

\begin{figure*}
    \centering
    \subfloat{{\includegraphics[scale=0.65]{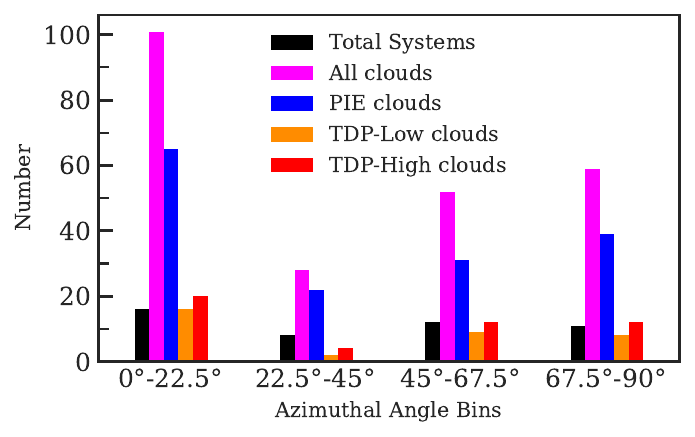} }}%
    \qquad
    \subfloat{{\includegraphics[scale=0.5]{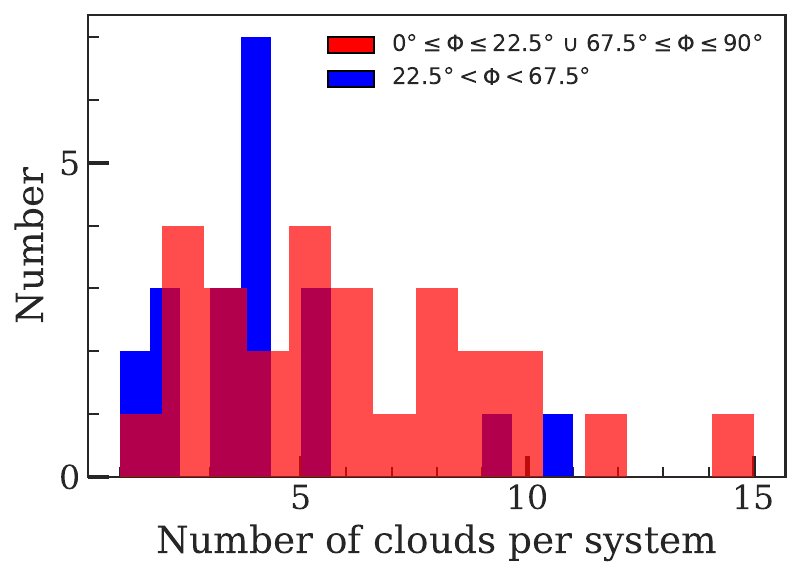} }}%
    \caption{(a): Barplot showing the total number of systems (black), clouds (magenta), PIE clouds (blue), TDP--Low clouds (orange), and TDP--High clouds (red), in different azimuthal angle bins for different cloud types. A greater number of clouds are seen along the major axis compared with other azimuthal angles. (b): The number of clouds per absorption system seen in all the phases along major and minor axis (red), and along interjacent angles (blue).
    }%
    \label{fig:numcomps}%
\end{figure*}

\begin{table*}
\begin{center}
\caption{\bf The bootstrapped mean metallicity and the error on the mean for the different types of clouds for the four azimuthal angle bins}
\label{tab:meanmetallicity_azi}

\begin{tabular}{ |c|c|c|c|c| } \hline

 Cloud type & $\Phi \leq 22.5\degree$ & $ 22.5\degree < \Phi \leq 45\degree$ & $ 45\degree < \Phi \leq 67.5\degree$ & $ \Phi > 67.5\degree$ \\\hline
  PIE & $-0.53$ $\pm$ 0.09 & $-0.25$ $\pm$ 0.19 & $-0.46$ $\pm$ 0.15 & $-0.66$ $\pm$ 0.16\\\hline
  TDP--Low & $-0.60$ $\pm$ 0.16 & $-0.89$ $\pm$ 0.76 & $-0.85$ $\pm$ 0.22 & $-0.63$ $\pm$ 0.28\\\hline
  TDP--High & $-1.35$ $\pm$ 0.11 & $-1.05$ $\pm$ 0.27 & $-1.37$ $\pm$ 0.10 & $-1.34$ $\pm$ 0.13\\\hline

\end{tabular} \\
\end{center}
\end{table*}

 In Fig.~\ref{fig:numcomps}a, we show the azimuthal angle, $\phi$, distributions for the number of clouds. We find that there are more PIE clouds along the major axis compared with other angles, with additional enhancement along the minor axis. We create four equal-width azimuthal angle bins: 0\degree--22.5\degree, 22.5\degree--45\degree, 45\degree--67.5\degree, and 67.5\degree--90\degree. Our choice of binning is to ensure that there are a sufficient number of data points in each bin, and the number of systems is
nearly evenly distributed in these bins. The total number of absorption systems is 16, 8, 12, and 11 in each of these azimuthal angle bins. The total number of clouds is
101, 28, 52, and 59 clouds in these four bins. We compare the number of clouds per absorption system along the major and minor axis ($0\degree \leq \Phi \leq 22.5\degree$ $\cup$ $67.5\degree \leq \Phi \leq 90\degree$) with the number of clouds per absorption system along the interjacent angles ($22.5\degree < \Phi < 67.5\degree$), and show their distributions in Fig.~\ref{fig:numcomps}b. We see that along the major and minor axes, there is a relatively higher frequency of observing more than $\sim$5 clouds per absorption system, compared with the interjacent angles. We compare their distributions using a 2-sample Anderson-Darling test~\citep{adtest} to test the null hypothesis that the distribution of the number of clouds per system along the minor and major axis is different from the distribution of the number of clouds per system along the interjacent angles. This test is particularly suitable for sample sizes of less than 25 observations each. We find statistically significant evidence to reject the null hypothesis ($p$-value = 0.03). \citet{Berg2023}, in their bimodal absorption system imaging campaign (BASIC) survey, aimed at characterizing the galaxy environments of pLLSs and LLSs at $z \lesssim 1$, and found that metal-enriched absorbers ({\xh} $>$ $-1.4$) are preferentially found along the major and minor axes, consistent with our observations. We note that a significant percentage, $89_{-5}^{+4}$\%, of our galaxy-selected pLLSs and LLSs absorbers have metallicities, {\metallicity} $>$ $-1.4$.


\begin{figure*}
    \centering

 \subfloat{{\includegraphics[scale=0.505]
 {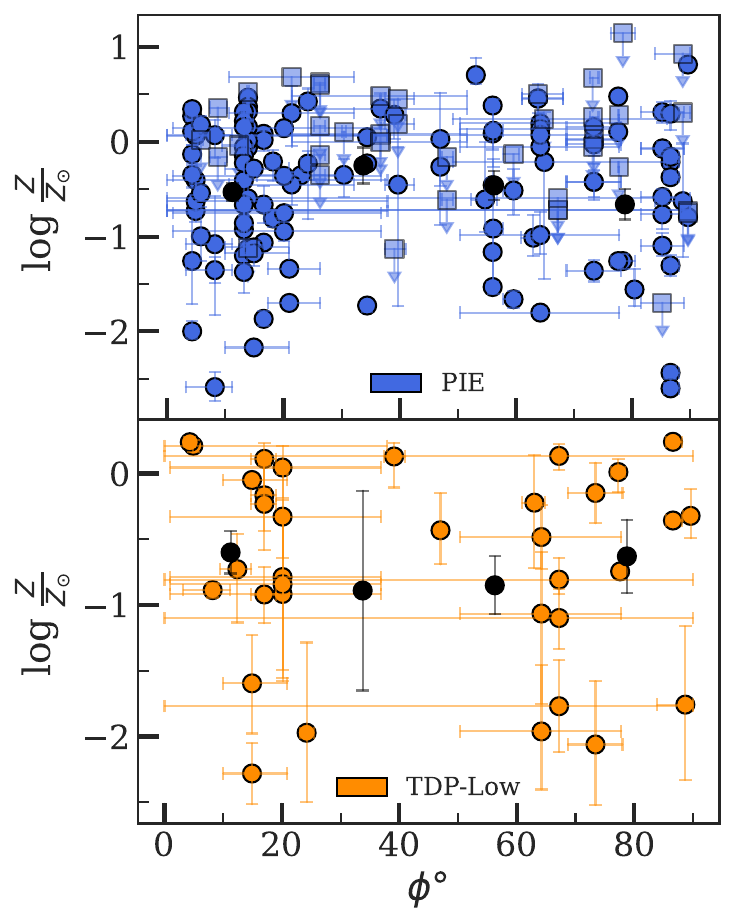} }}%
    \qquad
    \subfloat{{\includegraphics[scale=0.505]{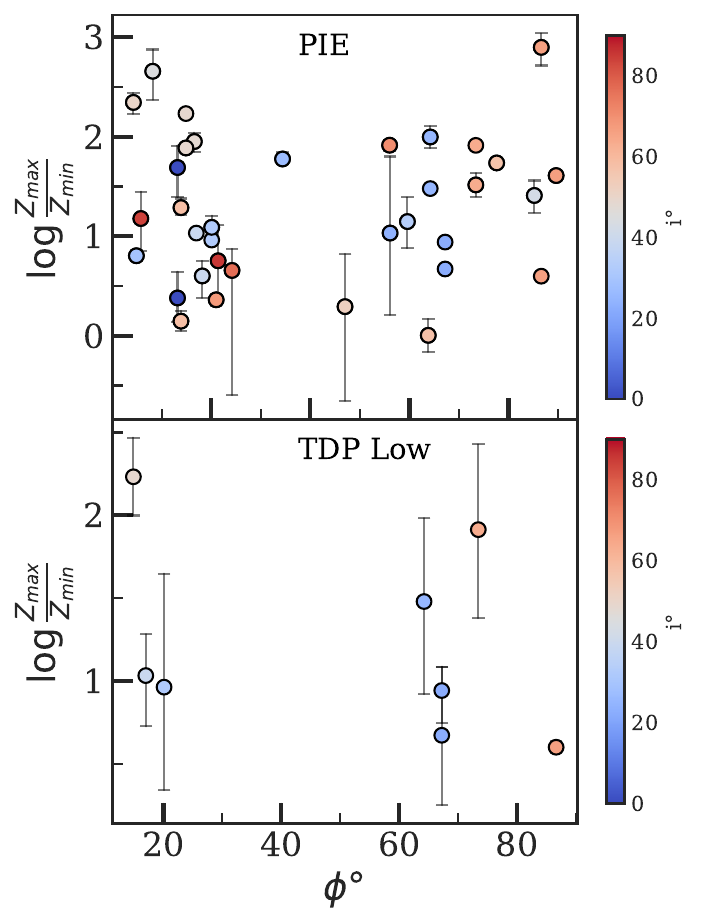} }}%
 \caption{(a): The metallicity as a function of azimuthal angle for PIE and TDP--Low cloud types. The metallicity measurements are shown as filled circles, while the upper limits on metallicity are shown as filled squares with downward pointing arrows. The bootstrapped mean of metallicity and the standard error of the mean within the four azimuthal angle bins are shown in black. The mean values across the various azimuthal angle bins are
 similar for these clouds. (b): The difference between maximum and minimum metallicities in a given absorption system as a function of azimuthal angle considering all cloud types and color-coded by the inclination angle.}%
    \label{fig:metallicity_azimuthal}%
\end{figure*}

\smallskip

We next examine the spatial distribution of CGM metallicities in Fig.~\ref{fig:metallicity_azimuthal}a with the bootstrapped mean metallicities in the various azimuthal angle bins overplotted in black. We observe a wide range of metallicities for all azimuthal angles, and observe a similar dispersion in metallicity for all the cloud types for all angles. We find that the mean metallicities are consistent with one another within their uncertainties (standard error on the mean) for each of the cloud types. The bootstrapped metallicity mean and its uncertainty are tabulated in Table~\ref{tab:meanmetallicity_azi}. We also perform a log-rank test, accounting for the non-detections, to test the null hypothesis of no difference in metallicities between any two chosen azimuthal angle bins for the three different types of clouds. The inferred $p-$values from these tests suggest that the null hypothesis of no difference in the metallicity distribution of gas along the minor axis, major axis, and interjacent angles cannot be rejected for all three cloud types.

\smallskip

\citet{Veilleux2005} suggested that outflows should have large cloud-cloud variations in metallicity. Such signals are also observed in very strong {\mgii} absorbers~\citep{Bond2001} and at lower column densities~\citep[e.g.,][]{Rauch2002,zonak2004absorption}. In Fig.~\ref{fig:metallicity_azimuthal}b, we examine this by plotting the metallicity variation, the difference between maximum and minimum metallicities in a given absorption system along a given sightline, with points coloured by their inclination angle.
We only plot points corresponding to an absorption system with more than one cloud. We do not observe significant differences in the metallicity variation along the minor or major axis. In the case of TDP--Low clouds, the number of clouds is too low to observe a trend. It is possible that the dependency on other galaxy or gas properties is the cause of scatter in the metallicity distribution as a function of the azimuthal angle. We investigate any such dependency while taking into account the {\hi}-column density, inclination angle, impact parameter scaled by virial radius, $B-K$ galaxy colour, absorber velocity scaled by the galaxy circular velocity, and the halo mass of the galaxy.

\smallskip

\begin{figure*}
  \includegraphics[scale=0.5]{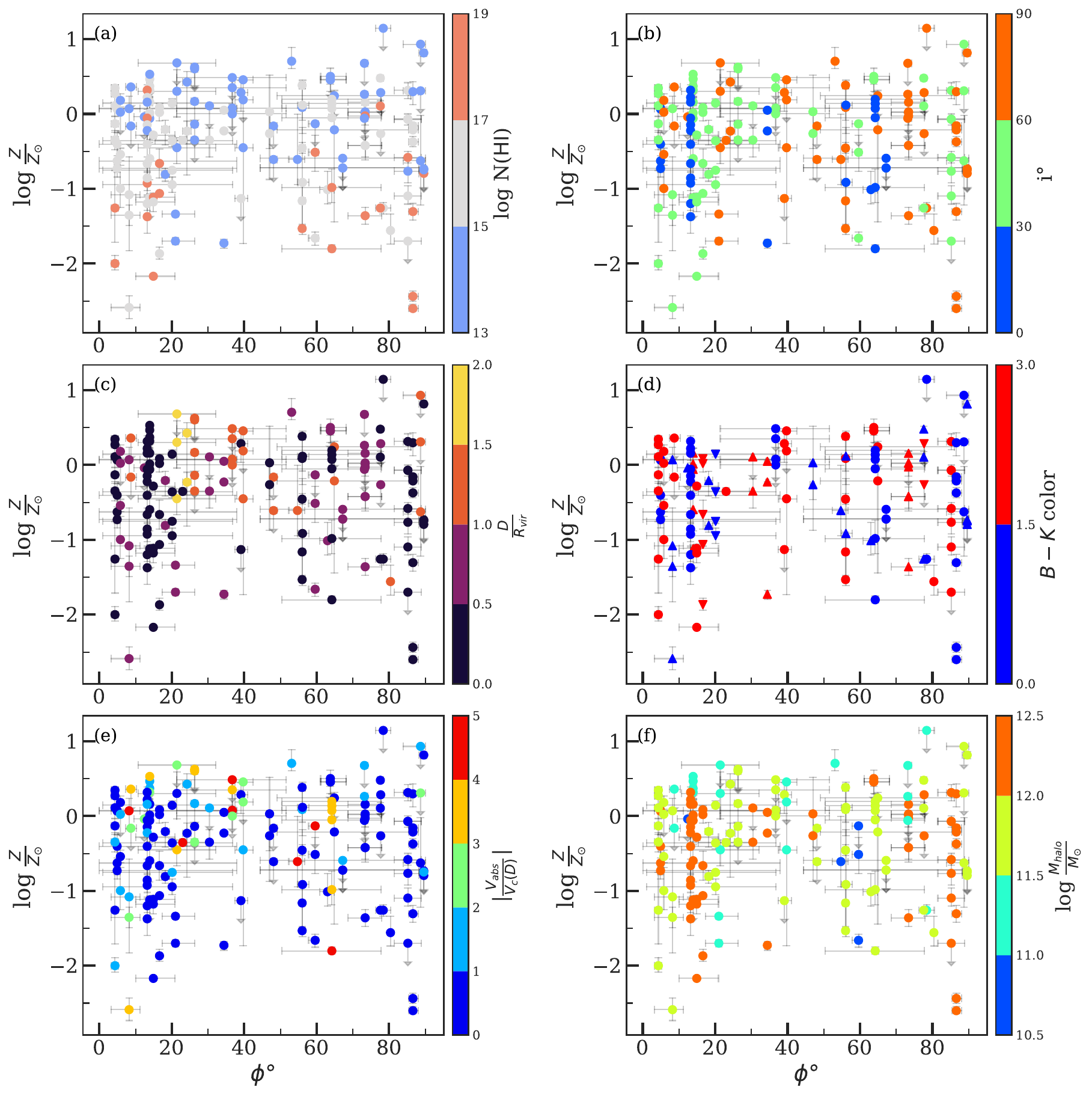}
\caption{The metallicity as a function of azimuthal angle color-coded by different galaxy properties for PIE clouds. The upper limits on metallicity are indicated by downward pointing arrows. Color-coded by (a): neutral hydrogen column density; (b): inclination angle; (c): normalized impact parameter; (d): galaxy $B-K$ color with marker shapes of upward triangles indicating star-forming galaxies, downward triangles indicating passive galaxies, and circles for which the nature of galaxy is not determined; (e): absorber velocity scaled by the galaxy circular velocity; (f): galaxy halo mass. Error bars reflect the 1$\sigma$ uncertainty.
}%
  \label{fig:metallicity_azimuthal_galaxy_pie}
  \end{figure*}

    \begin{figure*}

  \includegraphics[scale=0.5]{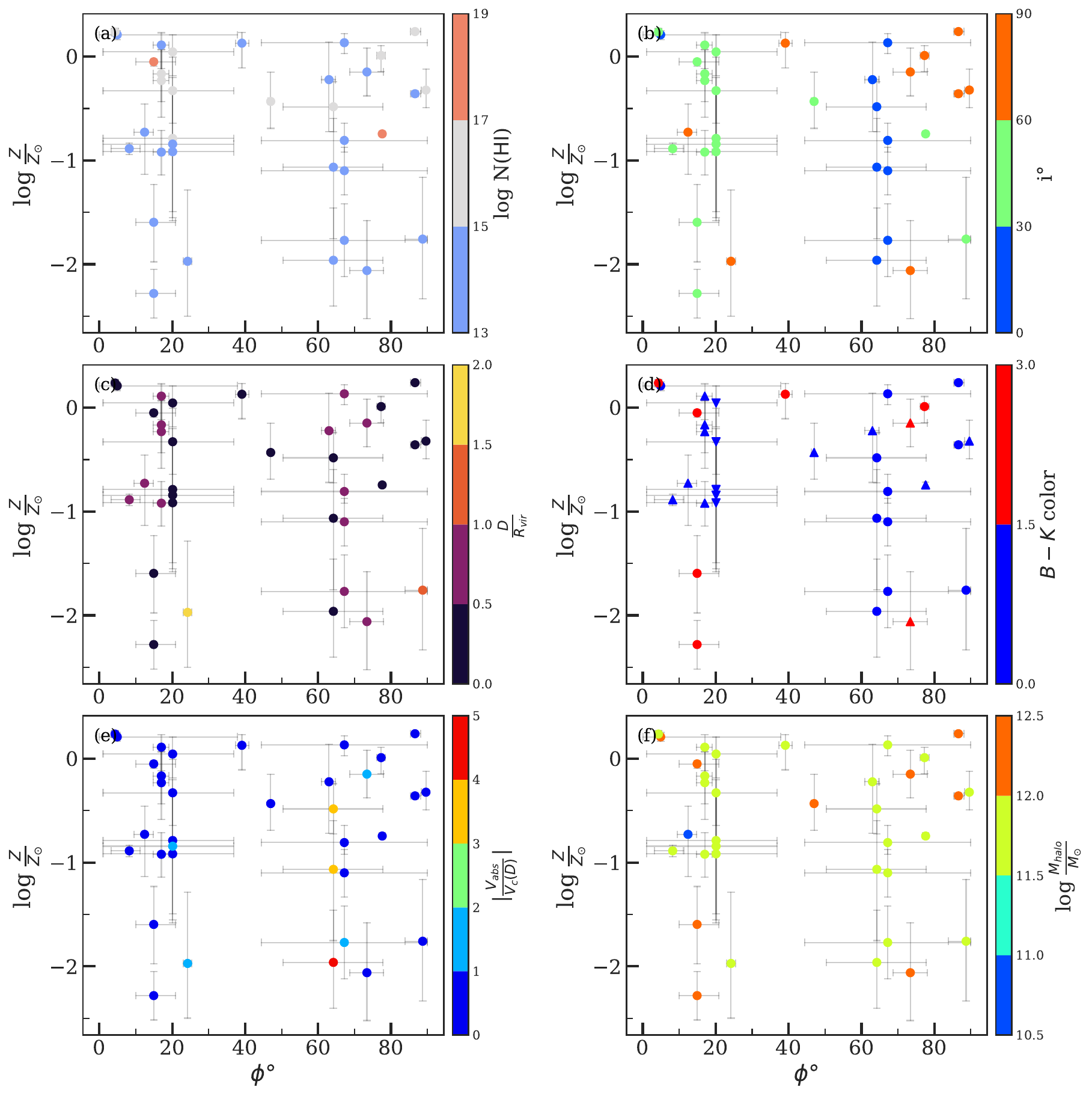}
\caption{Same as in Fig.~\ref{fig:metallicity_azimuthal_galaxy_pie} but for TDP--Low clouds.}%
  \label{fig:metallicity_azimuthal_galaxy_noncbla}
  \end{figure*}

In Figs.~\ref{fig:metallicity_azimuthal_galaxy_pie} and~\ref{fig:metallicity_azimuthal_galaxy_noncbla}, we present the relationship between metallicity and azimuthal angle for PIE clouds and TDP--Low clouds where the data points are coloured by different galaxy properties. In Figure~\ref{fig:metallicity_azimuthal_galaxy_pie}a, the data points are coloured by neutral hydrogen column density. We observe that the
lowest neutral hydrogen column densities are often upper limits on metallicity, and these low hydrogen column densities are seen at all azimuthal angles. There is also no discernible trend in the association between metallicity and azimuthal angle among the absorbers chosen by {\hi} column density. A metallicity bimodality is expected to be strongest for pLLS due to inflows and outflows~\citep{Wotta2019}. We compare the metallicities of pLLSs for the minor and major axis groupings, however, we do not find evidence of a significant difference in their metallicity distributions. This finding suggests that a generalization that metal-poor accretion occurs along the major axis and metal-enriched outflows along the minor axis is too simplistic. Instead, this indicates that the CGM is evenly mixed across the entire range of azimuthal angle and for all {\hi} column densities. We also do not see any evidence of a trend for TDP--Low as well (Fig.~\ref{fig:metallicity_azimuthal_galaxy_noncbla}a). 

\smallskip

Furthermore, we take into account how the inclination angle affects the correlation between metallicity and azimuthal angle. Assuming the simple picture of the CGM, if the galaxy were edge-on, outflows and inflows would be easier to identify since the cross sections of the gas flows on the sky are minimal and do not overlap. It becomes increasingly challenging to establish whether the quasar sight-line probes the major or minor axis as the galaxy's inclination becomes more face-on, when the cross sections of the gas flows grow, and there are potentially more structures along the line of sight~\citep{Churchill2015,Kacprzak2019ApJ,peeples2019figuring}. An edge-on galaxy sample maximizes the prospect of finding a metallicity-azimuthal angle bimodality. Therefore, it is crucial to account for the galaxy's inclination angle when determining if metallicity and azimuthal angle show any correlation. In Figure~\ref{fig:metallicity_azimuthal_galaxy_pie}b, the metallicity is plotted as a function of azimuthal angle with points coloured by the corresponding galaxy inclination angle. We observe that edge-on galaxies, $i \gtrsim$ 60\degree, show a range of metallicities, with no apparent trend between metallicity and azimuthal angle. We also do not see any evidence of a trend for TDP--Low (Fig.~\ref{fig:metallicity_azimuthal_galaxy_noncbla}b). 

\smallskip

We also consider how the impact parameter might affect the metallicity and azimuthal angle relationship. Previous works~\citep{Bordoloi2011,lan2018} have shown that for impact parameters less than 50--100 kpc, the equivalent width of {\mgii} absorbers was strongest along the minor axis, suggesting that the azimuthal distribution of the CGM metallicities is potentially bimodal at smaller impact parameters. Consequently, absorption clouds with lower impact parameters could have a metallicity dependence on the azimuthal angle. In Figure~\ref{fig:metallicity_azimuthal_galaxy_pie}c, the metallicity is plotted as a function of azimuthal angle with points coloured by the corresponding normalized impact parameter. We observe that clouds at low impact parameters $\frac{D}{R_{vir}} \lesssim 0.5$, show a range of metallicities, with no apparent trend between metallicity and azimuthal angle. We also do not see any evidence of a trend for TDP--Low (Figure~\ref{fig:metallicity_azimuthal_galaxy_noncbla}c). 

\smallskip

Previous works have also found a bimodality in the {\mgii} absorber azimuthal-angle distribution for blue star-forming galaxies~\citep{Bordoloi2011,kacprzak2012,lan2018}. This finding indicates that blue, star-forming galaxies may be where the simple CGM model holds up best. In Figure~\ref{fig:metallicity_azimuthal_galaxy_pie}d, we plot the metallicity as a function of azimuthal angle coloured by $B-K$ colour with markers indicating whether a galaxy is star-forming (upward triangles), passive (downward triangles), undetermined (circles). The blue star-forming galaxies, $B-K < 1.5$, in our sample have quasar sight-lines probing all azimuthal angles, but we observe similar metallicity spread across all orientations. Even along the minor axis, where it is anticipated that high metallicities due to outflows would be seen, we find clouds with the two lowest metallicities in our sample. A similar observation is made for TDP--Low (Figure~\ref{fig:metallicity_azimuthal_galaxy_noncbla}d. Thus, a metallicity and azimuthal angle relationship does not seem to exist for blue, star-forming galaxies. Though many of our galaxies are star-forming, their specific star-formation rates (sSFRs) are more than an order of magnitude lower than starburst galaxies. Starburst galaxies can experience star formation rates of 100 {\msun}/yr~\citep{schneider2006extragalactic} compared with an average of only few times {\msun}/yr in our sample. Starburst galaxies are expected to have active galactic winds that would produce high metallicities along their minor axes. We also do not see any evidence of a trend for TDP--Low clouds (Figure~\ref{fig:metallicity_azimuthal_galaxy_noncbla}d). 
 
\smallskip

The TNG50 simulations~\citep{peroux2020} predict that on average higher metallicities are expected to be linked with rapidly moving outflows emanating along the minor axis. Conversely, gas with lower metallicities is linked with slower-moving processes involving either accretion into the galaxy or inflow. In Figure~\ref{fig:metallicity_azimuthal_galaxy_pie} (e), we plot the metallicity as a function of the azimuthal angle coloured by the absorber velocity scaled by the galaxy circular velocity measured at the impact parameter. While the large majority of the absorbers have velocities within the circular velocity of the galaxy, we observe that these absorbers show a wide range of metallicities for all azimuthal angles. A minority of the absorbers that have velocities greater than the circular velocity of the galaxy neither show a preference for high metallicities nor are predominantly found along the minor axis. A similar observation is made for TDP--Low (Figure~\ref{fig:metallicity_azimuthal_galaxy_noncbla}e).

\smallskip

The TNG50 simulations~\citep{peroux2020} also find that the enhancement of metallicities along the minor axis compared with the major axis is strongest for halo masses {\mhalo}$\sim$ 11.5. In Figure~\ref{fig:metallicity_azimuthal_galaxy_pie}f, we plot the metallicity as a function of the azimuthal angle coloured by the halo mass. We find a range of metallicities for all azimuthal angles in various bins of galaxy halo masses. A similar observation is made for TDP--Low (Figure~\ref{fig:metallicity_azimuthal_galaxy_noncbla}f). 

\smallskip

 \citet{Berg2023} also did not find any strong correlations between the metallicities or {\hi} column densities of the gas and galaxy properties, except for stellar mass of the galaxies. For the cool, photoionized gas, they found that low-metallicity ({\xh} $\leq$ $-1.4$) systems have a probability of $0.39_{-0.15}^{+0.16}$ for being associated with a host galaxy with {\mstar} $\geq$ 9.0 within 1.5{\rvir}, while higher metallicity absorbers ({\xh} $>$ $-1.4$) have a probability of $0.78_{-0.13}^{+0.10}$.

\section{DISCUSSION}
\label{sec:discussion}

We applied the CMBM method to the absorbers from the ``Multiphase Galaxy Halos'' Survey to investigate the properties of the different regions/clouds of gas along the quasar sightlines through the CGM of known galaxies. The focus of our study was to examine if there is a relationship between the metallicity of the CGM clouds and the azimuthal angle for sightlines through $z < 0.7$ $L^{\star}$ isolated galaxies. Our methodology provides a much more detailed view of differing cloud properties along the sightlines, as compared with~\citetalias{Pointon2019}, who only computed an average metallicity for each galaxy/absorption line system.  With our CMBM analysis, we determined that CGM sightlines pass through multiple phases of gas, some consistent with photoionization equilibrium and some requiring time-dependent photoionization to explain the line ratios and profile shapes. The 240 absorbing clouds that we found in 47 absorption systems are found to be broadly consistent with three different underlying populations when characterized by their intrinsic properties.

\begin{landscape}
\begin{figure}
\hspace*{-0.5cm}\includegraphics[scale=0.45]{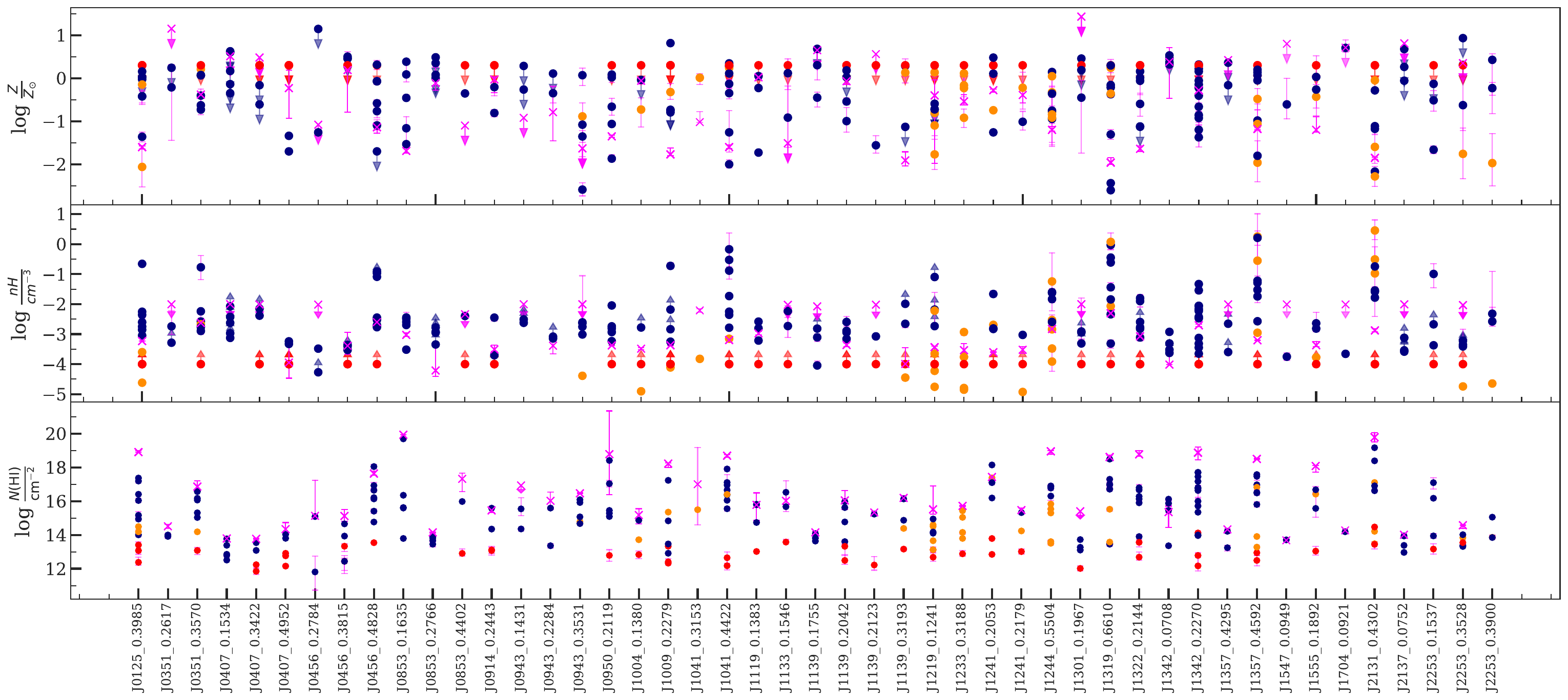}
  \caption{A comparison of the inferred properties \textit{viz.} {\metallicity}, {\hden}, and {\colden}, obtained using the CMBM for the different clouds along each sightline with the values from \citetalias{Pointon2019}. The PIE clouds are shown in navy blue, TDP--Low in dark orange, and TDP--High clouds in red. The values from \citetalias{Pointon2019} are shown as crosses in magenta. The upper limits on metallicity are shown as downward-pointing arrows, and the lower limits on density are shown as upward-pointing arrows. A single average value is not particularly representative of any of the gas clouds.}
   \label{fig:comparisonP19}
\end{figure}
\end{landscape}

\subsection{Interpreting the Absence of Metallicity/Azimuthal Angle Relationship}

In the introduction, we discussed the expected trends in galaxy metallicities, anticipating higher metallicities along the minor axis and lower metallicities along the major axis. 
This expectation is based on both intuition and simulations, particularly those presented in the study of \citet{peroux2020}. We also pointed out that no such trend was identified in the analysis conducted by \citetalias{Pointon2019} on the same galaxy sample as the one examined in the present paper. \citetalias{Pointon2019} delved into several potential reasons for this lack of agreement with simple models of the CGM, emphasizing that simple models of the CGM do not apply, those with bipolar outflows and cold-mode accretion in the disk plane. For instance, regions with high metallicity gas can arise close to the major axis due to the recycling of metal-enriched gas from outflows. Additionally, large wind opening angles might dilute metallicities along the minor axis. Furthermore, they pointed out that the signature of outflows would likely be more prominent at higher redshifts than in our current sample, which is limited to $z \lesssim 0.7$. \citetalias{Pointon2019} also suggested that low-metallicity gas might accrete along filaments that are not aligned with the major axis of the galaxy. Despite the various plausible explanations for the lack of a trend, the results of \citetalias{Pointon2019} were limited by their characterization of each sightline by just a single metallicity.

\smallskip

To compare the \citetalias{Pointon2019} results with the CMBM method, we show the derived metallicities, densities, and $N({\hi})$ for the different clouds along each sightline in Fig.~\ref{fig:comparisonP19}. The single-valued properties or limits derived for each system by \cite{Pointon2019} are shown as a magenta cross. A typical sightline through a galaxy has several clouds in PIE and one TDP--High cloud, and perhaps half also have one or more TDP--Low clouds. The ranges of metallicities and densities can be up to two orders of magnitude along a single galaxy sightline. This reflects the fact that most sightlines pass through different parts of a galaxy, potentially having contributions in absorption from outer disk, halo, outflows, recycled accretion, tidal streams, satellite galaxies, etc. We find that in most cases, the \citetalias{Pointon2019} metallicities and densities are consistent within the range of our values.
However, the single average metallicities are not systematically larger or smaller than the range of multi-cloud modelling results, showing that the situation is complicated. The single average value is not particularly representative of any of the gas clouds. The total system $N({\hi})$ values from \citetalias{Pointon2019} are often consistent with the value for the highest $N({\hi})$ cloud in the system using CMBM. The main cause for the differences in derived metallicities between our approach and \citetalias{Pointon2019} is that it is not typical for all of the {\hi} to be associated with all of the metal line column.

\smallskip

\begin{figure}
\centerline{
\includegraphics[scale=0.55]{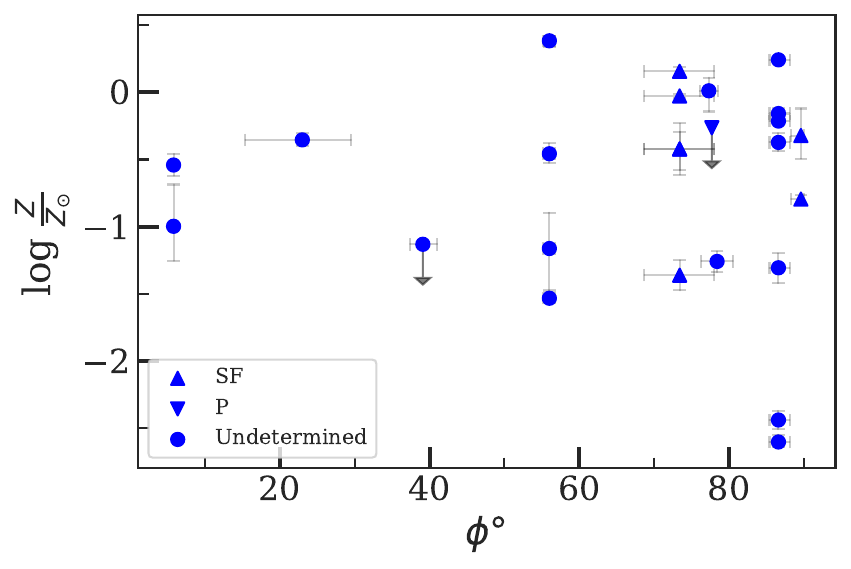}
}
\caption{The trend between metallicity and azimuthal angle for a chosen subsample, {\colden} $\geq$ 15.0, inclination angle $\geq$ 60$\degree$, and impact parameter $\leq R_{\rm vir}$, most likely to show difference in metallicity between major ($\phi$ = 0\degree) and minor axis ($\phi$ = 90\degree). The clouds associated with star-forming galaxies are shown as upward triangles, passive galaxies are shown as downward triangles, and the clouds without information on the galaxy type are shown as circles.}
\label{fig:metallicity_azi_chosen}%
\end{figure}

Even with the detailed cloud information about multiple phases, there is still no indication of a relationship between azimuthal angle and metallicity as seen in Figs~\ref{fig:metallicity_azimuthal_galaxy_pie} and~\ref{fig:metallicity_azimuthal_galaxy_noncbla}. There is no apparent trend for the highest or for the lowest metallicity cloud in a system, nor is there a trend for galaxies' near-edge-on inclinations, 
where the largest difference between the major and minor axis is expected. Both high and low metallicity clouds are found along the same sightlines. To investigate this further, we chose a subsample of sightlines most likely to show differences between major and minor axes ({\colden} $\geq$ 15.0, inclination angle $\geq$ 60$\degree$, and impact parameter $\leq R_{\rm vir}$).  A plot of the metallicity vs. azimuthal angle for the subsample, comprising of PIE and TDP--Low clouds, is shown in Fig.~\ref{fig:metallicity_azi_chosen}.  There are very few sightlines close to the major axis that satisfy all the subsample criteria, but it is worth noting that the full range of metallicities is apparent along the minor axis, including high values expected for outflow clouds, as well as lower values that could be contributed by dwarf satellites and their stripped material. The simplest interpretation is that the gas in these $z<0.7$ galaxies is often well mixed, meaning that gas originating from different processes is now located in various parts of the galaxy.

\smallskip

A substantial insight into that interpretation is provided by the results of the FIRE--2 simulations \citep{Hafen2019}. The trajectories of different particles, characterized by their origins in IGM accretion, winds, or satellite winds, were followed for the 1~Gyr prior to $z=0.25$ for galaxy halos of mass $10^{12}$~{M$_{\odot}$}, $10^{11}$~{M$_{\odot}$}, and $10^{10}$~{M$_{\odot}$}. The higher mass halos correspond to those in our present sample. The gas of different origins in the $10^{12}$~{M$_{\odot}$} halos appears to be relatively well mixed at $z=0.25$, with all three particle origins seen over the full halo volume.  When only particles which originated recently, between 1~Gyr and 0.5~Gyr ago, are considered, we do see specific regions that the three types of particles prefer.  The wind particles are close to the central galaxy, the satellite wind particles are mostly associated with three halo substructures and the accretion avoids some directions in the halo. When origins at all times are included, however, this asymmetry is no longer apparent. In conclusion, based on the results of \citet{Hafen2019}, we should not have expected a relationship between metallicity and azimuthal angle in our sample. Gas at $z=0.25$ is well mixed, with gas from several origins present at all impact parameters and azimuthal angles.

\smallskip

At higher redshifts the contribution of outflows is larger~(e.g., \citealt{Muratov2015,Mitchell2018}), and it is more likely that their signatures will be observable in absorption. \citet{Hafen2019} found that winds provided more than 40\% of the inner halo mass at $z=2$, and that the contributions of winds, satellite winds, and IGM accretion are all quite inhomogeneous and not well mixed.  Thus if we were to perform this same experiment at high redshift we would be more likely to see relationships between azimuthal angle and metallicity for some of the clouds.

\smallskip

Though we did not see any trends of cloud properties with galaxy properties, we should note that the galaxies in our sample are relatively isolated, with no neighbors within 100 kpc and 500~{\kms}. This would imply fewer active interactions in these galaxies and less wind activity than for a more crowded environment~\citepalias{Pointon2019}.

\subsection{Nature and Origin of Different Cloud Types}

In this section, we discuss the nature and origin of different types of clouds. One key principle to keep in mind is that every location along a sightline has a specific density, temperature, and metallicity. It is also important to note that together the thicknesses that we have derived for the clouds must sum to something of order a halo size or less.

\smallskip

In Fig.~\ref{fig:metallicity_temp}a, the three types of clouds found in our modelling, with the blue symbols representing cooler clouds in thermal equilibrium, the gold ones representing clouds of intermediate temperature that are modelled with time-dependent photoionization, and the red ones with detected {\ovi} as hotter clouds also modelled with time-dependent photoionization, but as we discuss below, they are found to be in collisional ionization equilibrium.

\subsubsection{PIE Clouds}

First, we will discuss the PIE clouds. Fig.~\ref{fig:metallicity_size}
showed that the majority of them have densities between  $-3.5 \leq$ {\hdenu} $\leq -2.5$, metallicities greater than solar, and small thicknesses less than a kpc and even sometimes sub-parsec. \citet{werk2013cos} argued that the large scatter in the low-ion absorption strengths indicates that the cool CGM is patchy and hence reveals a wide range of ionization conditions. Many of the PIE clouds have weak, low ionization absorption detected, and their properties thus coincide with the metal-rich weak {\mgii} absorbers studied by \citet{rigby2002population}, \citet{narayanan2005survey}, and \citet{muzahid2018cos}.  While it is tempting to characterize this population of absorbers as having a single origin, we must remember that gas of density {\hden} = $-3$ is surely not rare in galaxies and will occur in disk material, in outflows, in tidal debris, and in various other regions of the CGM.  This population is common, with a typical sightline passing through a few of them (see Fig.~\ref{fig:comparisonP19}).

\smallskip

\begin{figure}
\centerline{
\includegraphics[scale=0.55]{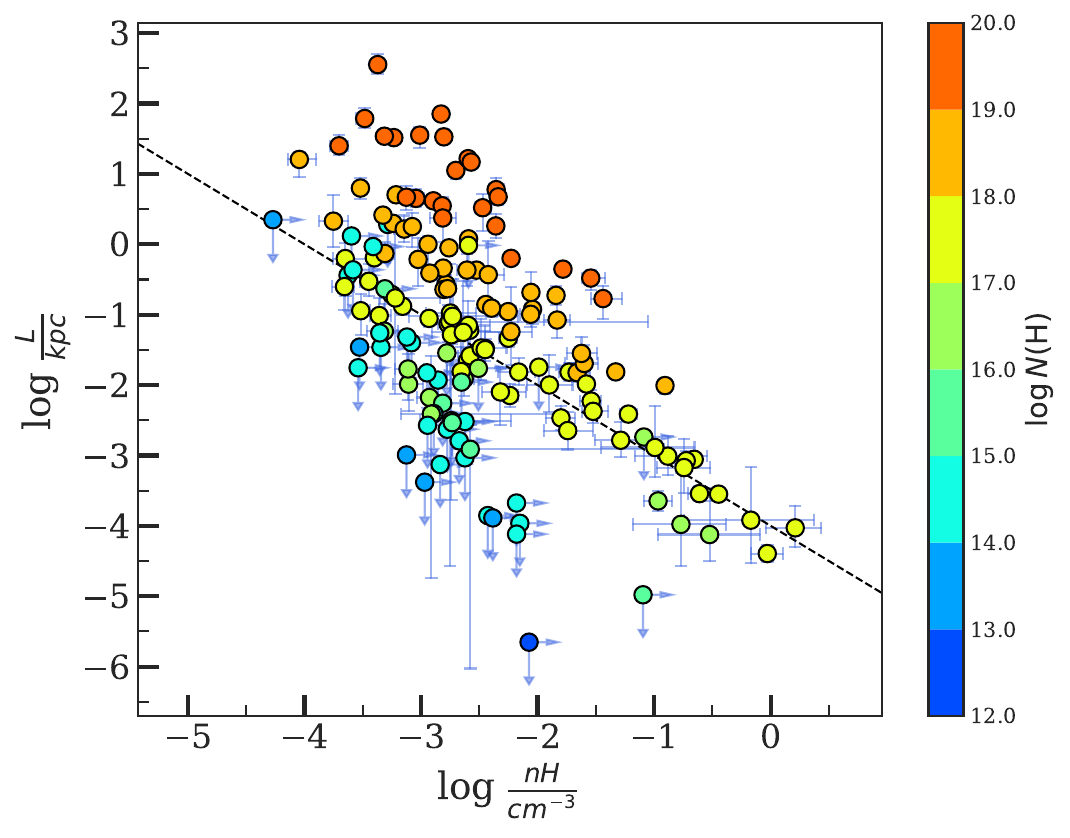}
}
\caption{Relationship between the sizes and hydrogen densities for the PIE clouds. The size scale of cool CGM clouds formed by the process of shattering is expected to follow $L \sim 0.1$pc/$n_{\rm H}$~\citep{McCourt2018MNRAS}, and shown as a dashed line on the plot. The limits are indicated by attached arrows.}
\label{fig:size_nH}%
\end{figure}

\smallskip

One plausible scenario for the origin of cool clouds is gas cooling out of a hot, ambient medium, mediated by thermal instability~\citep[e.g.,][]{Maller2004}. When gas begins to cool below $T \lesssim 10^{6}\,K$, line emission expedites the cooling to $T \sim 10^{4}\,K$, resulting in a two-phase medium - a hot phase at $T \sim 10^{6}\,K$ and a cooler phase at $T \sim 10^{4}\,K$. If the cooling is isochoric, maintaining the cloud size or density, the cooling cloud is out of pressure balance, contracting on a much longer sound-crossing timescale only after achieving pressure equilibrium. A faster route to achieving pressure equilibrium for the cooling cloud would be to cool isobarically, never leaving equilibrium, and forming fragmented clouds. The fragmented clouds at $T \gtrsim 10^{4}\,K$ undergo further shattering reaching a size scale where $c_{s}t_{cool}$ is minimized, where $c_{s}$ is the sound speed and $t_{cool}$ is the cooling time. The characteristic length scale is then given by the relationship $L_{cloud}$ $\sim$ min($c_{s}t_{cool}$) $\sim$ 0.1\,pc/$n_{\rm H}$~\citep{McCourt2018MNRAS}.

\smallskip

This relationship is shown as a dashed line on the plot of $\log L$ vs. $\log n_H$ in Fig.~\ref{fig:size_nH}. We find that the line is consistent with being a lower bound in PIE cloud thickness at every {\hden}, suggesting that fragmentation plays a role in their origin. The total column densities of PIE clouds hovering around the theoretical line are also consistent with $N_{\rm H}$ $\sim$ 10$^{17}$ {\cmsq} as expected from the $L = \frac{N_{\rm H}}{n_{\rm H}}$ relation. The apparent large spread at $\sim${\hden} $= -3$ is due to the limits, with no metals detected at the 3$\sigma$ detection limit, which could still be consistent with the theoretical line. The clouds with {\totalcolden} $\gtrsim$ 18 and with larger sizes could likely be being subject to shattering. The clouds with even higher {\totalcolden}, $\gtrsim$ 19, may undergo collapse due to self-gravity~\citep{Jeans1901,Low1976}. Despite the usefulness of such an observational comparison with theoretical prediction, we note that the lack of data points at the lower-left quadrant could be because of the lack of sensitivity to very low column densities, {\colden} $\lesssim$ 12.5.


\smallskip

\smallskip
\begin{figure*}
  \includegraphics[scale=0.55]{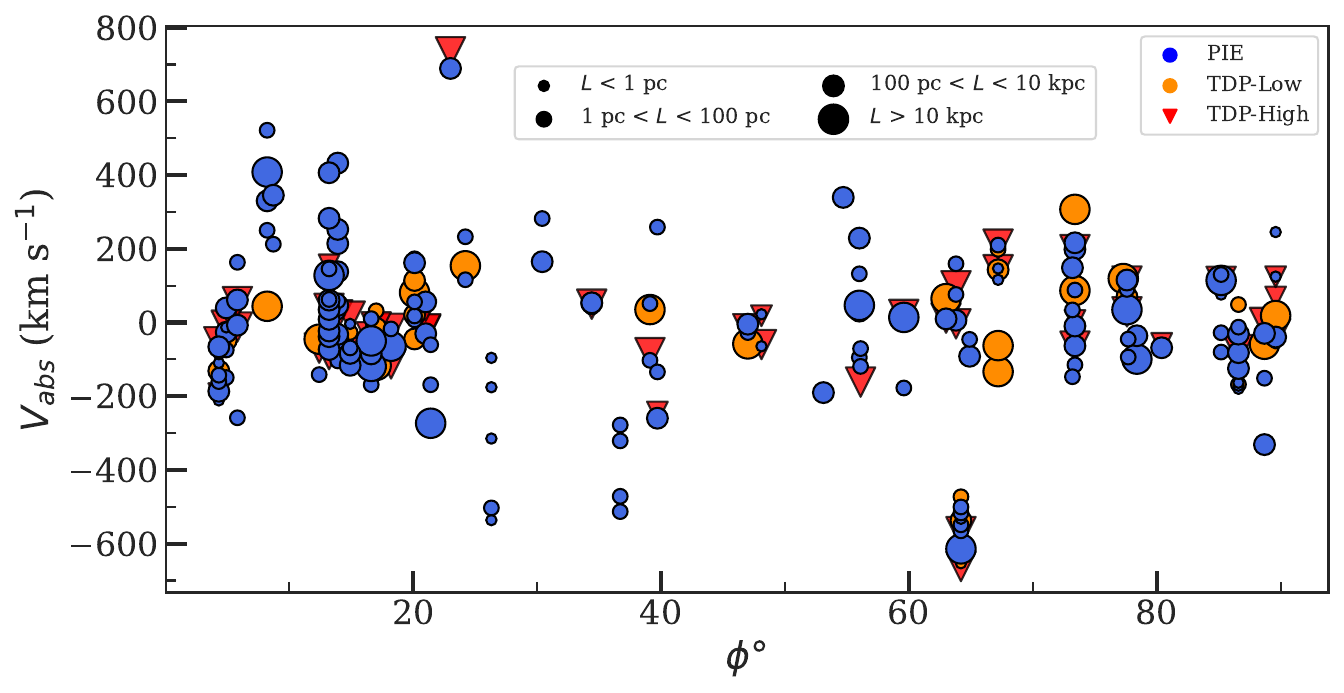}
  \caption{Absorber velocity as a function of azimuthal angle for the different cloud types. The marker size is indicative of the cloud size. The PIE clouds are shown in blue circles, TDP--Low in gold circles, and TDP--High clouds in red triangles. In some cases, clouds of different sizes are present along the same sightline suggesting that smaller clouds form by fragmentation from a parent cloud, either this same parent or a similar one which is not intercepted along this particular line of sight.
  }
   \label{fig:Vabs_Phi}
\end{figure*}

Several high-resolution numerical simulations also show small structures pervasive in the CGM.  For example, the {\mgii} column density distribution in the TNG50 simulation of \citet{Nelson2020} forms, through thermal instability, tens of thousands of fragments that are each a few 100 parsecs in size. \citet{Schneider2020} find many small fragments in the Cholla Outflow Simulation, which has a cell size of 5 pc. The expected CGM can be characterized as a galactic mist, with fragments from a larger cloud that are moving through a hot halo~\citep[e.g.,][]{Gronke2020}. The fragments/droplets are not stable, but the larger cloud continues to produce smaller ones so they are ubiquitous. We plot the cloud velocities seen along different sightlines in our sample in Fig.~\ref{fig:Vabs_Phi} as a function of the azimuthal angle; this figure shows that both larger clouds and several smaller clouds are present along the same sightline.  The larger clouds are, however, not always consistent in velocity with the small ones, but this is to be expected that the parent cloud would not always be intercepted along the same line of sight.

\smallskip

 The idea of a galactic mist is often connected to an origin in outflows, which at face value would predict higher metallicities along the minor axis. Since we do not see such a trend between azimuthal angle and metallicity in our sample, that might suggest galactic mist is not an important process. On the other hand, if mixing is efficient and clouds can survive, this would not be a concern. Furthermore, the idea of fragmentation of cooler clouds moving through a hot halo is more universal than the specific idea of galactic mist. 
The Milky Way Galaxy does not have active outflows on large scales~\citep{Fox2019}, but it does have a hot halo and multiphase high-velocity clouds passing through the halo.  This could lead to fragmentation and the type of PIE clouds we are observing in these absorption systems. 

\smallskip

\subsubsection{TDP--Low Clouds}

\begin{figure*}
\centerline{
\includegraphics[scale=0.625]{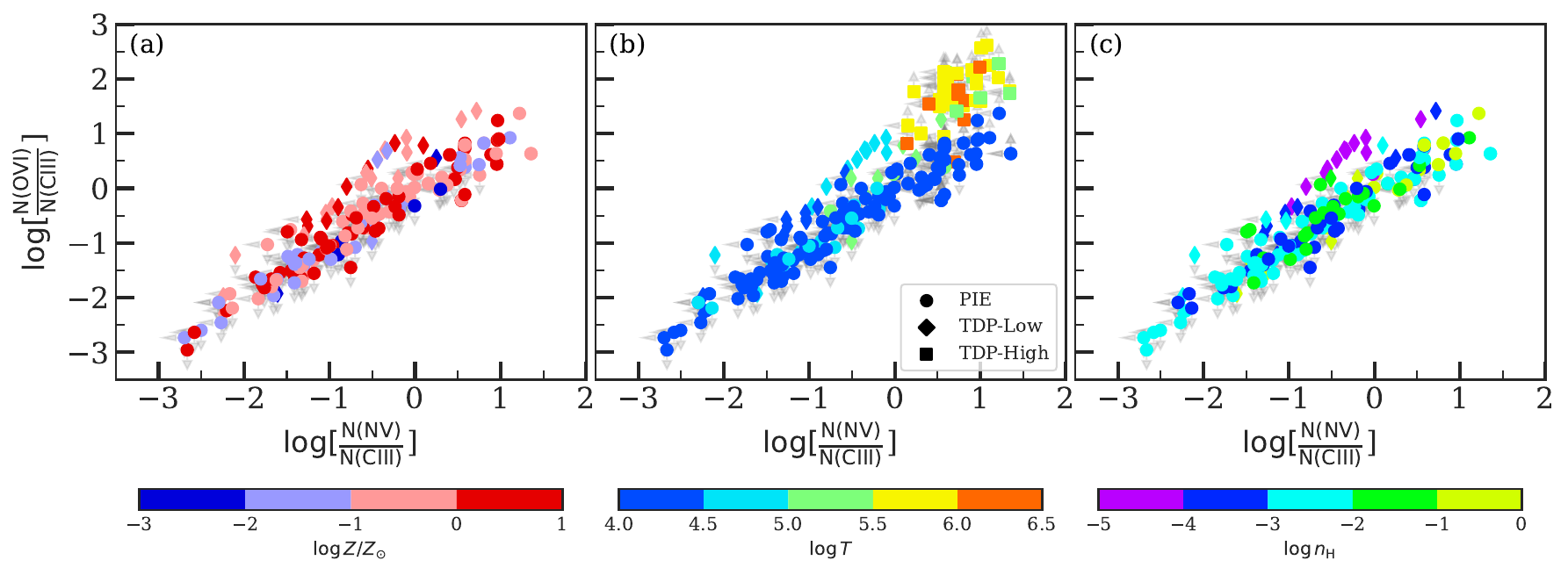}
}
\caption{Column density ratio of {\ovi}/{\ciii} plotted against {\nv}/{\ciii} for the three different cloud types. PIE clouds are shown as filled circles, TDP--Low clouds are shown as filled diamonds, and TDP--High clouds are shown as squares, and the data points are coloured by their (a) metallicity; (b) temperature; and (c) hydrogen number density. The TDP--Low clouds, with temperatures in the range $4.7 <$ {\temp} $< 5.1$ and metallicities in the range $-2.0 <$ {\metallicity} $< 0$,
have the lowest densities in our sample with {\hden} $< -4$. For such low-density clouds, the time to achieve collisional ionization equilibrium is longer, and this gives rise to the need for TDP.  We do not show the TDP--High clouds on the plots coloured by metallicity and density because these properties are not constrained for these clouds. The arrows are indicative of the upper limits on the column densities.
}
\label{fig:columnratios_cloudtypes}%
\end{figure*}

The TDP--Low clouds have considerable overlap with PIE clouds in the phase space of metallicity, density, and temperature. It is interesting to consider in what way they are out of thermal equilibrium in order to gain insight into their origins. Roughly half of the TDP--Low clouds have temperatures $4.7 \lesssim$ {\tempu} $\lesssim 5.1$, overlapping with PIE clouds but on the higher side, and metallicities $-2 <$ {\metallicity} $< 0$, also consistent with PIE clouds.  This can be seen in Fig.~\ref{fig:metallicity_temp}. The biggest difference between these TDP--Low clouds, with higher temperatures, and the PIE clouds is that the densities of many of these TDP--Low clouds are the lowest of clouds in our sample, only $-5 <$ {\hden} $< -4$ (see Fig.~\ref{fig:metallicity_doppler}b).  These densities correspond to large cloud thicknesses, in the range $10$--$100$~kpc, among the largest in our sample (see Fig.~\ref{fig:metallicity_size}). The Doppler parameters tend to be higher than those of the PIE sample because of higher temperatures (see Fig.~\ref{fig:metallicity_temp} and \ref{fig:metallicity_doppler}).  These larger Doppler parameters were constrained in our modelling process by the relative widths of lines from different ions.  Perhaps the higher temperatures are responsible for the non-equilibrium conditions.

\smallskip

In Fig.~\ref{fig:columnratios_cloudtypes}, we use the line of ratios of key intermediate and high ionization transitions as another way to distinguish the TDP--Low clouds from the PIE clouds. Most of the TDP--Low clouds have $\log$({\ovi}/{\ciii}) ratio $\sim0.5$--$1$ dex higher than the PIE clouds on a plot of $\log$({\ovi}/{\ciii}) vs. $\log$({\nv}/{\ciii}).  Fig.~\ref{fig:columnratios_cloudtypes}c shows that the TDP--Low clouds with the largest column densities of {\nv} and {\ovi} have the lowest densities, {\hden} $ < -4$, lower than any other clouds in our sample.  For such low-density clouds, the time to achieve collisional ionization equilibrium is longer, and this gives rise to the need for TDP. This same low-density TDP--Low population includes those clouds with the highest temperatures among the TDP--Low clouds, in the interval $4.7 \lesssim$ {\tempu} $\lesssim 5.1$. Especially at these intermediate temperatures, cooling occurs rapidly, often without sufficient time to attain equilibrium. The other TDP--Low clouds have temperatures similar to PIE clouds, which implies some of them may be in or close to PIE.  They were modelled with TDP only because they had detected {\civ}, but we found that in many of them the temperature needed to produce observed line ratios turned out to be consistent with equilibrium conditions. 

\subsubsection{TDP--High Clouds}

The TDP--High clouds are distinguished from the other classes by their broad {\ovi} profiles and higher derived temperatures in the range $5.5 <$ {\temp} $< 6.0$.  
In a few clouds, a discernable contribution to the {\lya} profile is seen, but in most, the contribution to {\lya} is negligible or dominated by other clouds in a saturated velocity range. The metallicities, densities, and sizes of these clouds are not constrained, as we discuss below.  

\smallskip

Though our models allowed for non-equilibrium collisional ionization and for contributions from time-dependent photoionization, most of these {\ovi} clouds are in collisional ionization equilibrium.  In order to have a departure from equilibrium, the collisional ionization time scale would have to be short compared with the timescale for recombination. The resulting non-equilibrium, ``overionized'' condition occurs only for low densities, {\hden} $< -4$ for collisional ionization timescales longer than 10 million years~\citep{Oppenheimer2013}. In Fig.~\ref{fig:ionfracs}, we show that the ionization fraction of {\ovi} diverges from the CIE expectation only at {\temp}$ \lesssim 5.2$, and especially for high metallicities ($Z/Z_{\odot} > 1$). At those lower temperatures, the cooling times are short enough to leave the gas ``overionized''.  In Fig.~\ref{fig:columnratios_cloudtypes}b, we see that {\ovi}/{\ciii} for the TDP--High clouds is elevated by roughly an order of magnitude over the values for PIE clouds with the same large values of {\nv}/{\ciii}.  This is also consistent with the origin of {\ovi} in collisionally ionized material at a higher temperature. 

\smallskip

In the absence of constraints on cloud sizes, an important question is whether the collisionally ionized {\ovi} absorbing gas found through the TDP--High clouds is characteristic of the entire galaxy halo or whether this {\ovi} arises in substructures within the halo. Such substructures could be small regions, heated by local processes, or could occupy a larger fraction of the halo (i.e., tens to 100~kpc). To understand the origin of these higher temperature regions giving rise to the TDP--High clouds, we can compare to numerical simulations and to known structures in the Milky Way galaxy and local universe. 

\smallskip

Fig.~2 in~\citet{Hafen2024} shows the temperature-vs-density distribution of CGM gas in FIRE-2 cosmological simulations of Milky Way-like galaxies.  A substantial phase space density is present at the temperatures found for the collisionally ionized {\ovi} clouds, and these clouds are at over an order of magnitude higher temperatures than given by the balance of heating and cooling.  \citet{Appleby2023} have studied the physical conditions of CGM gas at low redshift using the SIMBA simulation~\citet{dave2019}.  They find that {\ovi} absorbers arise in gas with a temperature distribution that peaks at $\log (T/{\rm K}) = 5.5$, with at least 80\% collisionally ionized, though there is a tail to lower temperatures with an origin in photoionized gas. They find that the majority of the {\ovi} absorption arises from the inner CGM and characterize it as arising in WHIM or hot halo gas. This would imply larger, more diffuse structures.

\smallskip

\begin{figure*}
\centerline{
\includegraphics[scale=0.7]{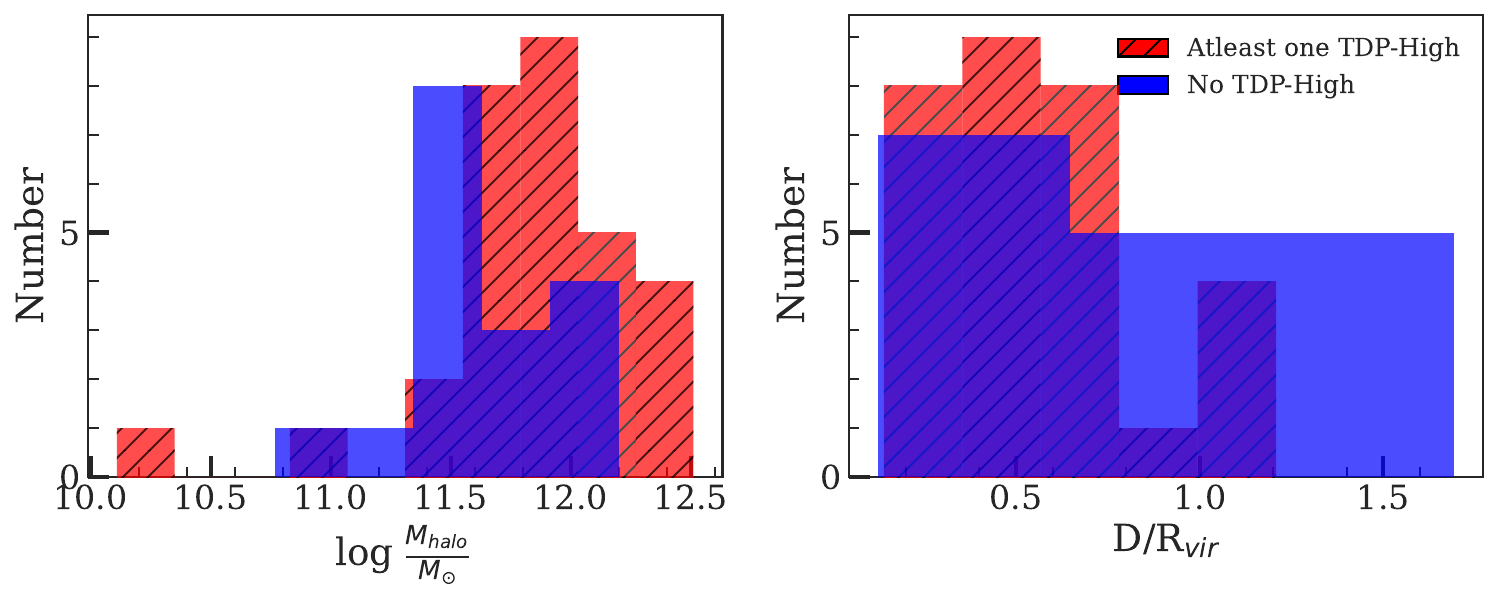}
}
\caption{Histograms showing (a): halo masses for absorption systems with at least one TDP--High (red) and systems with no TDP--High clouds (blue), (b): normalized impact parameter for absorption systems with at least one TDP--High (red) and systems with no TDP--High clouds (blue).}
\label{fig:tdphigh_hists}%
\end{figure*}

A favoured temperature for the presence of {\ovi} absorption predicts a dependence of {\ovi} on halo mass, as seen in a version of the \textsc{eagle} simulations that includes non-equilibrium ionization~\citep{Oppenheimer2016}. These simulations show maximum {\ovi} column densities for $L^{\star}$ galaxies with halo mass around $10^{12}$~{\msun}.  More massive galaxies or groups have higher virial temperatures which promotes much of the oxygen to {\ovii} and {\oviii}. This halo mass preference for {\ovi} absorption is seen observationally in the analyses of \citet{Pointon2017} and \citet{ng2019}, which compare galaxy properties to the measured column densities of {\ovi}.  \citep{nielsen2017highly,Kacprzak2019ApJ} also confirm that there are no obvious kinematic signatures in the {\ovi} profiles that would signal contributions of outflow, inflow, or co-rotation.  This is consistent with {\ovi} arising in gas distributed over much of the galaxy halo.

\smallskip

We can more definitively confirm the presence of {\ovi} halos around $L^{\star}$ galaxies by examining the galaxy properties for TDP--High clouds in our sample. Fig.~\ref{fig:tdphigh_hists}a compares the mass distributions for sightlines through the 30 galaxies with one or more TDP--High cloud to the 17 sightlines through galaxies that have no detected TDP--High absorption. The interquartile range for the mass of galaxies without any TDP--High cloud is $\log M_{\rm halo}/M_{\odot} = 11.4-11.9$, lower than the range of 11.7--12.1 for galaxies with at least one TDP--High cloud. We perform a 2-sample Anderson-Darling test to test the null hypothesis that the galaxies with detected TDP--High absorption and the galaxies without detected TDP--High absorption are drawn from the same population. We find that the null hypothesis can be rejected at the 5\% level ($p=0.022$). Fig.~\ref{fig:tdphigh_hists} (b) shows the distributions of impact parameters comparing galaxies with at least one TDP--High cloud to galaxies without TDP--High clouds. We find that these two samples are also significantly different from one another ($p=0.038$). This shows that several of the sightlines without TDP--High clouds are passing outside of the virial radius of their galaxies. We conclude that it is more likely for isolated, low redshift, more massive ($\sim L^{\star}$) galaxies to be surrounded by collisionally ionized {\ovi} halos extending to $0.6$--$1.2$ virial radii. 

\smallskip

Using semi-analytic models, \citet{richter2020} explored expectations for coronal broad {\lya} lines from galaxy halos and predicted $\log (N({\hi})/{\rm cm}^{-2}) = 12.4-13.4$ and $b({\hi})$ of 70--200~{\kms}.  Although the {\lya} is not often detected because of the presence of other phases and the $S/N$ of the data, these values are consistent with the expectations for our TDP--High clouds (see Figs.~\ref{fig:metallicity_NHI_D}a and \ref{fig:metallicity_doppler}a).

\smallskip

Within the coronal {\ovi} halos of $L^{\star}$ galaxies, and in the lower mass galaxies in our sample, there can also be substructures that produce {\ovi} absorption.  
In the specialized simulations of~\citet{Mandelker2020}, the {\ovi} absorption arises in the turbulent mixing layer formed as a cold filamentary stream falls through the hot CGM.  In such cases, the {\ovi} is thus arising in kpc-scale structures, and the surrounding medium is too hot to give rise to {\ovi} absorption.  These would be more analogous to the TDP--Low clouds in our sample, those with $\log (T/{\rm K}) \sim 5$ (Fig.~\ref{fig:metallicity_temp}a), which have {\ovi} along with lower ionization transitions due to over-ionization.  

\smallskip

There is also evidence that the TDP--High clouds are large substructures within halos.  They must be large because, for massive enough galaxies within $0.75 R_{\rm vir}$, the covering factor for TDP--High clouds is close to unity in our sample.  In addition, we find that 15 of the 30 sightlines with one or more TDP--High clouds have two or more of these clouds along the sightline. So these galaxies, with multiple components, have kinematically distinct substructures within the hot halos that give rise to the broad {\ovi} lines.  In other cases, there could be kinematically overlapping substructures that produce the {\ovi} absorption from different regions within the halo.  Fundamentally, there are likely to be many segments of gas along the sightline through the whole halo, with relatively low densities and temperatures consistent with the virial temperature, that shape the TDP--High {\ovi} profiles that we observe.

\smallskip

The most detailed view that we have of structures like those that may be responsible for the {\ovi} clouds (TDP--High and TDP--Low) is in the Local Group. For example,~\citet{Krishnarao2022} published a focused study of the corona of the Large Magellanic Cloud (LMC). Their analysis provides evidence for a {\temp}$\sim 5.5$ corona around the LMC within the larger {\temp}$\sim 6$ Milky Way corona. 
Cooler clouds ({\temp} $\sim 4$) lie within the LMC corona, and {\civ} and {\siiv} absorption is produced in interface layers between these cool clouds and the LMC corona.  
Generalizing those measurements of the LMC corona, there are likely to be many substructures in the CGM of the galaxies in our sample in the form of dwarf galaxies, and other infalling structures. There is also an extended {\ovi} corona surrounding Milky Way (and Andromeda) analogues in the High-resolution Environmental Simulations of The Immediate Area (HESTIA) model, with predicted {\ovi} similar to the galaxies in our sample \citet{Damle2022}.

\smallskip

We conclude that the TDP--High clouds are collisionally ionized.  They are prevalent in $\sim L^{\star}$ galaxies and absent in some of the lower-mass galaxies in our sample.  Their temperatures are consistent with the virial temperature for these halos, and their kinematics are consistent with expectations for one or two substructures of order the halo size.  Our CMBM analysis has made it possible to detect the diffuse halo of most isolated, low redshift galaxies, separating it from other phases through the broad {\ovi} absorption that it produces.

\section{CONCLUSION}
\label{sec:summary}

We inferred CGM absorption properties of 47, $z < 0.7$, isolated galaxy-absorber pairs from the ``Multiphase Galaxy Halos'' Survey using a cloud-by-cloud multiphase Bayesian ionization modelling approach. The metallicity of the CGM was compared with intrinsic absorber properties, including neutral hydrogen column density, temperature, Doppler parameter, hydrogen number density, and inferred line of sight thickness. We then compared the metallicity of the CGM to galaxy properties such as the galaxy inclination, azimuthal angle, impact parameter, halo mass, colours, and redshift. The galaxy inclinations and azimuthal angles were obtained by modelling the galaxies in {\hst} images with GIM2D by \citetalias{Pointon2019}. Our main findings are summarized here:

\begin{enumerate}
    \item We identify three main populations of CGM absorbers based on the type of models needed to best explain the observed multiphase absorption. The 47 absorption systems are found to comprise 240 clouds. The temperature distribution of the 157 cool PIE clouds is right-skewed with a median of $T \sim$ 10$^{4.1}$ K and a tail extending to $T \sim$ 10$^{5}$ K. The temperature distribution of the 35 warm-hot TDP--Low and 48 hotter TDP--High clouds appears to be flat between $T \sim $10$^{4.2-5.2}$ K) and ($T \sim $10$^{5.4-6.0}$ K).

    \smallskip

    \item The metallicity for the PIE clouds ranges between  $-2.6 \lesssim$ {\metallicity} $\lesssim 0.8$, with a median metallicity of {\metallicity} = $-0.41 \pm 0.09$. The TDP--Low clouds show a median metallicity of $-0.59 \pm 0.19$, and their metallicity ranges between $-2.3 \lesssim$ {\metallicity} $\lesssim 0.2$. The TDP--High clouds are found to be in CIE and thus have unconstrained metallicities. 

    \smallskip

    \item The {\hi} column density range of cool clouds (PIE) spans several decades 11.8 $\lesssim$  {\colden} $\lesssim 19.7$, and decreases as the impact parameter increases. The {\hi} column density of warm-hot (TDP--Low) clouds ranges between $12.8 \lesssim$ {\colden} $\lesssim 17.2$, and the {\hi} column density of hotter phases (TDP--High) ranges between $12.0 \lesssim$ {\colden} $\lesssim 15.2$.

    \smallskip

    \item The cool clouds are well explained by models that assume photoionization equilibrium (PIE). In agreement with theoretical predictions of cool clouds forming by the process of fragmentation, we find that the $L \sim 0.1$pc/$n_{\rm H}$ relationship is consistent with being a lower bound in PIE cloud thickness at every {\hden}.

    \smallskip
    
    \item About half of the warm-hot clouds, referred to as TDP--Low clouds, are found to be affected by time-dependent photoionization. These clouds are different from PIE clouds in their large ratios (0.5$-$1 dex higher) of $\log$({\ovi}/{\ciii}). The remaining TDP--Low clouds have properties similar to PIE clouds, and these clouds may be in or close to PIE. 

    \smallskip
    
    \item The hotter clouds, referred to as TDP--High clouds, are distinguished from other classes by their broad {\ovi} profiles. Though our models allowed for non-equilibrium ionization, most of the {\ovi} clouds are found to be close to collisional ionization equilibrium at temperatures consistent with the virial temperatures of their galaxy halos.  The lowest mass galaxies and sightlines at high impact parameters do not produce TDP--High clouds, but the $T \sim$ 10$^{5.7}$ K gas that produces them is abundant in $L^{\star}$ galaxy halos.

\smallskip

\item A bimodality is observed in the azimuthal distribution of the total number of clouds and mean number of clouds per absorption system. More clouds are seen along minor and major axes ($\Phi \lesssim 22.5\degree$ $\cup$ $\Phi \gtrsim 67.5\degree$), compared with interjacent azimuthal angles ($22.5\degree \lesssim \Phi \lesssim 67.5\degree$).  

\medskip

\item The metallicity is not related to the azimuthal angle. Additionally, the inclination angle, impact parameter, $B-K$ colours, subpopulations of the {\hi} column density, or the absorber velocities with respect to the galaxy central velocity, have no bearing on this connection. The lack of a link between CGM metallicity and azimuthal angle in this low-redshift study challenges a simple model of the CGM with outflows along the minor axis and inflows along the major axis.  A variety of other processes (e.g., disk material and recycled accretion) are contributing and there has been time for mixing of the different cloud populations in these galaxies.

\end{enumerate}

\section*{Acknowledgements}

We thank the referee for the detailed and thoughtful comments that improved the manuscript. We thank Nicolas Lehner and Christopher Howk for helpful discussions during the course of this work. S, J.C.C, B.W, acknowledges the support for this work through the grant number HST-AR-16607. Computations for this research were performed on the Pennsylvania State University's Institute for Computational and Data Sciences' Roar supercomputer. The authors also acknowledge the Texas Advanced Computing Center (TACC) at The University of Texas at Austin for providing HPC resources that have contributed to the research results reported within this paper. Parts of this research were supported by the Australian Research Council Centre of Excellence for All Sky Astrophysics in 3 Dimensions (ASTRO 3D), through project number CE170100013. A.N acknowledges the support for this work through grant number EMR/2017/002531 from the Department of Science and Technology, Government of India. We are grateful to Gary Ferland and collaborators for
developing the {\CLOUDY} photoionization code. We also thank
Orly Gnat for making the computational data on radiatively
cooling models public. This research made use of Astropy, a community-developed core
Python package for Astronomy~\citep{astropy2013}, NumPy~\citep{harris2020}, \textsc{matplotlib}~\citep{Hunter2007}, and SciPy~\citep{virtanen2020}. This research has made use of
the HSLA database, developed and maintained at STScI, Baltimore,
USA. Some of the data presented herein were obtained at the W. M. Keck Observatory, which is operated as a scientific partnership among the California Institute of Technology, the University of California and the National Aeronautics and Space Administration. Observations were supported by Swinburne Keck programs 2014A\_W178E, 2014B\_W018E, 2015\_W018E and 2016A\_W056E. The Observatory was made possible by the generous financial support of the W. M. Keck Foundation.  The authors wish to recognize and acknowledge the very significant cultural role and reverence that the summit of Maunakea has always had within the indigenous Hawaiian community.  We are most fortunate to have the opportunity to conduct observations from this mountain. This research has made use of the NASA/IPAC Extragalactic
Database (NED), which is operated by the Jet Propulsion Laboratory,
California Institute of Technology, under contract with NASA. We acknowledge the work of people involved in the design, construction and deployment of the COS on-board the Hubble Space Telescope, and thank all those who obtained data for the sight-lines studied in this paper. We thank KITP for hosting the Fundamentals of
Gaseous Halos workshop (supported by the NSF under Grant No. NSF PHY-1748958) during which some of the ideas presented in this work were formulated.

\section*{Data Availability}

The data underlying this article will be shared on reasonable request to the corresponding author.



\bibliographystyle{mnras}
\bibliography{references} 

\bsp
 



\appendix
\label{appendix}
\clearpage
\section{Galaxy Observations and Properties}
\onecolumn
\begin{landscape}
\begin{center}
\scriptsize
\begin{longtable}{lllcccccccccccl}
\caption{Galaxy Observations and Properties \label{tab:obsgal}}\\
\hline
\multicolumn{1}{c}{(1)}&
\multicolumn{1}{c}{(2)}&
\multicolumn{1}{c}{(3)}&
\multicolumn{1}{c}{(4)}&
\multicolumn{1}{c}{(5)}&
\multicolumn{1}{c}{(6)}&
\multicolumn{1}{c}{(7)}&
\multicolumn{1}{c}{(8)}&
\multicolumn{1}{c}{(9)}&
\multicolumn{1}{c}{(10)}&
\multicolumn{1}{c}{(11)}&
\multicolumn{1}{c}{(12)}&
\multicolumn{1}{c}{(13)}&
\multicolumn{1}{c}{(14)}&
\multicolumn{1}{c}{(15)}\\
\multicolumn{1}{c}{J-Name}&
\multicolumn{1}{c}{Cross ID}&
\multicolumn{1}{c}{$z_{\rm gal}$}&
\multicolumn{1}{c}{Ref.\tnote{a}}&
\multicolumn{1}{c}{RA}&
\multicolumn{1}{c}{DEC}&
\multicolumn{1}{c}{$D$}&
\multicolumn{1}{c}{$B-K$}&
\multicolumn{1}{c}{$\Phi$}&
\multicolumn{1}{c}{$i$}&
\multicolumn{1}{c}{$\log M_{h}/M_{\odot}$}&
\multicolumn{1}{c}{{\rvir}}&
\multicolumn{1}{c}{$D/R_{vir}$}&
\multicolumn{1}{c}{$V_{c}(D)$}&
\multicolumn{1}{c}{{\lsfr}, Galaxy Type, Reference}  \\
\multicolumn{1}{c}{}&
 \multicolumn{1}{c}{}&
\multicolumn{1}{c}{}&
\multicolumn{1}{c}{}&
\multicolumn{1}{c}{(J2000)}&
\multicolumn{1}{c}{(J2000)}&
\multicolumn{1}{c}{(kpc)}&
\multicolumn{1}{c}{}&
\multicolumn{1}{c}{$({\degree})$}&
\multicolumn{1}{c}{$({\degree})$}&
\multicolumn{1}{c}{}&
\multicolumn{1}{c}{(kpc)}&
\multicolumn{1}{c}{}&
\multicolumn{1}{c}{({\kms})}&
\multicolumn{1}{c}{(yr$^{-1}$)}\\
\hline
J0125 & PKS0122-00 & $0.3985$ & $1$ & $01:25:27.67$ & $-00:05:31.39$ & $163.0 \pm 0.1$ & $1.8$ & $73.4_{-4.7}^{+4.6}$ & $63.2_{-2.6}^{+1.7}$ & $12.5_{-0.1}^{+0.2}$ & $270.2_{-18.0}^{+16.9}$ & $0.6_{-0.0}^{+0.0}$ & $242.0_{-28.6}^{+27.3}$ & -10.0, SF, MZ15\\[2pt]
J0351 & 3C95 & $0.2617$ & $2$ & $03:51:28.93$ & $-14:29:54.31$ & $188.6 \pm 0.3$ & $2.3$ & $64.9_{-15.8}^{+21.1}$ & $83.0_{-3.0}^{+2.0}$ & $11.6_{-0.2}^{+0.4}$ & $137.4_{-23.3}^{+11.1}$ & $1.4_{-0.1}^{+0.2}$ & $99.0_{-27.4}^{+16.0}$ & {\nodata}\\[2pt]
J0351 & 3C95 & $0.3570$ & $1$ & $03:51:27.89$ & $-14:28:57.88$ & $72.3 \pm 0.4$ & $0.3$ & $4.9_{-4.9}^{+33.0}$ & $28.5_{-12.5}^{+19.8}$ & $12.0_{-0.2}^{+0.3}$ & $185.7_{-21.5}^{+14.1}$ & $0.4_{-0.0}^{+0.0}$ & $174.8_{-35.3}^{+26.3}$ & {\nodata}\\[2pt]
J0407 & PKS0405-12 & $0.1534$ & $3$ & $04:07:43.93$ & $-12:12:08.49$ & $195.9 \pm 0.1$ & {\nodata} & $26.3_{-1.0}^{+0.9}$ & $49.5_{-0.7}^{+0.5}$ & $11.9_{-0.2}^{+0.3}$ & $191.3_{-23.6}^{+15.2}$ & $1.1_{-0.1}^{+0.1}$ & $142.6_{-29.1}^{+20.9}$ & {\nodata}, SF, J13\\[2pt]
J0407 & PKS0405-12 & $0.3422$ & $3$ & $04:07:48.48$ & $-12:12:11.13$ & $172.0 \pm 0.1$ & {\nodata} & $48.1_{-0.9}^{+1.0}$ & $85.0_{-0.4}^{+0.1}$ & $11.6_{-0.2}^{+0.4}$ & $139.5_{-22.7}^{+11.3}$ & $1.2_{-0.1}^{+0.2}$ & $107.3_{-28.6}^{+17.3}$ & {\nodata}, SF, J15\\[2pt]
J0407 & PKS0405-12 & $0.4952$ & $4$ & $04:07:49.02$ & $-12:11:20.76$ & $107.6 \pm 0.4$ & {\nodata} & $21.0_{-3.7}^{+5.3}$ & $67.2_{-7.5}^{+7.6}$ & $11.4_{-0.2}^{+0.5}$ & $111.8_{-19.1}^{+8.9}$ & $0.9_{-0.1}^{+0.1}$ & $97.3_{-27.4}^{+16.1}$ & {\nodata}, SF, J13\\[2pt]
J0456 & PKS0454-22 & $0.2784$ & $2$ & $04:56:09.66$ & $-21:59:03.930$ & $50.7 \pm 0.5$ & $0.5$ & $78.4_{-2.1}^{+2.1}$ & $71.2_{-2.6}^{+2.6}$ & $11.4_{-0.2}^{+0.5}$ & $124.5_{-23.7}^{+10.0}$ & $0.4_{-0.0}^{+0.1}$ & $116.5_{-37.2}^{+21.0}$ & {\nodata}\\[2pt]
J0456 & PKS0454-22 & $0.3815$ & $1$ & $04:56:08.82$ & $-21:59:27.400$ & $103.4 \pm 0.3$ & $1.8$ & $63.8_{-2.7}^{+4.3}$ & $57.1_{-2.4}^{+19.9}$ & $12.0_{-0.2}^{+0.3}$ & $183.9_{-21.3}^{+14.0}$ & $0.5_{-0.0}^{+0.1}$ & $167.2_{-33.1}^{+24.4}$ & {\nodata}\\[2pt]
J0456 & PKS0454-22 & $0.4828$ & $5$ & $04:56:08.91$ & $-21:59:29.000$ & $108.0 \pm 0.5$ & $1.7$ & $85.2_{-3.7}^{+3.7}$ & $42.1_{-3.1}^{+3.1}$ & $12.3_{-0.1}^{+0.2}$ & $219.1_{-17.0}^{+13.5}$ & $0.4_{-0.0}^{+0.0}$ & $207.1_{-28.5}^{+24.1}$ & {\nodata}\\[2pt]
J0853 & US1867 & $0.1635$ & $2$ & $08:53:33.38$ & $43:49:03.97$ & $26.2 \pm 0.1$ & $1.8$ & $56.0_{-0.8}^{+0.8}$ & $70.1_{-0.8}^{+1.4}$ & $11.9_{-0.2}^{+0.3}$ & $183.4_{-24.0}^{+14.5}$ & $0.2_{-0.0}^{+0.0}$ & $174.9_{-43.0}^{+30.8}$ & {\nodata}\\[2pt]
J0853 & US1867 & $0.2766$ & $2$ & $08:53:36.88$ & $43:49:33.32$ & $179.4 \pm 0.2$ & $0.6$ & $36.7_{-15.3}^{+14.9}$ & $32.8_{-6.7}^{+5.7}$ & $11.6_{-0.2}^{+0.4}$ & $139.8_{-23.2}^{+11.3}$ & $1.3_{-0.1}^{+0.2}$ & $103.3_{-28.0}^{+16.7}$ & {\nodata}\\[2pt]
J0853 & US1867 & $0.4402$ & $2$ & $08:53:35.16$ & $43:48:59.81$ & $58.1 \pm 0.4$ & $1.8$ & $23.0_{-7.6}^{+6.5}$ & $73.3_{-3.0}^{+3.8}$ & $11.9_{-0.2}^{+0.3}$ & $173.0_{-18.6}^{+12.4}$ & $0.3_{-0.0}^{+0.0}$ & $168.2_{-32.2}^{+24.3}$ & {\nodata}\\[2pt]
J0914 & SDSSJ091440.40+282330.0 & $0.2443$ & $1$ & $09:14:41.75$ & $+28:23:51.18$ & $105.9 \pm 0.1$ & $1.0$ & $18.2_{-1.0}^{+1.1}$ & $39.0_{-0.2}^{+0.4}$ & $11.9_{-0.2}^{+0.3}$ & $176.8_{-23.1}^{+14.0}$ & $0.6_{-0.0}^{+0.1}$ & $152.6_{-33.4}^{+23.3}$ & -9.4, SF, W13\\[2pt]
J0943 & SDSSJ094331.60+053131.0 & $0.1431$ & $6$ & $09:43:29.21$ & $+05:30:41.75$ & $154.2 \pm 0.1$ & $2.5$ & $77.7_{-0.1}^{+0.1}$ & $75.5_{-0.1}^{+0.1}$ & $12.2_{-0.2}^{+0.2}$ & $227.3_{-23.0}^{+16.6}$ & $0.8_{-0.1}^{+0.1}$ & $185.0_{-31.6}^{+24.7}$ & $< -12.5$, P, W13\\[2pt]
J0943 & SDSSJ094331.60+053131.0 & $0.2284$ & $6$ & $09:43:33.78$ & $+05:31:22.26$ & $123.3 \pm 0.1$ & $1.9$ & $30.4_{-0.4}^{+0.3}$ & $52.3_{-0.3}^{+0.3}$ & $12.2_{-0.2}^{+0.2}$ & $227.3_{-21.3}^{+15.7}$ & $0.6_{-0.0}^{+0.1}$ & $197.3_{-31.6}^{+25.3}$ & -10.1, SF, W13\\[2pt]
J0943 & SDSSJ094331.60+053131.0 & $0.3531$ & $1$ & $09:43:30.67$ & $+05:31:18.08$ & $96.5 \pm 0.3$ & $1.0$ & $8.2_{-5.0}^{+3.0}$ & $44.4_{-1.2}^{+1.1}$ & $11.7_{-0.2}^{+0.4}$ & $143.3_{-22.8}^{+11.7}$ & $0.7_{-0.1}^{+0.1}$ & $126.1_{-33.4}^{+20.9}$ & -9.9, SF, W13\\[2pt]
J0950 & HS0946+4845 & $0.2119$ & $1$ & $09:50:00.86$ & $+48:31:02.59$ & $93.6 \pm 0.2$ & $2.4$ & $16.6_{-0.1}^{+0.1}$ & $47.7_{-0.1}^{+0.1}$ & $12.4_{-0.2}^{+0.2}$ & $260.6_{-19.3}^{+17.2}$ & $0.4_{-0.0}^{+0.0}$ & $235.9_{-31.3}^{+28.7}$ & $< -11.7$, P, W13\\[2pt]
J1004 & PG1001+291 & $0.1380$ & $2$ & $10:04:02.35$ & $+28:55:12.50$ & $56.7 \pm 0.2$ & $0.8$ & $12.4_{-2.9}^{+2.4}$ & $79.1_{-2.1}^{+2.2}$ & $10.9_{-0.2}^{+0.6}$ & $84.6_{-19.3}^{+6.7}$ & $0.7_{-0.1}^{+0.2}$ & $69.7_{-25.9}^{+12.9}$ & {\nodata}, SF, J17\\[2pt]
J1009 & SDSSJ100902.10+071344.0 & $0.2279$ & $1$ & $10:09:01.57$ & $+07:13:28.00$ & $64.0 \pm 0.8$ & $1.4$ & $89.6_{-1.3}^{+0.4}$ & $66.3_{-0.9}^{+0.6}$ & $11.8_{-0.2}^{+0.4}$ & $162.2_{-23.5}^{+13.4}$ & $0.4_{-0.0}^{+0.1}$ & $149.0_{-36.7}^{+24.7}$ & -9.2, SF, W13\\[2pt]
J1041 & 4C06.41 & $0.3153$ & $7$ & $10:41:16.85$ & $+06:10:06.13$ & $54.0 \pm 0.5$ & $2.2$ & $77.3_{-1.2}^{+1.2}$ & $72.6_{-1.3}^{+1.3}$ & $11.6_{-0.2}^{+0.4}$ & $135.7_{-22.5}^{+11.5}$ & $0.4_{-0.0}^{+0.1}$ & $128.0_{-36.0}^{+22.8}$ & {\nodata}\\[2pt]
J1041 & 4C06.41 & $0.4422$ & $1$ & $10:41:17.80$ & $+06:10:18.97$ & $56.2 \pm 0.3$ & $2.8$ & $4.3_{-1.0}^{+0.9}$ & $49.8_{-5.2}^{+7.4}$ & $12.0_{-0.2}^{+0.3}$ & $178.2_{-18.5}^{+12.8}$ & $0.3_{-0.0}^{+0.0}$ & $173.5_{-32.4}^{+25.2}$ & {\nodata}\\[2pt]
J1119 & PG1116+215 & $0.1383$ & $8$ & $11:19:06.67$ & $+21:18:29.56$ & $138.0 \pm 0.2$ & $2.2$ & $34.4_{-0.4}^{+0.4}$ & $26.4_{-0.4}^{+0.8}$ & $12.2_{-0.2}^{+0.2}$ & $242.1_{-20.7}^{+16.8}$ & $0.6_{-0.0}^{+0.1}$ & $203.2_{-29.8}^{+25.4}$ & {\nodata}, SF, K17\\[2pt]
J1133 & SDSSJ113327.78+032719.1 & $0.1546$ & $4$ & $11:33:28.21$ & $+03:26:59.00$ & $55.6 \pm 0.1$ & $1.1$ & $56.1_{-1.3}^{+1.7}$ & $23.5_{-0.2}^{+0.4}$ & $11.6_{-0.2}^{+0.4}$ & $151.9_{-24.2}^{+12.4}$ & $0.4_{-0.0}^{+0.1}$ & $139.2_{-37.3}^{+23.5}$ & -9.8, SF, W13\\[2pt]
J1139 & HE1136-1334 & $0.1755$ & $2$ & $11:39:10.53$ & $-13:49:48.59$ & $163.0 \pm 0.5$ & {\nodata} & $21.4_{-10.7}^{+10.7}$ & $85.0_{-0.2}^{+0.2}$ & $11.2_{-0.2}^{+0.6}$ & $104.3_{-22.5}^{+8.1}$ & $1.7_{-0.1}^{+0.4}$ & $69.5_{-24.2}^{+12.0}$ & {\nodata}\\[2pt]
J1139 & HE1136-1334 & $0.2042$ & $1$ & $11:39:11.52$ & $-13:51:08.69$ & $93.2 \pm 0.3$ & $2.3$ & $5.8_{-0.5}^{+0.4}$ & $83.4_{-0.5}^{+0.4}$ & $11.7_{-0.2}^{+0.4}$ & $155.0_{-24.2}^{+12.7}$ & $0.6_{-0.1}^{+0.1}$ & $132.4_{-34.3}^{+21.7}$ & {\nodata}\\[2pt]
J1139 & HE1136-1334 & $0.2123$ & $1$ & $11:39:09.53$ & $-13:51:31.46$ & $174.8 \pm 0.1$ & $2.1$ & $80.4_{-0.5}^{+0.4}$ & $85.0_{-0.6}^{+5.0}$ & $11.7_{-0.2}^{+0.4}$ & $159.4_{-24.3}^{+13.1}$ & $1.2_{-0.1}^{+0.2}$ & $119.4_{-29.9}^{+19.0}$ & {\nodata}\\[2pt]
J1139 & HE1136-1334 & $0.2197$ & $4$ & $11:39:08.33$ & $-13:50:45.64$ & $122.0 \pm 0.2$ & $2.1$ & $44.9_{-8.1}^{+8.9}$ & $85.0_{-8.5}^{+5.0}$ & $11.0_{-0.2}^{+0.6}$ & $93.6_{-20.7}^{+7.2}$ & $1.4_{-0.1}^{+0.3}$ & $66.9_{-23.9}^{+11.7}$ & {\nodata}\\[2pt]
J1139 & HE1136-1334 & $0.3193$ & $1$ & $11:39:09.80$ & $-13:50:53.08$ & $73.3 \pm 0.4$ & $1.6$ & $39.1_{-1.7}^{+1.9}$ & $83.4_{-1.1}^{+1.4}$ & $11.9_{-0.2}^{+0.3}$ & $169.2_{-22.7}^{+13.4}$ & $0.4_{-0.0}^{+0.1}$ & $156.8_{-36.0}^{+24.8}$ & {\nodata}\\[2pt]
J1219 & PG1216+069 & $0.1241$ & $8$ & $12:19:23.46$ & $+06:38:19.84$ & $93.4 \pm 5.3$ & $1.2$ & $67.2_{-22.8}^{+22.8}$ & $22.0_{-21.8}^{+18.7}$ & $11.9_{-0.2}^{+0.3}$ & $183.1_{-24.6}^{+14.5}$ & $0.6_{-0.1}^{+0.1}$ & $157.2_{-35.9}^{+24.9}$ & {\nodata}, SF, K18\\[2pt]
J1233 & LBQS1230-0015 & $0.3188$ & $4$ & $12:33:04.08$ & $-00:31:40.20$ & $88.9 \pm 0.2$ & $1.1$ & $17.0_{-2.3}^{+2.0}$ & $38.7_{-1.8}^{+1.6}$ & $11.9_{-0.2}^{+0.3}$ & $175.9_{-22.3}^{+14.0}$ & $0.5_{-0.0}^{+0.1}$ & $159.5_{-34.4}^{+24.6}$ & -10.0, SF, W13\\[2pt]
J1241 & SDSSJ124154.00+572107.0 & $0.2053$ & $1$ & $12:41:53.73$ & $+57:21:00.94$ & $21.1 \pm 0.1$ & $1.2$ & $77.6_{-0.4}^{+0.3}$ & $56.4_{-0.5}^{+0.3}$ & $11.6_{-0.2}^{+0.4}$ & $149.2_{-23.7}^{+12.2}$ & $0.2_{-0.0}^{+0.0}$ & $144.9_{-42.9}^{+27.8}$ & -9.5, SF, W13\\[2pt]
J1241 & SDSSJ124154.00+572107.0 & $0.2179$ & $4$ & $12:41:52.41$ & $+57:20:43.28$ & $94.6 \pm 0.2$ & $1.3$ & $63.0_{-2.1}^{+1.8}$ & $17.4_{-1.6}^{+1.4}$ & $11.6_{-0.2}^{+0.4}$ & $146.2_{-23.8}^{+11.9}$ & $0.7_{-0.1}^{+0.1}$ & $123.7_{-33.3}^{+20.4}$ & -10.0, SF, W13\\[2pt]
J1244 & PG1241+176 & $0.5504$ & $2$ & $12:44:11.04$ & $+17:21:05.05$ & $21.2 \pm 0.3$ & $1.3$ & $20.1_{-19.1}^{+16.7}$ & $31.7_{-4.8}^{+16.2}$ & $11.8_{-0.2}^{+0.3}$ & $149.8_{-18.3}^{+11.2}$ & $0.1_{-0.0}^{+0.0}$ & $144.4_{-37.4}^{+26.9}$ & {\nodata}, P, Z19\\[2pt]
J1301 & PG1259+593 & $0.1967$ & $2$ & $13:01:20.12$ & $+59:01:35.72$ & $135.5 \pm 0.1$ & $1.6$ & $39.7_{-2.2}^{+2.8}$ & $80.7_{-3.2}^{+4.3}$ & $11.4_{-0.2}^{+0.5}$ & $120.7_{-24.2}^{+9.6}$ & $1.2_{-0.1}^{+0.2}$ & $89.2_{-29.0}^{+15.3}$ & {\nodata}\\[2pt]
J1319 & TON153 & $0.6610$ & $9$ & $13:19:55.77$ & $+27:27:54.84$ & $103.9 \pm 0.5$ & $1.4$ & $86.6_{-1.2}^{+1.5}$ & $65.8_{-1.2}^{+1.2}$ & $12.2_{-0.1}^{+0.2}$ & $184.5_{-14.2}^{+11.2}$ & $0.5_{-0.0}^{+0.0}$ & $183.3_{-25.0}^{+21.2}$ & {\nodata}\\[2pt]
J1322 & SDSSJ132222.70+464535.0 & $0.2144$ & $1$ & $13:22:22.47$ & $+46:45:45.98$ & $38.6 \pm 0.2$ & $1.7$ & $13.9_{-0.2}^{+0.2}$ & $57.9_{-0.2}^{+0.1}$ & $12.1_{-0.2}^{+0.2}$ & $216.5_{-21.9}^{+15.8}$ & $0.2_{-0.0}^{+0.0}$ & $203.8_{-38.7}^{+30.9}$ & -11.0, SF, W13\\[2pt]
J1342 & HE1340-0038 & $0.0708$ & $6$ & $13:42:50.00$ & $-00:53:28.88$ & $39.4 \pm 0.5$ & {\nodata} & $13.9_{-0.2}^{+0.2}$ & $57.7_{-0.3}^{+0.3}$ & $11.4_{-0.2}^{+0.5}$ & $126.1_{-25.3}^{+10.0}$ & $0.4_{-0.0}^{+0.1}$ & $117.5_{-39.3}^{+21.5}$ & {\nodata}\\[2pt]
J1342 & HE1340-0038 & $0.2013$ & $6$ & $13:42:52.23$ & $-00:53:43.10$ & $31.8 \pm 0.2$ & $2.1$ & $44.5_{-0.3}^{+0.1}$ & $71.6_{-0.2}^{+0.3}$ & $11.7_{-0.2}^{+0.4}$ & $151.7_{-24.2}^{+12.4}$ & $0.2_{-0.0}^{+0.0}$ & $147.0_{-41.3}^{+26.6}$ & -10.1, SF, W13\\[2pt]
J1342 & HE1340-0038 & $0.2270$ & $1$ & $13:42:51.86$ & $-00:53:54.07$ & $35.3 \pm 0.2$ & $1.3$ & $13.2_{-0.4}^{+0.5}$ & $0.1_{-0.1}^{+0.6}$ & $12.4_{-0.2}^{+0.2}$ & $265.2_{-18.6}^{+17.5}$ & $0.1_{-0.0}^{+0.0}$ & $239.8_{-35.3}^{+33.9}$ & -11.0, P, W13\\[2pt]
J1357 & PKS1354+19 & $0.4295$ & $2$ & $13:57:03.29$ & $+19:18:44.41$ & $157.9 \pm 1.5$ & $1.7$ & $8.7_{-1.4}^{+1.6}$ & $85.0_{-1.7}^{+5.0}$ & $11.5_{-0.2}^{+0.4}$ & $122.0_{-20.1}^{+9.8}$ & $1.2_{-0.1}^{+0.2}$ & $96.4_{-26.0}^{+15.5}$ & {\nodata}\\[2pt]
J1357 & PKS1354+19 & $0.4592$ & $2$ & $13:57:04.53$ & $+19:19:15.15$ & $45.5 \pm 0.7$ & $1.4$ & $64.2_{-13.8}^{+13.6}$ & $24.7_{-6.5}^{+5.7}$ & $11.7_{-0.2}^{+0.3}$ & $143.9_{-19.1}^{+11.2}$ & $0.3_{-0.0}^{+0.0}$ & $142.6_{-33.5}^{+23.5}$ & {\nodata}\\[2pt]
J1547 & 3C323.1 & $0.0949$ & $2$ & $15:47:45.56$ & $+20:51:41.37$ & $79.8 \pm 0.5$ & $1.0$ & $54.7_{-2.4}^{+2.0}$ & $80.9_{-2.0}^{+1.8}$ & $10.8_{-0.2}^{+0.6}$ & $79.5_{-18.0}^{+6.3}$ & $1.1_{-0.1}^{+0.3}$ & $58.1_{-21.2}^{+10.4}$ & {\nodata}, SF, J17\\[2pt]
J1555 & SDSSJ155504.40+362847.0 & $0.1892$ & $1$ & $15:55:05.29$ & $+36:28:48.46$ & $33.4 \pm 0.1$ & $1.2$ & $47.0_{-0.8}^{+0.3}$ & $51.8_{-0.7}^{+0.7}$ & $12.1_{-0.2}^{+0.3}$ & $208.7_{-22.7}^{+15.1}$ & $0.2_{-0.0}^{+0.0}$ & $196.7_{-40.3}^{+30.6}$ & -9.9, SF, W13\\[2pt]
J1704 & 3C351.0 & $0.0921$ & $2$ & $17:04:34.33$ & $+60:44:47.59$ & $93.6 \pm 0.5$ & {\nodata} & $53.1_{-0.6}^{+0.6}$ & $72.0_{-0.5}^{+0.5}$ & $11.5_{-0.2}^{+0.5}$ & $137.3_{-25.2}^{+11.0}$ & $0.8_{-0.1}^{+0.1}$ & $110.5_{-33.2}^{+18.6}$ & {\nodata}\\[2pt]
J2131 & PKS2128-12 & $0.4302$ & $1$ & $21:31:35.63$ & $-12:06:58.56$ & $48.4 \pm 0.2$ & $2.1$ & $14.9_{-4.9}^{+6.0}$ & $48.3_{-3.7}^{+3.5}$ & $12.0_{-0.2}^{+0.2}$ & $186.1_{-18.7}^{+13.4}$ & $0.2_{-0.0}^{+0.0}$ & $180.9_{-33.5}^{+26.8}$ & {\nodata}\\[2pt]
J2137 & PKS2135-14 & $0.0752$ & $2$ & $21:37:45.08$ & $-14:32:06.27$ & $70.9 \pm 0.7$ & {\nodata} & $73.2_{-0.5}^{+1.0}$ & $71.0_{-1.0}^{+0.9}$ & $11.4_{-0.2}^{+0.5}$ & $129.8_{-25.6}^{+10.4}$ & $0.6_{-0.1}^{+0.1}$ & $109.3_{-35.4}^{+19.1}$ & {\nodata}\\[2pt]
J2253 & 3C454.3 & $0.3528$ & $1$ & $22:54:00.41$ & $+16:09:06.82$ & $203.2 \pm 0.5$ & $1.3$ & $88.7_{-4.8}^{+1.3}$ & $36.7_{-4.6}^{+6.9}$ & $11.9_{-0.2}^{+0.3}$ & $176.3_{-22.4}^{+14.0}$ & $1.1_{-0.1}^{+0.1}$ & $138.6_{-29.2}^{+20.6}$ & {\nodata}\\[2pt]\\
\hline
\end{longtable}
 \footnotesize
Columns (1): SDSS J-Name of the quasar, (2): Cross ID of the quasar,
(3): Galaxy redshift, (4): Galaxy redshift reference: $(1)$ \citet{Kacprzak2019ApJ}, $(2)$ \citet{chen2001origin}, $(3)$ \citet{Johnson2013}, $(4)$ P19, $(5)$ \citet{Kacprzak2011}, $(6)$ \citet{werk2012}, $(7)$ \citet{lanzetta1995}, $(8)$ \citet{Prochaska2011}, $(9)$ \citet{Kacprzak2012b}, (5) \& (6): RA and DEC in SDSS J2000 equatorial coordinates for the quasar, (7): Impact parameter of the sightline, (8): Galaxy color in $B - K$ magnitude, (9): Azimuthal angle of the galaxy, $\phi$ = 0$\degree$ corresponds to the major axis of the galaxy. We note that the uncertainties in the azimuthal angles of face-on galaxies ($i<20${\degree}) are statistical errors derived from modelling in GIM2D. It is possible that the systematic errors are larger, (10): Inclination angle of the galaxy, $i$ = 0$\degree$ implies the galaxy is oriented face-on along our line of sight, (11): Halo mass of the galaxy in log units of solar mass, (12): Virial radius of the galaxy, (13): Impact parameter scaled by the virial radius, (14): Galaxy circular velocity measured at the impact parameter, (15):  when available we give the {\lsfr}, Galaxy Type, and reference for this information. For galaxy type, SF refers to a star-forming galaxy, and P refers to a passive galaxy. A galaxy is considered star-forming when its {\lsfr} is $\geq -11$~\citep{tumlinson2011large}. The references are MZ15 -- \citet{muzahid2015extreme}, J13 -- \citet{Johnson2013}, J15 -- \citet{johnson2015possible}, J17 -- \citet{johnson2017}, W13 -- \citet{werk2013cos}, K17 -- \citet{keeney2017characterizing}, K18 -- \citet{keeney2018galaxy}, Z19 -- \citet{zahedy2019}. A portion of this table (columns 1, 3--11) is reproduced from \citetalias{Pointon2019} for the convenience of the reader. 
\end{center}

\end{landscape}
\twocolumn

\section{Notes on Ionization Modelling}
\label{appendix:notes_ionizationmodelling}

\subsection{Table of Cloud-by-cloud properties of the absorption systems}
\label{appendix:cloudbycloud}
\onecolumn
\begin{landscape}
\begin{center}
\begin{longtable}{llllllllllllc}
\caption{Cloud-by-cloud properties of the absorption systems}\\
\hline
      \multicolumn{1}{c}{(1)}     &
          \multicolumn{1}{c}{(2)}     &
          \multicolumn{1}{c}{(3)}     &
          \multicolumn{1}{c}{(4)}     &
          \multicolumn{1}{c}{(5)}     &
          \multicolumn{1}{c}{(6)}	    &
          \multicolumn{1}{c}{(7)}	    &
          \multicolumn{1}{c}{(8)}     &
           \multicolumn{1}{c}{(9)}     &
            \multicolumn{1}{c}{(10)}    &
            \multicolumn{1}{c}{(11)}     &
             \multicolumn{1}{c}{(12)}     &
            \multicolumn{1}{c}{(13)}     \\
          \multicolumn{1}{c}{J-Name}                  &
          \multicolumn{1}{c}{$z_{\rm gal}$}           &
          \multicolumn{1}{c}{$V$}           &
          \multicolumn{1}{c}{{\metallicity}}           &
          \multicolumn{1}{c}{{\hdenu}}           &
        \multicolumn{1}{c}{{\totalcoldenu}}           &
          \multicolumn{1}{c}{{\coldenu}}           &
          \multicolumn{1}{c}{{\tempu}}           &
          \multicolumn{1}{c}{{\thicknessu}} &
          \multicolumn{1}{c}{{\bturb}} &
          \multicolumn{1}{c}{{\btherm}} &
          \multicolumn{1}{c}{{\bnet}} &
          \multicolumn{1}{c}{{Ionization Model}} \\
          \multicolumn{1}{c}{}                &
          \multicolumn{1}{c}{}                &
        \multicolumn{1}{c}{{\kms}}                &
          \multicolumn{1}{c}{}                &
          \multicolumn{1}{c}{}                &
          \multicolumn{1}{c}{}                &
          \multicolumn{1}{c}{}                &
          \multicolumn{1}{c}{}                &
          \multicolumn{1}{c}{} & 
          \multicolumn{1}{c}{{\kms}} & 
          \multicolumn{1}{c}{{\kms}} & 
          \multicolumn{1}{c}{{\kms}} &
          \multicolumn{1}{c}{} \\
\hline
\end{longtable}
\footnotesize
 Properties of the different gas phases that contribute to the absorption towards different sightlines. Notes: (1) J-Name of the quasar; (2) Redshift of the galaxy; (3) Velocity of the cloud w.r.t the galaxy, sorted in increasing order; (4) cloud metallicity; (5) cloud total hydrogen volume density; (6) cloud total hydrogen column density; (7) cloud hydrogen column density; (8) cloud temperature; (9) cloud line of sight thickness; (10) cloud non-thermal Doppler broadening parameter (11) cloud thermal Doppler broadening parameter measured for {\hi}; (12) cloud total Doppler broadening parameter measured for {\hi}; (13) type of ionization model for the cloud; PIE - Photoionized Equilibrium, TDP - Time Dependent Photoionized (see Section~\ref{sec:methodology}). The marginalised posterior values of model parameters are given as the median along with the upper and lower bounds corresponding to the 16--84 percentiles. The median and the 1-sigma credible interval are derived from fractional weighted quantiles. The weighting is done by sample probability following \citet{Buchner2016}. The lower and upper limits correspond to $\mu_{1/2} - 2\sigma$ and $\mu_{1/2} + 2\sigma$, respectively.
\end{center}
\end{landscape}
\twocolumn

\subsection{The absorption associated with the z = 0.3985 galaxy towards the quasar J0125}
\label{sec:J0125_0.3985}

A system plot of this absorber is presented in Fig.~\ref{fig:J0125_0.3985}. This system shows absorption in the low ionization transitions of {\siii}, {\mgii}, {\cii}, and {\nii}, and the intermediate ionization transition of {\siiii}. We also see absorption in the higher ionization transitions of {\nv} and {\ovi}. The {\civ} and {\siiv} are covered by the G225M grating but do not provide very useful constraints because of the poor {\sn} of the data. 

\smallskip

The {\cii}~$\lambda$1036 is blended with {\niv}$\lambda$765 line of the $z \approx$ 0.8950 absorber at $\approx$ 150 {\kms}. The contributions of {\cii}~$\lambda$1036 and {\ovi}~$\lambda$1037 to one another within this system are masked. The {\siii}~$\lambda$1190 is contaminated with {\siiii} of the $z \approx$ 0.3791 absorber at $\approx$ $-$160 {\kms}. The {\siii}~$\lambda$1193 is contaminated with {\lya} of the $z \approx$ 0.3726 absorber at $\approx$ $-$25 {\kms}, and by the Galactic {\alii} absorption between $\approx$ 200--300 {\kms}. The {\siii}~$\lambda$1260 is contaminated with {\ovi} of the $z \approx$ 0.6981 absorber between $\approx$ $-$150 -- $-$100 {\kms}, and with {\hi}~$\lambda$930 of the $z \approx$ 0.8950 absorber between $\approx$ 200 -- 300 {\kms}. The {\nii}~$\lambda$1083 is contaminated with {\siiii} of the $z \approx$ 0.2565 absorber at $\approx$ 0 {\kms}. The {\siiv}~$\lambda$1393 is contaminated with {\lya} of the $z \approx$ 0.6044 and 0.6048 absorbers between $\approx$ 150 -- 350 {\kms}. The {\civ}~$\lambda$1548 is contaminated with {\lya} of the $z \approx$ 0.7797, 0.7802, 0.7813, and 0.7820 absorbers between $\approx$ $-$200 -- 200 {\kms}. The {\civ}~$\lambda$1550 is contaminated with {\lya} of the $z \approx$ 0.7844 absorber at 150 {\kms}. The {\nv}~$\lambda$1238 is contaminated with {\lya} of the $z \approx$ 0.4267 at $\approx$ 325 {\kms}. The {\nv}~$\lambda$1242 is contaminated with {\siii} of the $z \approx$ 0.3791 absorber at $\approx$ 37 {\kms}, and by the {\lya} of the $z \approx$ 0.4314 at $\approx$ 350 {\kms}.

\smallskip

The low ionization absorption traced by {\mgii} resolves into three components at $\approx$ $-$10, 184, and 198 {\kms} in the high-resolution UVES data. The cloud at $\approx$ $-$10 {\kms} is a low metallicity cloud with {\metallicity} $\approx$ $-$1.4 and a density of $\approx$ $-$2.4. The clouds at 190 {\kms} and 200 {\kms} have metallicities consistent with solar {\metallicity} $\approx$ 0 and 0.15, respectively, but contrasting densities of {\hden} $\approx$ $-$0.6 and $-$2.3, respectively. The densities of these clouds are well constrained owing to coincident absorption seen in {\mgi}, {\nii}, {\cii}, {\siii}, and {\siiii}, and the metallicities are constrained from the profile shapes of {\lyb} and {\lya}. The overproduction of absorption in {\caii} suggests depletion due to dust, however, it was not used as a constraint. 

\smallskip

The clouds at $\approx$ $-$63 {\kms} and $\approx$ 88 {\kms} are found to probe intermediate ionization gas traced by {\siiii}. The former cloud is determined to have a metallicity {\metallicity} $\approx$ $-$0.4 and a density {\hden} $\approx$ $-$2.8. The properties of this cloud are constrained by the non-detections in low ionization species and the profile shape of {\lyb}. The latter cloud is determined to have a metallicity of {\metallicity} $\approx$ 0 and a density {\hden} $\approx$ $-$2.6. There is also a cloud traced by the {\siiii} absorption at $\approx$ 216 {\kms} with a metallicity, {\metallicity} $\approx$ $-$0.4 and a density {\hden} $\approx$ $-$3.0. The metallicity of this cloud is constrained by the shape of the {\lyb} profile, and its density is constrained by the ratio of {\siii} and {\siiii}, and coincident non-detections of other low ionization species.

\smallskip

The higher ionization gas phase traced by the {\ovi} absorption resolves into five components centred at $\approx$ $-$64 {\kms}, 5 {\kms}, 87  {\kms}, 198 {\kms}, and 307 {\kms}. The components at $-$64 {\kms}, 5 {\kms}, and 198 {\kms} contribute as weak broad {\lya} absorbers with temperatures between {\temp} $\sim$ 5.5--5.8. The metallicities and densities of these clouds are unconstrained as indicated by their platykurtic posterior distributions. We adopt upper limits and lower limits for metallicity and density, and set them to the upper and lower bound of their priors, respectively. The {\ovi} clouds at 87 {\kms} and 307 {\kms} are at relatively lower temperatures {\temp} $\approx$ 5.2 and 4.6, respectively, owing to their narrow profile shape and also produce absorption seen in the noisy {\civ} profiles.

\smallskip

There is a {\lya}-only cloud with no associated metals at $-$115 {\kms}. There properties of this cloud are not well constrained owing to the lack of detected metal absorption. We adopt an upper limit for its metallicity and a lower limit for its density.

\subsection{The absorption associated with the z = 0.2617 galaxy towards the quasar J0351}
\label{sec:J0351_0.2617}

A system plot of this absorber is presented in Fig.~\ref{fig:J0351_0.2617}. This system does not show any metal line absorption except in {\ciii}. We mainly see absorption in {\hi} {\lya}. The absorption in {\hi} resolves into two components centred at $\approx$ $-91$ {\kms} and $-46$ {\kms}. The {\lyb} is contaminated by {\siiii} arising in the $z \approx$ 0.07192, 0.07218, and 0.07252 absorbers. The {\lyg} is also contaminated by {\cii}~$\lambda$903.6 and $\lambda$903.9 arising in the $z \approx$ 0.35667 absorber. The {\cii}~$\lambda$1334 line is blended with the {\lya} absorber at $z \approx$ 0.38517 at $\approx$ 20 {\kms}.

\smallskip

The metallicity and density of these clouds are unconstrained due to the lack of metal line absorption. The {\hi} column densities are relatively well constrained because of the access to unsaturated higher-order Lyman series lines.

\subsection{The absorption associated with the z = 0.3570 galaxy towards the quasar J0351}
\label{sec:J0351_0.3570}

A system plot of this absorber is presented in Fig.~\ref{fig:J0351_0.3570}. This system shows absorption in the low ionization transitions of {\mgii}, {\cii}, {\siii}, and {\nii}, and in the intermediate ionization transition of {\siiii}. We also see absorption in the high ionization transitions of {\civ}, {\nv}, and {\ovi}. 

\smallskip

The {\cii}~$\lambda$903.9 line is presented, and is overlapping with the {\cii}~$\lambda$903.6 line which is masked out. The {\cii}~$\lambda$1036 line is contaminated by the {\ciii}~$\lambda$977 line from the $z \approx$ 0.43981 absorber at $\approx$ 100 {\kms}. The  {\niii}~$\lambda$989 line is overlapping with the {\siiii}~$\lambda$989 at $\approx$ 70 {\kms}. {\nv}~$\lambda$1238 is contaminated by {\lya} from the $z \approx$ 0.38212 absorber at $-$160 {\kms}. The {\ovi}~$\lambda$1031 line is contaminated with {\lyg} of the $z \approx$ 0.43981 absorber between $-$70 {\kms} -- 70 {\kms}. The {\siii}~$\lambda$1190 is contaminated with {\lya} of the $z \approx$ 0.32849 absorber at $-$65 {\kms}.

\smallskip

The high resolution spectrum of {\mgii} resolves into three components at $\approx$ $-$75 {\kms}, $-$13 {\kms}, and 40 {\kms}. The properties of the gas phases traced by these three {\mgii} components are metallicities of {\metallicity} $\approx$ 0, $-$0.6, and 0, and densities of {\hden} $\approx$ $-2.2$, $-0.8$, and $-2.9$. These gas phases also explain all of the absorption seen in the {\siii}, {\cii}, and some of the absorption in {\SiIV} and {\civ}. The Lyman series constrains the metallicities of these phases, while the {\siiii} and other intermediate ionization states constrain the density of the gas. The strongest {\siiii} absorption seen at $-$26 {\kms} is found to arise in a separate phase and is needed to fill in the absorption seen in higher order Lyman series. The properties of this gas phase are \metallicity $\approx$ $-$0.7 and {\hden} $\approx$ $-$2.8. There is also some absorption seen in {\siiii} at $-$149 {\kms} that is needed to explain the coincident absorption seen in the Lyman series with a metallicity, {\metallicity} $\approx -0.4$ and density, {\hden} $\approx -2.6$. The {\ovi} at $\approx$ $-$48 {\kms} is at a temperature of {\temp} $\approx$ 5.1 and this gas phase also produces the absorption seen in {\civ} and {\nv} at coincident velocities. The metallicity of this cloud is {\metallicity} $\approx$ 0.2 and a density of {\hden} $\approx$ $-$2.7. The {\ovi} at $\approx$ $-$7 {\kms} produces weak broad absorption in {\lya}; this phase is at a higher temperature of {\temp} $\approx$ 5.7, and both its metallicity and density are unconstrained.

\subsection{The absorption associated with the z = 0.1534 galaxy towards the quasar J0407}
\label{sec:J0407_0.1534}

A system plot of this absorber is presented in Fig.~\ref{fig:J0407_0.1534}. We only see absorption in the {\lya} and {\lyb} lines, and do not see any metal line absorption. The absorption resolves into five {\lya} clouds. The metallicity and density are not constrained for these clouds, and we adopt upper limits and lower limits, respectively, for these parameters.

\subsection{The absorption associated with the z = 0.3422 galaxy towards the quasar J0407}
\label{sec:J0407_0.3422}

A system plot of this absorber is presented in Fig.~\ref{fig:J0407_0.3422}. This absorption system does not show any absorption in the low or intermediate ionization species. We only see absorption in {\ovi}, which resolves into two components, a broad component at $-$60 {\kms} and a relatively narrower component at 18 {\kms}. We also see absorption in {\lya} centred at $-65$ {\kms} and 22 {\kms}. We determine that the {\ovi} and {\hi} arise in separate phases. For {\hi}-only clouds, we adopt upper limits on their metallicities and lower limits on densities due to the absence of associated metal line absorption. The {\ovi} clouds are temperatures of {\temp} $\approx$ 5.8 and 5.4, with unconstrained metallicities and densities.

\subsection{The absorption associated with the z = 0.4952 galaxy towards the quasar J0407}
\label{sec:J0407_0.4952}

A system plot of this absorber is presented in Fig.~\ref{fig:J0407_0.4952}. This absorption system does not show any absorption in the low ionization species including {\mgii}, {\siii}, and {\oii}. We also do see absorption in the intermediate ionization species including {\oiii} and {\ciii}, and in the higher ionization species of {\oiv}, {\oxyv}, {\ovi}, and also in {\neviii}. The {\lya} in this system is covered by the G230-L grating, thus, the weak absorption does not show prominently in its spectrum.

\smallskip

The cloud centred at $-$30 {\kms} produces all of the absorption seen in intermediate ionization transitions of {\ciii} and {\oiii}, and some of the absorption in {\oiv}. The properties of this cloud are constrained to be {\metallicity} $\approx$ $-$1.3 and {\hden} $\approx$ $-$3.2. The metallicity is well constrained due to the absorption seen in {\lyb} and {\lyg}, and access to several higher Lyman series lines, and the non-detections constrain the density in lower ionization states. 

\smallskip

The absorption in {\ovi} resolves into two components at $-$17 {\kms} and 33 {\kms}. The cloud at $-$17 {\kms} also explains the absorption seen in {\oxyv} and most of the absorption seen in {\oiv}. The metallicity and density of these clouds are unconstrained with temperatures of {\temp} $\approx$ 5.4 and 5.5, respectively. 

\smallskip

There is modest absorption in {\neviii}, centred at $-$20 {\kms}, and is found to arise in a different phase from the {\ovi} phase; its metallicity and density are also unconstrained. The temperature of this phase is quite high at {\temp} $\approx$ 6.5. 

\smallskip

We also see absorption in {\lya} centred at 56 {\kms} which also produces modest absorption in {\oiv}. The properties of this cloud are metallicity, {\metallicity} $\approx$ $-1.7$ and density, {\hden} $\approx$ $-3.3$.

\subsection{The absorption associated with the z = 0.2784 galaxy towards the quasar J0456}
\label{sec:J0456_0.2784}

A system plot of this absorber is presented in Fig.~\ref{fig:J0456_0.2784}. This system does not show any absorption in the low or intermediate ionization states. However, we do see absorption in the higher ionization phase of {\civ}. The absorption resolves into two components at $\approx$ $-$100 {\kms} and $\approx$ $-$36 {\kms}. The blueward component is well constrained and has a metallicity, {\metallicity} $\approx$ $-$1.3, density {\hden} $\approx$ $-$3.5, and a high temperature of {\temp} $\approx$ 5.0. This phase also produces modest absorption in {\nv}. The redward component is needed to explain the asymmetry in the {\lya} profile. However, its properties are not well constrained, thus we adopt an upper limit on its metallicity and a lower limit on its hydrogen density.

\subsection{The absorption associated with the z = 0.3815 galaxy towards the quasar J0456}
\label{sec:J0456_0.3815}

A system plot of this absorber is presented in Fig.~\ref{fig:J0456_0.3815}. This system does not show absorption in low ionization transitions. We see absorption in intermediate and high ionization transitions. {\siiii} and {\civ} show absorption centred at $\approx$ 6 {\kms} and $\approx$ 76 {\kms}. The former component has a metallicity of {\metallicity} $\approx$ 0.5 and a density of {\hden} $\approx$ $-$3.4. The latter cloud also has very similar properties to the former one. 

\smallskip

The absorption in {\ovi} shows two components at $\approx$ $-4$ {\kms} and $\approx$ 100 {\kms}. The metallicity and density of both these components are not well constrained, while the temperatures are {\temp} $\approx$ 5.75 and 5.65 in the blueward and redward {\ovi} clouds, respectively.  

\smallskip

We also see a {\lya}-only cloud at $\approx$ 160 {\kms} whose properties are not well constrained. The metallicity is adopted as an upper limit, and hydrogen density is adopted as a lower limit. 

\subsection{The absorption associated with the z = 0.4828 galaxy towards the quasar J0456}
\label{sec:J0456_0.4828}

A system plot of this absorber is presented in Fig.~\ref{fig:J0456_0.4828}. This system shows absorption in low, intermediate, and high ionization transitions. The {\lyb} is contaminated by {\ovi} absorption arising in the $z$ = 0.47466 absorber at $\approx$ 10 {\kms}. The {\lya} from GHRS G200M grating has poor {\sn} hence it is entirely masked. It is displayed to indicate that the model is consistent with the data. The {\lya} is contaminated by {\di} of this same system at $\approx$ $-$150 {\kms}. 

\smallskip

The high resolution spectrum of {\mgii} resolves it into three components at $\approx$ 76 {\kms}, 109 {\kms}, and 130 {\kms}. This phase of gas also explains all of the absorption seen in {\cii} and {\siii}. The properties of the cloud at $\approx$ 75 {\kms} are {\metallicity} $\approx$ $-$0.1 and {\hden} $\approx$ $-$1.0. The cloud at 109 {\kms} is the strongest component among the three and is found to have a {\metallicity} $\approx$ $-$0.6 and {\hden} $\approx$ $-$0.9. It also produces absorption seen in coincident {\siii}, {\cii}, and {\feii}. The cloud at 130 {\kms} is found to have a {\metallicity} $\approx$ 0.3 and {\hden} $\approx$ $-$2.5, and also explains the concomitant absorption seen in {\cii} and {\siii}. 

\smallskip

The intermediate ionization phase traced by {\ciii} resolves into two components at $-$28 {\kms} and 107 {\kms}. The low ionization phase cannot entirely explain these two clouds of gas. The cloud at $-$28 {\kms} is found to explain the asymmetry in the profile of the Lyman series in the blueward {\hi} component. The properties of this gas phase are determined to be {\metallicity} $\approx$ $-$0.8 and {\hden} $\approx$ $-$2.8. The cloud at 107 {\kms} produces the saturated core of the Lyman series in the component centred at this velocity and also explains the asymmetric structure of {\civ} on the blueward side of its profile. The properties of this cloud are determined to be {\metallicity} $\approx$ 0 and {\hden} $\approx$ $-$2.5.

\smallskip

The {\civ} absorption was found to not exist in the same phase as {\ciii} in the component at 107 {\kms}, leading us to invoke an additional cloud found to be centred at $\approx$ 114 {\kms}. The {\siiv} which exists in a noisy part of the spectrum confirms the need for this cloud. The properties of this cloud are determined to be {\metallicity} $\approx$ $-$1.1 and {\hden} $\approx$ $-$2.7. 

\smallskip

The {\ovi} absorption centred at 112 {\kms} arises in a hot collisionally ionized phase at a temperature of {\temp} $\approx$ 6. The metallicity and density are not constrained for this gas phase. This absorption manifests as a broad {\lya} absorber.

\subsection{The absorption associated with the z = 0.1635 galaxy towards the quasar J0853}
\label{sec:J0456_0.1635}

A system plot of this absorber is presented in Fig.~\ref{fig:J0853_0.1635}. This system shows absorption in low and intermediate ionization transitions. We do not see absorption in {\civ} and {\ovi}. The {\lyb} is contaminated by molecular hydrogen H2:1~$\lambda$1108.6 line from the same system at $ z \approx$ $-$160 {\kms}. The {\ovi}~$\lambda$1031 line is contaminated by the Galactic {\NI} lines. The {\ovi}~$\lambda$1037 is contaminated by the Galactic {\siiii} line. 

\smallskip

The low ionization absorption resolves into three components centred at 31 {\kms}, 47 {\kms}, and 132 {\kms}, and is seen in {\cii}, {\siii}, {\nii}, and {\feii}. The metallicities and densities of these low ionization clouds are well constrained owing to the availability of several of the higher-order Lyman series lines. The properties of the cloud at 31 {\kms} are {\metallicity} $\approx$ $-$0.5 and {\hden} $\approx$ $-$2.4. In addition to producing some absorption in the low ionization transitions, this cloud also produces the bulk of the absorption seen in the coincident intermediate ionization transition of {\ciii}. The cloud at 47 {\kms} shows the strongest absorption in low ionization transitions and also explains the asymmetry of the {\ciii} profile shape. This cloud is also found to produce absorption in the neutral transitions of {\oi} and {\nitri} and is a sub-DLA with {\colden} $\approx$ 19.7. This is a low metallicity cloud with {\metallicity} $\approx$ $-$1.5 and {\hden} $\approx$ $-$2.6. The cloud at 132 {\kms} is found to produce all of the low and intermediate ionization transitions at the coincident velocity. The properties of this cloud are determined to be {\metallicity} $\approx$ 0.4 and {\hden} $\approx$ $-$2.7. 

\smallskip

There are also two clouds flanking the above three absorption components and are centred at $-$95 {\kms} and 229 {\kms}. These clouds are found to produce just the intermediate ionization gas traced by {\ciii}. The metallicity of the blueward cloud is {\metallicity} $\approx$ $-$1.2 and its density is $\approx$ $-$2.5. The redward cloud is found to have a metallicity of {\metallicity} $\approx$ 0.1, and its density is $\approx$ $-$3.5. 

\subsection{The absorption associated with the z = 0.2766 galaxy towards the quasar J0853}
\label{sec:J0853_0.2766}

A system plot of this absorber is presented in Fig.~\ref{fig:J0853_0.2766}. The {\lyb} is contaminated by {\lyg} of the $z =$ 0.34528 absorber between $-$320 {\kms} to $-$220 {\kms}. The {\ciii} is contaminated by {\lya} of the $z = $0.02452 absorber. All other blends are confidently masked out based on access to other transitions of a doublet.

\smallskip

This system shows absorption in intermediate ionization transitions of {\ciii} and {\siiii} only. The {\lya} resolves into four components centred at $\approx$ $-$513 {\kms}, $-$472 {\kms}, $-$321 {\kms}, and $-$278 {\kms}. The column densities of these four clouds are well constrained due to the access to several higher-order Lyman lines. The component at $-$321 {\kms} also shows absorption in {\ciii} and {\siiii}. The properties of this cloud are determined to be {\metallicity} $\approx$ 0.4 and density {\hden} $\approx$ $-$2.9. The properties of the other three clouds are upper limits on metallicity and lower limits on density. 

\subsection{The absorption associated with the z = 0.4402 galaxy towards the quasar J0853}
\label{sec:J0853_0.4402}

A system plot of this absorber is presented in Fig.~\ref{fig:J0853_0.4402}. The {\lya} is contaminated by a blend at 650 {\kms}. The {\ciii} is blended with {\lya} of the $z = $0.16033 absorber between 700 {\kms} to 775 {\kms}.  

\smallskip

This system shows absorption in low, intermediate, and high ionization transitions. The low and intermediate ionization transitions together tightly constrain the properties of the absorber at $\approx$ 690 {\kms}. With access to consecutive ionization states of oxygen, the density of this cloud is determined to be {\hden} $\approx$ $-$2.4 and its metallicity is constrained to be {\metallicity} $\approx$ $-$0.4, by access to several higher-order Lyman lines. We also see broad absorption in {\ovi} and weak absorption in {\neviii}, which manifests as a very weak coronal broad {\lya} absorber at a temperature of {\temp} $\approx$ 6.2, with unconstrained metallicity and density. 

\subsection{The absorption associated with the z = 0.2443 galaxy towards the quasar J0914}
\label{sec:J0914_0.2443}

A system plot of this absorber is presented in Fig.~\ref{fig:J0914_0.2443}. The {\ovi}~$\lambda$1031 line is blended with {\lya} from the $z = $0.05551 absorber at $-$150 {\kms}. The {\lyg} is blended with the {\lyb} at $z = $0.17906 between $\approx$ $-$250 
{\kms} -- $-$115{\kms}. We completely blank out the {\feii} and {\mgii} as they have very poor {\sn} coverage, but present here for completeness.

\smallskip

The strongest absorption seen in the Lyman series is centred at $\approx$ $-$17 {\kms}. This cloud produces absorption in the intermediate ionization transition of {\siiii} and also modest absorption in {\cii}. The metallicity of this cloud is determined to be {\metallicity} $\approx$ $-$0.2 and it has a density of $\approx$ $-$2.5. There is also absorption seen in {\niii} at $\approx$ $-$64 {\kms}, which is needed to explain the asymmetric absorption profiles of {\lya} and {\lyb}. The properties of this cloud are determined to be {\metallicity} $\approx$ $-$0.8 and it has a density of $\approx$ $-$3.7. 

\smallskip

The absorption in {\ovi} resolves into two components, centred at $-$112 {\kms} and $-$16 {\kms}. Both these clouds show up as weak {\lya} absorbers. The temperature for these clouds is {\temp} $\approx$ 5.8 and 5.5, respectively; their metallicities and densities, however, are unconstrained.

\subsection{The absorption associated with the z = 0.1431 galaxy towards the quasar J0943}
\label{sec:J0943_0.1431}

A system plot of this absorber is presented in Fig.~\ref{fig:J0943_0.1431}. This system is a {\hi} only absorber that shows absorption in {\lyb} that resolves into two components centred at $-$94 {\kms} and $-$45 {\kms}. The {\lya} is completely blanked out as it blended with {\lyb} absorption arising in the $z = $0.35421, 0.35456, and 0.35546 absorbers. Both the components seen in {\lyb} have upper limits in metallicity and lower limits in hydrogen number density.

\subsection{The absorption associated with the z = 0.2284 galaxy towards the quasar J0943}
\label{sec:J0943_0.2284}

A system plot of this absorber is presented in Fig.~\ref{fig:J0943_0.2284}. The {\ciii} is contaminated by Galactic {\nitri} between 100 {\kms} -- 160 {\kms}. 

\smallskip

The absorption in {\lya} shows two components centred at 165 {\kms} and 282 {\kms}. The blueward cloud shows absorption in low and intermediate ionization transitions. With access to multiple Lyman series lines, and detections in singly and doubly ionized states, the properties of this cloud are constrained to be {\metallicity} $\approx$ $-$0.4 and hydrogen density {\hden} $\approx$ $-$3.1. The redward cloud is an {\hi}-only cloud with an upper limit on metallicity and a lower limit on hydrogen number density. 

\smallskip

\subsection{The absorption associated with the z = 0.3531 galaxy towards the quasar J0943}
\label{sec:J0943_0.3531}

A system plot of this absorber is presented in Fig.~\ref{fig:J0943_0.3531}. The {\ciii} line is blended with {\lya} of the $z =$ 0.08814 absorber at $\approx$ 190 {\kms}. 

\smallskip

We do not see any absorption in the low ionization metal transitions. The absorption in {\ciii} resolves into five components. Based on our modelling, we find that a cloud at $\approx$ 250 {\kms|, which is partly blended with the {\lya} contamination, is needed to self-consistently explain the absorption seen in the Lyman series. The five absorption components are centred at $\approx$ 44 {\kms}, 250 {\kms}, 330 {\kms}, 408 {\kms}, and 522 {\kms}. We also see absorption in {\nv} and {\ovi} that is coincident in velocity with the blueward component in {\ciii}. The properties of these five clouds are {\metallicity} $\approx$ $-$0.9, $-$1.1, $-$1.4, $-$2.6, and 0, and their densities are $-$4.4, $-$2.8, $-$2.6, $-$3.0, and $-$2.6.

\subsection{The absorption associated with the z = 0.2119 galaxy towards the quasar J0950}
\label{sec:J0950_0.2119}

A system plot of this absorber is presented in Fig.~\ref{fig:J0950_0.2119}. This system shows absorption in the low, intermediate, and high ionization transitions. The {\ovi} $\lambda$1031 line is blended with the Galactic {\SI} line at $\approx$ 0 {\kms}. We mask out blending from the {\siii}$\lambda$989 to the redward side of the {\niii}$\lambda$989 line. 

\smallskip

The absorption in {\mgii} resolves into two components centred at $\approx$ $-$85 {\kms} and $-$50 {\kms}. Both these clouds are found to be contributing to the absorption in low and intermediate ionization states. The properties of the blueward cloud in {\mgii} are determined to be {\metallicity} $\approx$ $-$0.7 and {\hden} $\approx$ $-$2.0. The redward cloud shows {\metallicity} $\approx$ $-$1.1 and {\hden} $\approx$ $-$2.8. The redward cloud is also found to produce absorption in {\oi}. 

\smallskip

In addition to the clouds tracing {\mgii}, we also find absorption arising only in intermediate ions of {\ciii}, {\siiii}, and {\niii}. These clouds are most conspicuous in {\ciii} and centred at $\approx$ $-$169 {\kms}, $-$119 {\kms}, and 11 {\kms}. The flanking clouds show quite similar properties. The metallicities are {\metallicity} $\approx$ 0.1 and 0, and densities are {\hden} $\approx$ $-$2.9 and $-$2.7, respectively. The cloud at $-$119 {\kms} shows a {\metallicity} $\approx$ $-$1.9 and {\hden} $\approx$ $-$3.2.

\smallskip

We also observe absorption in {\ovi}, centred at $-$39 {\kms} at a temperature of {\temp} $\approx$ 5.7. The metallicity and density of this cloud are not constrained.

\subsection{The absorption associated with the z = 0.1380 galaxy towards the quasar J1004}
\label{sec:J1004_0.1380}

A system plot of this absorber is presented in Fig.~\ref{fig:J1004_0.1380}. The {\siiii}~$\lambda$1206 line is contaminated by {\lya} of the $z =$ 0.12916 at $-$65 {\kms}. 

\smallskip

This system does not show any absorption in low or intermediate ionization states. We see absorption in {\hi} transitions centred at $\approx$ $-$141 {\kms}. The properties of this cloud are not constrained, and we adopt an upper limit on its metallicity and a lower limit on its density. 

\smallskip

The higher ionization transition of {\ovi} shows absorption that resolves into two components, a broad one centred at $\approx$ $-$70 {\kms}, and a relatively narrow cloud at $-$45 {\kms}. The broad one is a high temperature cloud at {\temp} $\approx$ 5.8, but its metallicity and density are unconstrained. The narrow one is also found to produce modest amounts of intermediate ionization {\ciii} and explains the {\hi} component seen in {\lya} at the coincident velocity. The metallicity of this component is consistent with {\metallicity} $\approx$ $-$0.7, while its density is constrained to be {\hden} $\approx$ $-$4.9. 

\subsection{The absorption associated with the z = 0.2279 galaxy towards the quasar J1009}
\label{sec:J1009_0.2279}

A system plot of this absorber is presented in Fig.~\ref{fig:J1009_0.2279}. The {\cii}$\lambda$1036 line is blended with the {\feii}$\lambda$1142 line of the $z =$ 0.11408 absorber at 50 {\kms}. The {\cii}$\lambda$1334 line is blended with the {\civ}$\lambda$1548 line of the $z =$ 0.05867 absorber at 70 {\kms}. The {\siiii}~$\lambda$1206 line is blended with the {\lya} of the $z = $0.21913 absorber at $\approx$ 140 {\kms}. 

\smallskip

This system shows absorption in low, intermediate, and high ionization transitions. The high resolution spectrum of {\mgii} shows a very narrow component, $b_{\rm {\hi}} = $ 3 {\kms}, centred at $\approx$ $-$57 {\kms}. This cloud produces modest absorption in {\cii} and {\siii}. The properties of this cloud are {\metallicity} $\approx$ $-$0.8 and {\hden} $\approx$ $-$0.7. The absorption in {\siii} and {\cii} is not completely explained by the {\mgii} phase. Thus, we include an additional cloud in the low ionization phase to account for it. This cloud is centred at $-$40 {\kms} and also accounts for the absorption seen in {\siiii} and {\siiv} at coincident velocities. The properties of this cloud are {\metallicity} $\approx$ 0.8, {\hden} $\approx$ $-$3.3. 

\smallskip

The {\ovi} resolves into four components centred at $\approx$ $-$31 {\kms}, 19 {\kms}, 69 {\kms}, and 124 {\kms}. The blueward component at $\approx$ $-$31 {\kms} manifests as a CBLA with {\temp} $\approx$ 5.4, and also produces absorption in {\nv}. The cloud at 19 {\kms} is found to also produce {\cii}, {\siiii}, and {\nv}, and has the properties {\metallicity} $\approx$ $-$0.3, {\hden} $\approx$ $-$4.1, and a lower temperature of {\temp} $\approx$ 4.5. The clouds at 69 {\kms} and 124 {\kms} are seen in {\ovi} only with no other metals observed; both have a {\temp} $\approx$ 5.5, but their metallicities and densities are unconstrained.

\smallskip

We also see two clouds in {\lya} only centred at $\approx$ 125 {\kms} and 245 {\kms} with no associated metals. We adopt upper limits on their metallicity and lower limits on density.

\subsection{The absorption associated with the z = 0.3153 galaxy towards the quasar J1041}
\label{sec:J1041_0.3153}

A system plot of this absorber is presented in Fig.~\ref{fig:J1041_0.3153}. This system does not show low ionization metal absorption. We see absorption in the intermediate ionization transition of {\siiii} and in the high ionization transition of {\nv}. The properties of this cloud are determined to be metallicity, {\metallicity} $\approx$ 0, and density, {\hden} $\approx$ $-$3.8.

\subsection{The absorption associated with the z = 0.4422 galaxy towards the quasar J1041}
\label{sec:J1041_0.4422}

A system plot of this absorber is presented in Fig.~\ref{fig:J1041_0.4422}. The {\cii}$\lambda$1036 line is contaminated by {\oiii}$\lambda$833 absorption from the $z = $ 0.79310 system at $\approx$ 200 {\kms}, and the {\lya} is affected by the {\ciii} line from that same system at $-$150 {\kms}, however, the predicted model suggests that the contaminant is not affecting the profile shape significantly. The {\ovi} $\lambda$1037 is contaminated by {\cii}$\lambda\lambda$903.6, 903.9 lines arising in the $z =$ 0.65537 absorber between $-$100 {\kms} and 0~{\kms}. 

\smallskip

This system shows absorption in low, intermediate, and high ionization transitions. The high resolution spectrum of {\mgii} resolves it into seven components centred at $-$211 {\kms}, $-$187 {\kms}, $-$160 {\kms}, $-$143 {\kms}, $-$110 {\kms}, $-$86 {\kms}, and $-$66 {\kms}. The properties of these clouds vary from cloud to cloud. The metallicities are determined to be {\metallicity} $\approx$ $-$0.4, $-$2.0, 0.3, 0.3, $-$1.3, $-$0.1, and 0.1. The densities are determined to be {\hden} $\approx$ $-0.5$, $-2.4$, $-1.7$, $-2.3$, $-0.2$, $-0.9$, and $-2.8$. The densities of these clouds are constrained due to access to {\mgi} and other low ionization transitions. Although the {\hi} series is saturated, the $b$-parameters of {\mgii} constrain the $b$-parameters of {\hi} and yield well-constrained metallicities for many of these clouds. 

\smallskip

The intermediate ionization transitions of {\ciii}, {\siiii}, and {\niii} are found to arise in a predominantly different phase centred at $-131$ {\kms}. This phase of gas is also found to explain the {\ovi} absorption coincident in velocity. The properties of this cloud are constrained to be {\metallicity} $\approx$ 0.2, {\hden} $\approx$ $-3.2$, and a relatively lower temperature of {\temp} $\approx$ 4.4, accommodating both the intermediate and high ionization ions.

\smallskip

We find two more phases of gas flanking the intermediate ionization phase discussed above. These clouds are found to produce just the {\ovi} centred at $-193$ {\kms} and $-52$ {\kms}. The metallicity and density of these clouds are not constrained. These clouds are temperatures of {\temp} $\approx$ 5.5 and 5.7, respectively.

\subsection{The absorption associated with the z = 0.1383 galaxy towards the quasar J1119}
\label{sec:J1119_0.1383}

A system plot of this absorber is presented in Fig.~\ref{fig:J1119_0.1383}. The {\hi}$\lambda$923 line is contaminated by the Galactic molecular hydrogen H$_{2}$ L 4–0 P(1)$\lambda$1051.03 line at $\approx$ 20 {\kms}. The {\hi}$\lambda$949 line is contaminated by the Galactic molecular hydrogen H$_{2}$ L 2–0 P(2) $\lambda$1081.265 line at $\approx$ 5 {\kms}. The {\oiii}$\lambda$832 line is contaminated by H$_{2}$ L 14$-$0 R(2)$\lambda$948.47 line beyond $\approx$ 65 {\kms}, and we completely blank it out due to ambiguity in the amount of its contribution to the absorption of interest. 

\smallskip

This system shows absorption in low, intermediate, and high ionization states. The {\mgii} resolves into two components centred at $42$ {\kms} and 53 {\kms}. The properties of these clouds are well constrained and determined to be {\metallicity} $\approx$ 0.1 and $-0.2$, and their densities are {\hden} $\approx$ $-$2.6 and $-$2.8. These two clouds explain the absorption seen in all the low ionization transitions. 

\smallskip

The intermediate ionization transitions of {\siiii} and {\ciii}, however, are not fully explained by the low ionization phases. An additional gas phase centred at $\approx$ 53 {\kms} was needed to explain the absorption seen in the intermediate ionization transitions. This phase also seems to satisfactorily predict the {\oiii} at the coincident velocity, although it was completely blanked out. The properties of this phase are determined to be {\metallicity} $\approx$ $-$1.7 and {\hden} $\approx$ $-$3.2. 

\smallskip

We also see broad absorption in {\ovi}. The metallicity and density of this cloud are unconstrained; it is at a high temperature of {\temp} $\approx$ 6.0. This cloud manifests as a weak CBLA.

\subsection{The absorption associated with the z = 0.1546 galaxy towards the quasar J1133}
\label{sec:J1133_0.1546}

A system plot of this absorber is presented in Fig.~\ref{fig:J1133_0.1546}. This system shows absorption in low, intermediate, and high ionization transitions. The absorption is found to be traced by single-phase gas clouds centred at $\approx$ $-$119 {\kms} and $\approx$ $-$71 {\kms}. The low and intermediate transitions are well explained by these clouds. The properties of these clouds are weakly constrained, {\metallicity} $\approx$ $-$0.9 and 0.1, and their densities are {\hden} $\approx$ $-$2.1 and $-$2.7, respectively. 

\smallskip

We also see absorption in {\ovi} which is found to arise in a hot phase at {\temp} $\approx$ 6.1, and it manifests as a CBLA. The metallicity and density of this cloud are unconstrained.

\subsection{The absorption associated with the z = 0.1755 galaxy towards the quasar J1139}
\label{sec:J1139_0.1755}

A system plot of this absorber is presented in Fig.~\ref{fig:J1139_0.1755}. The {\lyb} is contaminated by {\cii}$\lambda$$\lambda$903.6, 903.9 lines of the $z =$ 0.33292 absorber between $-$365 {\kms}  and $-$170 {\kms}. The {\cii}$\lambda$1036 line is contaminated by the {\hi}$\lambda$923 line of the $z = $0.31947 absorber at $\approx$ $-$40 {\kms}.

\smallskip

This system shows modest metal line absorption in the intermediate ionization state of {\ciii} only. The absorption is seen in three distinct components in {\lya} centred at $-$273 {\kms}, $-$169 {\kms}, and $-$60 {\kms}. The blueward cloud is consistent with {\metallicity} $\approx$ -0.5 and {\hden} $\approx$ $-$4.0. The central cloud has an upper limit on metallicity and a lower limit on its density. The redward cloud has a metallicity consistent with solar, and its density is {\hden} $\approx$ $-$3.1.

\subsection{The absorption associated with the z = 0.2042 galaxy towards the quasar J1139}
\label{sec:J1139_0.2042}

A system plot of this absorber is presented in Fig.~\ref{fig:J1139_0.2042}. The {\hi}$\lambda$949 line is contaminated by Galactic {\feii} line at $\approx$ $-$110 {\kms}. The {\lyb} is contaminated by the  {\hi}$\lambda$926.2 line of the $z =  0.33292$ at $-$140 {\kms}. The {\ovi}$\lambda$1031 line is contaminated by the {\lyb} of the $z$ = 0.21202 absorber at $\approx$ 130 {\kms}. The {\ovi}$\lambda$1037 line is contaminated by the {\hi}$\lambda$937.8 line of the $z$ = 0.33266 absorber at $\approx$135 {\kms}.

\smallskip

This system shows intermediate and high ionization transitions. We also see modest absorption in {\cii}. The absorption in Lyman series lines is mainly arising in a phase traced by the intermediate {\ciii}, {\siiii}, and {\niii} ions. The absorption in {\ciii} resolves into three components centred at $-$7 {\kms}, 61 {\kms}, and 163 {\kms}. The metallicities of these clouds are {\metallicity} $\approx$ $-$0.5, 0.2, and $-$1.0, respectively. Their hydrogen densities are {\hden} $\approx$ $-$2.9, $-$3.2, and $-$2.6. The blueward cloud also produces weak absorption in {\cii}. The density of this cloud is constrained by the detections in low and intermediate ions, while its metallicity is constrained from access to {\lya} through {\lyd}. The central cloud shows absorption in the intermediate ions. The metallicity of this cloud is constrained by the absorption seen in the {\hi} component at $\approx$ 61 {\kms} and its density from low and intermediate ionization transitions. The redward cloud shows absorption in {\ciii} and {\siiii} without any concomitant absorption in lower ionization transitions. 

\smallskip

The absorption in {\ovi} resolves into two components at $\approx$ $-13$ {\kms} and 57 {\kms}. The cloud centred at 57 {\kms} manifests as a CBLA with a temperature of {\temp} $\approx$ 5.9. The cloud at $-13$ {\kms} has a temperature of {\temp} $\approx$ 5.6. The metallicity and density of these clouds are unconstrained. 

\smallskip

\subsection{The absorption associated with the z = 0.2123 galaxy towards the quasar J1139}
\label{sec:J1139_0.2123}

A system plot of this absorber is presented in Fig.~\ref{fig:J1139_0.2123}. The {\lyb} is contaminated by {\ovi} from the $z$ = 0.20415 absorber between $\approx$ $-$230 {\kms} and $-$ 120 {\kms}. The {\cii}$\lambda$1036 line is contaminated by the {\lya} of the $z$ = 0.03333 absorber at $-$25{\kms}. The {\ovi}$\lambda$1031 line is contaminated by Galactic {\SII} at $\approx$ $-$ 60{\kms}. The {\niii}$\lambda$989 is swamped by contamination.

\smallskip

This system does not show absorption in low ionization metal transition. We see absorption in {\ciii} and {\siiii} centred at $-$69 {\kms}. The properties of this phase are {\metallicity} $\approx$ $-$1.6 and {\hden} $\approx$ $-$3.1. We also see absorption in {\ovi} with {\temp} $\approx$ 5.4, and unconstrained metallicity and density.

\subsection{The absorption associated with the z = 0.3193 galaxy towards the quasar J1139}
\label{sec:J1139_0.3193}

A system plot of this absorber is presented in Fig.~\ref{fig:J1139_0.3193}. The {\cii}~903.9 line is contaminated by {\niii} from the $z$ = 0.20415 absorber at $\approx$ $-$100 {\kms}. The {\hi}$\lambda$926 and the {\hi}$\lambda$930 lines are contaminated by the {\hi}$\lambda$916.4 line and the {\hi}$\lambda$920.9 lines of the $z$ = 0.33292 absorber at $\approx$ $-$100 {\kms}. The {\ovi}$\lambda$1037 line is contaminated by the {\lya} of the $z$ = 0.40684 absorber at $\approx$ $-$100 {\kms}. 

\smallskip

This system shows modest absorption in the low ionization transition of {\cii}, but no absorption is seen in {\nii}. We see absorption in the intermediate ionization transition of {\ciii} and in the high ionization transition of {\ovi}. The absorption in {\ciii} resolves into two components centred at $\approx$ $-$103 {\kms} and $\approx$ 35 {\kms}. The blueward cloud has the properties of {\metallicity} $\approx$ 0.3 and {\hden} $\approx$ $-$2.7, and this cloud harbors most of the {\hi} seen at coincident velocity. The redward cloud has the properties of {\metallicity} $\approx$ 0.1 and {\hden} $\approx$ $-$4.5. The low density of the redward cloud is needed to explain both the {\ciii} and relatively narrow {\ovi} absorption at a temperature of {\temp} $\approx$ 4.6. 

\smallskip

We also see absorption in the high ionization transition of {\ovi} centred at $\approx$ $-$82 {\kms}. The blueward component manifests as a CBLA and has a temperature of {\temp} $\approx$ 5.9, while its density and metallicity are not constrained. The redward cloud showing absorption in {\ovi} is in the same phase as the {\ciii} discussed earlier. 

\smallskip

We also see absorption in {\hi} series with no associated metals centred at $\approx$ 51 {\kms}. The metallicity and density are upper and lower limits, respectively. 

\subsection{The absorption associated with the z = 0.1241 galaxy towards the quasar J1219}
\label{sec:J1219_0.1241}

A system plot of this absorber is presented in Fig.~\ref{fig:J1219_0.1241}. The {\nii}$\lambda$1083 line is contaminated by the {\lyd} of the $z$ = 0.28198 system blueward of $\approx$ $-$100 {\kms}. The {\lyg} and {\ciii}$\lambda$977 transitions are covered by the $FUSE$ instrument, however, with a comparatively low {\sn}. The {\lyg} is masked because of the availability of higher {\sn} {\lya} and {\lyb} profiles, but {\ciii} is not masked to better constrain the density of gas. The {\cii}$\lambda$1334 line is contaminated by the {\lya} of the $z$ = 0.23442 absorber at $\approx$ 90 {\kms}. 

\smallskip

This system shows absorption in intermediate and higher ionization transitions. The intermediate ionization transitions of {\ciii} and {\siiii} are found to arise in the same phase as {\civ}. The absorption in {\civ} resolves into four components centred at $-$130 {\kms}, $-$60 {\kms}, 145 {\kms}, and 200 {\kms}. The properties of these phases are determined to be {\metallicity} $\approx$ 0.1, $-$0.9, $-$1.2, and $-$1.9 and density $\approx$ $-$4.2, $-$4.8, $-$3.7, and $-$2.4. The {\ovi} in components centred at $-$130 {\kms} and $-$60 {\kms} also arises in the same phase as the {\civ}. 

\smallskip

The {\ovi} centred at 145 {\kms} and 200 {\kms} is found to arise in a much hotter phase with temperatures {\temp} $\approx$ 6.4 and 5.7, respectively, but their metallicity and density are not constrained. 

\smallskip

We find that in order to best explain the absorption seen in the Lyman series additional gas phases must exist. These phases of gas do not produce any noticeable metal line absorption. The properties of these {\hi}-only clouds are not well constrained. 

\subsection{The absorption associated with the z = 0.3188 galaxy towards the quasar J1233}
\label{sec:J1233_0.3188}

A system plot of this absorber is presented in Fig.~\ref{fig:J1233_0.3188}. The {\niii}$\lambda$989 line falls on the {\lya} emission of the Galaxy. The {\siii}$\lambda$1260 line is contaminated by {\lya} of the $z = $0.36757 absorber at $\approx$ 50 {\kms}. 

\smallskip

This system shows absorption only in the intermediate and high ionization transitions. The absorption in {\ciii}$\lambda$977 resolves into four components centred at $\approx$ $-$117 {\kms}, $-$73 {\kms}, $-$16 {\kms}, and 32 {\kms}. The properties of these components are determined to be {\metallicity} $\approx$ $-$0.9, 0.1, $-$0.2, and $-$0.2, and their densities are {\hden} $\approx$ $-$4.9, $-$4.8, $-3.8$, and $-$3.0, respectively. The low densities of the blueward clouds at $-$117 {\kms} and $-$73 {\kms} allow for the narrow {\ovi} absorption to arise in the same phase as {\ciii}. 

\smallskip

The {\ovi} shows broad absorption at a temperature of $\approx$ 5.6 which manifests as a weak CBLA. Neither its density nor its metallicity are constrained. 

\subsection{The absorption associated with the z = 0.2053 galaxy towards the quasar J1241}
\label{sec:J1241_0.2053}

A system plot of this absorber is presented in Fig.~\ref{fig:J1241_0.2053}. The {\hi}$\lambda$949 line is contaminated by Galactic {\feii}$\lambda$1144 at $-$50 {\kms}. The {\hi}$\lambda$926 line is contaminated by {\lyg} of the $z = $0.14765 absorber blueward of $\approx$ $-$20 {\kms}. The {\cii}$\lambda$1334 line is contaminated by Galactic {\feii}$\lambda$1608 blueward of $\approx$ 5 {\kms}. The {\niii}$\lambda$989 line is contaminated by the Galactic {\siii}$\lambda$1193 line at 10 {\kms} and also from the {\siii}$\lambda$989 line of the same system. The complexity of the contributions of these contaminants prevents us from using it as a constraint. 

\smallskip

The {\sn} of the {\mgii}$\lambda$$\lambda$2796, 2803 transitions is poor, making it challenging to resolve the component structure. Based on our modelling, we find that the low ionization phase needs four clouds to best explain the absorption seen in {\oi}, {\siii}, and {\cii}. These four clouds also explain the absorption seen in the intermediate ionization transitions of {\ciii}, {\siiii}, and {\siiv}. The clouds are centred at 35 {\kms}, 70 {\kms}, 90 {\kms}, and 115 {\kms}. The properties of these clouds are determined to be {\metallicity} $\approx$ $-$1.3, $-$0.7, 0.1, and 0.5, and the densities are {\hden} $\approx$ $-$2.8, $-$2.7, $-$1.7, and $-$2.8, respectively. 

\smallskip

The absorption in {\ovi} resolves into two components centred at $\approx$29 {\kms} and $\approx$112 {\kms}. The blueward component manifests as a CBLA and explains the asymmetric wing of the {\lya}, while the redward component manifests as a relatively weaker CBLA. The temperatures of these clouds are {\temp} $\approx$ 6.2 and 5.5, respectively. The metallicities and densities of these two clouds are not constrained. 

\subsection{The absorption associated with the z = 0.2179 galaxy towards the quasar J1241}
\label{sec:J1241_0.2179}

A system plot of this absorber is presented in Fig.~\ref{fig:J1241_0.2179}. The {\lyb} is contaminated by the {\cii}$\lambda$1036 line of the $z = $0.20545 absorber, blueward of $\approx$ 10{\kms}. The {\ciii}$\lambda$977 line is swamped by contamination from the Galactic {\siii}$\lambda$1190 line. 

\smallskip

This system shows absorption in intermediate and high ionization transitions only. The {\siiii}$\lambda$1206 line shows absorption centred at $\approx$ 10{\kms}. With access to several of the higher order Lyman series lines, the metallicity is constrained to be {\metallicity} $\approx$ $-$1.0 and density, {\hden} $\approx$ $-$3. 

\smallskip

The absorption in {\ovi} resolves into two components centred at $\approx$ 12 {\kms} and 65 {\kms}. The blueward cloud manifests as a weak CBLA at a temperature of {\temp} $\approx$ 5.5, while the redward one is relatively narrower and hence at a lower temperature of {\temp} $\approx$ 4.6.

\subsection{The absorption associated with the z = 0.5504 galaxy towards the quasar J1244}
\label{sec:J1244_0.5504}

A system plot of this absorber is presented in Fig.~\ref{fig:J1244_0.5504}. Many of the {\hi}-Lyman series lines and metal lines are contaminated by molecular hydrogen arising in the same system. The {\hi}$\lambda$919 line is contaminated by the H2:2$\lambda$919.4 and by the H2:3$\lambda$919.5 lines from $\approx$ 40 {\kms} -- 100{\kms}. The {\hi}$\lambda$923 line is contaminated by an unidentified blend at 110{\kms}. The {\hi}$\lambda$930 line is contaminated by the {\oi}$\lambda$930 line, the H2:1$\lambda$930.5 lines, and the H2:2$\lambda$930.4 line between $\approx$ $-$ 75{\kms}and $-$20 {\kms}. It is also contaminated by H2:0$\lambda$931 at $\approx$ 105 {\kms}. The {\hi}$\lambda$949 line is contaminated by the H2:3$\lambda$950.3 line at $\approx$ 210 {\kms}. The {\hi}$\lambda$972 line is contaminated by the H2:1$\lambda$973.3 line at $\approx$ 255 {\kms}. The {\cii}$\lambda$1036 line is contaminated by H2:0$\lambda$1036.5 line at $\approx$ 70 {\kms}. The {\ciii}$\lambda$977 line is contaminated by H2:3$\lambda$976.5 line at $\approx$ 120 {\kms}. The {\civ}$\lambda$1548 line is contaminated by {\lya} of the $z = $ 0.97426 absorber at $\approx$ $-$35 {\kms}. The {\niii}$\lambda$989 line is contaminated by the H2:3$\lambda$989.7 line at $\approx$ 20 {\kms}. The {\ovi}$\lambda$1037 line is contaminated by the  Galactic {\feii}$\lambda$1608 lines between $-$120 and 20 {\kms}, and by the H2:1$\lambda$1037.1 line at $\approx$  165 {\kms}. The {\siii}$\lambda$1526 line is contaminated by the $z = $0.94667 absorber blueward of $\approx$ 15 {\kms}. 

\smallskip

This system shows absorption in the low, intermediate, and high ionization transitions. The high resolution spectrum of {\mgii} shows four components centred at $\approx$ 7 {\kms}, 18 {\kms}, 56 {\kms}, and 162 {\kms}. The properties of these clouds are constrained to be {\metallicity} $\approx$ $-$0.4, 0.1, $-$1, and $-$0.8, and their densities are {\hden} $\approx$ $-$1.8, $-1.6$, $-1.6$, and $-2.6$, respectively. These phases also explain the absorption seen in {\feii} and {\siii} but slightly overpredict the absorption in {\alii} and {\caii} suggesting depletion due to dust in these ions. Some absorption in {\cii} and in intermediate ionization transitions of {\ciii} and {\siiii} also arises in these gas clouds. 

\smallskip

The absorption in {\civ} resolves into five components centred at $\approx$ $-$ 45 {\kms}, 21 {\kms}, 81 {\kms}, 114 {\kms}, and 173 {\kms}. The properties of these clouds are constrained to be {\metallicity} $\approx$ $-$0.3, $-$0.8, $-$0.9, 0, and $-$0.8, and their densities are {\hden} $\approx$ $-$3.5, $-2.5$, $-3.9$, $-2.8$, and $-1.2$, respectively. The low density component at 81 {\kms} predicts {\ovi}, while affected by contamination and masked, it is found to be consistent with the observed data. Most of the absorption seen in {\cii}, {\ciii}, and {\siiii} also arises in these clouds. 

\subsection{The absorption associated with the z = 0.1967 galaxy towards the quasar J1301}
\label{sec:J1301_0.1967}

A system plot of this absorber is presented in Fig.~\ref{fig:J1301_0.1967}. The {\hi}$\lambda$937 line is completely contaminated by the Galactic {\feii}. The {\lyb} is contaminated by {\lyd} of the $z = 0.29237$ at $\approx$ $-$25 {\kms}. The {\ciii}$\lambda$977 line falls near an emission line at $\approx$ $-$135 {\kms}. The {\nv}$\lambda$1238 line is contaminated by {\lya} of the $z = 0.21911$ and 0.21947 absorbers between $-$165 {\kms} and 110 {\kms}. The {\nv}$\lambda$1242 line is contaminated by the {\lya} of the $z = $0.22311 absorber at $\approx$ $-$80 {\kms}. 

\smallskip

This system shows metal absorption in {\ovi}--only, centred at $-$243 {\kms}. The metallicity and density of this cloud are not well constrained, and its temperature is consistent with {\temp} $\approx$ 5.6. We also see absorption in {\hi}--only centred at $\approx$ $-$260 {\kms}, $-$134 {\kms}, and 259 {\kms}. The metallicities and densities of these clouds are unconstrained due to the lack of associated metal-line absorption. We adopt upper and lower limits for metallicity and density for these clouds. 

\subsection{The absorption associated with the z = 0.6610 galaxy towards the quasar J1319}
\label{sec:J1319_0.6610}

A system plot of this absorber is presented in Fig.~\ref{fig:J1319_0.6610}. The {\hi}$\lambda$920 line is contaminated by {\lyb} of the $z = $0.49020 absorber at  $-$250 {\kms} -- $-$200 {\kms}. The {\hi}$\lambda$972 line is contaminated by the  {\hi}$\lambda$930.7 line of the $z =$ 0.73386 absorber at $\approx$ $-$300 {\kms}. The {\cii}$\lambda$1036 line is contaminated by  the {\cii}$\lambda$1334 line of the $z =$ 0.28881, 0.28900, and 0.28914 absorbers between $\approx$ $-$300 {\kms} and  $\approx$ $-$150 {\kms}. The {\ciii}$\lambda$977 line is contaminated by the {\lya} of the $z = $0.33435 absorber at $\approx$ $-$150 {\kms}, and by the {\lyg} of the $z = $0.66906 absorber between 100 -- 150 {\kms}. The {\ovi}$\lambda$1031 line is contaminated by the {\lyb} of the $z = $0.67127 absorber at $\approx$ 100 {\kms}. 

\smallskip

This system shows metal line absorption in low, intermediate, and high ionization transitions. The high resolution spectrum of {\mgii} shows six components centred at $\approx$ $-180$ {\kms}, $-171$ {\kms}, $-163$ {\kms}, $-125$ {\kms}, $-82$ {\kms}, and $-32$ {\kms}. The properties of these clouds are constrained to be {\metallicity} $\approx$ $-$0.2, $-0.2$, $-0.4$, $-$2.4, $-1.3$, and $-2.6$, and their densities are {\hden} $\approx$ $-$0.5, $-0.6$, $0$, $-2.3$, $-1.8$, and $-1.4$, respectively. 

\smallskip

The phases traced by the {\mgii} do not produce any absorption in the higher ionization transition of {\civ}. The absorption in {\civ} resolves into three components centred at $-168$ {\kms}, $-13$ {\kms}, and 49 {\kms}. The properties of these clouds are constrained to be {\metallicity} $\approx$ 0.2, $0.3$, and $-0.4$, and their densities are {\hden} $\approx$ $-$2.1, $-3.3$, and $0.1$, respectively. The extreme blueward and redward of these clouds also produce the modest narrow absorption seen in the {\ovi}. However, the stronger absorption in {\ovi} centred at $\approx$ $-74$ {\kms} is found to arise in a much hotter phase at {\temp} $\approx$ 6, and it manifests as a weak CBLA in {\lya}. 

\subsection{The absorption associated with the z = 0.2144 galaxy towards the quasar J1322}
\label{sec:J1322_0.2144}

A system plot of this absorber is presented in Fig.~\ref{fig:J1322_0.2144}. The {\ovi}$\lambda$1031 is corrected for contamination by the Galactic {\siii} at $\approx$ 50 {\kms}, and the deblended profile is shown. 

\smallskip

The low ionization transition of {\mgii} in the noisy \textsc{hires} spectrum shows two components centred at $\approx$ 28 {\kms} and 59 {\kms}. These two clouds also account for some of the absorption seen in {\siii} and {\cii} at coincident velocities. The properties of these clouds are determined to be {\metallicity} $\approx$ $-0.6$ and $0$, and their densities are determined to be {\hden} $\approx$ $-1.9$ and $-1.8$, respectively. We also find the need for a {\hi}-only cloud at $\approx$ $-32$ {\kms} that explains the blueward absorption seen in higher-order Lyman lines. The properties of this are not well constrained due to the lack of associated metals. We adopt an upper limit on its metallicity and a lower limit on its density. 

\smallskip 

The intermediate ionization transitions of {\ciii} and {\siiii} resolve into three components, a weak one and two stronger ones,  centred at $\approx$ $-104$ {\kms}, $-43$ {\kms}, and $46$ {\kms}. The stronger components also produce absorption in {\nii}, {\siii}, {\cii}, and {\niii}. The blueward, weaker component is not very well constrained in its metallicity; its metallicity is consistent with $\approx$ solar, and its density is {\hden} $\approx$ $-3.1$. The properties of the stronger components are determined to be {\metallicity} $\approx$ $0.2$ and $-1.1$, and their densities are determined to be {\hden} $\approx$ $-2.8$ and $-2.8$, respectively. 

\smallskip

The absorption in {\ovi} resolves into two components centred at $-58$ {\kms} and 21 {\kms}. The blueward cloud at the temperature of {\temp} $\approx$ 6, manifests as a CBLA, while the redward one at the temperature of {\temp} $\approx$ 5.6 shows up as a very weak CBLA. Both these clouds have unconstrained densities and metallicities.

\subsection{The absorption associated with the z = 0.0708 galaxy towards the quasar J1342}
\label{sec:J1342_0.0708}

A system plot of this absorber is presented in Fig.~\ref{fig:J1342_0.0708}. The {\lya} is contaminated by Galactic {\oi} absorption blueward of $\approx$ 150 {\kms}. The {\nii}$\lambda$1083 line is contaminated by H2:0$\lambda$946.4 line of the $z = $ 0.22710 absorber at $\approx$ 150 {\kms}. The {\siii}$\lambda$1190 line is contaminated by the {\oi}$\lambda$1039 line of the $z = 0.22710$ absorber at $\approx$ 135 {\kms}. 

\smallskip

This system shows absorption in low, intermediate, and high ionization transitions. The low ionization absorption traced by {\siii} resolves into three components centred at $\approx$ 138 {\kms}, 214 {\kms}, and 253 {\kms}. It is found that the intermediate ionization absorption seen in {\siiii} and the high ionization absorption seen in {\civ} arise in the same gas phases traced by the {\siii}. The properties of these clouds are constrained to have metallicities of {\metallicity} $\approx$ 0.5, 0.4, and 0.3, and their densities are {\hden} $\approx$ $-$2.9, $-3.5$, and $-3.3$, respectively. We also see absorption in {\hi} only without any associated metals at $\approx$ 433 {\kms}. We adopt an upper limit and lower limit for its metallicity and density, respectively.

\subsection{The absorption associated with the z = 0.2270 galaxy towards the quasar J1342}
\label{sec:J1342_0.2270}

A system plot of this absorber is presented in Fig.~\ref{fig:J1342_0.2270}. The {\lyg} is contaminated by the Galactic {\siii} absorption at $\approx$ $-$130 {\kms}. The {\lyb} is contaminated by the Galactic {\SII}$\lambda$1258 line at $\approx$ 220 {\kms}, and by the Galactic {\siii}$\lambda$1260 line at $\approx$ 300 {\kms}. The {\ciii}$\lambda$977 line is contaminated by the Galactic {\nitri}1199 line at $\approx$ 180 {\kms, 350, and 480 {\kms}. The {\siii}$\lambda$1260 line is contaminated by the {\lya} of the $z = 0.27287$ absorber at $\approx$ 150 {\kms}. 

\smallskip

This system shows absorption in the low, intermediate, and high ionization transitions. The absorption in {\mgii} is best fit with 8 clouds centred at $\approx$ $-38$ {\kms}, $-23$ {\kms}, 9 {\kms}, 35 {\kms}, 61 {\kms}, 62 {\kms}, 105{\kms}, and 145 {\kms}. We make use of the unsaturated components in other low-ionization lines to establish the low-ionization component structure because of the poor {\sn} $\approx$ 4 of {\mgii}. The properties of these clouds are found to be {\metallicity} $\approx$ $-0.7$, $-1.4$, $0.3$, $-0.4$, $-0.9$, $0.2$, $0$, and $-0.2$, in increasing order of velocity. The densities are found to be {\hden} $\approx$ $-2.1$, $-2.5$, $-2.2$, $-2.4$, $-2.1$, $-2.2$, $-1.3$, and $-1.5$, also in increasing order of velocity. These clouds also explain the absorption seen in the neutral transitions of {\oi}, {\ci}, and {\nitri}, also the low ionization transitions of {\feii}, {\nii}, and {\siii}, and partly account for the absorption in {\cii}, {\ciii}, {\siiii}.

\smallskip

We needed to invoke additional clouds centred at $-73$ {\kms}, 128 {\kms}, 283 {\kms}, and 406 {\kms} to fully explain the absorption seen in the intermediate ionization transitions of {\ciii} and {\siiii}. The properties of these clouds are {\metallicity} $\approx$ $-1.2$, $-0.9$, $-0.2$, and $0.2$, respectively. The densities are {\hden} $\approx$ $-3.1$, $-3.3$, $-3.5$, and $-3.7$, respectively. 

\smallskip

The absorption in {\ovi} is found to be explained using three clouds centred at $\approx$ $-87$ {\kms}, 38 {\kms}, and 156 {\kms}. The metallicities and densities of these clouds are not constrained. Their temperatures are consistent with {\temp} $\approx$ 5.9, 6.5, and 5.5, respectively. The middle one among these clouds manifests as a CBLA and is needed to explain the broad wing of {\lya} on its left side of the profile.

\subsection{The absorption associated with the z = 0.4295 galaxy towards the quasar J1357}
\label{sec:J1357_0.4295}

A system plot of this absorber is presented in Fig.~\ref{fig:J1357_0.4295}. The {\lyg} is contaminated by the {\lya} of the $z = 0.1447$ absorber at $\approx$ 300 {\kms}. The {\lya} is contaminated by {\siii}$\lambda$ of the $z = 0.45662$ absorber blueward of 150 {\kms}. The {\ciii}$\lambda$977 line is contaminated by the {\lyd} of the $z = 0.47199$ absorber at $\approx$ 300 {\kms} and by the {lya} of the $z = 0.15044$  absorber at $\approx$ 400 {\kms}. 

\smallskip

This system shows absorption in {\hi} only in two components centred at 212 {\kms} and 345 {\kms}. The metallicity and density of these clouds are unconstrained. We adopt upper limits and lower limits on metallicity and density for these clouds. 

\subsection{The absorption associated with the z = 0.4592 galaxy towards the quasar J1357}
\label{sec:J1357_0.4592}

A system plot of this absorber is presented in Fig.~\ref{fig:J1357_0.4592}. The {\lyg} line is contaminated by the {\hi}$\lambda$930.7 line of the $z =$ 0.52168 absorber at $\approx$ $-650$ {\kms}. The {\cii}$\lambda$1036 line is contaminated by the {\lyb} of the $z = 0.47199$ absorber between $-$525 -- $-450$ {\kms}. The {\niii}$\lambda$989 line is blended with the {\siii}$\lambda$989 line of the same system. The {\ovi}$\lambda$1031 line is contaminated by the {\lya} of the $z = 0.23763$ absorber at $\approx$ $-475$ {\kms}. The {\siii}$\lambda$1193 is contaminated by the {\lya} of the $z = 0.43054$ and 0.43116 absorbers redward of $-500$ {\kms}. 

\smallskip
 
This system shows absorption in low, intermediate, and high ionization transitions. The high resolution spectrum of {\mgii} reveals six components centred at $-614$ {\kms}, $-565$ {\kms}, $-549$ {\kms}, $-533$ {\kms}, $-517$ {\kms}, and $-501$ {\kms}. The properties of these clouds are constrained to be {\metallicity} $\approx$ $-1.8$, 0, $0.2$, $-1$, $0.1$, and 0, and their densities are {\hden} $\approx$ $-2.6$, $-1.2$, $-1.7$, $0.2$, $-1.5$, and $-1.3$, respectively. The absorption in other low ionization transitions of {\oi}, {\cii}, {\siii}, and {\nii} is partly accounted for by these phases of gas. We find that modest absorption is predicted in {\caii}, indicative of depletion due to dust for this transition. 

\smallskip

The absorption in intermediate ionization transitions of {\ciii}, {\siiii}, and {\niii} is found to be much stronger to arise just from the phases traced by the {\mgii}. We invoke additional phases of gas centred at $-653$ {\kms}, $-537$ {\kms}, and $-473$ {\kms}. The higher ionization transition of {\civ} is also found to arise in the same phase as the intermediate ionization transitions. The properties of these clouds are found to be {\metallicity} $\approx$ $-2$, $-0.5$, and $-1.1$, and their densities are found to be {\hden} $\approx$ 0.3, $-$3, and $-0.6$, respectively. 

\smallskip

We also see absorption in {\ovi} which resolves into two components centred at $-662$ {\kms} and $-568$ {\kms}. While the metallicity and density are less certain for these clouds, the temperatures are {\temp} $\approx$  5.7 and 5.3, respectively. These clouds manifest as CBLAs in {\lya}. 

\subsection{The absorption associated with the z = 0.0949 galaxy towards the quasar J1547}
\label{sec:J1547_0.0949}

A system plot of this absorber is presented in Fig.~\ref{fig:J1547_0.0949}. This system shows absorption in {\lya} only centred at $\approx$ 340 {\kms}. The inferred properties are an upper limit on metallicity and a lower limit on hydrogen number density. 

\subsection{The absorption associated with the z = 0.1892 galaxy towards the quasar J1555}
\label{sec:J1555_0.1892}

A system plot of this absorber is presented in Fig.~\ref{fig:J1555_0.1892}. This system shows absorption in low, intermediate, and high ionization transitions. The low ionization transition of {\mgii} shows two components centred at $\approx$ $-27$ {\kms} and $-5$ {\kms}. Although the data is quite noisy in the {\mgii}$\lambda\lambda$2796, 2803 transitions, there is evidence for a component at $-27$ in both the {\cii} and {\siiii} transitions. The properties of the component at $-27$ are not well constrained, the metallicity is found to be consistent with solar and its density is {\hden} $\approx$ $-2.7$. The component at $-5$ has the properties of {\metallicity} $\approx$ $-0.3$ and {\hden} $\approx$ $-2.8$. Both these clouds show up in absorption in {\cii}, {\nii}, {\siii}, {\siiii}, and {\niii}. 

\smallskip

There is strong absorption seen in the intermediate ionization transitions of {\ciii}, {\siiii}, and {\niii} centred at $\approx$ $-58$ {\kms}, offset in velocity from the absorption seen in low ions discussed above. This cloud of gas also explains the coincident absorption seen in other ionization states including {\cii}, {\siii}, {\siiv}, {\nv}, and also the {\ovi}. The properties of this cloud are constrained to be {\metallicity} $\approx$ $-0.4$, {\hden} $\approx$ $-3.8$, and {\temp} $\approx$ 4.2. 

\smallskip

There is also absorption in {\ovi} that is found to arise in a much hotter phase at {\temp} $\approx$ 5.7, whose metallicity and density are not well constrained. This cloud manifests as a weak CBLA. 

\subsection{The absorption associated with the z = 0.0921 galaxy towards the quasar J1704}
\label{sec:J1704_0.0921}

A system plot of this absorber is presented in Fig.~\ref{fig:J1704_0.0921}. The {\lyb} is contaminated by the {\hi}$\lambda$916.4 line and by the {\hi}$\lambda$917.1 line of the $z = $ 0.22106 absorber, blueward of $-250$ {\kms} and redward of $-150$ {\kms}, respectively. The {\lya} is contaminated by the {\lyg} of the $z = 0.36442$ and 0.36474 absorbers redward of $-200$ {\kms}. The {\cii}$\lambda$1334 line is contaminated by the {\siii}$\lambda$1193 line of $z = 0.22093$ absorber at $-100$ {\kms}. The {\civ}$\lambda$1548 line is contaminated by the {\nv}$\lambda$1238 line of $z = 0.36442$ absorber redward of $-200$ {\kms}. The {\siiii}$\lambda$1206 line is contaminated by the {\lyb} of the $z = 0.28366$ and 0.28395 absorbers at $-200$ {\kms} and $-140$ {\kms}. Although we have completely blanked out the {\siiii}, the predicted amount of {\siiii} suggests that the effect of contamination at $-200$ {\kms} is not significant.  

\smallskip

This system does not show any absorption in the low ionization transitions. There is modest, broad absorption seen in the {\siiv} and {\civ} transitions at $\approx -190$ {\kms}. The inferred metallicity is found to be consistent with a {\metallicity} $\approx 0.7$, albeit with large uncertainty, and density is {\hden} $\approx$ $-3.7$. We do not see any absorption in {\ovi}. 

\subsection{The absorption associated with the z = 0.4302 galaxy towards the quasar J2131}
\label{sec:J2131_0.4302}

A system plot of this absorber is presented in Fig.~\ref{fig:J2131_0.4302}. Several of the Lyman series lines and metal lines are affected by blending due to molecular hydrogen arising in the same system. The {\hi}$\lambda$923 line is contaminated by the {\lya} of the $z = 0.08522$ absorber at $\approx$ $-200$ {\kms}. The {\hi}$\ lambda$926 line is contaminated by the {\lya} of the $z = 0.08887$ absorber blueward of $-100$ {\kms}. The {\hi}$\lambda$930 line is contaminated by the {\oi}$\lambda$930 line of the $z = 0.42985$ absorber between $-200$ -- $-150$ {\kms}. The {\hi}$\lambda$937 is contaminated by an unidentified blend between $-200$ -- $-150$ {\kms}, by the {\lya} of the $z$ = 0.10357 absorber and by the H2:0$\lambda$938.4 line of the $z =$ 0.42985 absorber redward of 0 {\kms}. The {\hi}$\lambda$949 line is contaminated by H2:2$\lambda$949.3 and H2:2$\lambda$949.6 lines blueward of $-150$ {\kms}. The {\hi}$\lambda$972 line is contaminated by the H2:0$\lambda$971.9 line of the $z = 0.42985$ absorber between $-200$ -- $-150$ {\kms}. The {\lyb} is contaminated by the H2:1$\lambda$1024.9 line of the $z = 0.42985$ absorber blueward of $-150$ {\kms}. The {\niii}$\lambda$989 line is contaminated by the H2:3$\lambda$989.7 line of the $z = 0.42985$ absorber. The {\ovi}$\lambda$1031 line is contaminated by the {\lyb} of the $z = 0.43930$ absorber redward of 50 {\kms}. The {\ovi}$\lambda$1037 line is contaminated by the H2:1 line of the $z = 0.42985$ absorber at $\approx$ 75 {\kms}.

\smallskip

This system shows absorption in the low, intermediate, and high ionization transitions. The spectrum of {\mgii} resolves three components centred at $-116$ {\kms}, $-84$ {\kms}, and $-69$ {\kms}. The properties of these clouds are determined to be {\metallicity} $\approx$ $-2.2$, $-1.1$, and $-0.3$, and their densities are $-1.5$, $-1.8$, and $-1.6$. The absorption in other low ions such as {\cii}, {\siii}, and {\nii} is only partially explained by the three components seen in {\mgii}. An additional cloud centred at $\approx$ $-$77 {\kms} is found to explain the stronger absorption component seen in the other low ions. This cloud has the properties of {\metallicity} $\approx$ 0 and {\hden} $\approx$ 0.5. There is also modest absorption seen in {\siii} at $\approx$ $-4$ {\kms}. This cloud is needed to explain the absorption seen in the higher-order Lyman lines. The properties of this cloud are determined to be {\metallicity} $\approx$ $=1.2$ and {\hden} $\approx$ $-0.7$.

\smallskip

The absorption in intermediate ionization transitions {\siiii} and {\ciii} is not centred on the phases tracing the low ions. We invoke a cloud centred at $-27$ {\kms} that explains the absorption seen in the intermediate ions. The properties of this cloud are determined to be {\metallicity} $\approx$ $-2.3$ and {\hden} $\approx$ $-0.5$. 

\smallskip

The absorption in higher ionization transition of {\civ} is found to arise in a separate phase centred at $-78$ {\kms}. The metallicity and density of this cloud are determined to be {\metallicity} $\approx$ $-1.6$ and {\hden} $\approx$ $-1$.

\smallskip

The absorption in {\ovi} is found to be best explained by two broad components. The metallicities and densities of these clouds are not constrained. These clouds are at temperatures of {\temp} $\approx$ 6 and 6.6, respectively, and manifest as CBLAs in {\lya}.

\subsection{The absorption associated with the z = 0.0752 galaxy towards the quasar J2137}
\label{sec:J2137_0.0752}

A system plot of this absorber is presented in Fig.~\ref{fig:J2137_0.0752}. The {\lyb} is contaminated by the {\hi}$\lambda$919.3 line of the $z = 0.19964$ absorber at $\approx$ 0 {\kms}. 

\smallskip

This system does not show any metal line absorption. We only see absorption in the {\lya} and {\lyb} lines in three components at $\approx$ $-147$ {\kms}, 34 {\kms}, and 149 {\kms}. The properties of these clouds are not constrained. We adopt upper limits on metallicity and lower limits on density for these clouds.

\subsection{The absorption associated with the z = 0.1537 galaxy towards the quasar J2253}
\label{sec:J2253_0.1537}

A system plot of this absorber is presented in Fig.~\ref{fig:J2253_0.1537}. The {\lya} line is contaminated by the {\ovi}$\lambda$1037 line of the $z = 0.35253$ absorber at 180 {\kms}. The {\civ}$\lambda$1548 line is contaminated by the {\lyb} of the $z = 0.74211$ absorber redward of $\approx$ 15 {\kms}. The {\nii}$\lambda$1083 line is contaminated by the Galactic {\SII$\lambda$1250 line at $\approx$ $-25$ {\kms}. The {\ovi}$\lambda$1031 line is contaminated by the Galactic {\siii}$\lambda$1190 line blueward of 0 {\kms}. The {\siii}$\lambda$1260 line is contaminated by the {\lya} of the $z = 0.19609$ absorber at $\approx$ $-40$ {\kms}. The {\siiii}$\lambda$1206 line is swamped by the Galactic {\siiv}$\lambda$1393 absorption at 0 {\kms}. The {\siiv}$\lambda$1393 line is contaminated by the Galactic {\feii}$\lambda$1608 line redward of $\approx$ 20 {\kms}. 

\smallskip

This system shows absorption in the low, intermediate, and high ionization transitions. The absorption in the low ionization transition of {\siii} is centred at $\approx$ 7 {\kms}, and this cloud also produces modest absorption in {\cii}. The properties of this cloud are determined to be {\metallicity} $\approx$ $-0.5$ and {\hden} $\approx$ $-1$. 

\smallskip

The absorption in the higher ionization transition of {\civ} is found to be arising in a different phase centred at 14 {\kms}. This cloud is also found to explain the absorption seen in {\siiv}, and also in {\siiii}, although it is completely blanked out. The properties of this cloud are determined to be {\metallicity} $\approx$ $-1.7$ and {\hden} $\approx$ $-3.4$.

\smallskip

There is also a {\hi} only cloud with no associated metals at $\approx -178$ {\kms}. Its properties are not constrained. 

\smallskip

The {\ovi} is found to arise in a much higher temperature phase at {\temp} $\approx$ 5.7, with unconstrained metallicity and density.

\subsection{The absorption associated with the z = 0.3528 galaxy towards the quasar J2253}
\label{sec:J2253_0.3528}

A system plot of this absorber is presented in Fig.~\ref{fig:J2253_0.3528}. The {\hi}$\lambda$949 line is contaminated by the {\lyg} of the $z = 0.32061$ absorber blueward of $-100$ {\kms}. The {\ciii}$\lambda$977 line is contaminated by the {\lyd} of the $z = 0.39074$ absorber at $\approx$ $-200$ {\kms}. The {\siiii}$\lambda$1206 line is contaminated by the {\lya} of the $z = 0.34140$ at $-250$ {\kms}. 

\smallskip

This system shows absorption in the intermediate and high ionization transitions. We do not see absorption in the low ionization transitions. The intermediate ionization transition of {\ciii} shows absorption centred at $\approx$ $-59$ {\kms}. The properties of this cloud are found to be {\metallicity} $\approx$ $-1.8$, {\hden} $\approx$ $-4.7$, and {\temp} $\approx$ 5. This low density gas is found to also produce the {\civ} and {\ovi} at the coincident velocity. The centering of this gas cloud implies that there needs to be an additional cloud to explain the absorption seen in the {\lya}. This additional cloud is found to be centred at $\approx$ $-30$ {\kms}, and its properties are {\metallicity} $\approx$ $-0.6$, {\hden} $\approx$ $-3.2$. This cloud also explains the asymmetry seen in the redward portion of the {\ciii} absorption profile. In addition to these clouds, we also see absorption in the {\lya} at $\approx$ $-151$ {\kms}, which is found to produce absorption in {\ciii}. However, due to the uncertain amounts of contribution from coincident blends we completely blanked this region of the {\ciii} spectrum. The properties of this cloud are not constrained, and we adopt an upper limit on its metallicity and a lower limit on its density.

\smallskip

We also see broad absorption in {\ovi} centred at $\approx$ 0 {\kms}. This cloud manifests as a CBLA and is at a temperature of {\temp} $\approx$ 6. 

\smallskip

We also see a cloud in {\lya} with no associated metals centred at $\approx$ $-331$ {\kms}. The inferred properties imply an upper limit on its metallicity and a lower limit on its density.

\subsection{The absorption associated with the z = 0.3900 galaxy towards the quasar J2253}
\label{sec:J2253_0.3900}

A system plot of this absorber is presented in Fig.~\ref{fig:J2253_0.3900}. The {\ciii}$\lambda$977 line is contaminated by {\lya} arising in the $z = 0.11758$ and 0.11735 absorbers blueward of 150 {\kms}. 

\smallskip

This system shows absorption in the intermediate and high ionization transitions. The intermediate ionization transitions of {\ciii} and {\oiii} show absorption in two components, a strong component centred at $\approx$ 154 {\kms}, and a relatively weaker one at 233 {\kms}. The stronger component is also found to explain the absorption seen in {\nv} and {\ovi}. The metallicity of this cloud is weakly constrained to be {\metallicity} $\approx$ $-2$, and its density is determined to be {\hden} $\approx$ $-4.7$. The weaker cloud at $\approx$ 232 {\kms} is poorly constrained in its properties. The metallicity is found to be {\metallicity} consistent with $\approx$ 0.4 and density, {\hden} $\approx$ $-2.6$. 

\smallskip

Neither of the clouds at 154 {\kms} and 233 {\kms} are centred on the absorption seen in the Lyman series lines. Thus, we invoke an additional cloud at $\approx$ 116 {\kms} needed to explain the {\hi} Lyman series. We also observe modest absorption in {\siii} and {\ciii} in this cloud. However, {\ciii} is blanked out at this location. The properties of this cloud are constrained to be {\metallicity} $\approx$ $-0.2$, and {\hden} $\approx$ $-2.3$. 

\section{System Plots for all the absorbers}
\label{appendix:systemplots}

 \begin{landscape}

\begin{figure}
\begin{center}
 
\includegraphics[width=\linewidth]{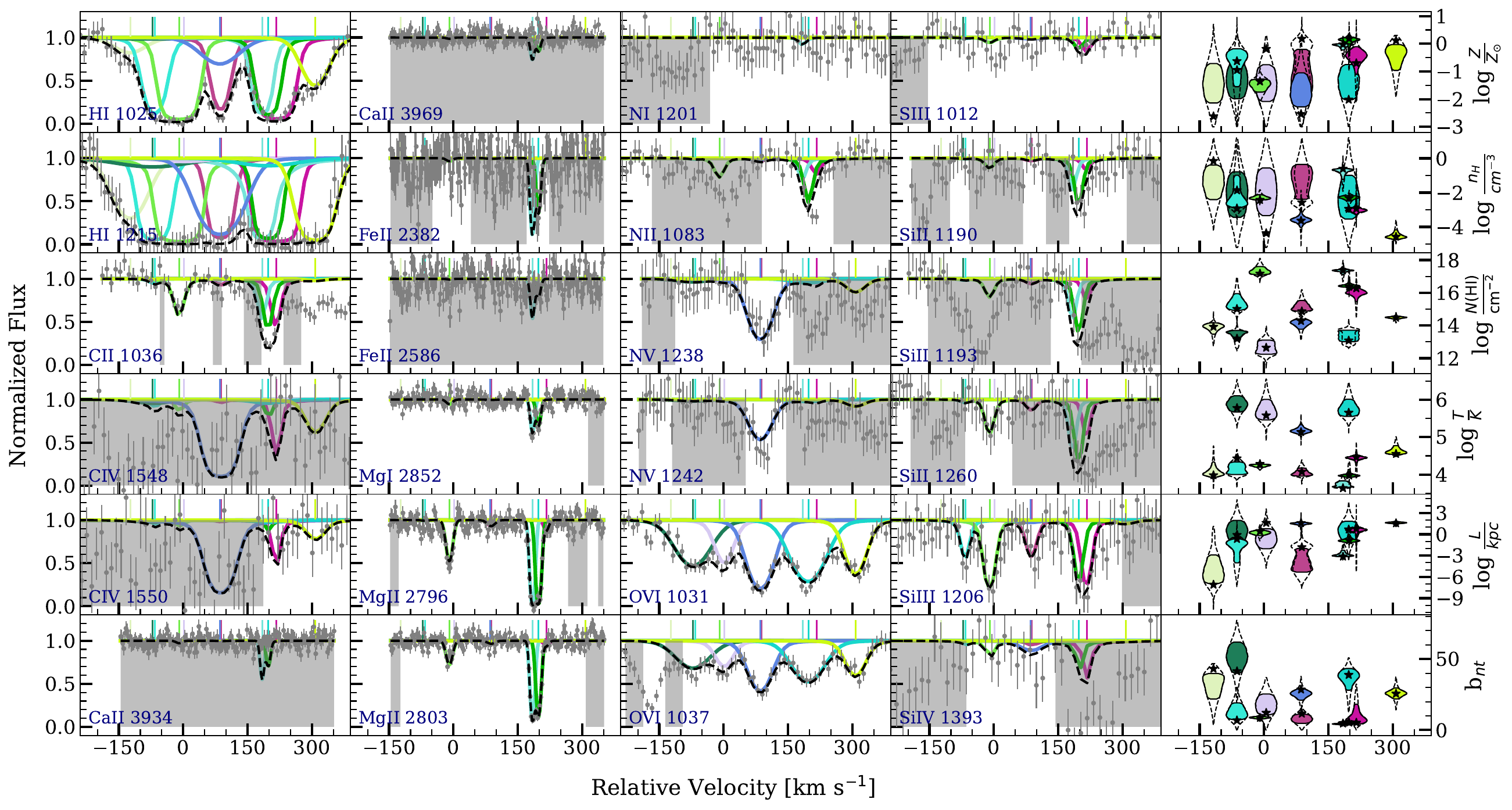}
\caption{\CLOUDY~models for the $z = 0.3985$ absorber towards the quasar J0125 obtained using the MLE values. The spectral data are shown in grey with 1$\sigma$~error. The centroids of absorption components for various constraining ions are indicated by the vertical tick marks on top of each panel. The superposition of all the models is shown by the black dashed curve. The region shaded in grey shows the pixels that were not used in the evaluation of the log-likelihood. The inferred properties \textit{viz.} {\metallicity}, {\hden}, {\colden}, {\temp}, {\thickness}, and {\bturb}, of the gas clouds are shown as violin plots in the right-most panel.}
\label{fig:J0125_0.3985}
\end{center}
\end{figure}

\clearpage

\begin{figure}
\begin{center}
\includegraphics[width=\linewidth]{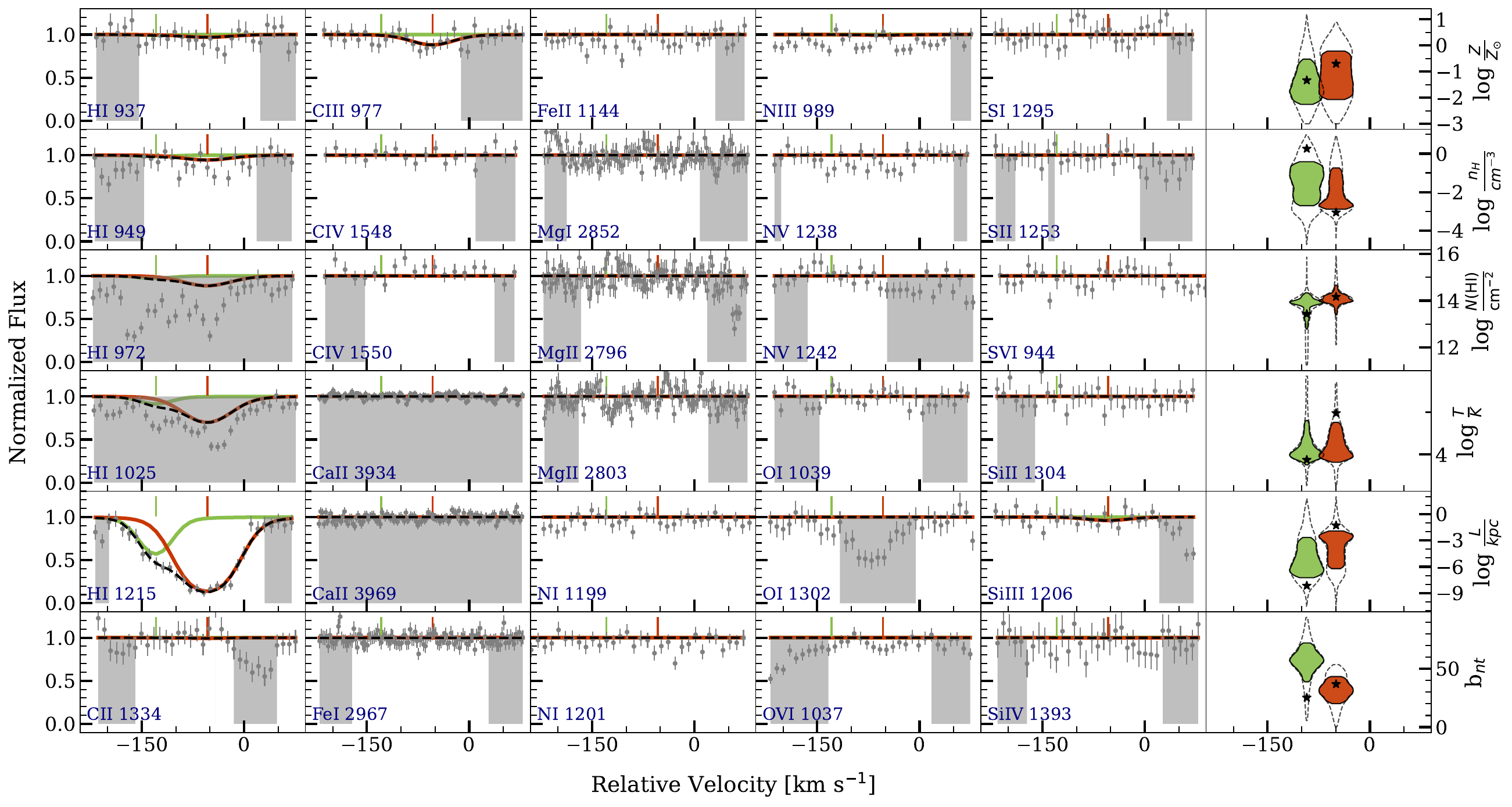}
\caption{Same as Fig.\ref{fig:J0125_0.3985} but for the absorber nearest to the $z$ = 0.2617 galaxy towards the background quasar J0351. The basic galaxy properties are $D \approx 189$ kpc, $\phi$ = $64.9_{-15.8}^{+21.1}$, $i$ = $83.0_{-3.0}^{+2.0}$}
\label{fig:J0351_0.2617}
\end{center}
\end{figure}

\clearpage

\begin{figure}
\begin{center}
\includegraphics[width=\linewidth]{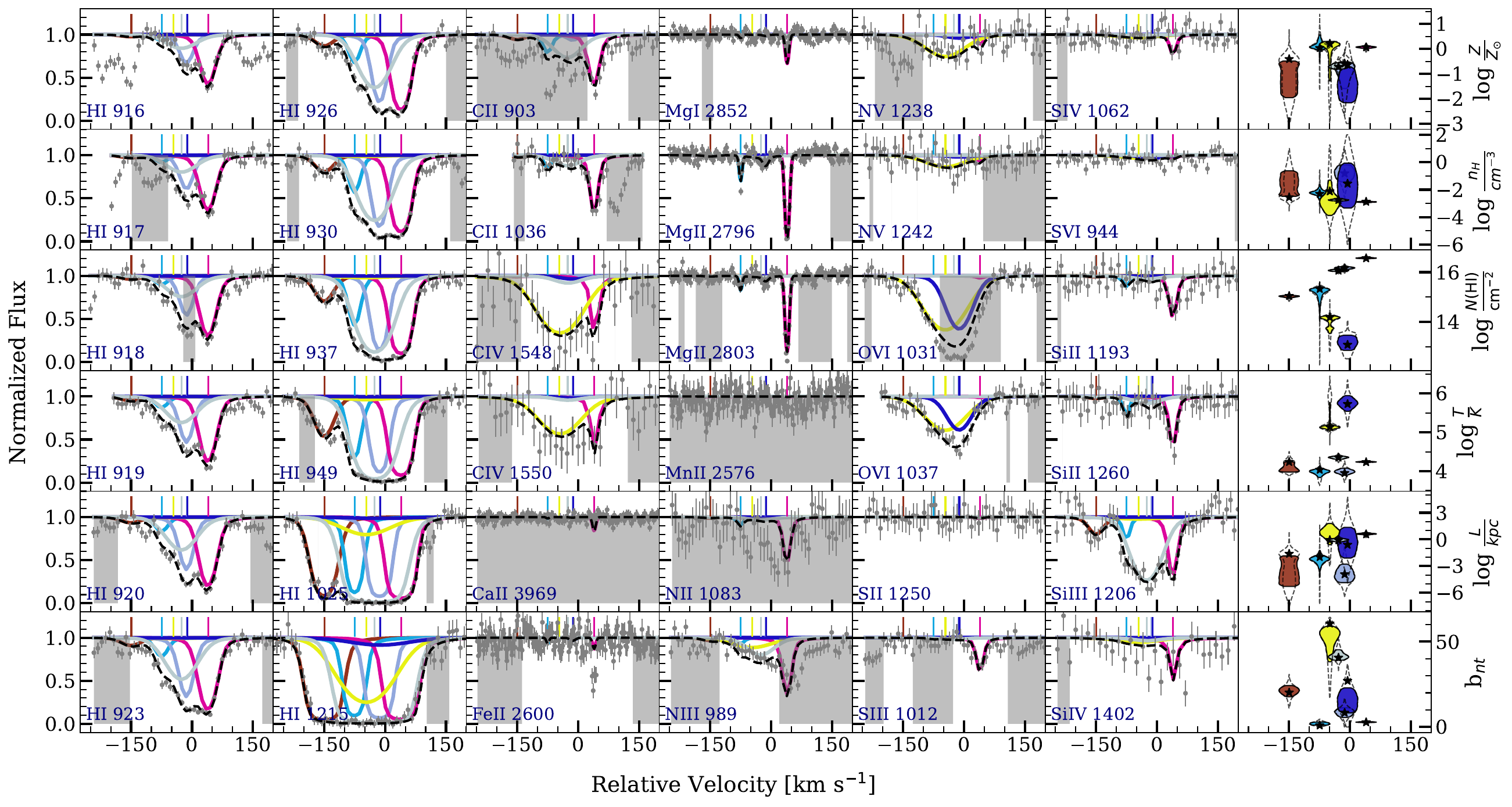}
\caption{Same as Fig.\ref{fig:J0125_0.3985} but for the absorber nearest to the $z$ = 0.3570 galaxy towards the background quasar J0351. The basic galaxy properties are $D \approx 72$ kpc, $\phi$ = $4.9_{-4.9}^{+33.0}$, $i$ = $28.5_{-12.5}^{+19.8}$}
\label{fig:J0351_0.3570}
\end{center}
\end{figure}

\clearpage

\begin{figure}
\begin{center}
\includegraphics[width=\linewidth]{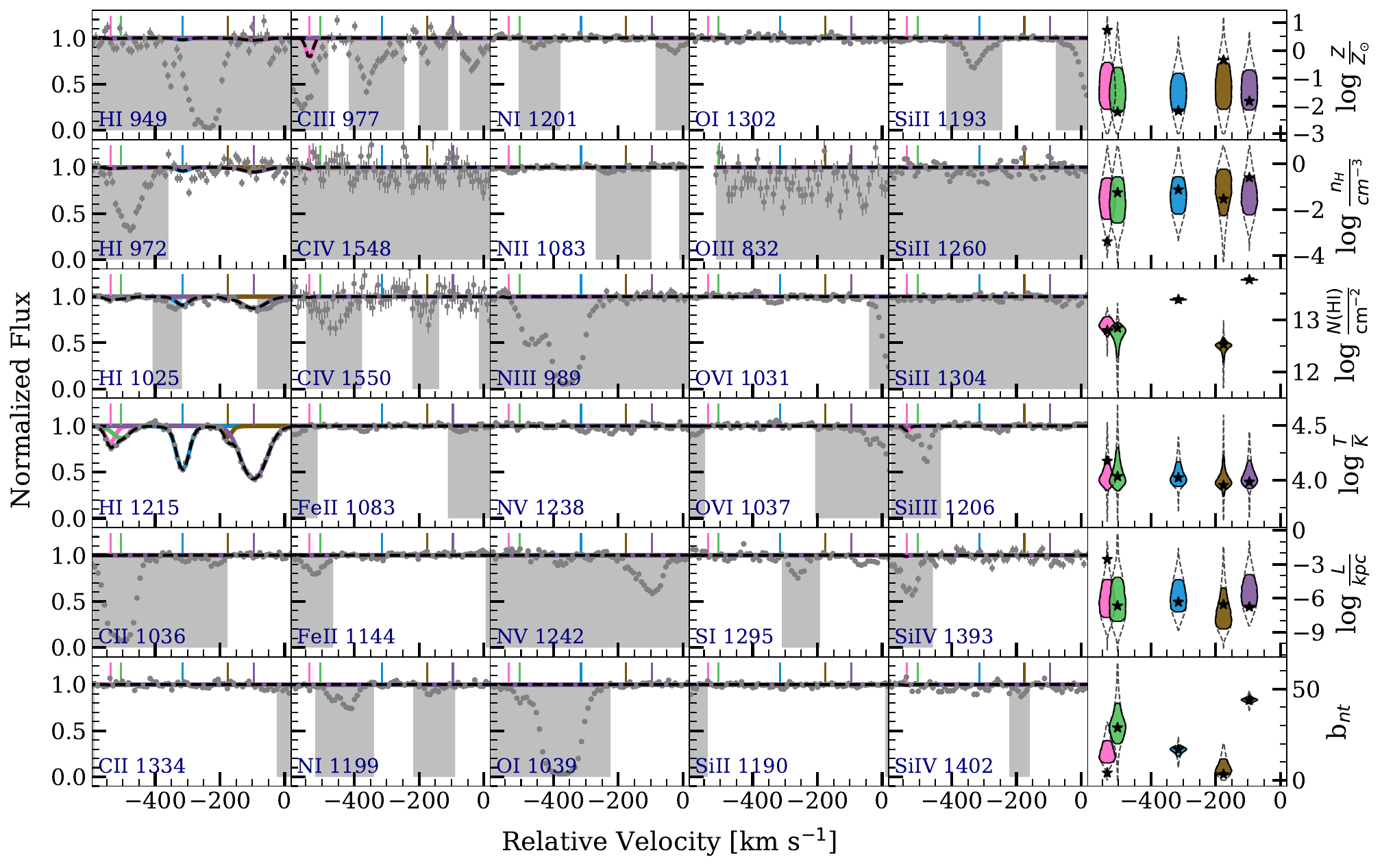}
\caption{Same as Fig.\ref{fig:J0125_0.3985} but for the absorber nearest to the $z$ = 0.1534 galaxy towards the background quasar J0407. The basic galaxy properties are $D \approx 196$ kpc, $\phi$ = $26.3_{-1.0}^{+0.9}$, $i$ = $49.5_{-0.7}^{+0.5}$}
\label{fig:J0407_0.1534}
\end{center}
\end{figure}

\clearpage

\begin{figure}
\begin{center}
\includegraphics[width=\linewidth]{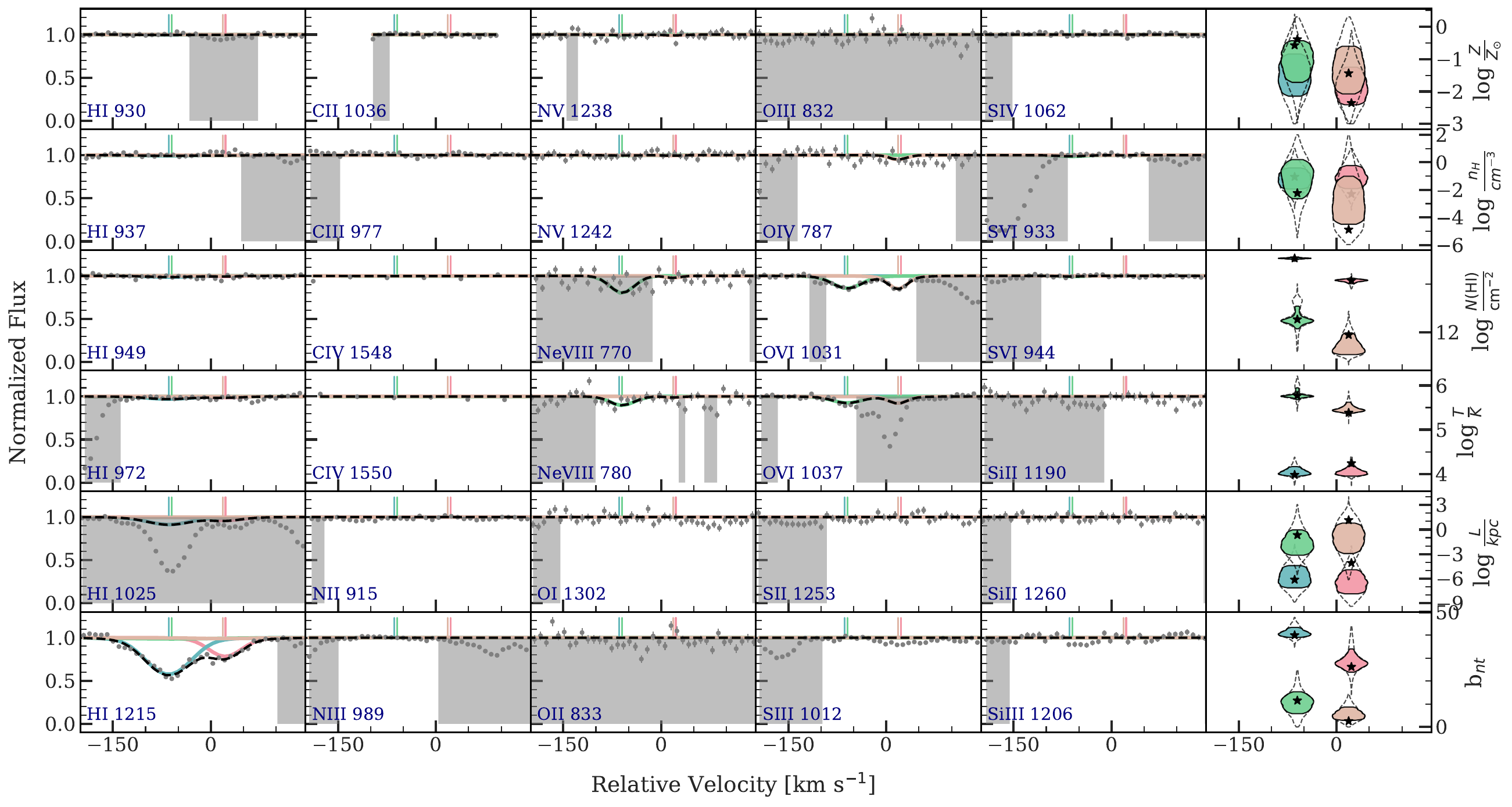}
\caption{Same as Fig.\ref{fig:J0125_0.3985} but for the absorber nearest to the $z$ = 0.3422 galaxy towards the background quasar J0407. The basic galaxy properties are $D \approx 172$ kpc, $\phi$ = $48.1_{-0.9}^{+1.0}$, $i$ = $85.0_{-0.4}^{+0.1}$}
\label{fig:J0407_0.3422}
\end{center}
\end{figure}

\clearpage

\begin{figure}
\begin{center}
\includegraphics[width=\linewidth]{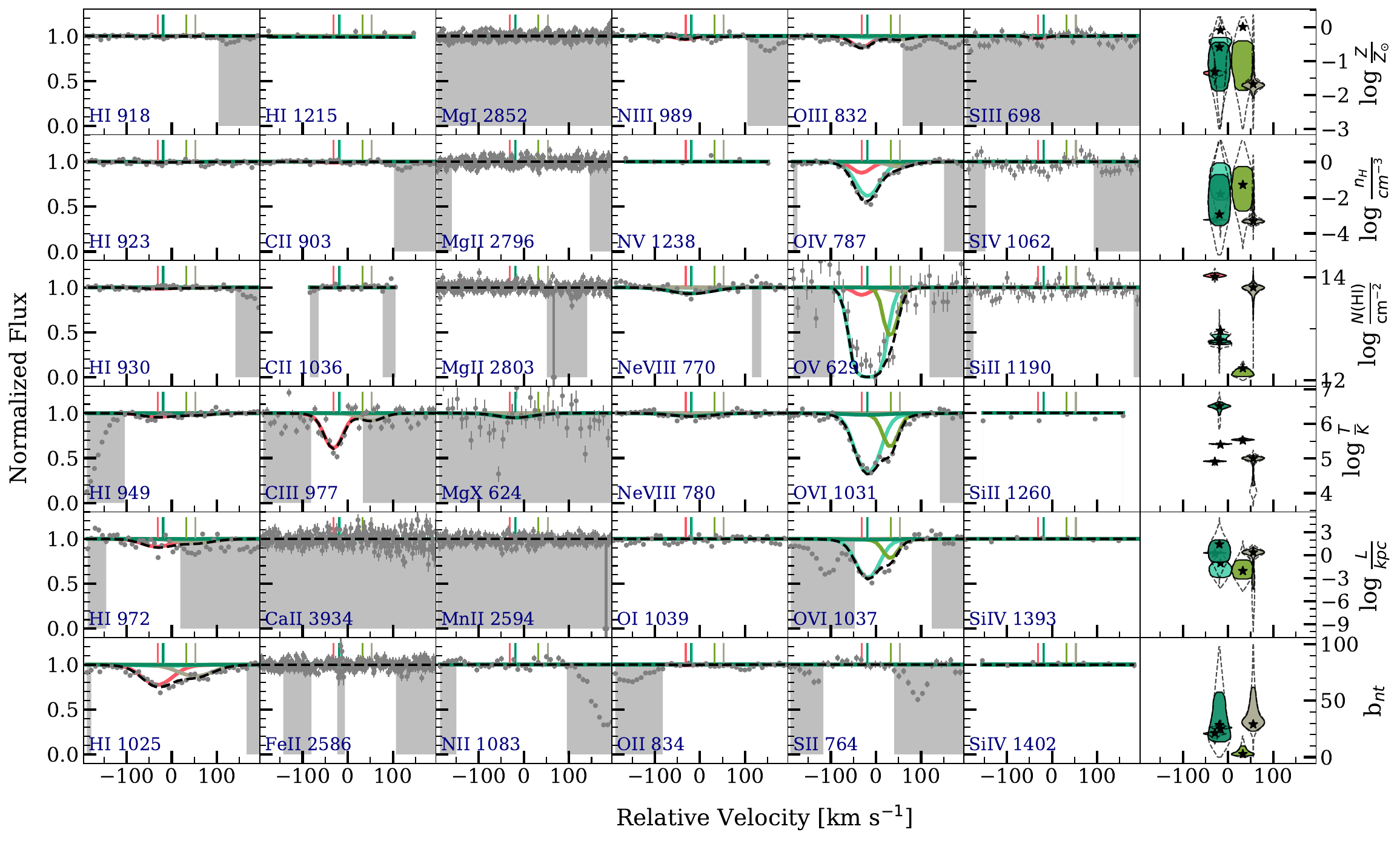}
\caption{Same as Fig.\ref{fig:J0125_0.3985} but for the absorber nearest to the $z$ = 0.4952 galaxy towards the background quasar J0407. The basic galaxy properties are $D \approx 108$ kpc, $\phi$ = $21.0_{-3.7}^{+5.3}$, $i$ = $67.2_{-7.5}^{+7.6}$}
\label{fig:J0407_0.4952}
\end{center}
\end{figure}

\clearpage

\begin{figure}
\begin{center}
\includegraphics[width=\linewidth]{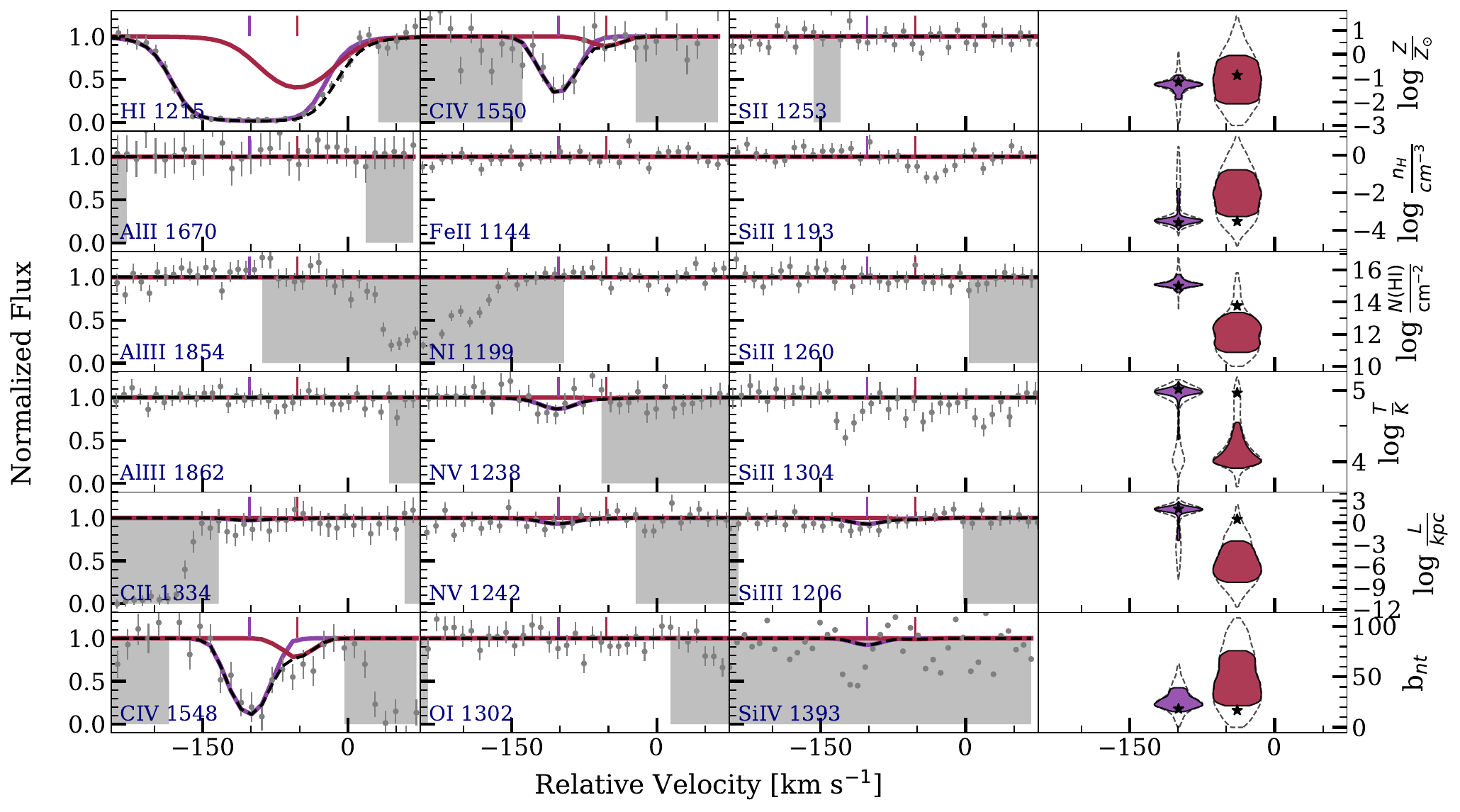}
\caption{Same as Fig.\ref{fig:J0125_0.3985} but for the absorber nearest to the $z$ = 0.2784 galaxy towards the background quasar J0456. The basic galaxy properties are $D \approx 51$ kpc, $\phi$ = $78.4_{-2.1}^{+2.1}$, $i$ = $71.2_{-2.6}^{+2.6}$}
\label{fig:J0456_0.2784}
\end{center}
\end{figure}

\clearpage

\begin{figure}
\begin{center}
\includegraphics[width=\linewidth]{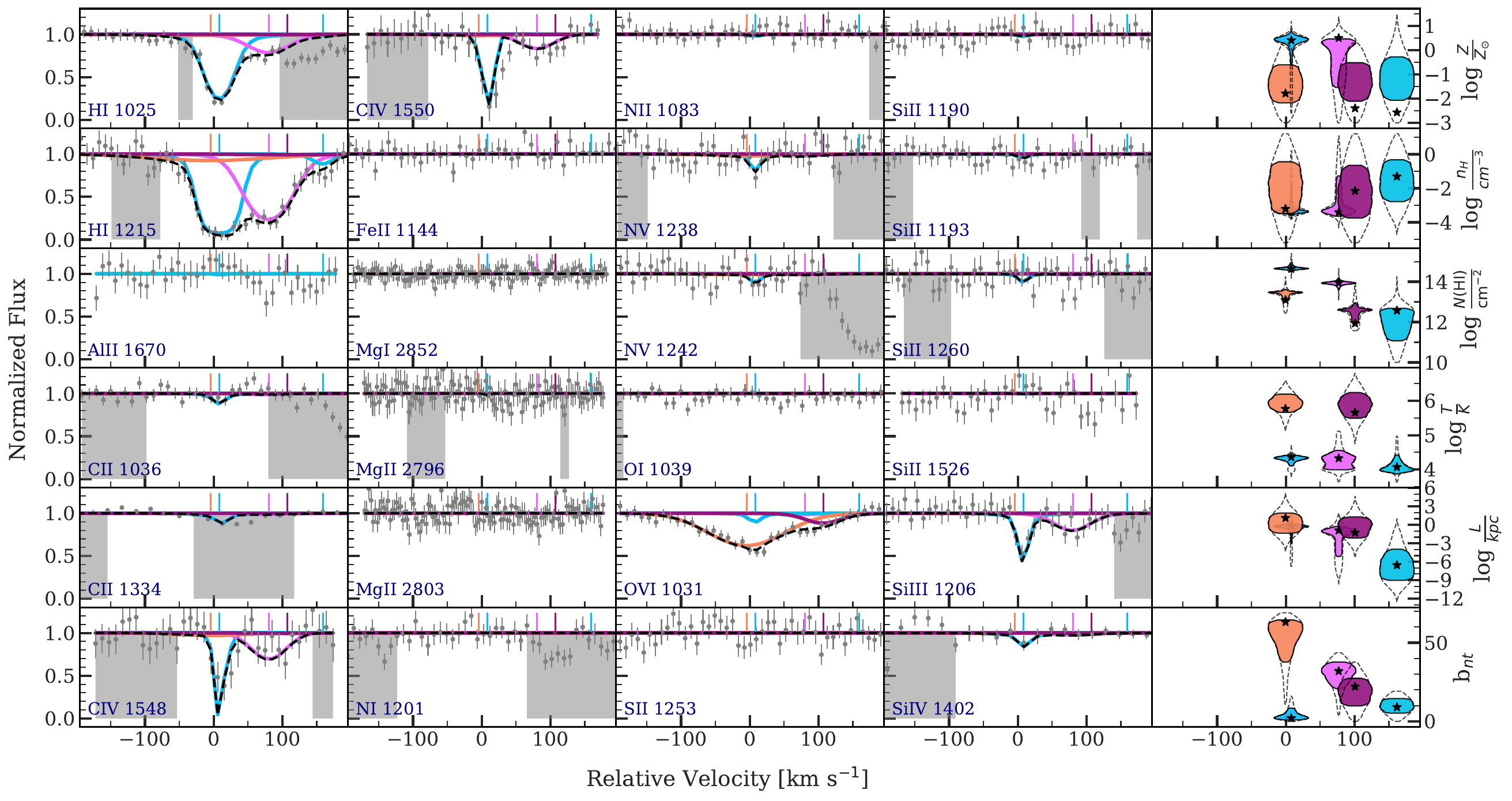}
\caption{Same as Fig.\ref{fig:J0125_0.3985} but for the absorber nearest to the $z$ = 0.3815 galaxy towards the background quasar J0456. The basic galaxy properties are $D \approx 103$ kpc, $\phi$ = $63.8_{-2.7}^{+4.3}$, $i$ = $57.1_{-2.4}^{+19.9}$}
\label{fig:J0456_0.3815}
\end{center}
\end{figure}

\clearpage

\begin{figure}
\begin{center}
\includegraphics[width=\linewidth]{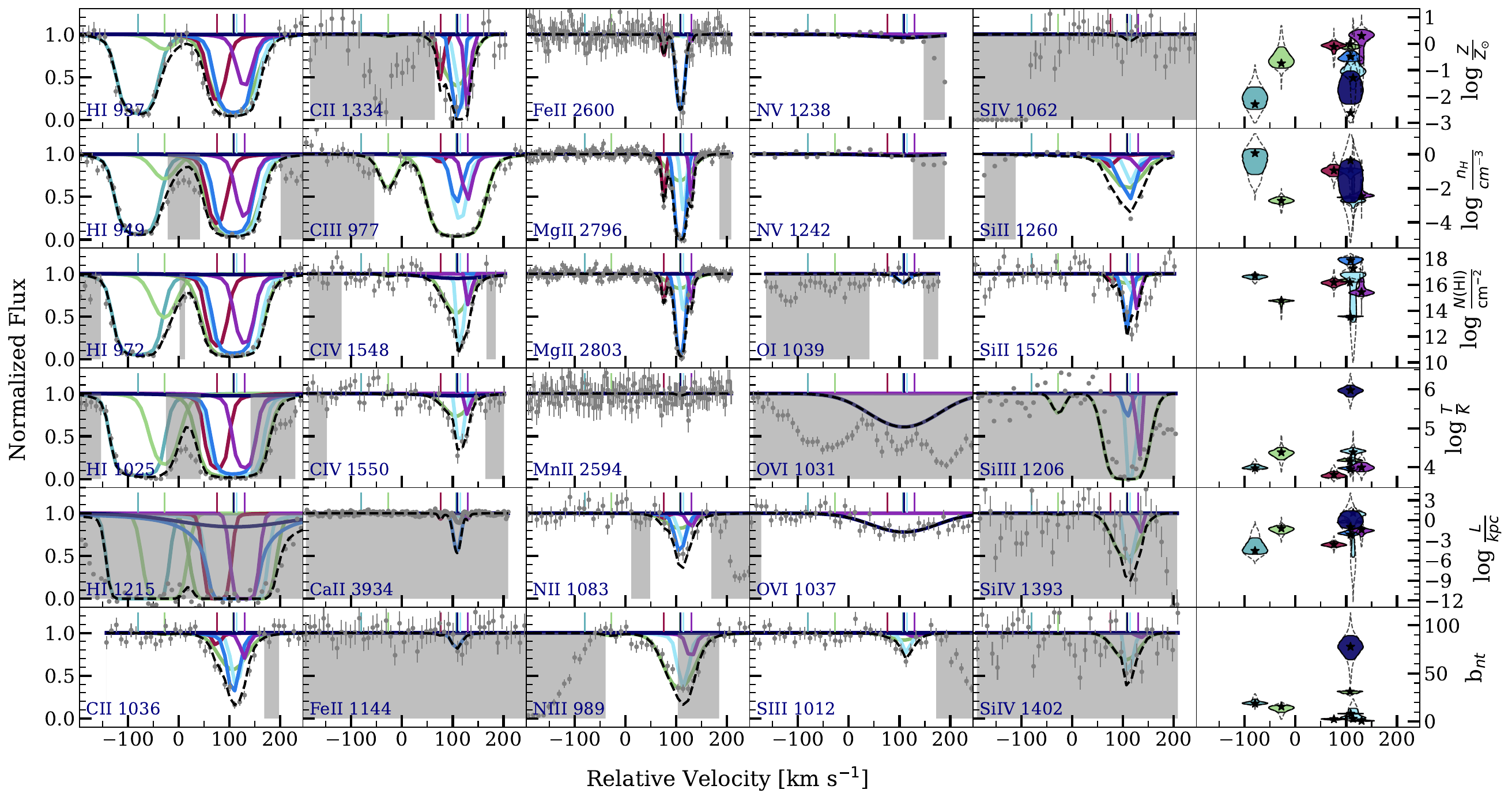}
\caption{Same as Fig.\ref{fig:J0125_0.3985} but for the absorber nearest to the $z$ = 0.4828 galaxy towards the background quasar J0456. The basic galaxy properties are $D \approx 108$ kpc, $\phi$ = $85.2_{-3.7}^{+3.7}$, $i$ = $42.1_{-3.1}^{+3.1}$}
\label{fig:J0456_0.4828}
\end{center}
\end{figure}

\clearpage

\begin{figure}
\begin{center}
\includegraphics[width=\linewidth]{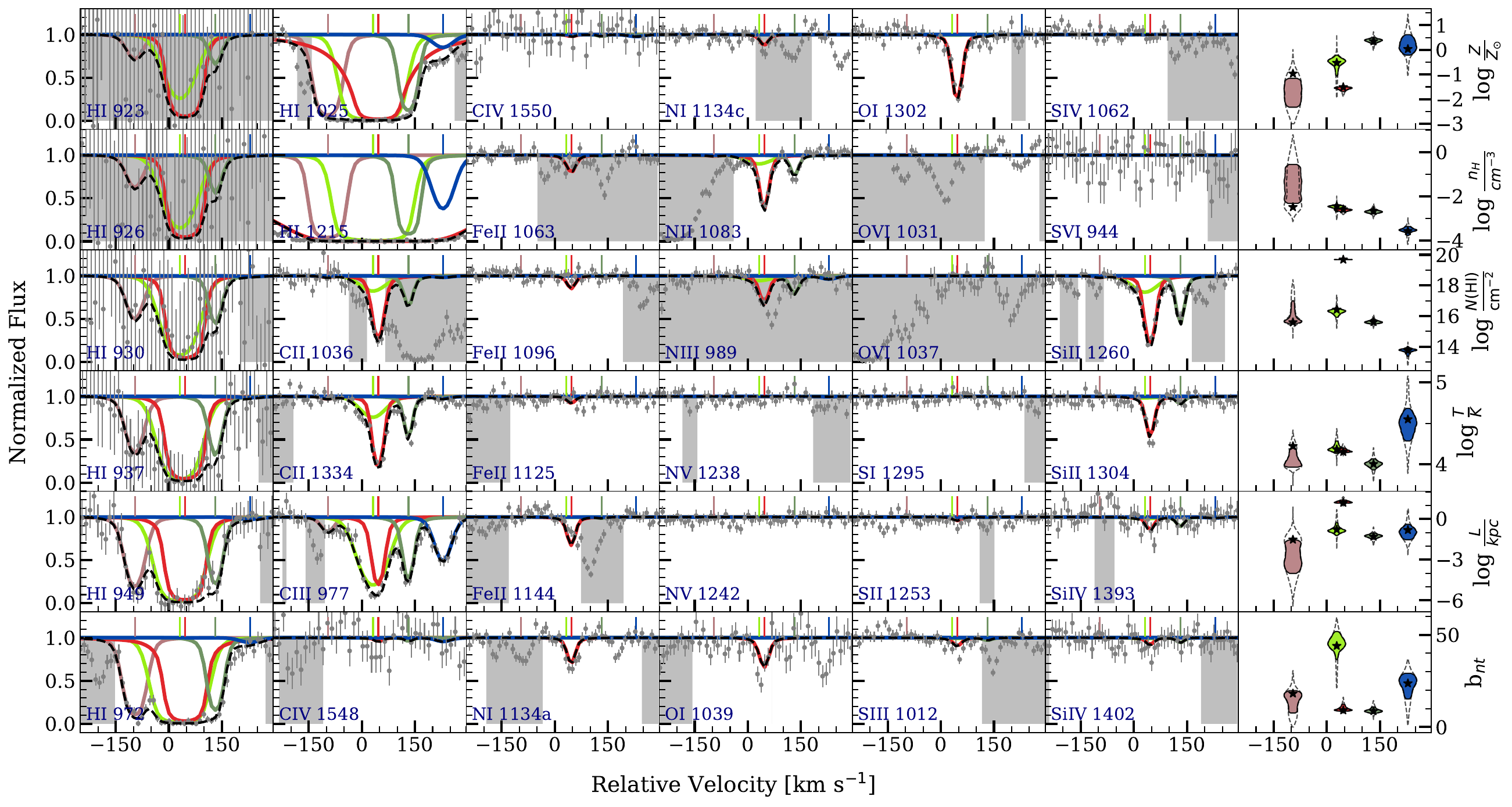}
\caption{Same as Fig.\ref{fig:J0125_0.3985} but for the absorber nearest to the $z$ = 0.1635 galaxy towards the background quasar J0853. The basic galaxy properties are $D \approx 26$ kpc, $\phi$ = $56.0_{-0.8}^{+0.8}$, $i$ = $70.1_{-0.8}^{+1.4}$}
\label{fig:J0853_0.1635}
\end{center}
\end{figure}

\clearpage

\begin{figure}
\begin{center}
\includegraphics[width=\linewidth]{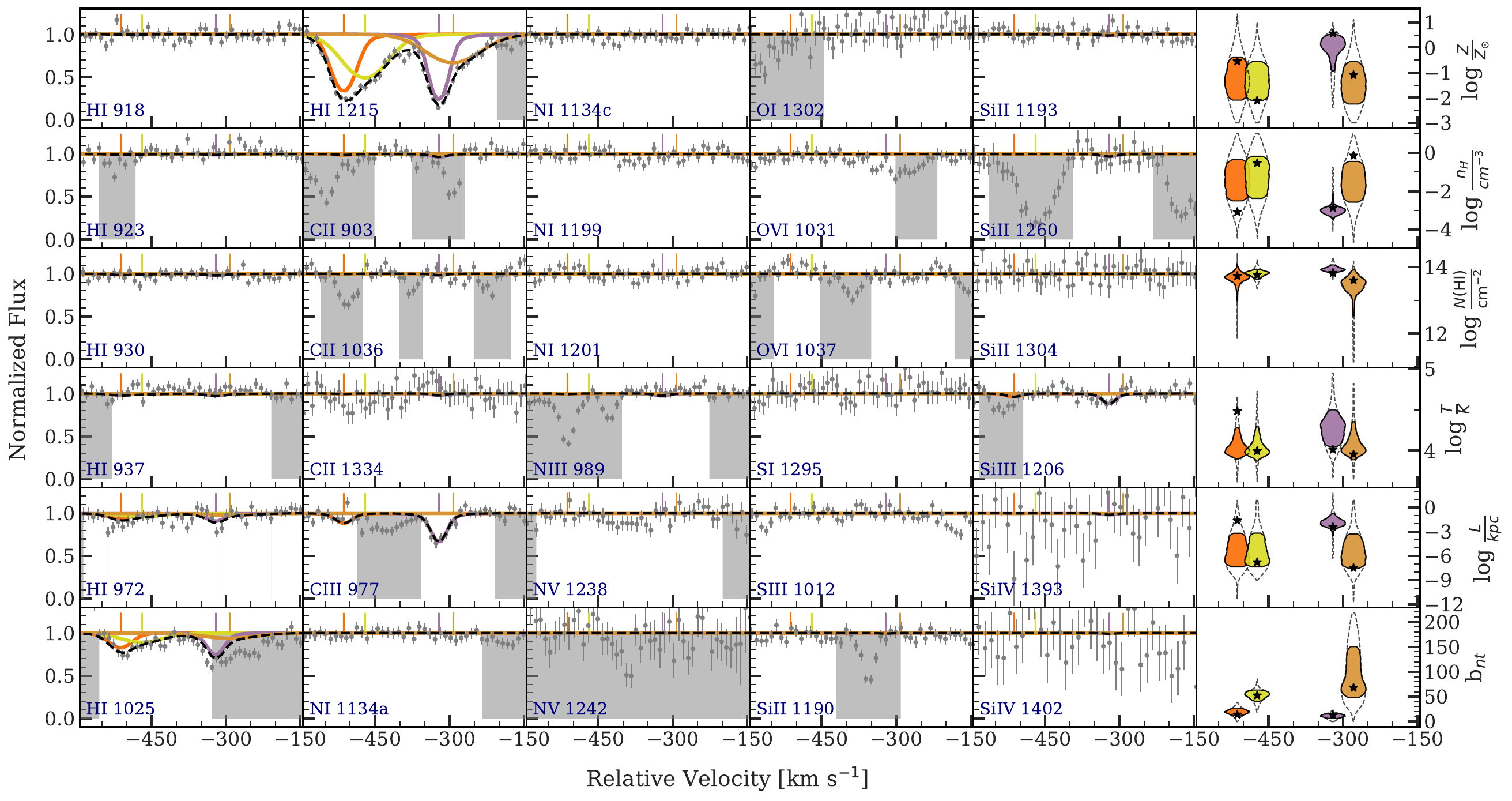}
\caption{Same as Fig.\ref{fig:J0125_0.3985} but for the absorber nearest to the $z$ = 0.2766 galaxy towards the background quasar J0853. The basic galaxy properties are $D \approx 179$ kpc, $\phi$ = $36.7_{-15.3}^{+14.9}$, $i$ = $32.8_{-6.7}^{+5.7}$}
\label{fig:J0853_0.2766}
\end{center}
\end{figure}

\clearpage

\begin{figure}
\begin{center}
\includegraphics[width=\linewidth]{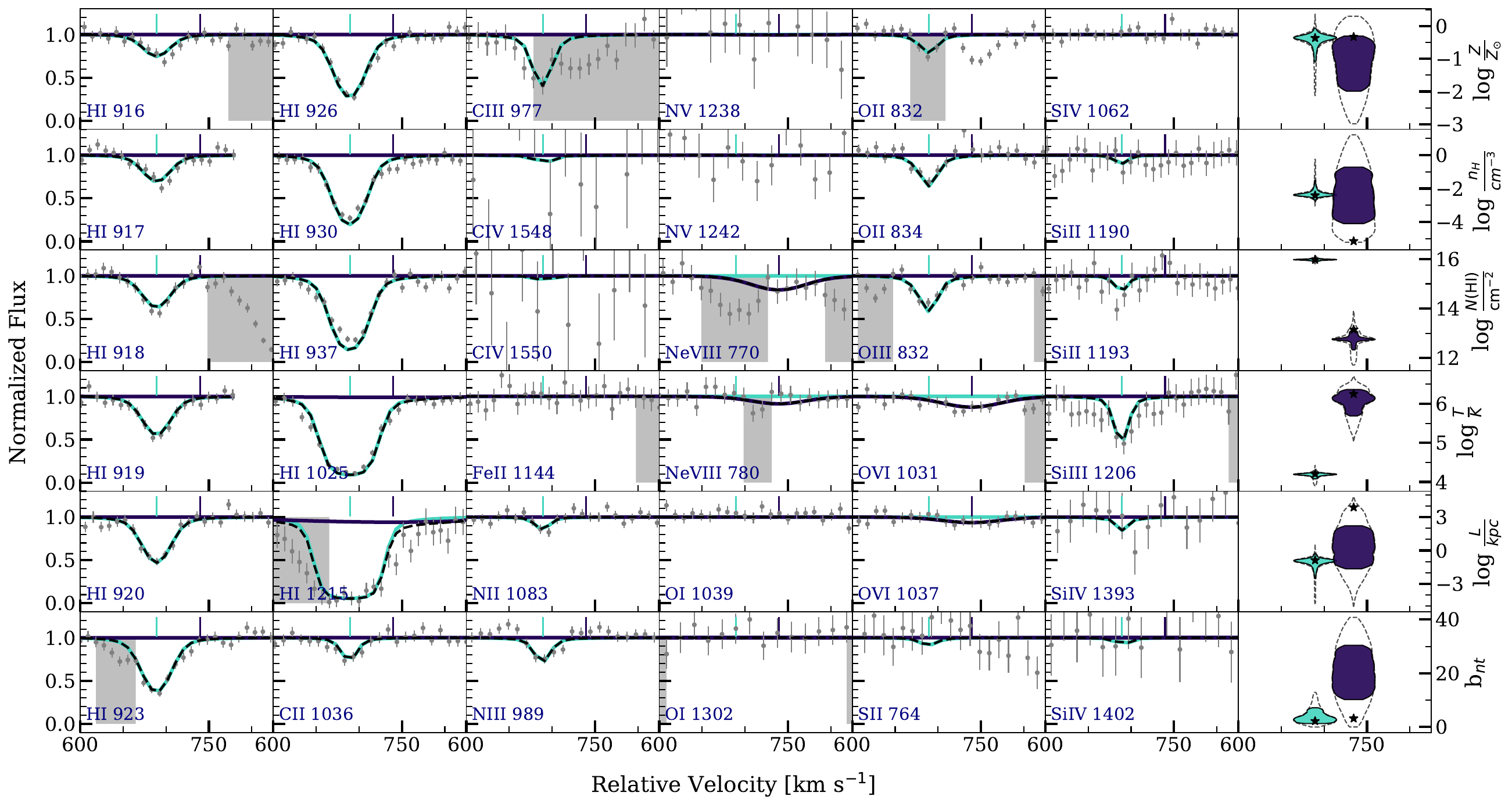}
\caption{Same as Fig.\ref{fig:J0125_0.3985} but for the absorber nearest to the $z$ = 0.4402 galaxy towards the background quasar J0853. The basic galaxy properties are $D \approx 58$ kpc, $\phi$ = $23.0_{-7.6}^{+6.5}$, $i$ = $73.3_{-3.0}^{+3.8}$}
\label{fig:J0853_0.4402}
\end{center}
\end{figure}

\clearpage

\begin{figure}
\begin{center}
\includegraphics[width=\linewidth]{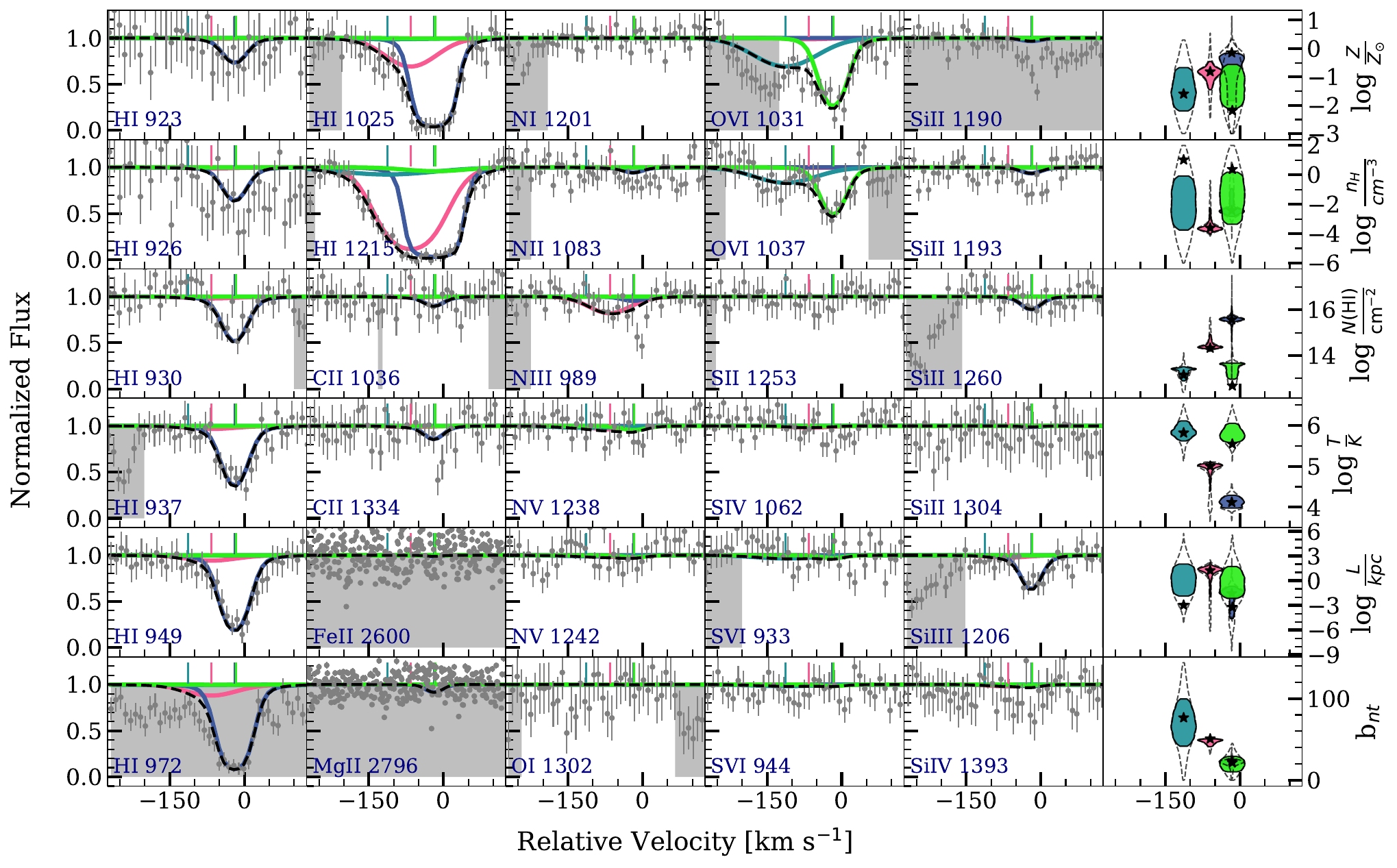}
\caption{Same as Fig.\ref{fig:J0125_0.3985} but for the absorber nearest to the $z$ = 0.2443 galaxy towards the background quasar J0914. The basic galaxy properties are $D \approx 106$ kpc, $\phi$ = $18.2_{-1.0}^{+1.1}$, $i$ = $39.0_{-0.2}^{+0.4}$}
\label{fig:J0914_0.2443}
\end{center}
\end{figure}

\clearpage

\begin{figure}
\begin{center}
\includegraphics[width=\linewidth]{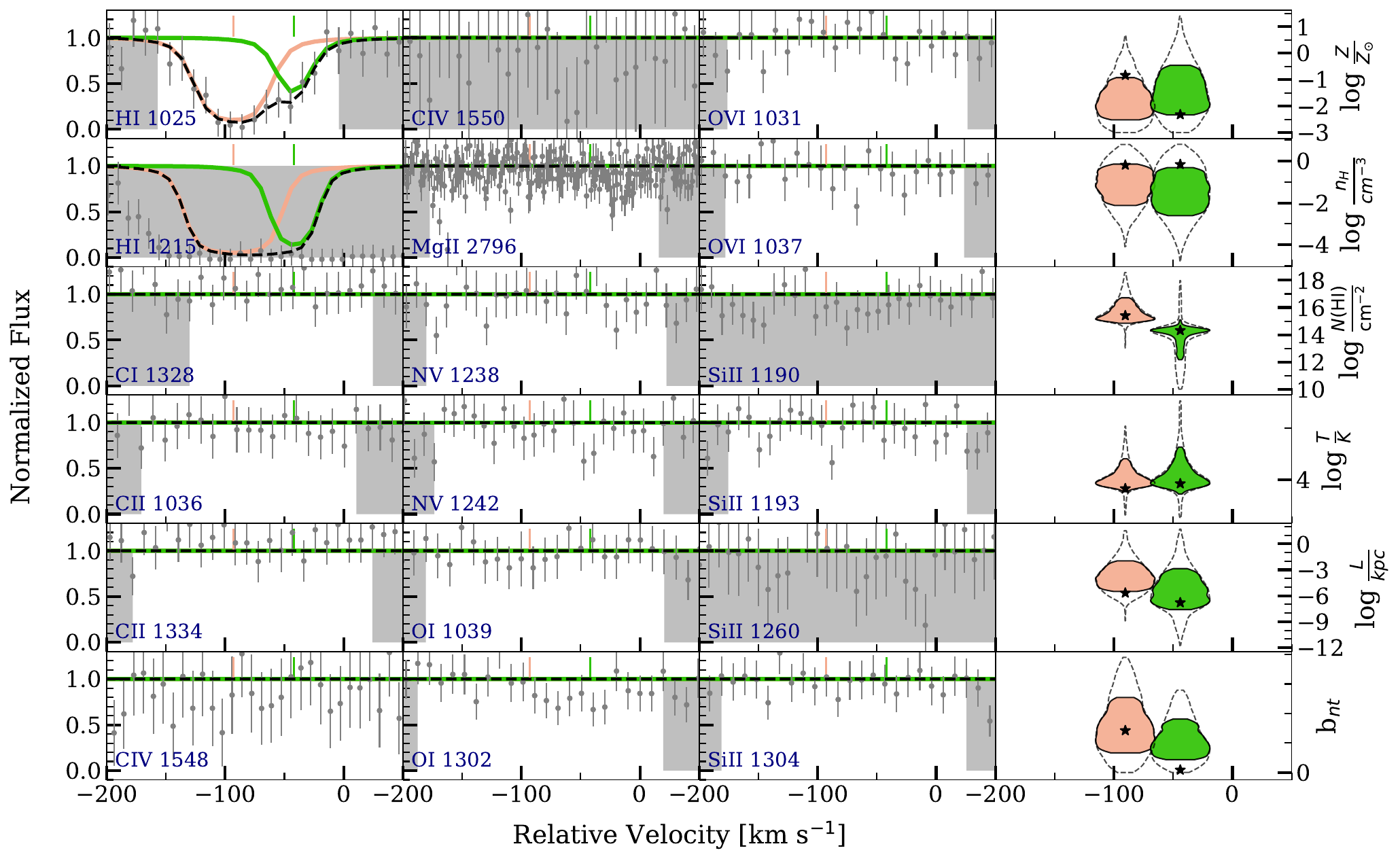}
\caption{Same as Fig.\ref{fig:J0125_0.3985} but for the absorber nearest to the $z$ = 0.1431 galaxy towards the background quasar J0943. The basic galaxy properties are $D \approx 154$ kpc, $\phi$ = $77.7_{-0.1}^{+0.1}$, $i$ = $75.5_{-0.1}^{+0.1}$}
\label{fig:J0943_0.1431}
\end{center}
\end{figure}

\clearpage

\begin{figure}
\begin{center}
\includegraphics[width=\linewidth]{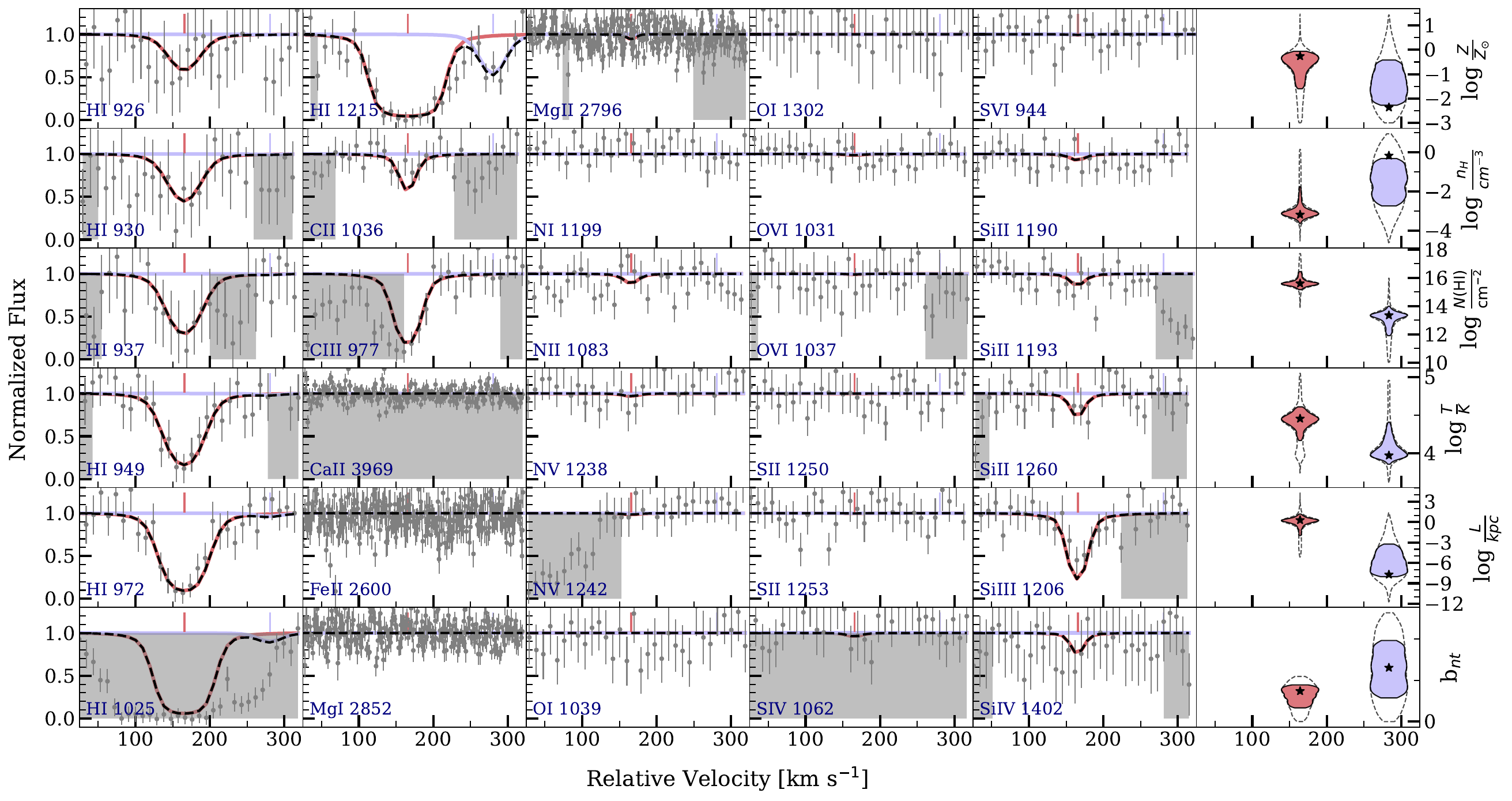}
\caption{Same as Fig.\ref{fig:J0125_0.3985} but for the absorber nearest to the $z$ = 0.2284 galaxy towards the background quasar J0943. The basic galaxy properties are $D \approx 123$ kpc, $\phi$ = $30.4_{-0.4}^{+0.3}$, $i$ = $52.3_{-0.3}^{+0.3}$}
\label{fig:J0943_0.2284}
\end{center}
\end{figure}

\clearpage

\begin{figure}
\begin{center}
\includegraphics[width=\linewidth]{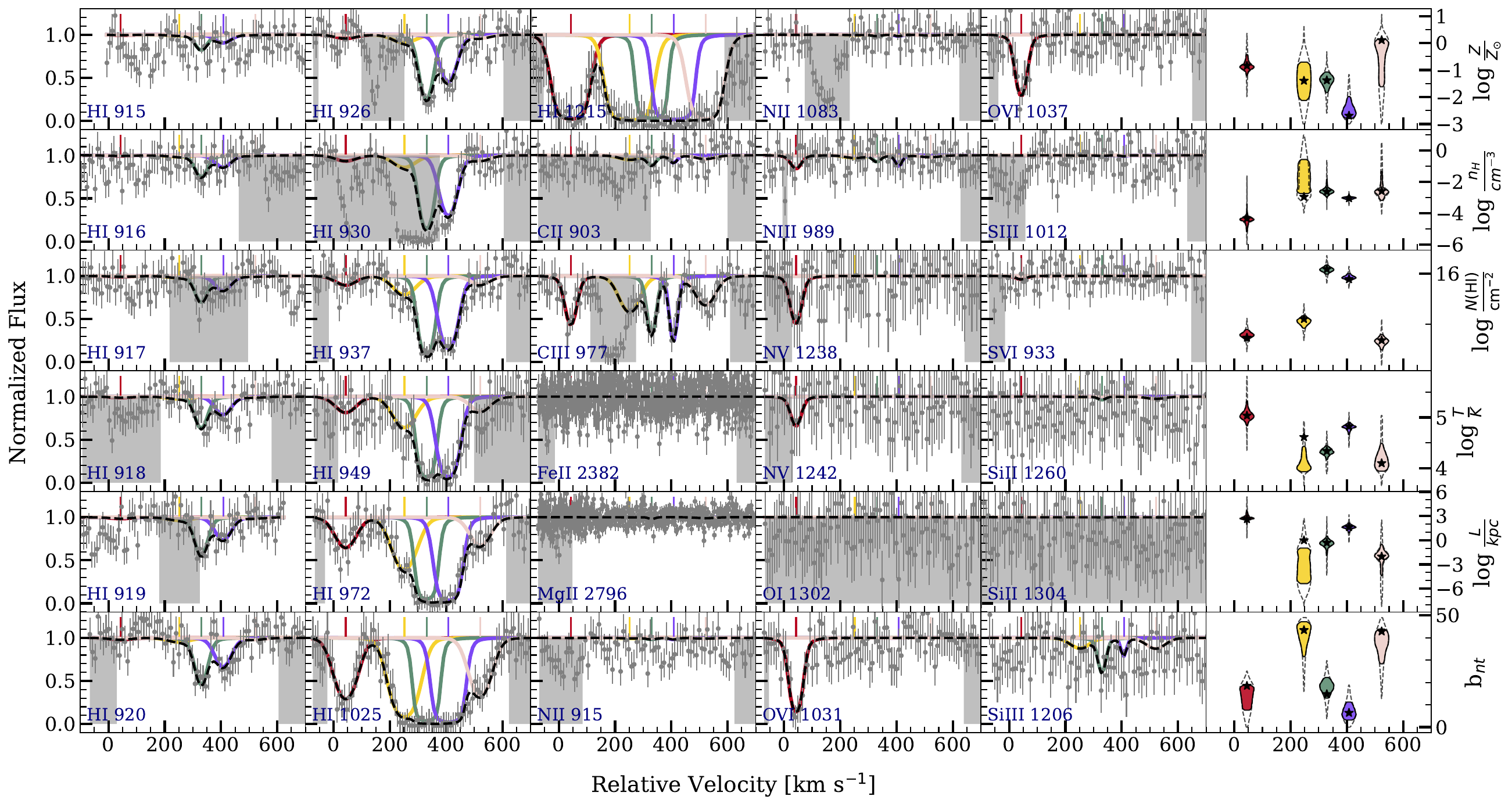}
\caption{Same as Fig.\ref{fig:J0125_0.3985} but for the absorber nearest to the $z$ = 0.3531 galaxy towards the background quasar J0943. The basic galaxy properties are $D \approx 96$ kpc, $\phi$ = $8.2_{-5.0}^{+3.0}$, $i$ = $44.4_{-1.2}^{+1.1}$}
\label{fig:J0943_0.3531}
\end{center}
\end{figure}

\clearpage

\begin{figure}
\begin{center}
\includegraphics[width=\linewidth]{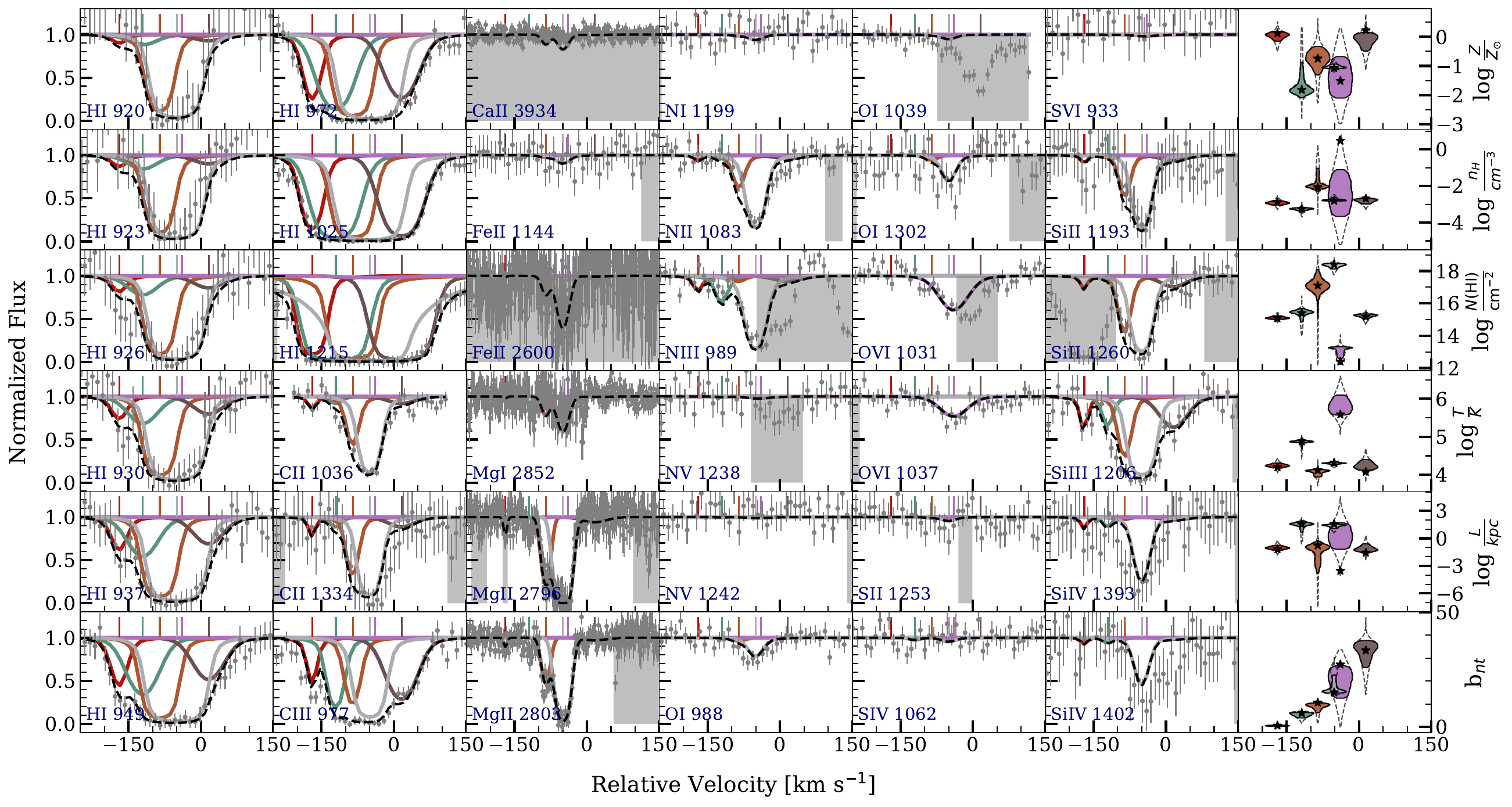}
\caption{Same as Fig.\ref{fig:J0125_0.3985} but for the absorber nearest to the $z$ = 0.2119 galaxy towards the background quasar J0950. The basic galaxy properties are $D \approx 94$ kpc, $\phi$ = $16.6_{-0.1}^{+0.1}$, $i$ = $47.7_{-0.1}^{+0.1}$}
\label{fig:J0950_0.2119}
\end{center}
\end{figure}

\clearpage

\begin{figure}
\begin{center}
\includegraphics[width=\linewidth]{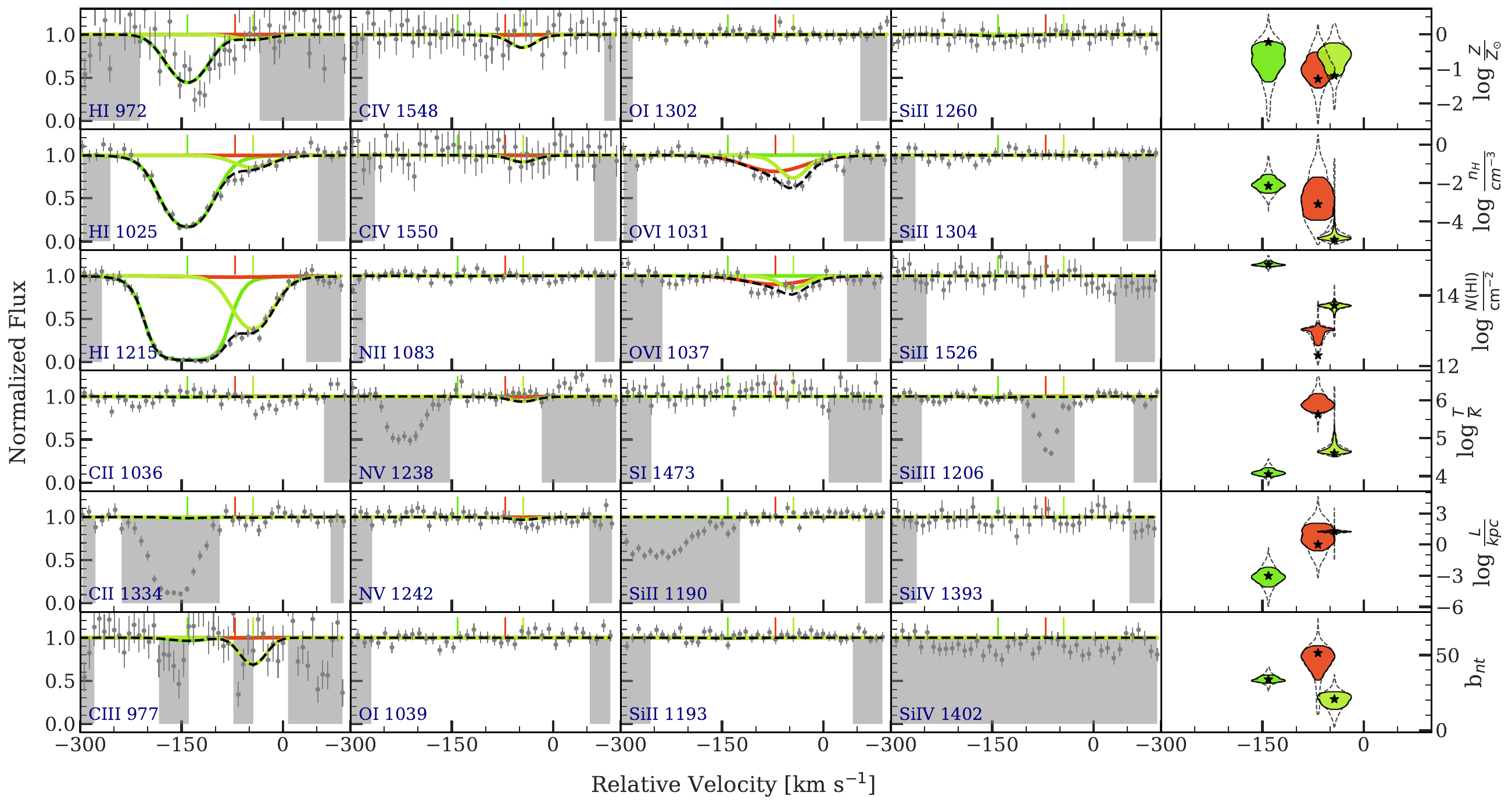}
\caption{Same as Fig.\ref{fig:J0125_0.3985} but for the absorber nearest to the $z$ = 0.1380 galaxy towards the background quasar J1004. The basic galaxy properties are $D \approx 57$ kpc, $\phi$ = $12.4_{-2.9}^{+2.4}$, $i$ = $79.1_{-2.1}^{+2.2}$}
\label{fig:J1004_0.1380}
\end{center}
\end{figure}

\clearpage

\begin{figure}
\begin{center}
\includegraphics[width=\linewidth]{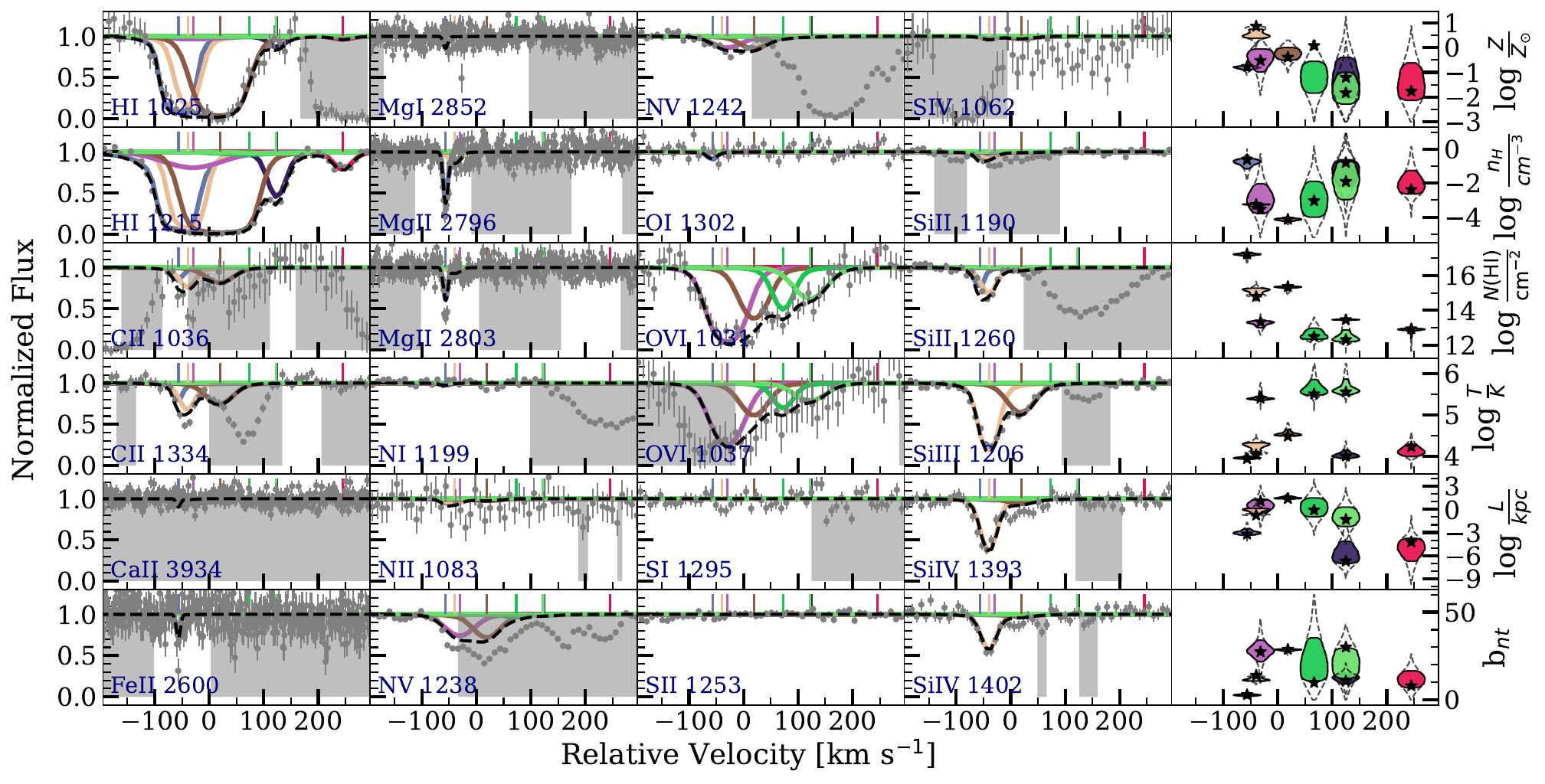}
\caption{Same as Fig.\ref{fig:J0125_0.3985} but for the absorber nearest to the $z$ = 0.2279 galaxy towards the background quasar J1009. The basic galaxy properties are $D \approx 64$ kpc, $\phi$ = $89.6_{-1.3}^{+0.4}$, $i$ = $66.3_{-0.9}^{+0.6}$}
\label{fig:J1009_0.2279}
\end{center}
\end{figure}

\clearpage

\begin{figure}
\begin{center}
\includegraphics[width=\linewidth]{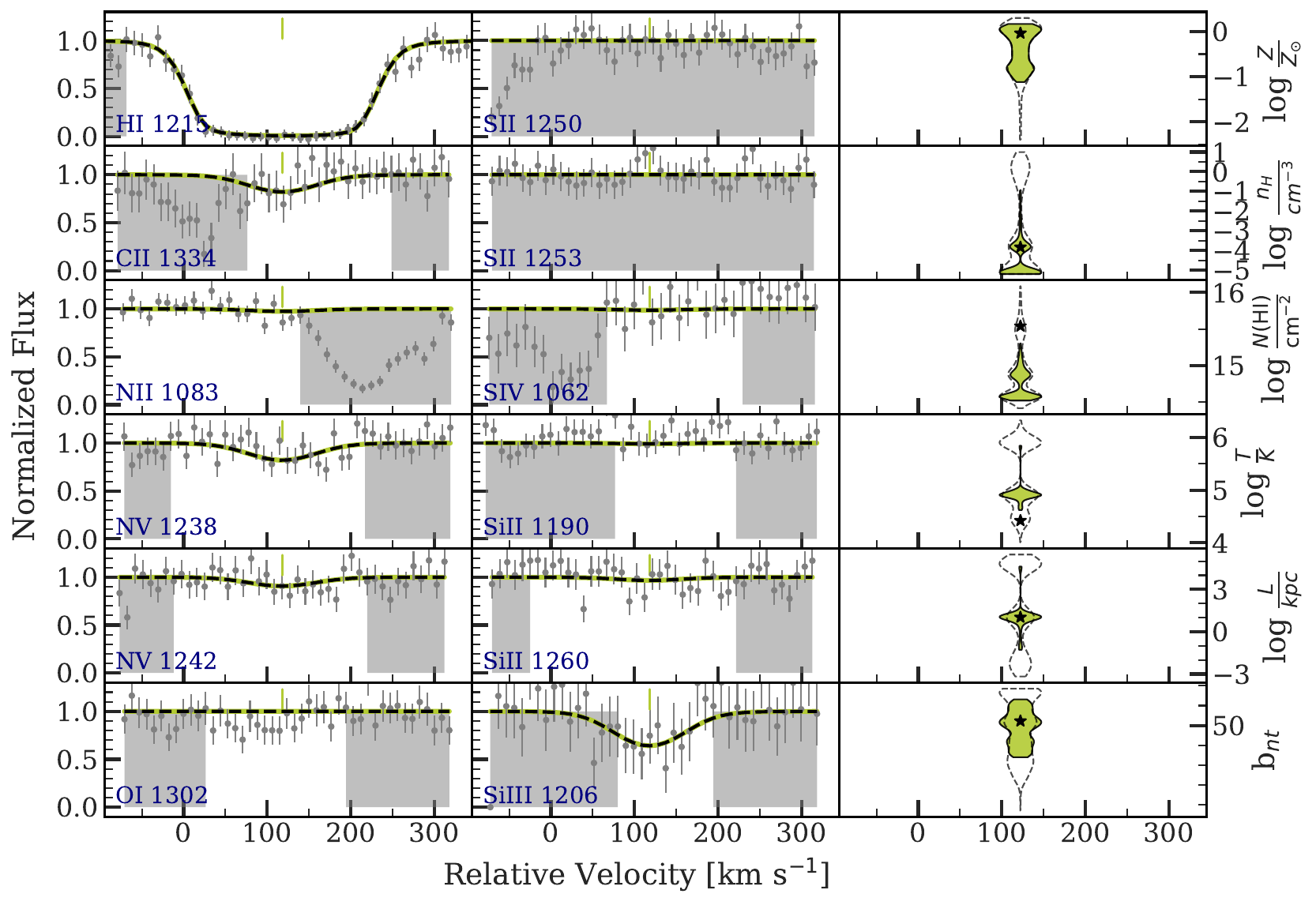}
\caption{Same as Fig.\ref{fig:J0125_0.3985} but for the absorber nearest to the $z$ = 0.3153 galaxy towards the background quasar J1041. The basic galaxy properties are $D \approx 54$ kpc, $\phi$ = $77.3_{-1.2}^{+1.2}$, $i$ = $72.6_{-1.3}^{+1.3}$}
\label{fig:J1041_0.3153}
\end{center}
\end{figure}

\clearpage

\begin{figure}
\begin{center}
\includegraphics[width=\linewidth]{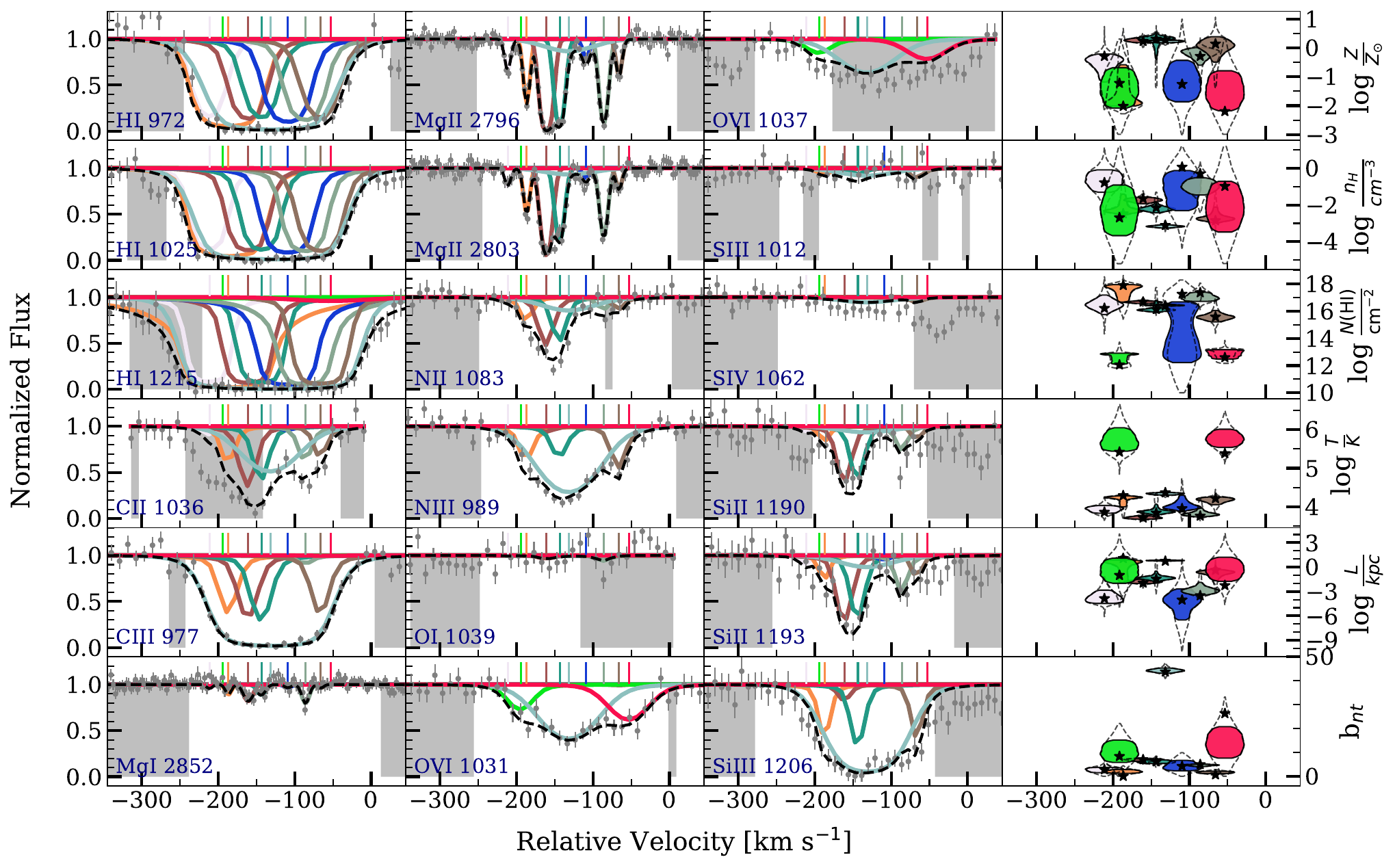}
\caption{Same as Fig.\ref{fig:J0125_0.3985} but for the absorber nearest to the $z$ = 0.4422 galaxy towards the background quasar J1041. The basic galaxy properties are $D \approx 56$ kpc, $\phi$ = $4.3_{-1.0}^{+0.9}$, $i$ = $49.8_{-5.2}^{+7.4}$}
\label{fig:J1041_0.4422}
\end{center}
\end{figure}

\clearpage

\begin{figure}
\begin{center}
\includegraphics[width=\linewidth]{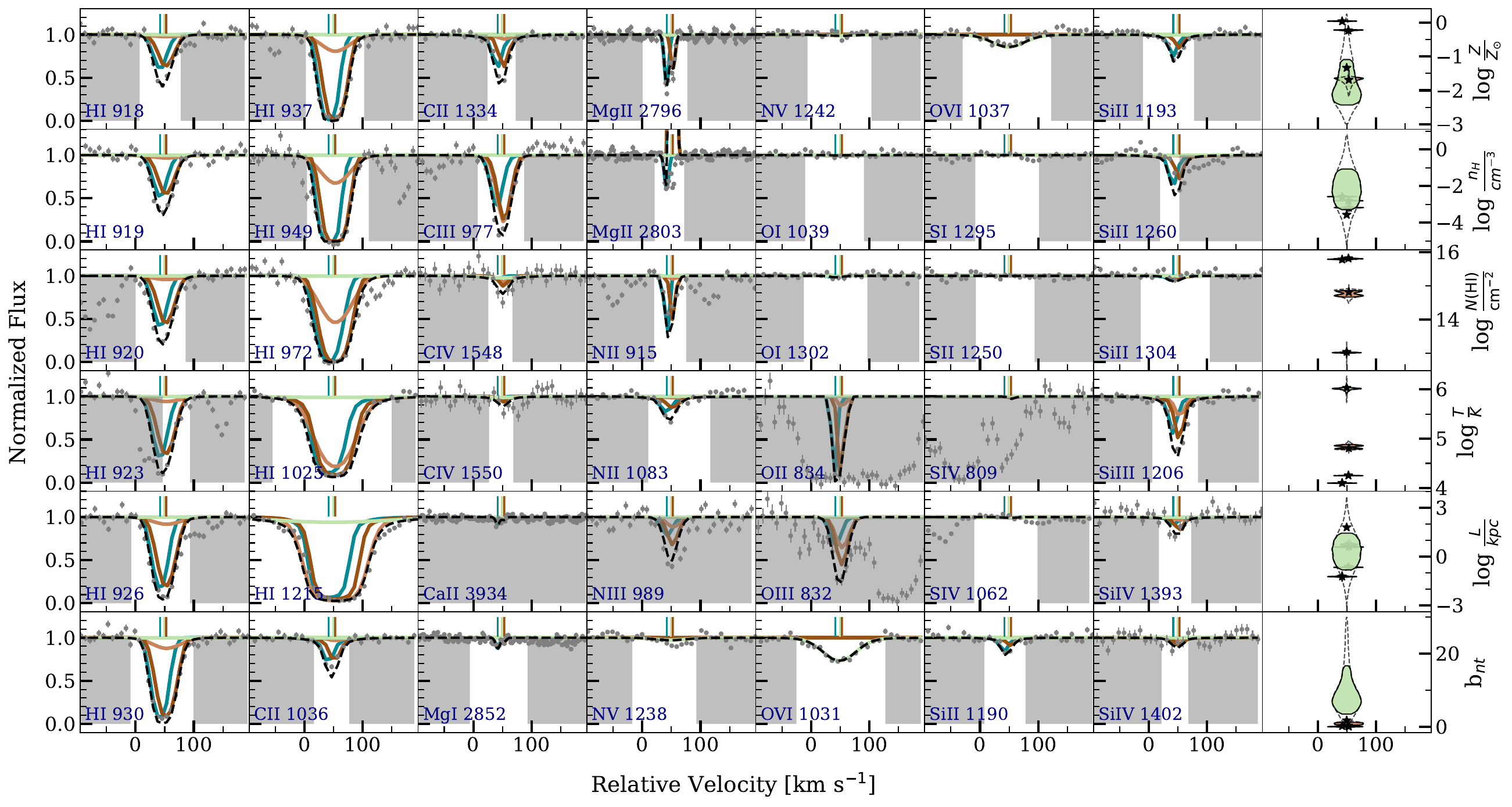}
\caption{Same as Fig.\ref{fig:J0125_0.3985} but for the absorber nearest to the $z$ = 0.1383 galaxy towards the background quasar J1119. The basic galaxy properties are $D \approx 138$ kpc, $\phi$ = $34.4_{-0.4}^{+0.4}$, $i$ = $26.4_{-0.4}^{+0.8}$}
\label{fig:J1119_0.1383}
\end{center}
\end{figure}

\clearpage

\begin{figure}
\begin{center}
\includegraphics[width=\linewidth]{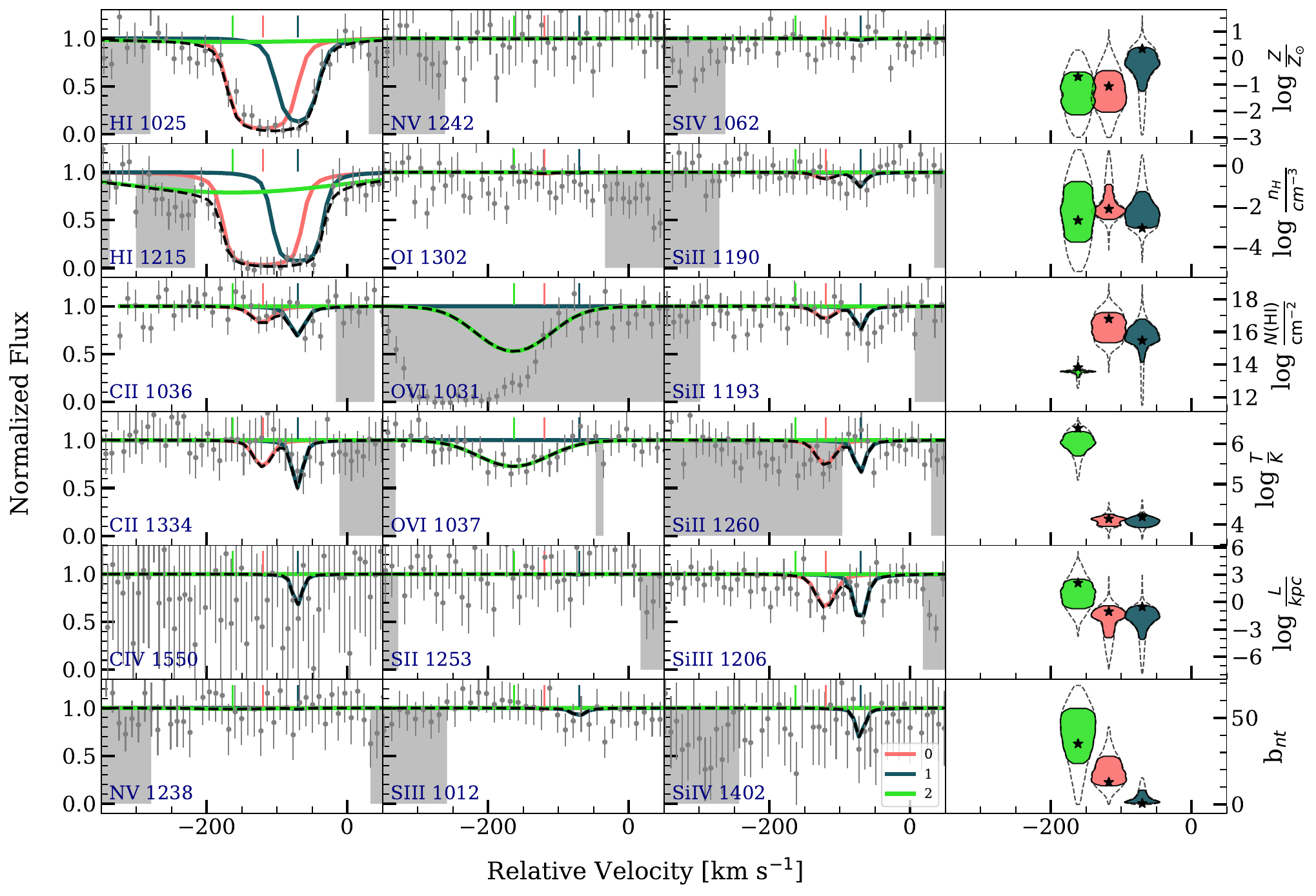}
\caption{Same as Fig.\ref{fig:J0125_0.3985} but for the absorber nearest to the $z$ = 0.1546 galaxy towards the background quasar J1133. The basic galaxy properties are $D \approx 56$ kpc, $\phi$ = $56.1_{-1.3}^{+1.7}$, $i$ = $23.5_{-0.2}^{+0.4}$}
\label{fig:J1133_0.1546}
\end{center}
\end{figure}

\clearpage

\begin{figure}
\begin{center}
\includegraphics[width=\linewidth]{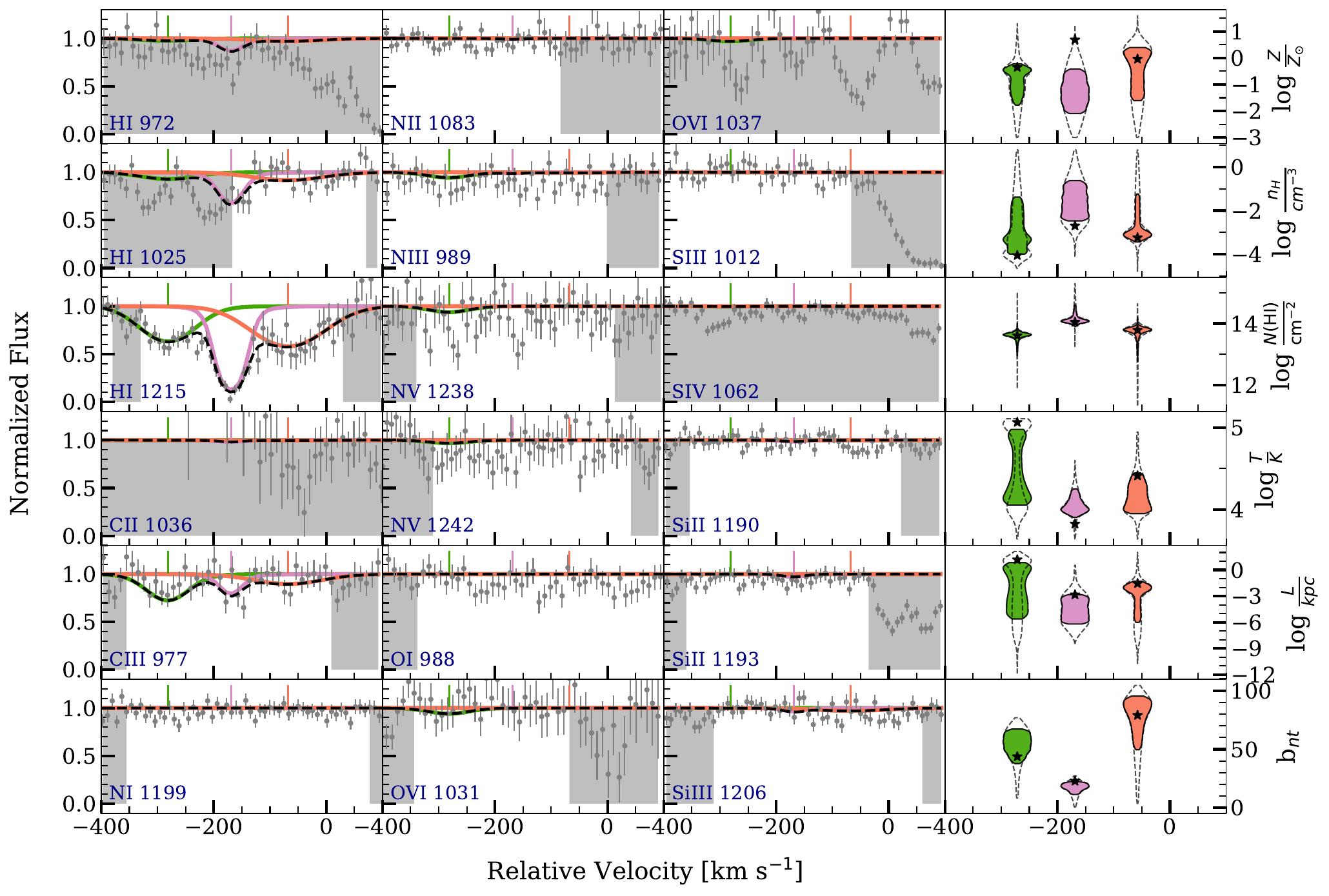}
\caption{Same as Fig.\ref{fig:J0125_0.3985} but for the absorber nearest to the $z$ = 0.1755 galaxy towards the background quasar J1139. The basic galaxy properties are $D \approx 163$ kpc, $\phi$ = $21.4_{-10.7}^{+10.7}$, $i$ = $85.0_{-0.2}^{+0.2}$}
\label{fig:J1139_0.1755}
\end{center}
\end{figure}

\clearpage

\begin{figure}
\begin{center}
\includegraphics[width=\linewidth]{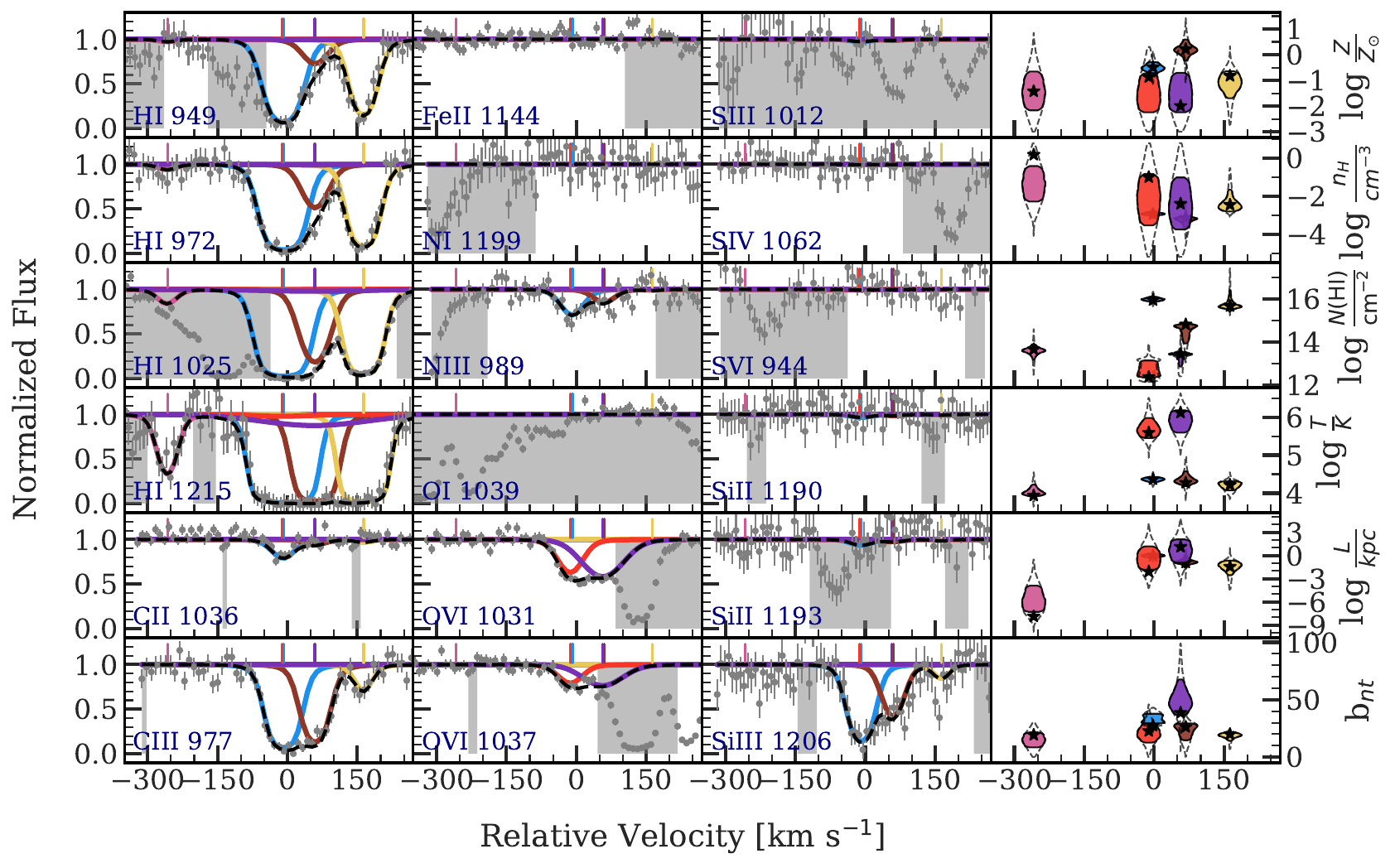}
\caption{Same as Fig.\ref{fig:J0125_0.3985} but for the absorber nearest to the $z$ = 0.2042 galaxy towards the background quasar J1139. The basic galaxy properties are $D \approx 93$ kpc, $\phi$ = $5.8_{-0.5}^{+0.4}$, $i$ = $83.4_{-0.5}^{+0.4}$}
\label{fig:J1139_0.2042}
\end{center}
\end{figure}

\clearpage

\begin{figure}
\begin{center}
\includegraphics[width=\linewidth]{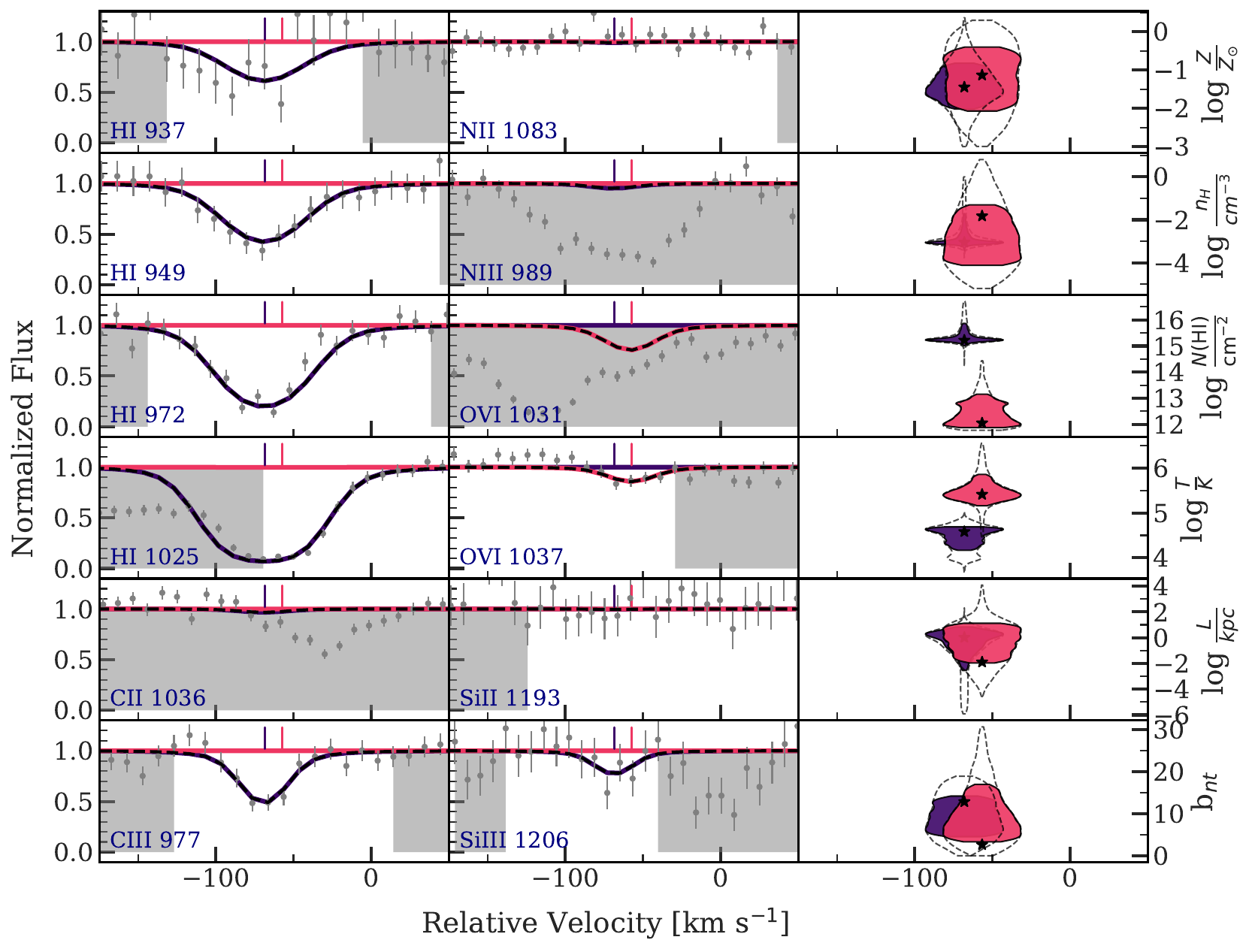}
\caption{Same as Fig.\ref{fig:J0125_0.3985} but for the absorber nearest to the $z$ = 0.2123 galaxy towards the background quasar J1139. The basic galaxy properties are $D \approx 175$ kpc, $\phi$ = $80.4_{-0.5}^{+0.4}$, $i$ = $85.0_{-0.6}^{+5.0}$}
\label{fig:J1139_0.2123}
\end{center}
\end{figure}

\clearpage

\begin{figure}
\begin{center}
\includegraphics[width=\linewidth]{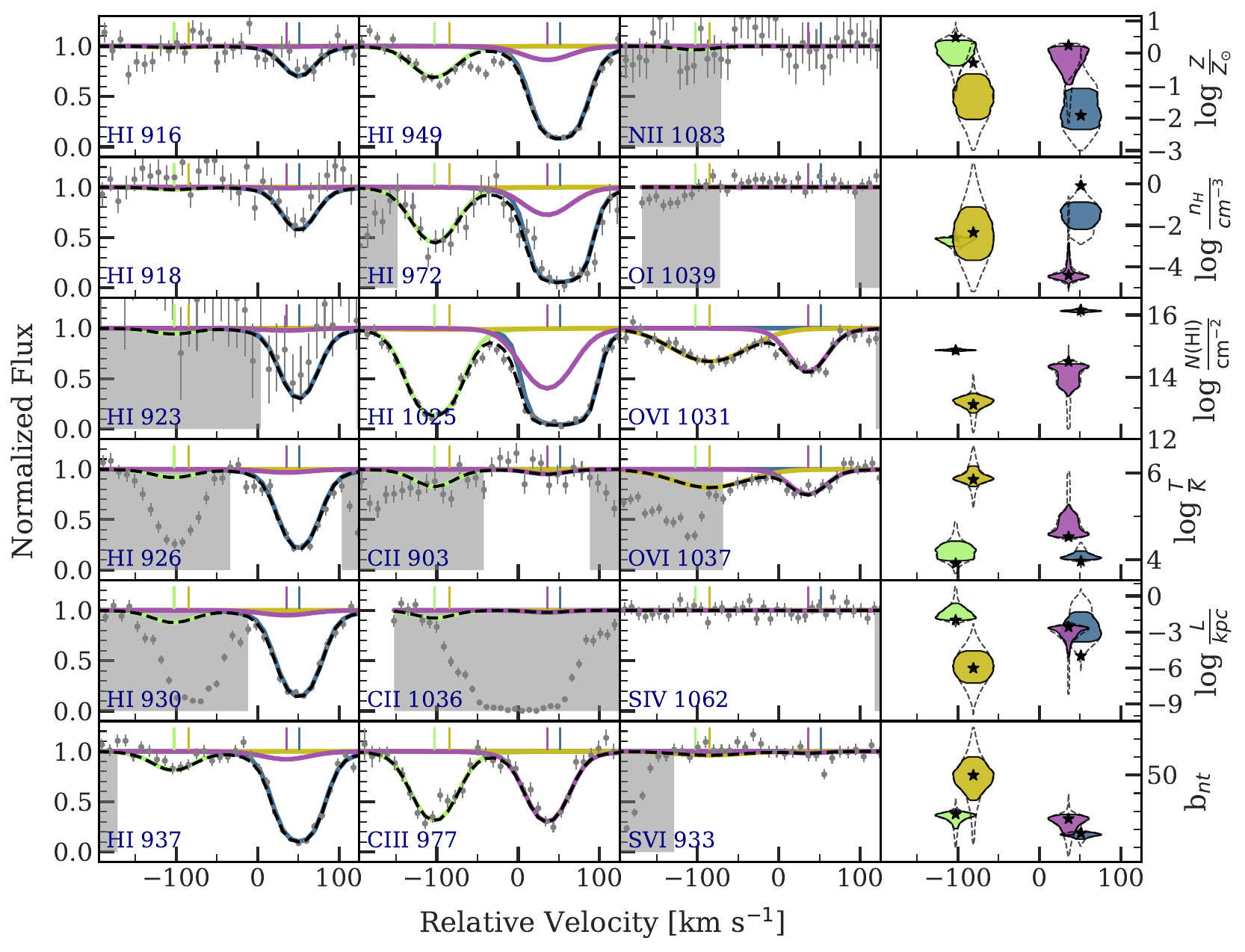}
\caption{Same as Fig.\ref{fig:J0125_0.3985} but for the absorber nearest to the $z$ = 0.3193 galaxy towards the background quasar J1139. The basic galaxy properties are $D \approx 73$ kpc, $\phi$ = $39.1_{-1.7}^{+1.9}$, $i$ = $83.4_{-1.1}^{+1.4}$}
\label{fig:J1139_0.3193}
\end{center}
\end{figure}

\clearpage

\begin{figure}
\begin{center}
\includegraphics[width=\linewidth]{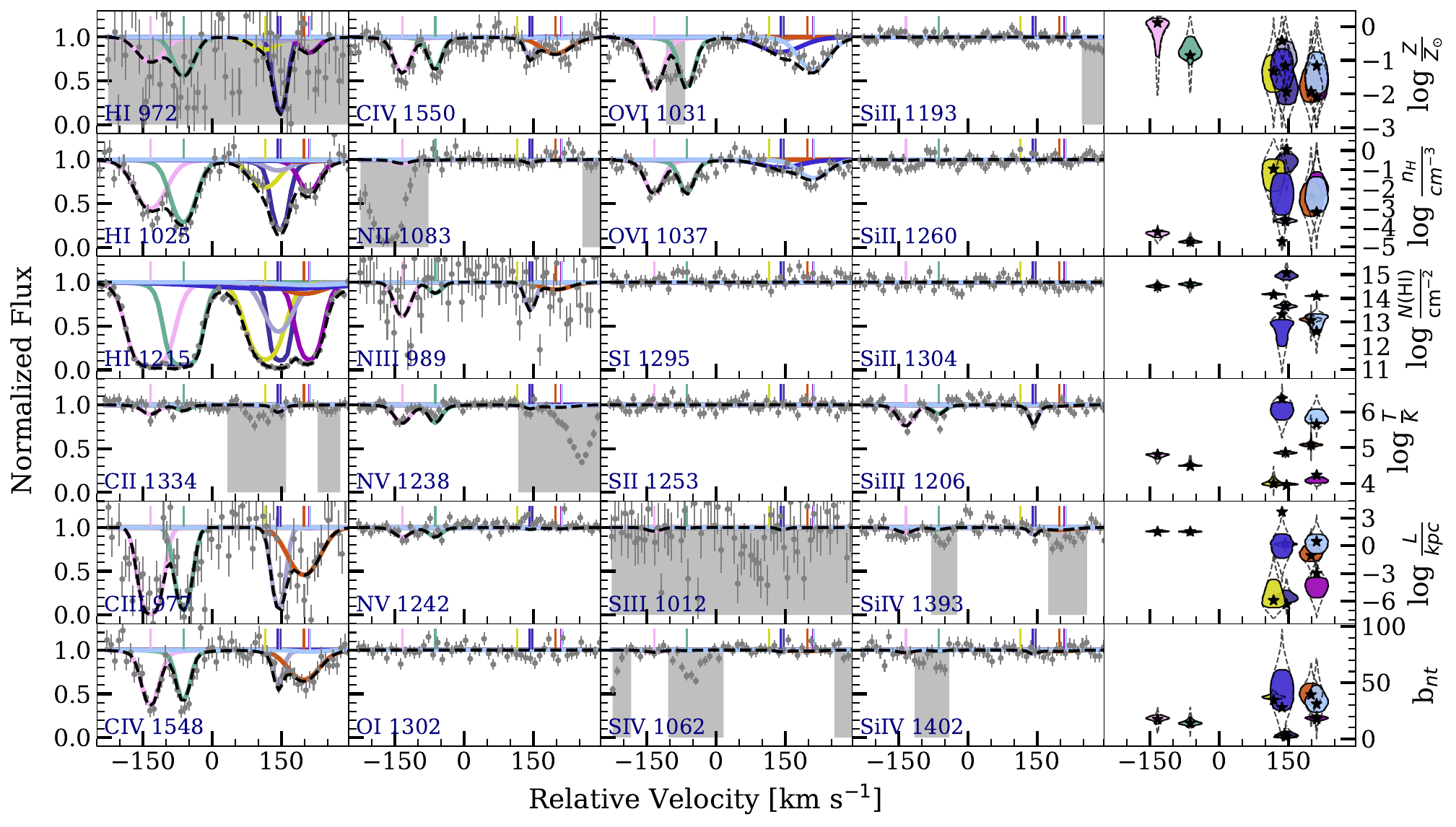}
\caption{Same as Fig.\ref{fig:J0125_0.3985} but for the absorber nearest to the $z$ = 0.1241 galaxy towards the background quasar J1219. The basic galaxy properties are $D \approx 93$ kpc, $\phi$ = $67.2_{-67.2}^{+22.8}$, $i$ = $22.0_{-21.8}^{+18.7}$}
\label{fig:J1219_0.1241}
\end{center}
\end{figure}

\clearpage

\begin{figure}
\begin{center}
\includegraphics[width=\linewidth]{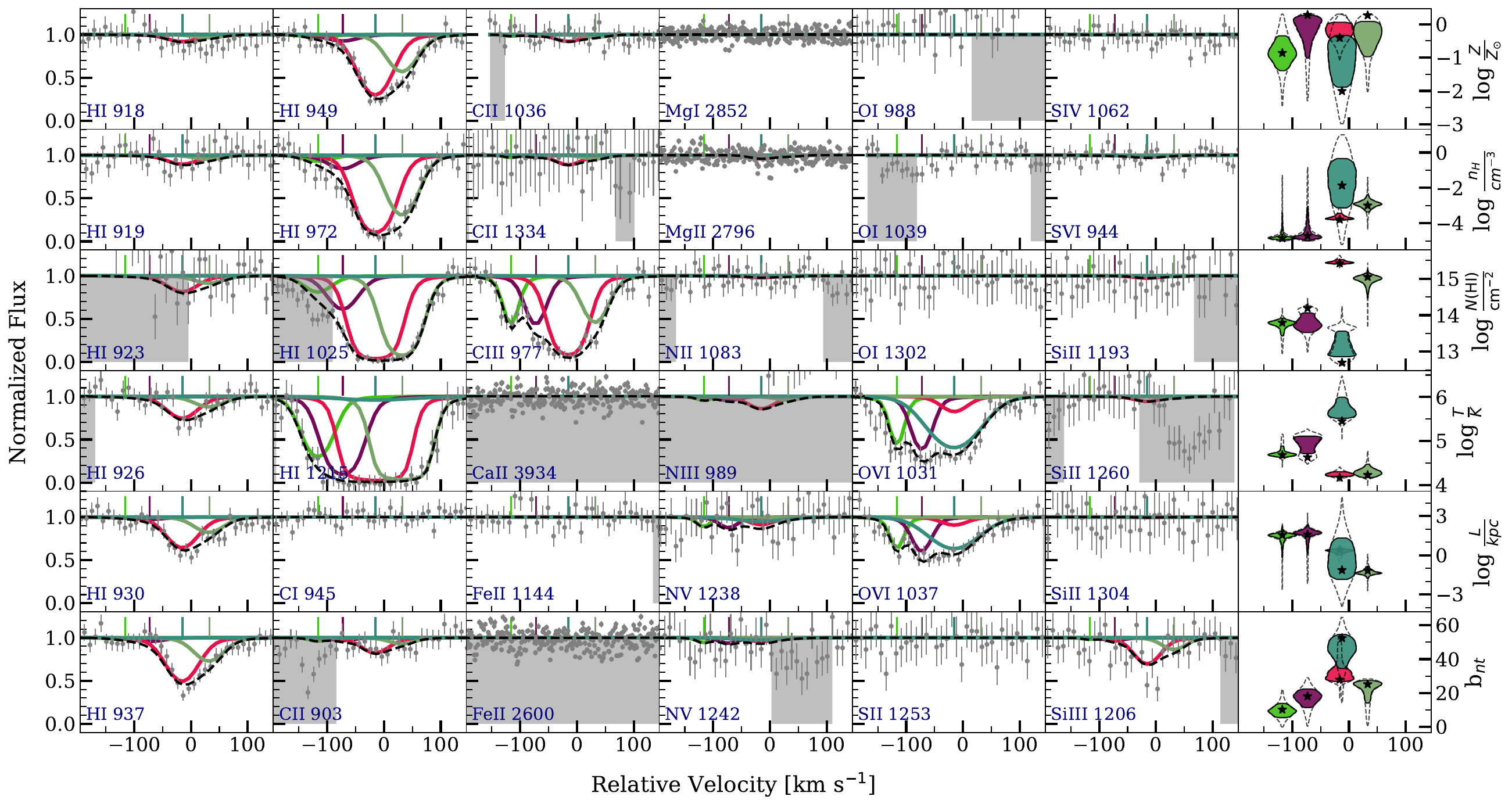}
\caption{Same as Fig.\ref{fig:J0125_0.3985} but for the absorber nearest to the $z$ = 0.3188 galaxy towards the background quasar J1233. The basic galaxy properties are $D \approx 89$ kpc, $\phi$ = $17.0_{-2.3}^{+2.0}$, $i$ = $38.7_{-1.8}^{+1.6}$}
\label{fig:J1233_0.3188}
\end{center}
\end{figure}

\clearpage

\begin{figure}
\begin{center}
\includegraphics[width=\linewidth]{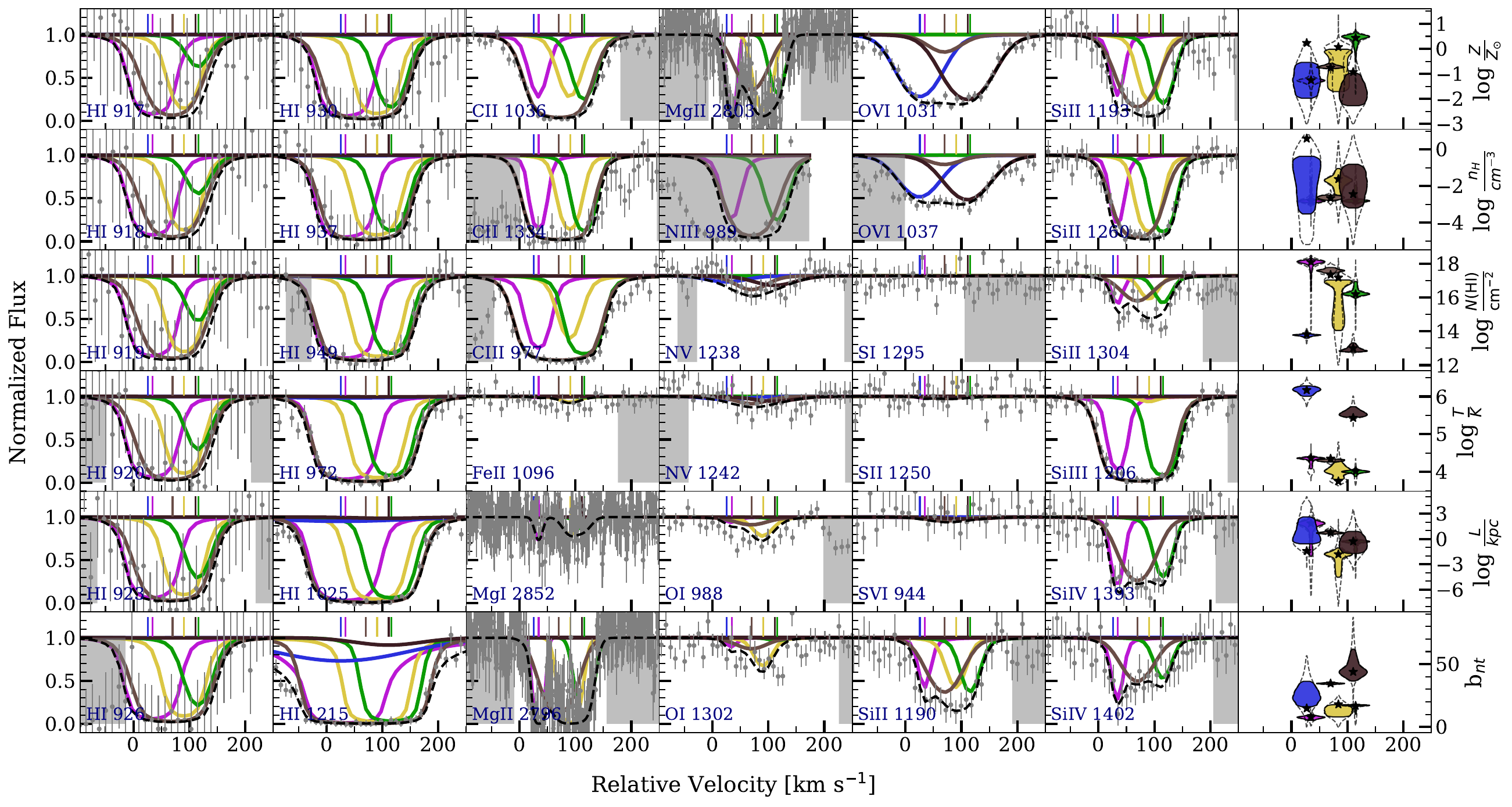}
\caption{Same as Fig.\ref{fig:J0125_0.3985} but for the absorber nearest to the $z$ = 0.2053 galaxy towards the background quasar J1241. The basic galaxy properties are $D \approx 21$ kpc, $\phi$ = $77.6_{-0.4}^{+0.3}$, $i$ = $56.4_{-0.5}^{+0.3}$}
\label{fig:J1241_0.2053}
\end{center}
\end{figure}

\clearpage

\begin{figure}
\begin{center}
\includegraphics[width=\linewidth]{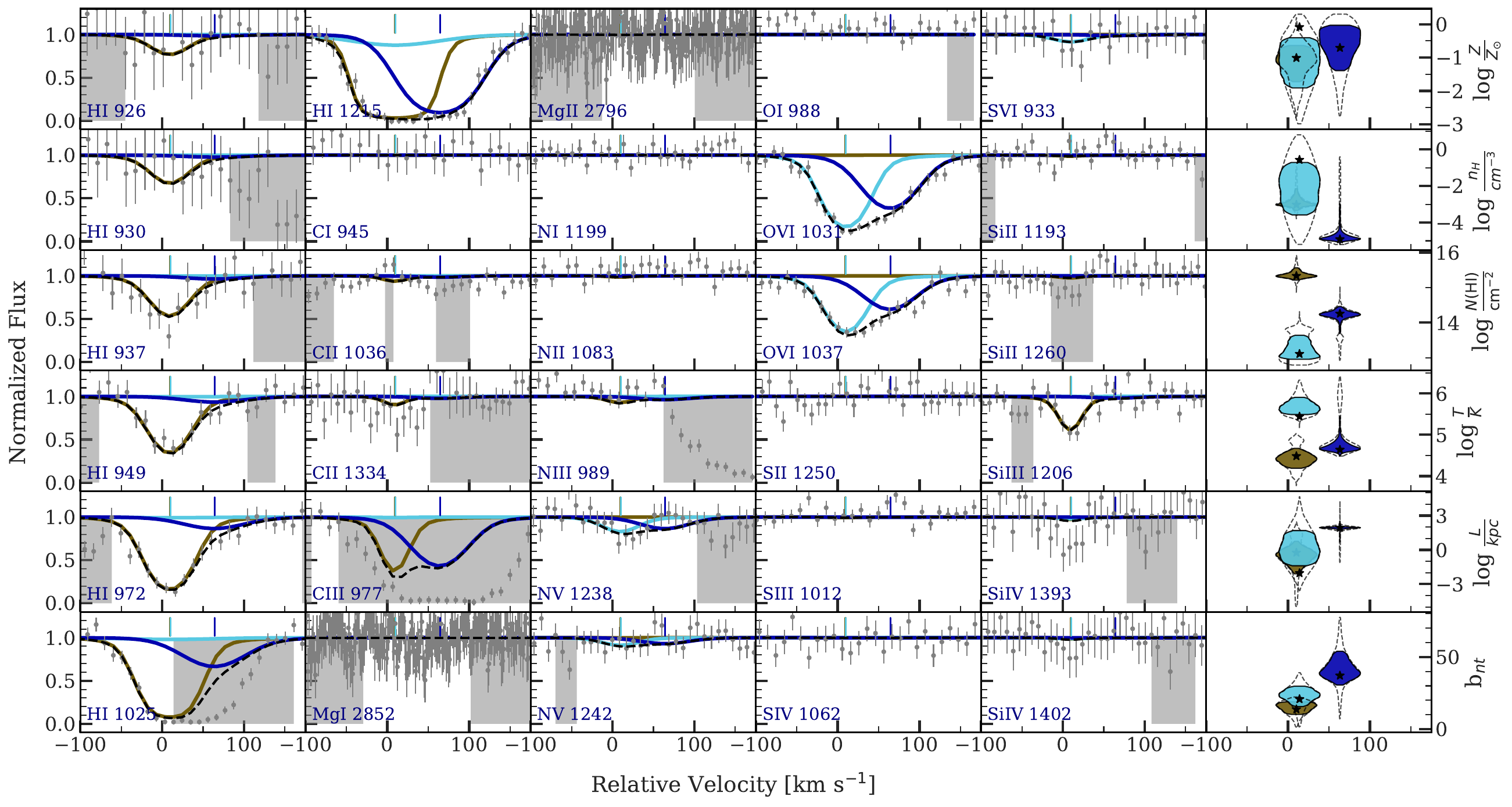}
\caption{Same as Fig.\ref{fig:J0125_0.3985} but for the absorber nearest to the $z$ = 0.2179 galaxy towards the background quasar J1241. The basic galaxy properties are $D \approx 95$ kpc, $\phi$ = $63.0_{-2.1}^{+1.8}$, $i$ = $17.4_{-1.6}^{+1.4}$}
\label{fig:J1241_0.2179}
\end{center}
\end{figure}

\clearpage

\begin{figure}
\begin{center}
\includegraphics[width=\linewidth]{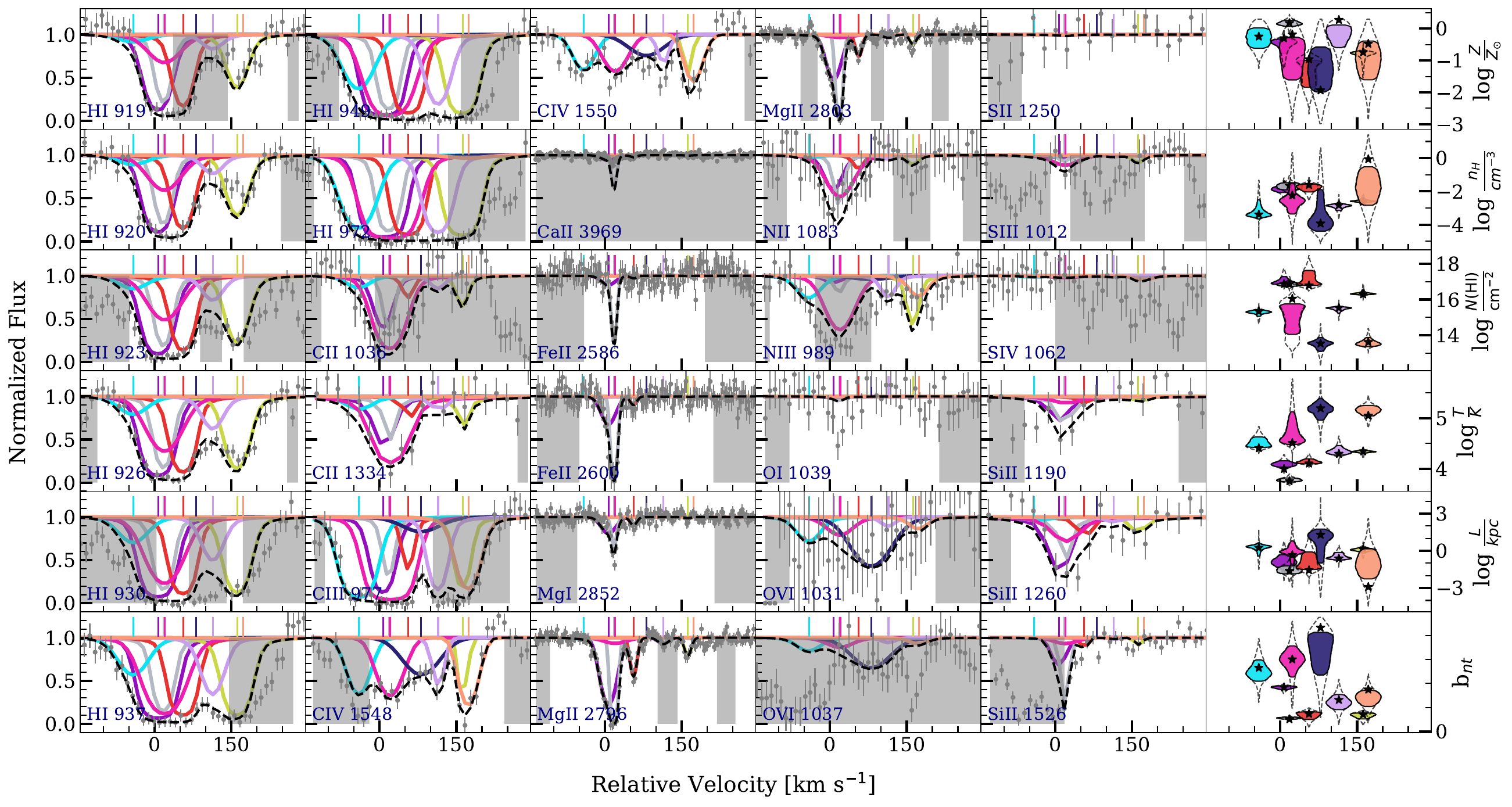}
\caption{Same as Fig.\ref{fig:J0125_0.3985} but for the absorber nearest to the $z$ = 0.5504 galaxy towards the background quasar J1244. The basic galaxy properties are $D \approx 21$ kpc, $\phi$ = $20.1_{-19.1}^{+16.7}$, $i$ = $31.7_{-4.8}^{+16.2}$}
\label{fig:J1244_0.5504}
\end{center}
\end{figure}

\clearpage

\begin{figure}
\begin{center}
\includegraphics[width=\linewidth]{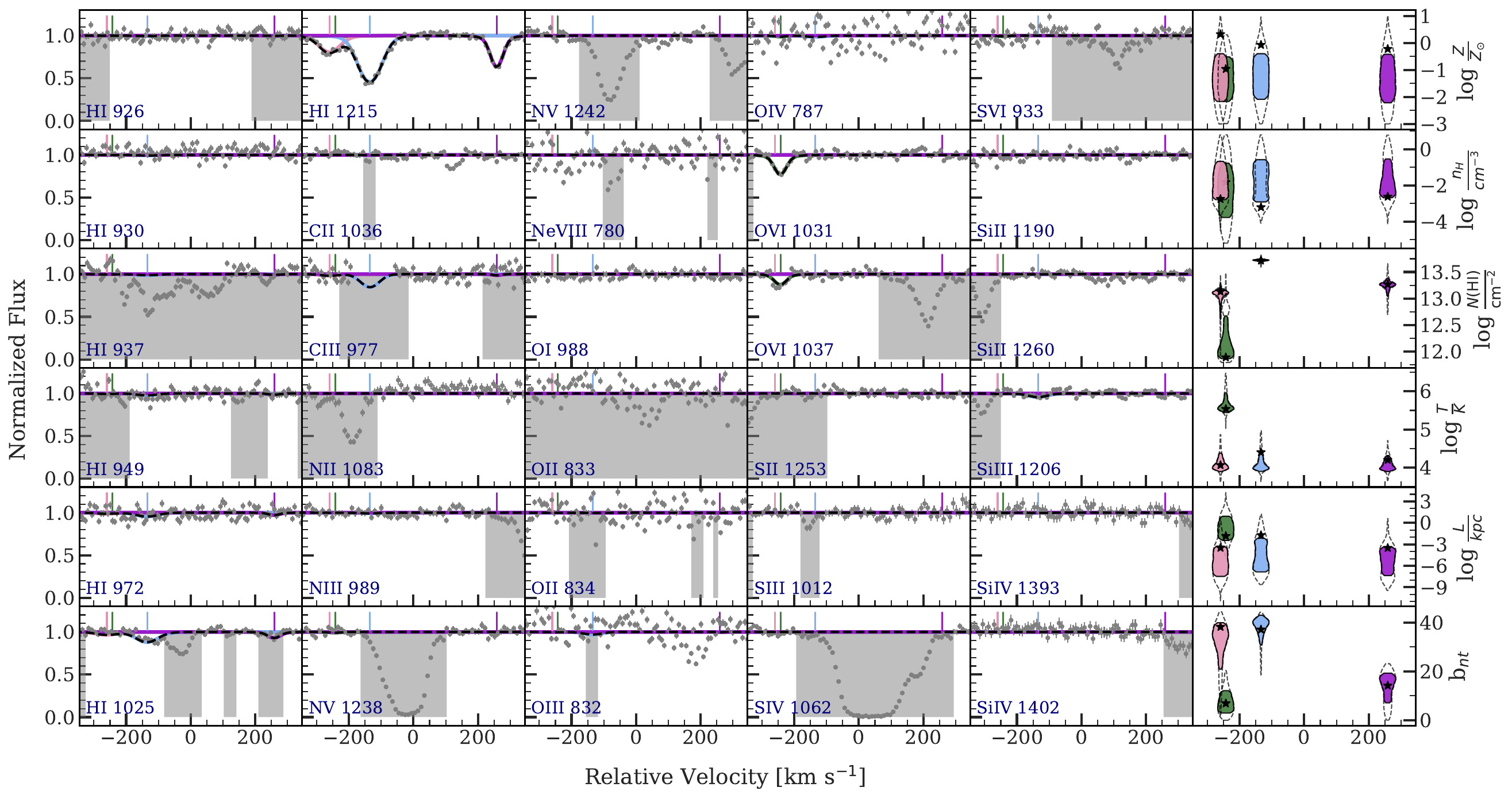}
\caption{Same as Fig.\ref{fig:J0125_0.3985} but for the absorber nearest to the $z$ = 0.1967 galaxy towards the background quasar J1301. The basic galaxy properties are $D \approx 136$ kpc, $\phi$ = $39.7_{-2.2}^{+2.8}$, $i$ = $80.7_{-3.2}^{+4.3}$}
\label{fig:J1301_0.1967}
\end{center}
\end{figure}

\clearpage

\begin{figure}
\begin{center}
\includegraphics[width=\linewidth]{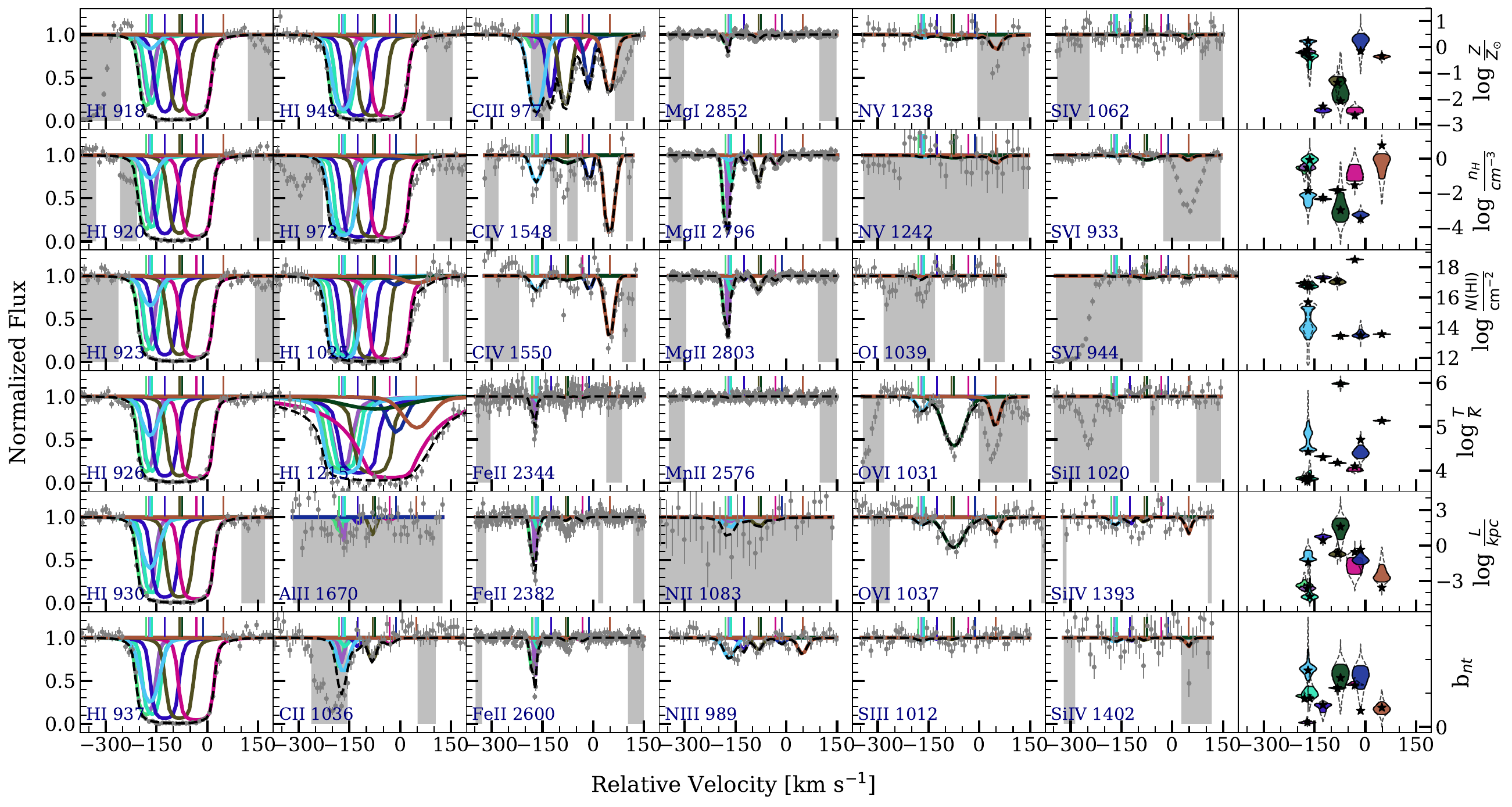}
\caption{Same as Fig.\ref{fig:J0125_0.3985} but for the absorber nearest to the $z$ = 0.6610 galaxy towards the background quasar J1319. The basic galaxy properties are $D \approx 104$ kpc, $\phi$ = $86.6_{-1.2}^{+1.5}$, $i$ = $65.8_{-1.2}^{+1.2}$}
\label{fig:J1319_0.6610}
\end{center}
\end{figure}

\clearpage

\begin{figure}
\begin{center}
\includegraphics[width=\linewidth]{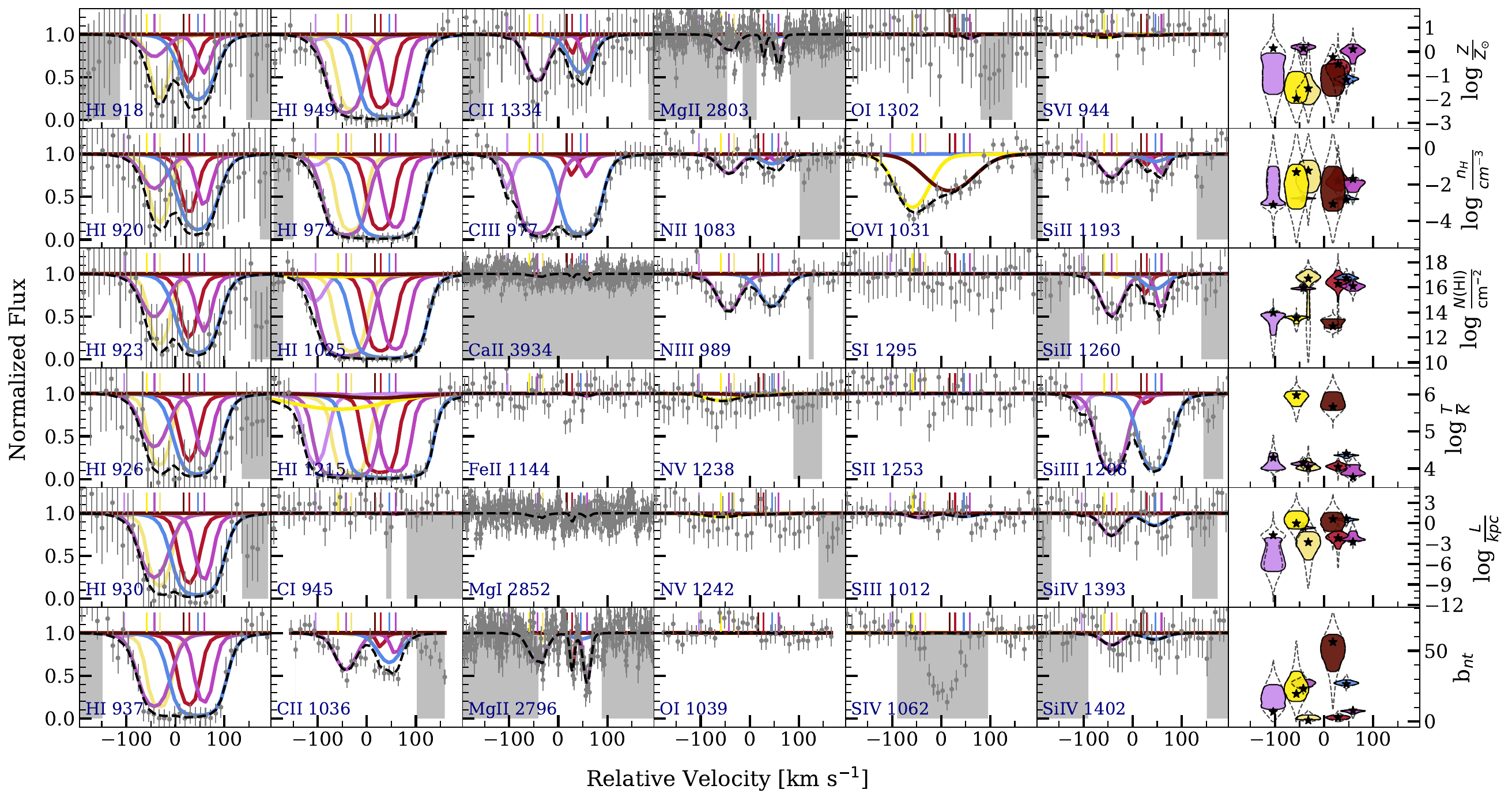}
\caption{Same as Fig.\ref{fig:J0125_0.3985} but for the absorber nearest to the $z$ = 0.2144 galaxy towards the background quasar J1322. The basic galaxy properties are $D \approx 39$ kpc, $\phi$ = $13.9_{-0.2}^{+0.2}$, $i$ = $57.9_{-0.2}^{+0.1}$}
\label{fig:J1322_0.2144}
\end{center}
\end{figure}

\clearpage

\begin{figure}
\begin{center}
\includegraphics[width=\linewidth]{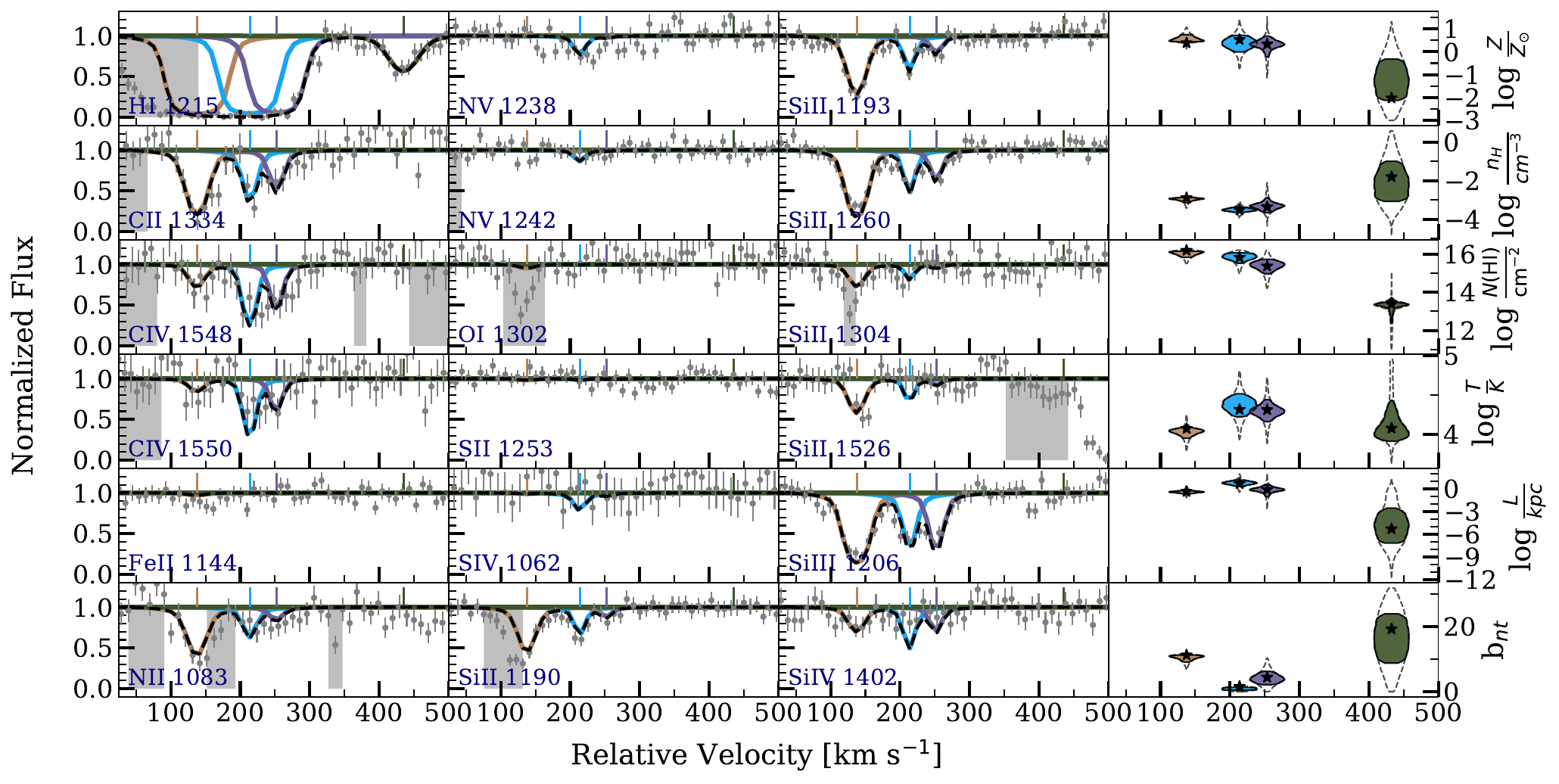}
\caption{Same as Fig.\ref{fig:J0125_0.3985} but for the absorber nearest to the $z$ = 0.0708 galaxy towards the background quasar J1342. The basic galaxy properties are $D \approx 39$ kpc, $\phi$ = $13.9_{-0.2}^{+0.2}$, $i$ = $57.7_{-0.3}^{+0.3}$}
\label{fig:J1342_0.0708}
\end{center}
\end{figure}

\clearpage

\begin{figure}
\begin{center}
\includegraphics[width=\linewidth]{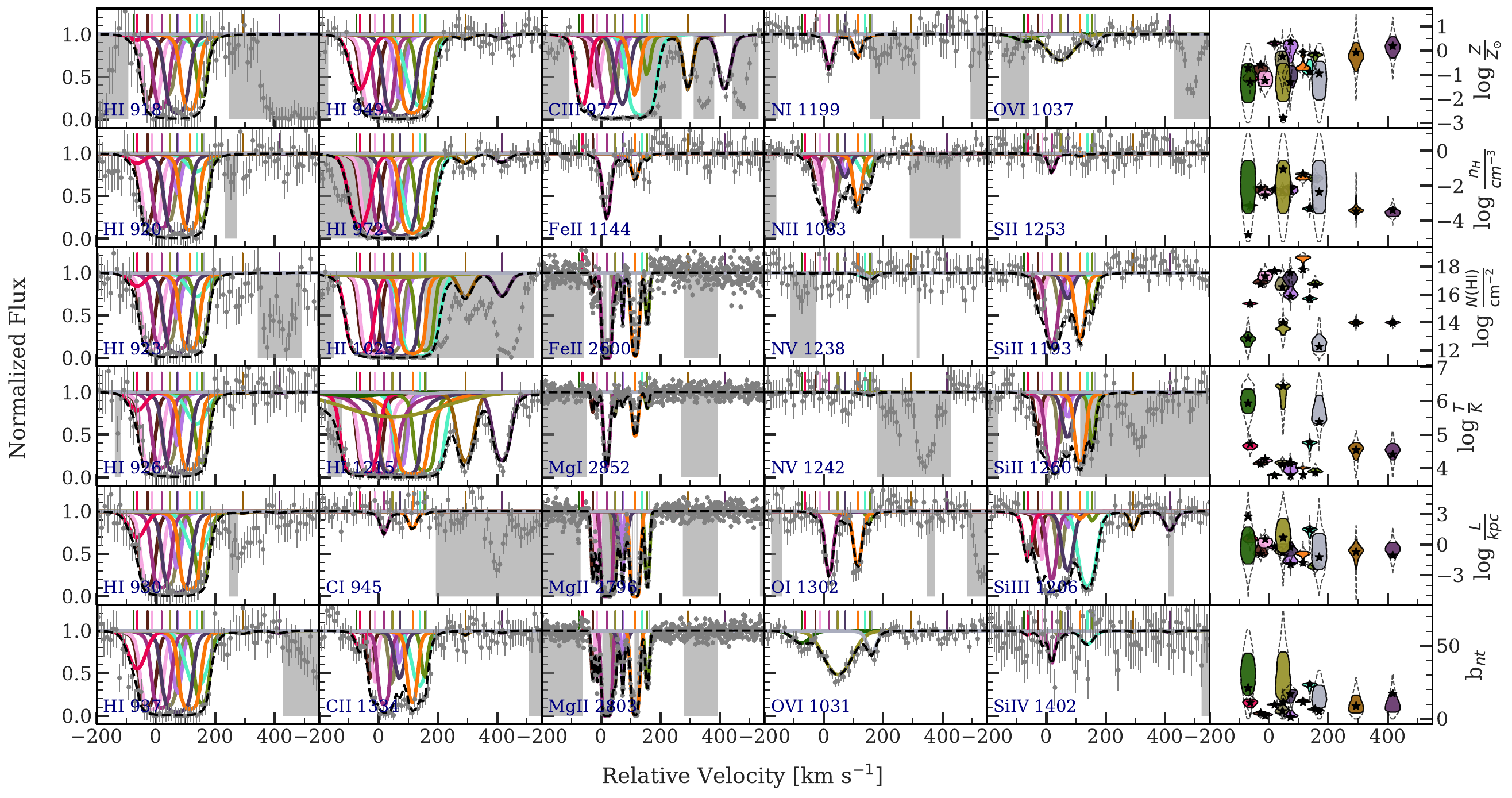}
\caption{Same as Fig.\ref{fig:J0125_0.3985} but for the absorber nearest to the $z$ = 0.2270 galaxy towards the background quasar J1342. The basic galaxy properties are $D \approx 35$ kpc, $\phi$ = $13.2_{-0.4}^{+0.5}$, $i$ = $0.1_{-0.1}^{+0.6}$}
\label{fig:J1342_0.2270}
\end{center}
\end{figure}

\clearpage

\begin{figure}
\begin{center}
\includegraphics[width=\linewidth]{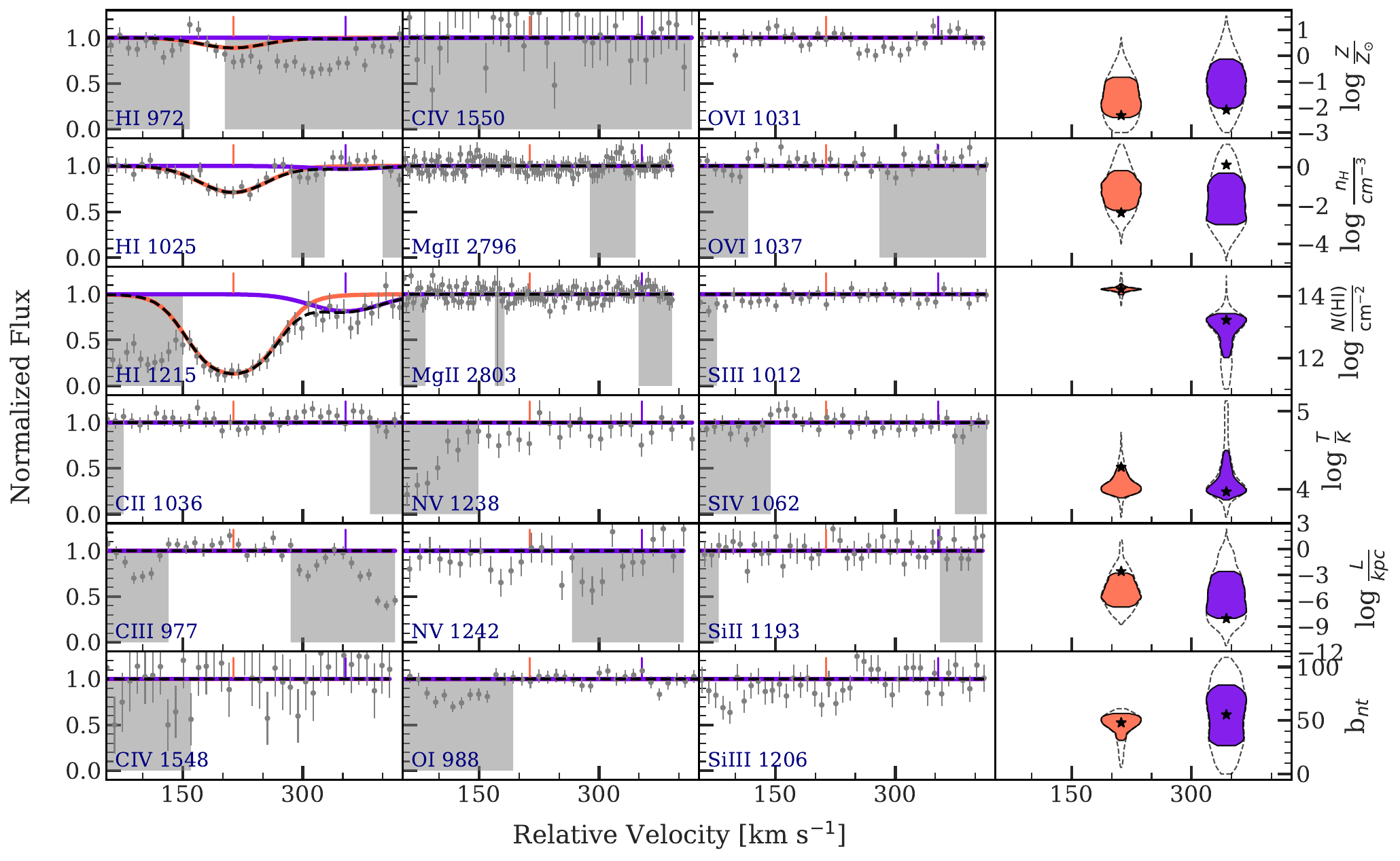}
\caption{Same as Fig.\ref{fig:J0125_0.3985} but for the absorber nearest to the $z$ = 0.4295 galaxy towards the background quasar J1357. The basic galaxy properties are $D \approx 158$ kpc, $\phi$ = $8.7_{-1.4}^{+1.6}$, $i$ = $85.0_{-1.7}^{+5.0}$}
\label{fig:J1357_0.4295}
\end{center}
\end{figure}

\clearpage

\begin{figure}
\begin{center}
\includegraphics[width=\linewidth]{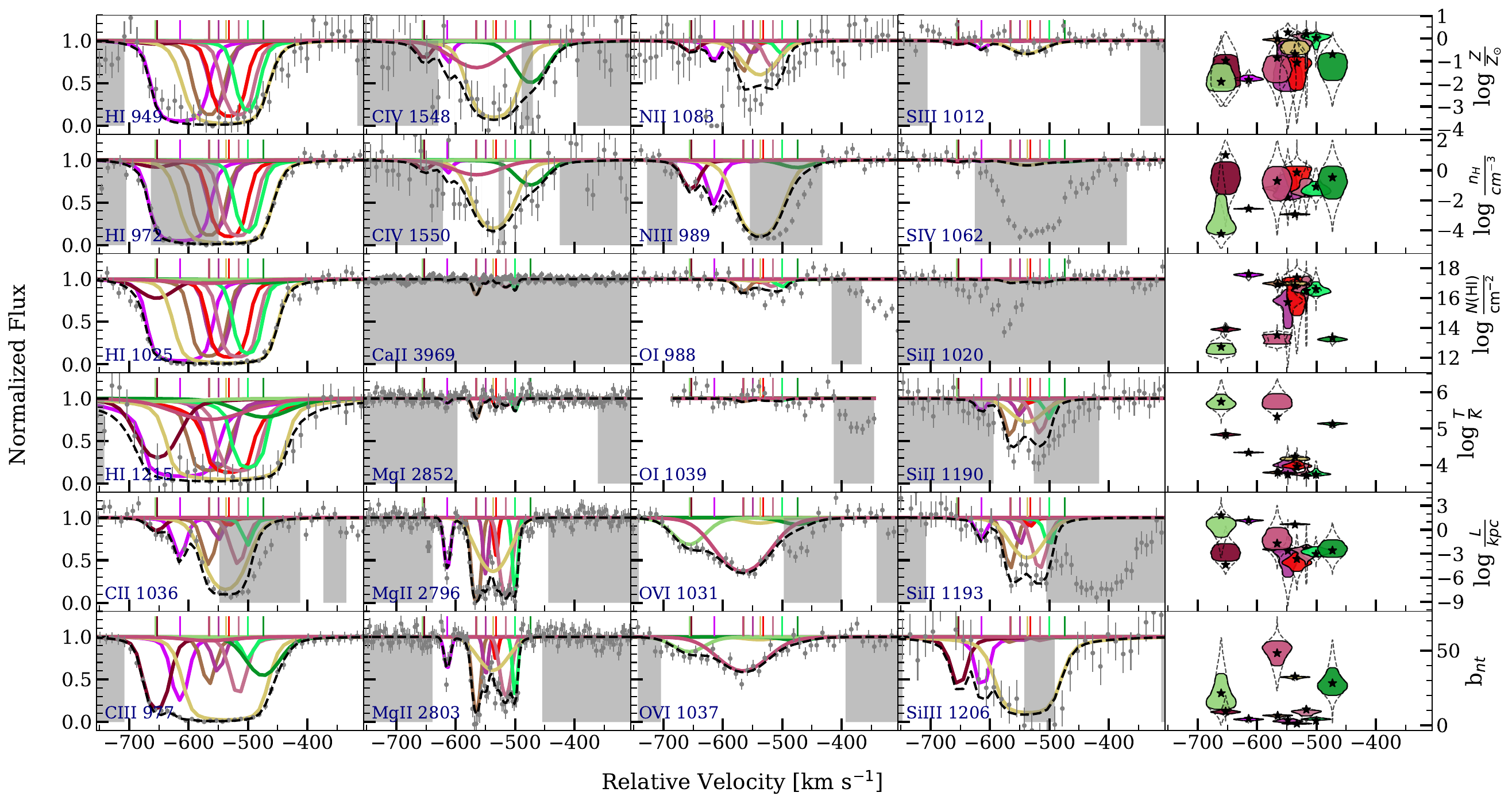}
\caption{Same as Fig.\ref{fig:J0125_0.3985} but for the absorber nearest to the $z$ = 0.4592 galaxy towards the background quasar J1357. The basic galaxy properties are $D \approx 46$ kpc, $\phi$ = $64.2_{-13.8}^{+13.6}$, $i$ = $24.7_{-6.5}^{+5.7}$}
\label{fig:J1357_0.4592}
\end{center}
\end{figure}

\clearpage

\begin{figure}
\begin{center}
\includegraphics[width=\linewidth]{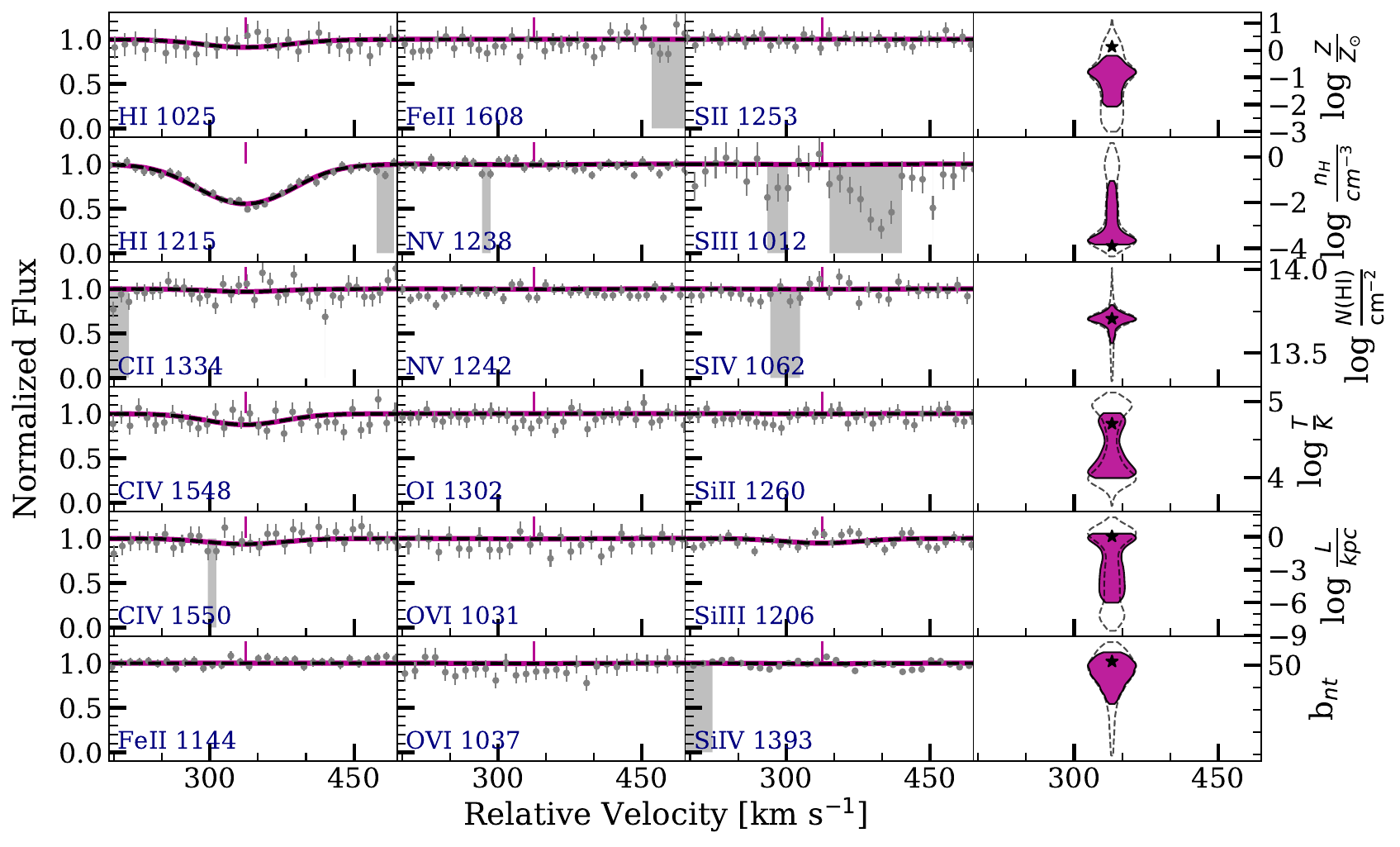}
\caption{Same as Fig.\ref{fig:J0125_0.3985} but for the absorber nearest to the $z$ = 0.0949 galaxy towards the background quasar J1547. The basic galaxy properties are $D \approx 80$ kpc, $\phi$ = $54.7_{-2.4}^{+2.0}$, $i$ = $80.9_{-2.0}^{+1.8}$}
\label{fig:J1547_0.0949}
\end{center}
\end{figure}

\clearpage

\begin{figure}
\begin{center}
\includegraphics[width=\linewidth]{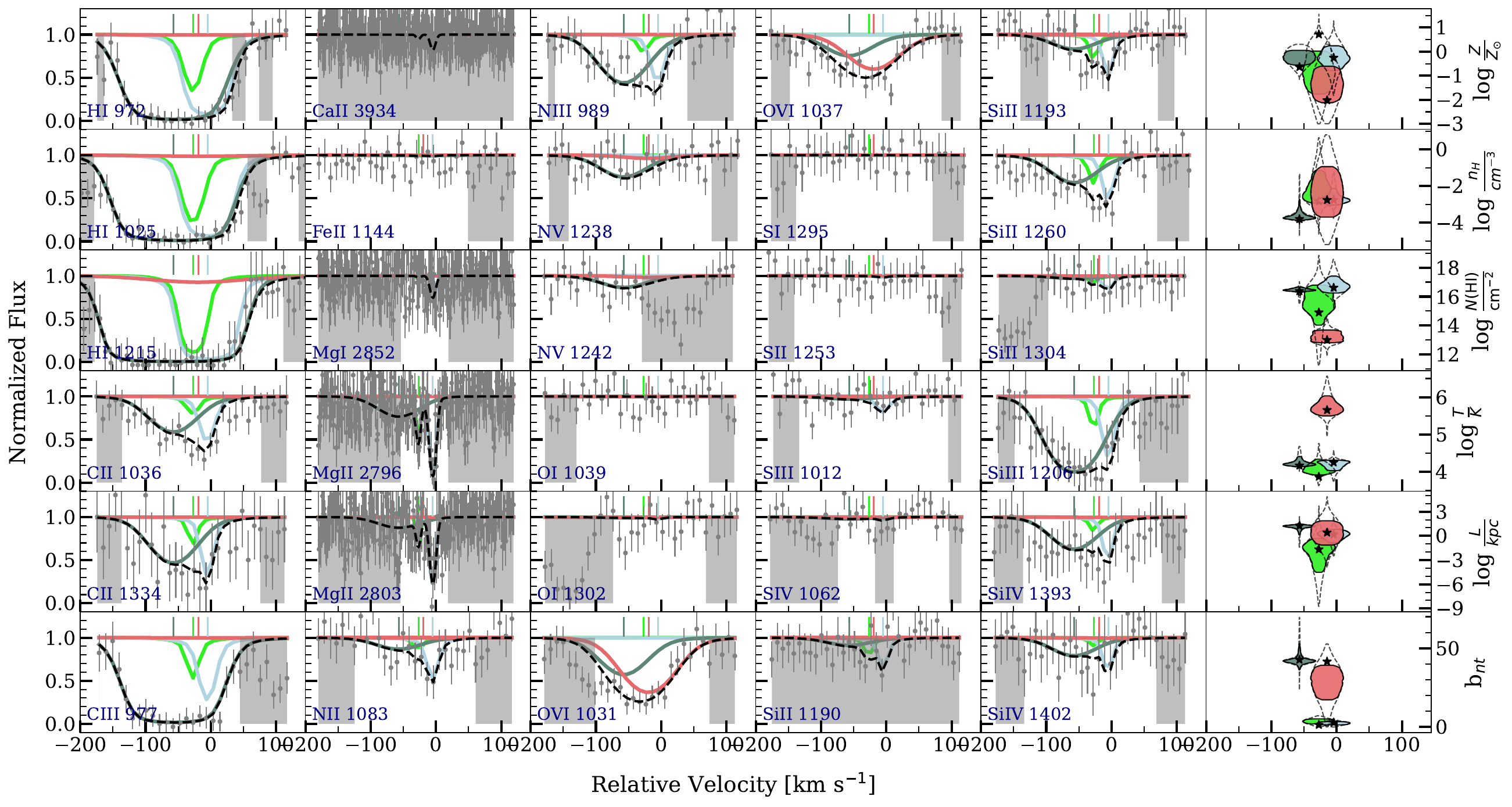}
\caption{Same as Fig.\ref{fig:J0125_0.3985} but for the absorber nearest to the $z$ = 0.1892 galaxy towards the background quasar J1555. The basic galaxy properties are $D \approx 33$ kpc, $\phi$ = $47.0_{-0.8}^{+0.3}$, $i$ = $51.8_{-0.7}^{+0.7}$}
\label{fig:J1555_0.1892}
\end{center}
\end{figure}

\clearpage

\begin{figure}
\begin{center}
\includegraphics[width=\linewidth]{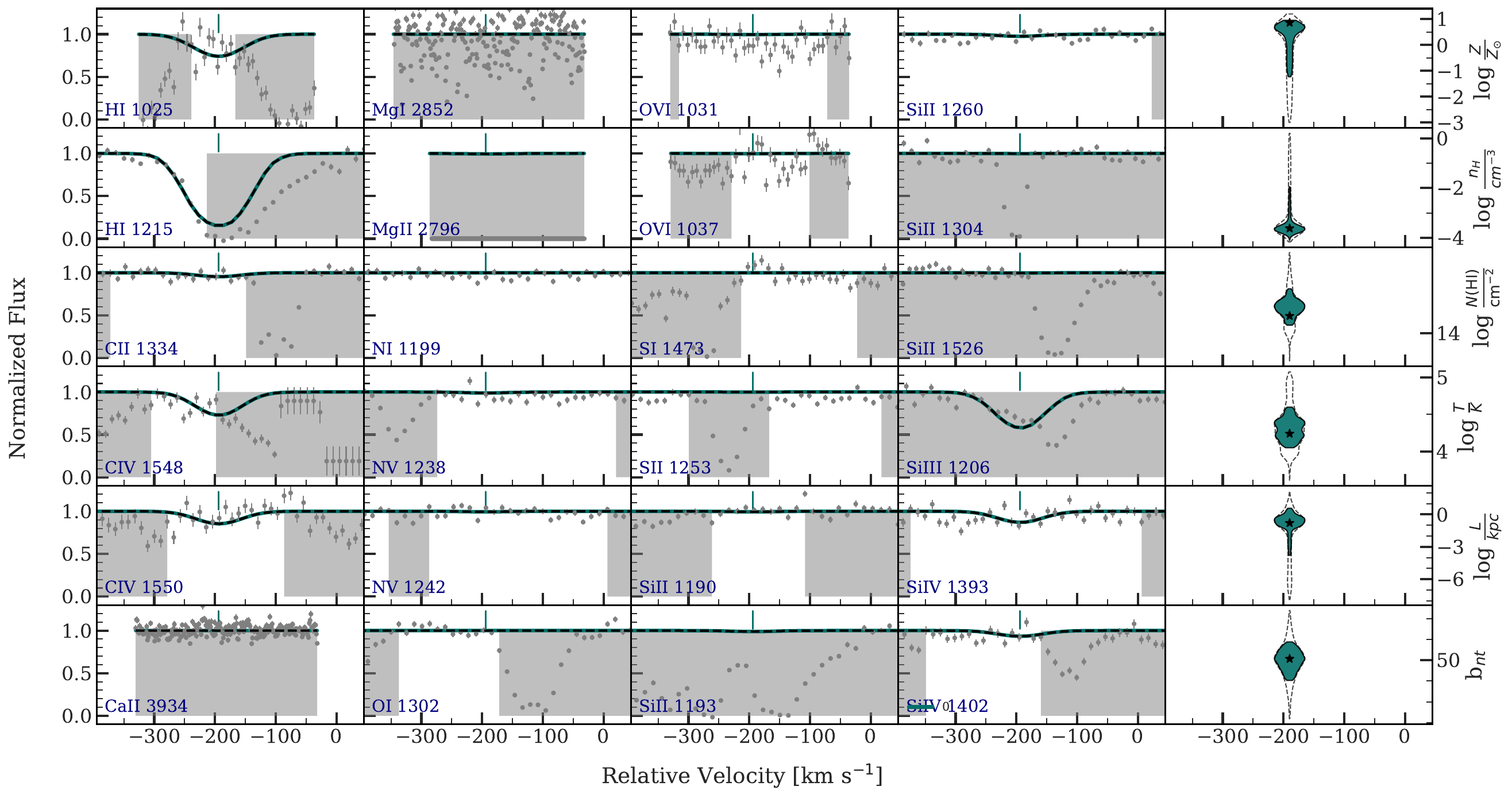}
\caption{Same as Fig.\ref{fig:J0125_0.3985} but for the absorber nearest to the $z$ = 0.0921 galaxy towards the background quasar J1704. The basic galaxy properties are $D \approx 94$ kpc, $\phi$ = $53.1_{-0.6}^{+0.6}$, $i$ = $72.0_{-0.5}^{+0.5}$}
\label{fig:J1704_0.0921}
\end{center}
\end{figure}

\clearpage

\begin{figure}
\begin{center}
\includegraphics[width=\linewidth]{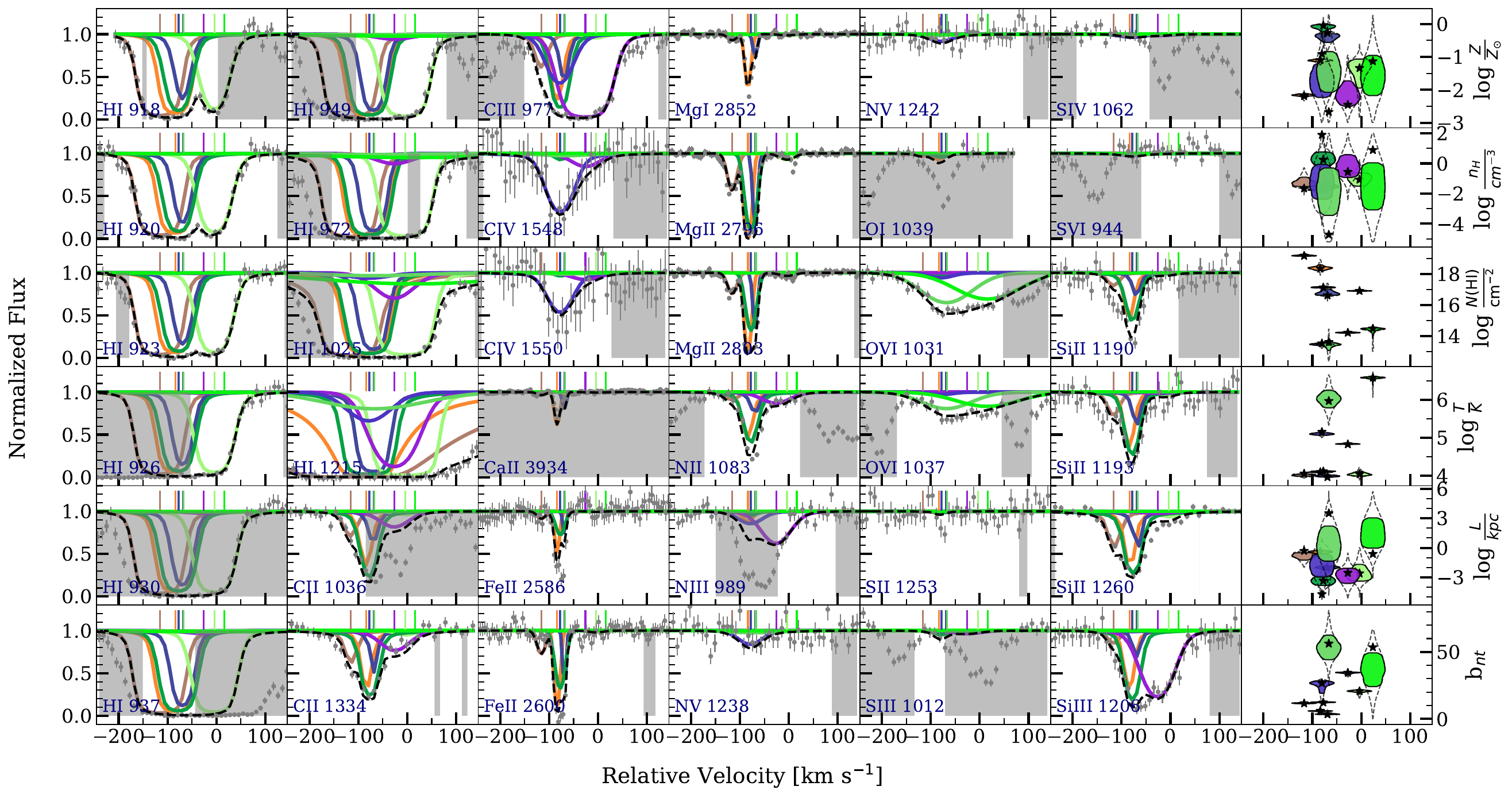}
\caption{Same as Fig.\ref{fig:J0125_0.3985} but for the absorber nearest to the $z$ = 0.4302 galaxy towards the background quasar J2131. The basic galaxy properties are $D \approx 48$ kpc, $\phi$ = $14.9_{-4.9}^{+6.0}$, $i$ = $48.3_{-3.7}^{+3.5}$}
\label{fig:J2131_0.4302}
\end{center}
\end{figure}

\clearpage

\begin{figure}
\begin{center}
\includegraphics[width=\linewidth]{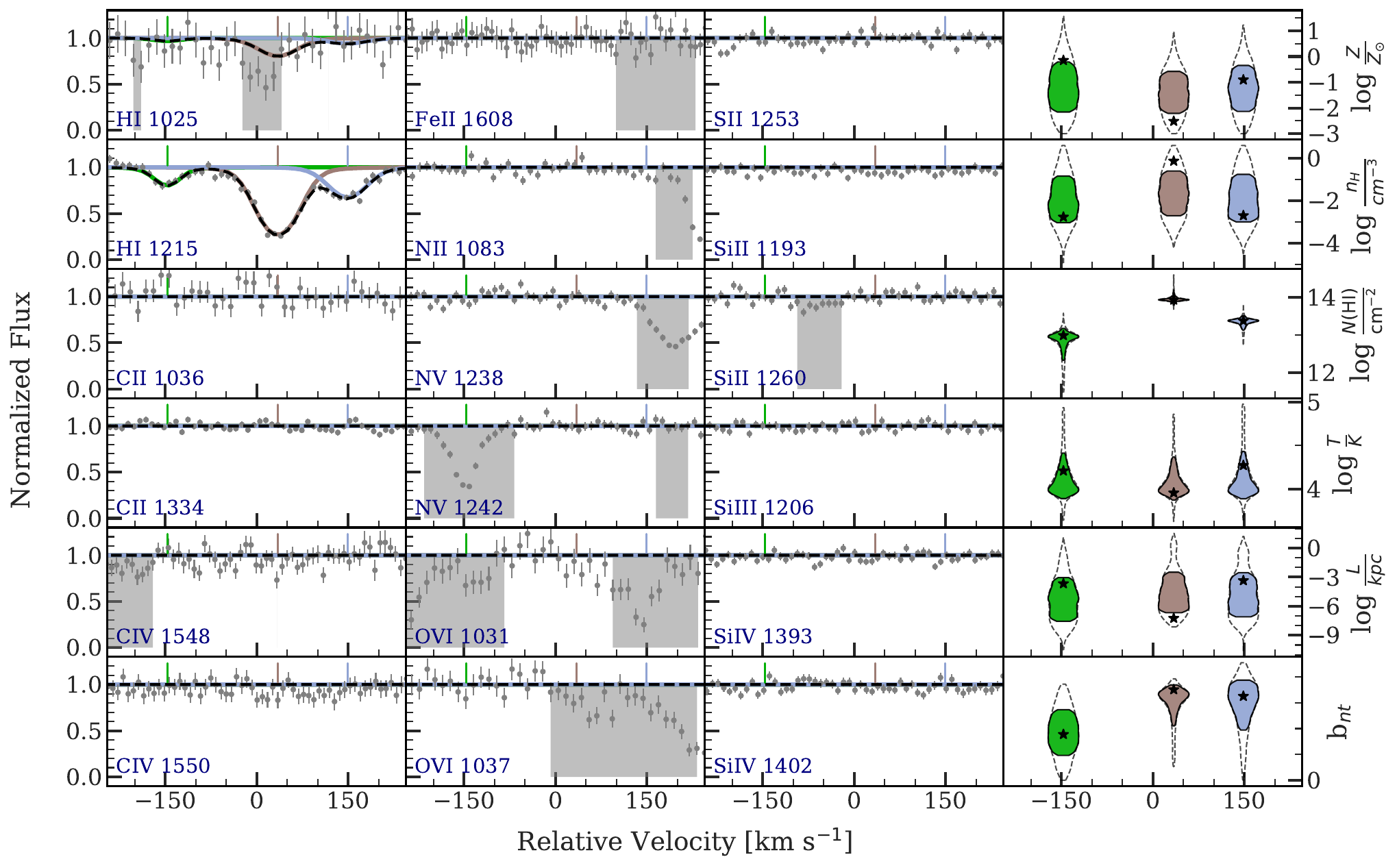}
\caption{Same as Fig.\ref{fig:J0125_0.3985} but for the absorber nearest to the $z$ = 0.0752 galaxy towards the background quasar J2137. The basic galaxy properties are $D \approx 71$ kpc, $\phi$ = $73.2_{-0.5}^{+1.0}$, $i$ = $71.0_{-1.0}^{+0.9}$}
\label{fig:J2137_0.0752}
\end{center}
\end{figure}

\clearpage

\begin{figure}
\begin{center}
\includegraphics[width=\linewidth]{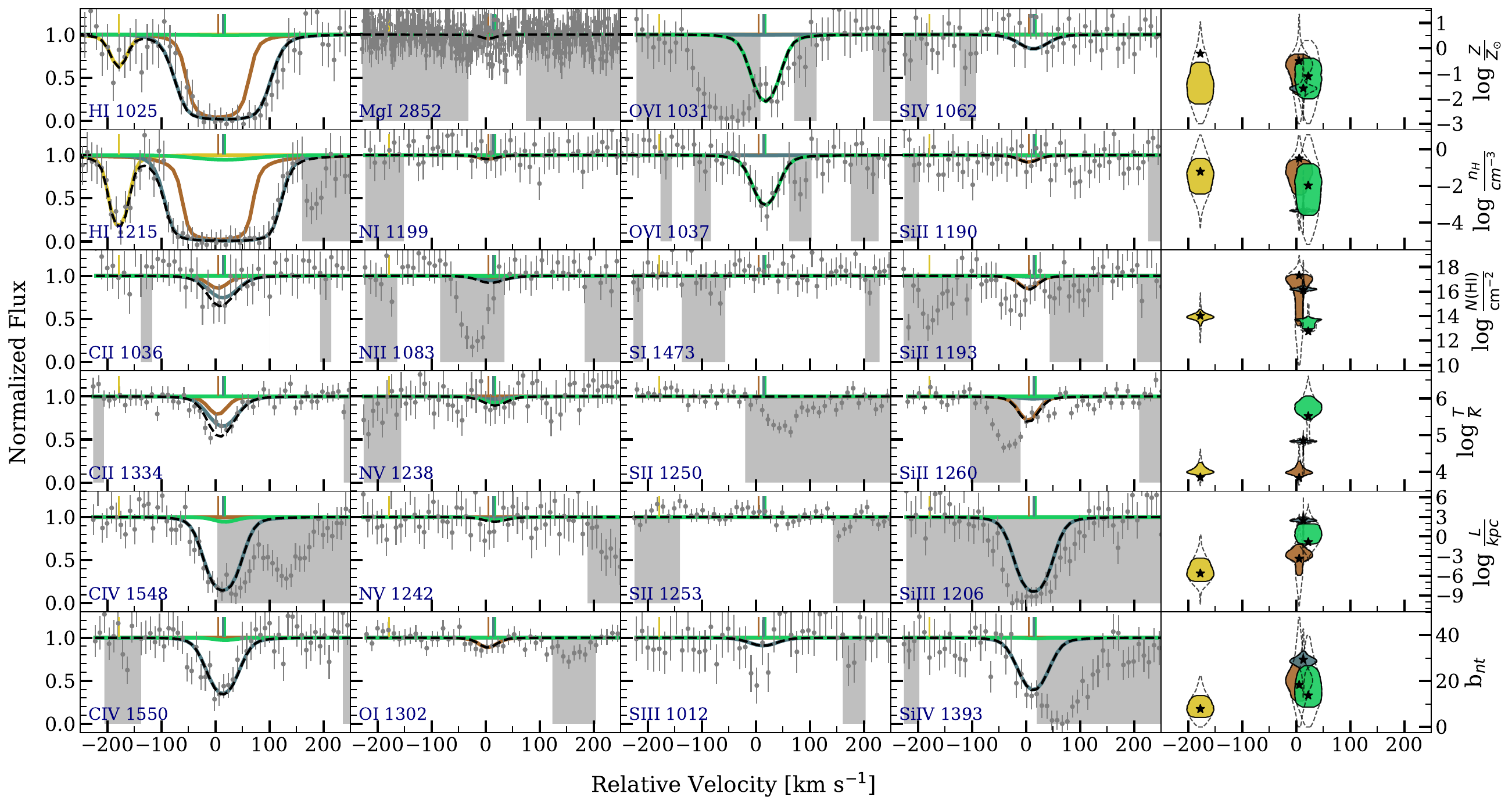}
\caption{Same as Fig.\ref{fig:J0125_0.3985} but for the absorber nearest to the $z$ = 0.1537 galaxy towards the background quasar J2253. The basic galaxy properties are $D \approx 32$ kpc, $\phi$ = $59.6_{-1.8}^{+0.9}$, $i$ = $33.3_{-2.0}^{+2.7}$}
\label{fig:J2253_0.1537}
\end{center}
\end{figure}

\clearpage

\begin{figure}
\begin{center}
\includegraphics[width=\linewidth]{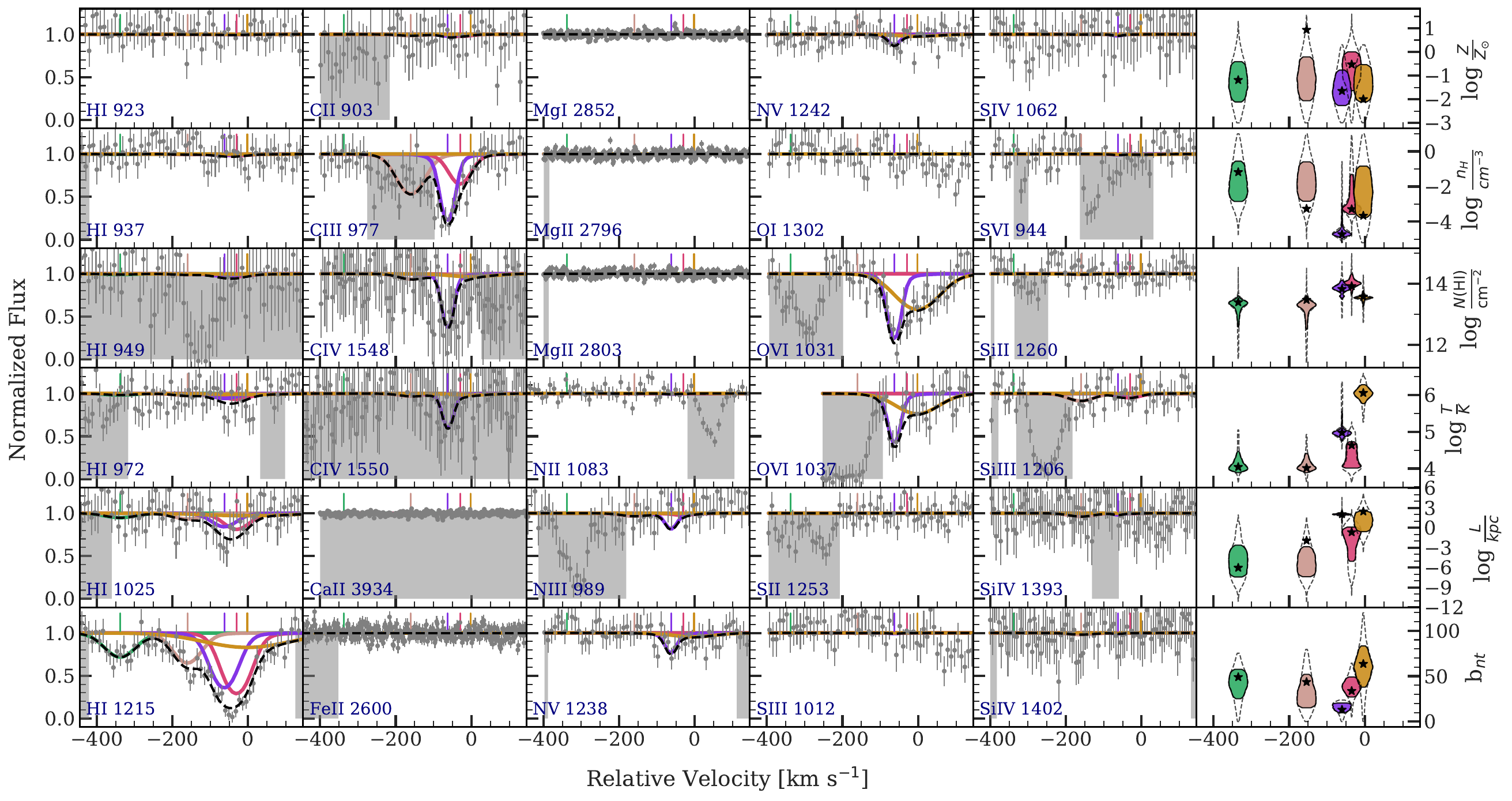}
\caption{Same as Fig.\ref{fig:J0125_0.3985} but for the absorber nearest to the $z$ = 0.3528 galaxy towards the background quasar J2253. The basic galaxy properties are $D \approx 203$ kpc, $\phi$ = $88.7_{-4.8}^{+1.3}$, $i$ = $36.7_{-4.6}^{+6.9}$}
\label{fig:J2253_0.3528}
\end{center}
\end{figure}

\clearpage

\begin{figure}
\begin{center}
\includegraphics[width=\linewidth]{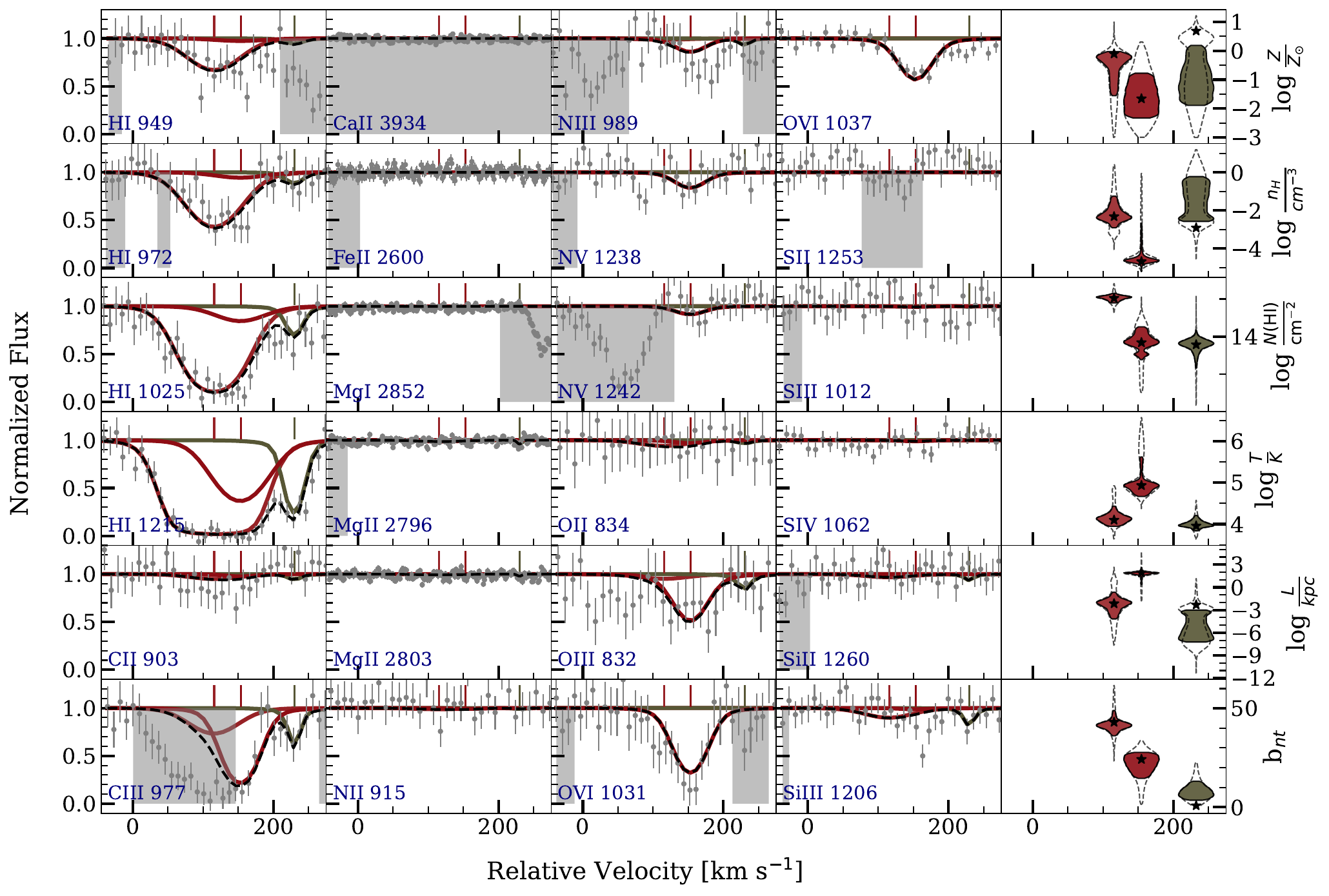}
\caption{Same as Fig.\ref{fig:J0125_0.3985} but for the absorber nearest to the $z$ = 0.3900 galaxy towards the background quasar J2253. The basic galaxy properties are $D \approx 276$ kpc, $\phi$ = $24.2_{-1.2}^{+1.2}$, $i$ = $76.1_{-1.2}^{+1.1}$}
\label{fig:J2253_0.3900}
\end{center}
\end{figure}

\clearpage

 \end{landscape}
 

\end{document}